\documentclass[longauth,traditabstract]{aa}

\usepackage[nonamebreak]{natbib}
\usepackage[stable]{footmisc}
\usepackage{amsmath}
\usepackage{txfonts}
\usepackage{placeins}
\usepackage{natbib}

\bibpunct{(}{)}{;}{a}{}{,}
\usepackage{graphicx}
\usepackage{subfig}
\usepackage{epstopdf}
\usepackage{ifthen}
\usepackage{mathtools}
\usepackage{fixltx2e}
\usepackage{url}
\usepackage[table,usenames,dvipsnames]{xcolor}
\usepackage{soul} 
\usepackage[export]{adjustbox}

\definecolor{linkcolor}{rgb}{0.6,0,0}
\definecolor{citecolor}{rgb}{0,0,0.75}
\definecolor{urlcolor}{rgb}{0.12,0.46,0.7}
\usepackage[breaklinks, colorlinks, urlcolor=urlcolor, linkcolor=linkcolor,citecolor=citecolor,pdfencoding=auto]{hyperref}
\hypersetup{linktocpage}

\graphicspath{{./fig/}} 

\newcommand{\citelowell}{\citetalias{planck2014-a10}}
\newcommand{\citehfimap}{\citetalias{planck2014-a09}}
\newcommand{\citeES}{\citetalias{planck2016-ES}}

\defcitealias{planck2014-a10}{LowEll2016}
\defcitealias{planck2014-a09}{HFImaps2015}
\defcitealias{planck2016-ES}{Explanatory Supplement}

\newcommand{\Nside}{\ensuremath{N_{\mathrm{side}}}}
\newcommand{\sroll}{{\tt SRoll}}
\def\microK{\ifmmode \,\mu$K$\else \,$\mu$\hbox{K}\fi}
\def\microKcarre{\ifmmode \,\mu$K$^2\else \,$\mu$\hbox{K}$^2$\fi}

\def\setsymbol#1#2{\expandafter\def\csname #1\endcsname{#2}}
\def\getsymbol#1{\csname #1\endcsname}

\def\Planck{\textit{Planck}}

\def\HeJT{$^4$He-JT}

\newbox\tablebox    \newdimen\tablewidth
\def\leaderfil{\leaders\hbox to 5pt{\hss.\hss}\hfil}
\def\endPlancktable{\tablewidth=\columnwidth 
    $$\hss\copy\tablebox\hss$$
    \vskip-\lastskip\vskip -2pt}
\def\endPlancktablewide{\tablewidth=\textwidth 
    $$\hss\copy\tablebox\hss$$
    \vskip-\lastskip\vskip -2pt}
\def\tablenote#1 #2\par{\begingroup \parindent=0.8em
    \abovedisplayshortskip=0pt\belowdisplayshortskip=0pt
    \noindent
    $$\hss\vbox{\hsize\tablewidth \hangindent=\parindent \hangafter=1 \noindent
    \hbox to \parindent{$^#1$\hss}\strut#2\strut\par}\hss$$
    \endgroup}
\def\doubleline{\vskip 3pt\hrule \vskip 1.5pt \hrule \vskip 5pt}

\def\L2{\ifmmode L_2\else $L_2$\fi}

\def\DeltaT{\ifmmode \Delta T\else $\Delta T$\fi}
\def\deltat{\ifmmode \Delta t\else $\Delta t$\fi}
\def\fknee{\ifmmode f_{\rm knee}\else $f_{\rm knee}$\fi}
\def\Fmax{\ifmmode F_{\rm max}\else $F_{\rm max}$\fi}
\def\solar{\ifmmode{\rm M}_{\mathord\odot}\else${\rm M}_{\mathord\odot}$\fi}
\def\Msolar{\ifmmode{\rm M}_{\mathord\odot}\else${\rm M}_{\mathord\odot}$\fi}
\def\Lsolar{\ifmmode{\rm L}_{\mathord\odot}\else${\rm L}_{\mathord\odot}$\fi}
\def\inv{\ifmmode^{-1}\else$^{-1}$\fi}
\def\mo{\ifmmode^{-1}\else$^{-1}$\fi}
\def\sup#1{\ifmmode ^{\rm #1}\else $^{\rm #1}$\fi}
\def\expo#1{\ifmmode \times 10^{#1}\else $\times 10^{#1}$\fi}
\def\,{\thinspace}
\def\lsim{\mathrel{\raise .4ex\hbox{\rlap{$<$}\lower 1.2ex\hbox{$\sim$}}}}
\def\gsim{\mathrel{\raise .4ex\hbox{\rlap{$>$}\lower 1.2ex\hbox{$\sim$}}}}

\def\simprop{\mathrel{\raise .4ex\hbox{\rlap{$\propto$}\lower 1.2ex\hbox{$\sim$}}}}
\def\deg{\ifmmode^\circ\else$^\circ$\fi}
\def\pdeg{\ifmmode $\setbox0=\hbox{$^{\circ}$}\rlap{\hskip.11\wd0 .}$^{\circ}
          \else \setbox0=\hbox{$^{\circ}$}\rlap{\hskip.11\wd0 .}$^{\circ}$\fi}
\def\arcs{\ifmmode {^{\scriptstyle\prime\prime}}
          \else $^{\scriptstyle\prime\prime}$\fi}
\def\arcm{\ifmmode {^{\scriptstyle\prime}}
          \else $^{\scriptstyle\prime}$\fi}
\newdimen\sa  \newdimen\sb
\def\parcs{\sa=.07em \sb=.03em
     \ifmmode \hbox{\rlap{.}}^{\scriptstyle\prime\kern -\sb\prime}\hbox{\kern -\sa}
     \else \rlap{.}$^{\scriptstyle\prime\kern -\sb\prime}$\kern -\sa\fi}
\def\parcm{\sa=.08em \sb=.03em
     \ifmmode \hbox{\rlap{.}\kern\sa}^{\scriptstyle\prime}\hbox{\kern-\sb}
     \else \rlap{.}\kern\sa$^{\scriptstyle\prime}$\kern-\sb\fi}
\def\ra[#1 #2 #3.#4]{#1\sup{h}#2\sup{m}#3\sup{s}\llap.#4}
\def\dec[#1 #2 #3.#4]{#1\deg#2\arcm#3\arcs\llap.#4}
\def\deco[#1 #2 #3]{#1\deg#2\arcm#3\arcs}
\def\rra[#1 #2]{#1\sup{h}#2\sup{m}}

\def\dots{\relax\ifmmode \ldots\else $\ldots$\fi}
\def\WHzsr{\ifmmode $W\,Hz\mo\,sr\mo$\else W\,Hz\mo\,sr\mo\fi}
\def\mHz{\ifmmode $\,mHz$\else \,mHz\fi}
\def\GHz{\ifmmode $\,GHz$\else \,GHz\fi}
\def\mKs{\ifmmode $\,mK\,s$^{1/2}\else \,mK\,s$^{1/2}$\fi}
\def\muKs{\ifmmode \,\mu$K\,s$^{1/2}\else \,$\mu$K\,s$^{1/2}$\fi}
\def\muKRJs{\ifmmode \,\mu$K$_{\rm RJ}$\,s$^{1/2}\else \,$\mu$K$_{\rm RJ}$\,s$^{1/2}$\fi}
\def\muKHz{\ifmmode \,\mu$K\,Hz$^{-1/2}\else \,$\mu$K\,Hz$^{-1/2}$\fi}
\def\MJysr{\ifmmode \,$MJy\,sr\mo$\else \,MJy\,sr\mo\fi}
\def\MJysrmK{\ifmmode \,$MJy\,sr\mo$\,mK$_{\rm CMB}\mo\else \,MJy\,sr\mo\,mK$_{\rm CMB}\mo$\fi}
\def\microns{\ifmmode \,\mu$m$\else \,$\mu$m\fi}

\def\muK{\ifmmode \,\mu$K$\else \,$\mu$\hbox{K}\fi}
\def\microK{\ifmmode \,\mu$K$\else \,$\mu$\hbox{K}\fi}
\def\muW{\ifmmode \,\mu$W$\else \,$\mu$\hbox{W}\fi}
\def\kms{\ifmmode $\,km\,s$^{-1}\else \,km\,s$^{-1}$\fi}
\def\kmsMpc{\ifmmode $\,\kms\,Mpc\mo$\else \,\kms\,Mpc\mo\fi}
\providecommand{\sorthelp}[1]{}

\begin{document}

\title{\vglue -10mm\Planck\ 2018 results. III. High Frequency Instrument data processing and frequency maps}

\author{\small
Planck Collaboration: N.~Aghanim\inst{47}
\and
Y.~Akrami\inst{49, 51}
\and
M.~Ashdown\inst{58, 5}
\and
J.~Aumont\inst{86}
\and
C.~Baccigalupi\inst{71}
\and
M.~Ballardini\inst{19, 35}
\and
A.~J.~Banday\inst{86, 8}
\and
R.~B.~Barreiro\inst{53}
\and
N.~Bartolo\inst{24, 54}
\and
S.~Basak\inst{78}
% ...
}
\institute{\small
AIM, CEA, CNRS, Universit\'{e} Paris-Saclay, F-91191 Gif sur Yvette, France. AIM, Universit\'{e} Paris Diderot, Sorbonne Paris Cit\'{e}, F-91191 Gif sur Yvette, France.\goodbreak
\and
APC, AstroParticule et Cosmologie, Universit\'{e} Paris Diderot, CNRS/IN2P3, CEA/lrfu, Observatoire de Paris, Sorbonne Paris Cit\'{e}, 10, rue Alice Domon et L\'{e}onie Duquet, 75205 Paris Cedex 13, France\goodbreak
\and
African Institute for Mathematical Sciences, 6-8 Melrose Road, Muizenberg, Cape Town, South Africa\goodbreak
\and
Aix Marseille Univ, CNRS, CNES, LAM, Marseille, France\goodbreak
\and
Astrophysics Group, Cavendish Laboratory, University of Cambridge, J J Thomson Avenue, Cambridge CB3 0HE, U.K.\goodbreak
\and
Astrophysics \& Cosmology Research Unit, School of Mathematics, Statistics \& Computer Science, University of KwaZulu-Natal, Westville Campus, Private Bag X54001, Durban 4000, South Africa\goodbreak
\and
CITA, University of Toronto, 60 St. George St., Toronto, ON M5S 3H8, Canada\goodbreak
\and
CNRS, IRAP, 9 Av. colonel Roche, BP 44346, F-31028 Toulouse cedex 4, France\goodbreak
\and
Cahill Center for Astronomy and Astrophysics, California Institute of Technology, Pasadena CA,  91125, USA\goodbreak
\and
California Institute of Technology, Pasadena, California, U.S.A.\goodbreak
\and
Centre for Theoretical Cosmology, DAMTP, University of Cambridge, Wilberforce Road, Cambridge CB3 0WA, U.K.\goodbreak
\and
Computational Cosmology Center, Lawrence Berkeley National Laboratory, Berkeley, California, U.S.A.\goodbreak
\and
D\'{e}partement de Physique Th\'{e}orique, Universit\'{e} de Gen\`{e}ve, 24, Quai E. Ansermet,1211 Gen\`{e}ve 4, Switzerland\goodbreak
\and
Departamento de Astrof\'{i}sica, Universidad de La Laguna (ULL), E-38206 La Laguna, Tenerife, Spain\goodbreak
\and
Departamento de F\'{\i}sica, Universidad de Oviedo, C/ Federico Garc\'{\i}a Lorca, 18 , Oviedo, Spain\goodbreak
\and
Department of Astrophysics/IMAPP, Radboud University, P.O. Box 9010, 6500 GL Nijmegen, The Netherlands\goodbreak
\and
Department of Mathematics, University of Stellenbosch, Stellenbosch 7602, South Africa\goodbreak
\and
Department of Physics \& Astronomy, University of British Columbia, 6224 Agricultural Road, Vancouver, British Columbia, Canada\goodbreak
\and
Department of Physics \& Astronomy, University of the Western Cape, Cape Town 7535, South Africa\goodbreak
\and
Department of Physics and Astronomy, University of Sussex, Brighton BN1 9QH, U.K.\goodbreak
\and
Department of Physics, Gustaf H\"{a}llstr\"{o}min katu 2a, University of Helsinki, Helsinki, Finland\goodbreak
\and
Department of Physics, Princeton University, Princeton, New Jersey, U.S.A.\goodbreak
\and
Department of Physics, University of Illinois at Urbana-Champaign, 1110 West Green Street, Urbana, Illinois, U.S.A.\goodbreak
\and
Dipartimento di Fisica e Astronomia G. Galilei, Universit\`{a} degli Studi di Padova, via Marzolo 8, 35131 Padova, Italy\goodbreak
\and
Dipartimento di Fisica e Scienze della Terra, Universit\`{a} di Ferrara, Via Saragat 1, 44122 Ferrara, Italy\goodbreak
\and
Dipartimento di Fisica, Universit\`{a} La Sapienza, P. le A. Moro 2, Roma, Italy\goodbreak
\and
Dipartimento di Fisica, Universit\`{a} degli Studi di Milano, Via Celoria, 16, Milano, Italy\goodbreak
\and
Dipartimento di Fisica, Universit\`{a} degli Studi di Trieste, via A. Valerio 2, Trieste, Italy\goodbreak
\and
Dipartimento di Fisica, Universit\`{a} di Roma Tor Vergata, Via della Ricerca Scientifica, 1, Roma, Italy\goodbreak
\and
European Space Agency, ESAC, Planck Science Office, Camino bajo del Castillo, s/n, Urbanizaci\'{o}n Villafranca del Castillo, Villanueva de la Ca\~{n}ada, Madrid, Spain\goodbreak
\and
European Space Agency, ESTEC, Keplerlaan 1, 2201 AZ Noordwijk, The Netherlands\goodbreak
\and
Gran Sasso Science Institute, INFN, viale F. Crispi 7, 67100 L'Aquila, Italy\goodbreak
\and
Haverford College Astronomy Department, 370 Lancaster Avenue, Haverford, Pennsylvania, U.S.A.\goodbreak
\and
Helsinki Institute of Physics, Gustaf H\"{a}llstr\"{o}min katu 2, University of Helsinki, Helsinki, Finland\goodbreak
\and
INAF - OAS Bologna, Istituto Nazionale di Astrofisica - Osservatorio di Astrofisica e Scienza dello Spazio di Bologna, Area della Ricerca del CNR, Via Gobetti 101, 40129, Bologna, Italy\goodbreak
\and
INAF - Osservatorio Astronomico di Padova, Vicolo dell'Osservatorio 5, Padova, Italy\goodbreak
\and
INAF - Osservatorio Astronomico di Trieste, Via G.B. Tiepolo 11, Trieste, Italy\goodbreak
\and
INAF, Istituto di Radioastronomia, Via Piero Gobetti 101, I-40129 Bologna, Italy\goodbreak
\and
INAF/IASF Milano, Via E. Bassini 15, Milano, Italy\goodbreak
\and
INFN - CNAF, viale Berti Pichat 6/2, 40127 Bologna, Italy\goodbreak
\and
INFN, Sezione di Bologna, viale Berti Pichat 6/2, 40127 Bologna, Italy\goodbreak
\and
INFN, Sezione di Ferrara, Via Saragat 1, 44122 Ferrara, Italy\goodbreak
\and
INFN, Sezione di Milano, Via Celoria 16, Milano, Italy\goodbreak
\and
INFN, Sezione di Roma 1, Universit\`{a} di Roma Sapienza, Piazzale Aldo Moro 2, 00185, Roma, Italy\goodbreak
\and
INFN, Sezione di Roma 2, Universit\`{a} di Roma Tor Vergata, Via della Ricerca Scientifica, 1, Roma, Italy\goodbreak
\and
Imperial College London, Astrophysics group, Blackett Laboratory, Prince Consort Road, London, SW7 2AZ, U.K.\goodbreak
\and
Institut d'Astrophysique Spatiale, CNRS, Univ. Paris-Sud, Universit\'{e} Paris-Saclay, B\^{a}t. 121, 91405 Orsay cedex, France\goodbreak
\and
Institut d'Astrophysique de Paris, CNRS (UMR7095), 98 bis Boulevard Arago, F-75014, Paris, France\goodbreak
\and
Institute Lorentz, Leiden University, PO Box 9506, Leiden 2300 RA, The Netherlands\goodbreak
\and
Institute of Astronomy, University of Cambridge, Madingley Road, Cambridge CB3 0HA, U.K.\goodbreak
\and
Institute of Theoretical Astrophysics, University of Oslo, Blindern, Oslo, Norway\goodbreak
\and
Instituto de Astrof\'{\i}sica de Canarias, C/V\'{\i}a L\'{a}ctea s/n, La Laguna, Tenerife, Spain\goodbreak
\and
Instituto de F\'{\i}sica de Cantabria (CSIC-Universidad de Cantabria), Avda. de los Castros s/n, Santander, Spain\goodbreak
\and
Istituto Nazionale di Fisica Nucleare, Sezione di Padova, via Marzolo 8, I-35131 Padova, Italy\goodbreak
\and
Jet Propulsion Laboratory, California Institute of Technology, 4800 Oak Grove Drive, Pasadena, California, U.S.A.\goodbreak
\and
Jodrell Bank Centre for Astrophysics, Alan Turing Building, School of Physics and Astronomy, The University of Manchester, Oxford Road, Manchester, M13 9PL, U.K.\goodbreak
\and
Kavli Institute for Cosmological Physics, University of Chicago, Chicago, IL 60637, USA\goodbreak
\and
Kavli Institute for Cosmology Cambridge, Madingley Road, Cambridge, CB3 0HA, U.K.\goodbreak
\and
LAL, Universit\'{e} Paris-Sud, CNRS/IN2P3, Orsay, France\goodbreak
\and
LERMA, CNRS, Observatoire de Paris, 61 Avenue de l'Observatoire, Paris, France\goodbreak
\and
LERMA/LRA, Observatoire de Paris, PSL Research University, CNRS, Ecole Normale Sup\'erieure, 75005 Paris, France\goodbreak
\and
Laboratoire AIM, CEA - Universit\'{e} Paris-Saclay, 91191 Gif-sur-Yvette, France\goodbreak
\and
Laboratoire de Physique Subatomique et Cosmologie, Universit\'{e} Grenoble-Alpes, CNRS/IN2P3, 53, rue des Martyrs, 38026 Grenoble Cedex, France\goodbreak
\and
Laboratoire de Physique Th\'{e}orique, Universit\'{e} Paris-Sud 11 \& CNRS, B\^{a}timent 210, 91405 Orsay, France\goodbreak
\and
Lawrence Berkeley National Laboratory, Berkeley, California, U.S.A.\goodbreak
\and
Low Temperature Laboratory, Department of Applied Physics, Aalto University, Espoo, FI-00076 AALTO, Finland\goodbreak
\and
Max-Planck-Institut f\"{u}r Astrophysik, Karl-Schwarzschild-Str. 1, 85741 Garching, Germany\goodbreak
\and
Mullard Space Science Laboratory, University College London, Surrey RH5 6NT, U.K.\goodbreak
\and
NAOC-UKZN Computational Astrophysics Centre (NUCAC), University of KwaZulu-Natal, Durban 4000, South Africa\goodbreak
\and
Nicolaus Copernicus Astronomical Center, Polish Academy of Sciences, Bartycka 18, 00-716 Warsaw, Poland\goodbreak
\and
SISSA, Astrophysics Sector, via Bonomea 265, 34136, Trieste, Italy\goodbreak
\and
San Diego Supercomputer Center, University of California, San Diego, 9500 Gilman Drive, La Jolla, CA 92093, USA\goodbreak
\and
School of Chemistry and Physics, University of KwaZulu-Natal, Westville Campus, Private Bag X54001, Durban, 4000, South Africa\goodbreak
\and
School of Physical Sciences, National Institute of Science Education and Research, HBNI, Jatni-752050, Odissa, India\goodbreak
\and
School of Physics and Astronomy, Cardiff University, Queens Buildings, The Parade, Cardiff, CF24 3AA, U.K.\goodbreak
\and
School of Physics and Astronomy, Sun Yat-sen University, 2 Daxue Rd, Tangjia, Zhuhai, China\goodbreak
\and
School of Physics and Astronomy, University of Nottingham, Nottingham NG7 2RD, U.K.\goodbreak
\and
School of Physics, Indian Institute of Science Education and Research Thiruvananthapuram, Maruthamala PO, Vithura, Thiruvananthapuram 695551, Kerala, India\goodbreak
\and
Simon Fraser University, Department of Physics, 8888 University Drive, Burnaby BC, Canada\goodbreak
\and
Sorbonne Universit\'{e}-UPMC, UMR7095, Institut d'Astrophysique de Paris, 98 bis Boulevard Arago, F-75014, Paris, France\goodbreak
\and
Space Research Institute (IKI), Russian Academy of Sciences, Profsoyuznaya Str, 84/32, Moscow, 117997, Russia\goodbreak
\and
Space Science Data Center - Agenzia Spaziale Italiana, Via del Politecnico snc, 00133, Roma, Italy\goodbreak
\and
Space Sciences Laboratory, University of California, Berkeley, California, U.S.A.\goodbreak
\and
The Oskar Klein Centre for Cosmoparticle Physics, Department of Physics, Stockholm University, AlbaNova, SE-106 91 Stockholm, Sweden\goodbreak
\and
UPMC Univ Paris 06, UMR7095, 98 bis Boulevard Arago, F-75014, Paris, France\goodbreak
\and
Universit\'{e} de Toulouse, UPS-OMP, IRAP, F-31028 Toulouse cedex 4, France\goodbreak
% ...
}

\titlerunning{\Planck\ 2018 results. HFI DPC.}
\authorrunning{Planck Collaboration}
\date{Accepted by A\&A.}

\abstract {This paper presents the High Frequency Instrument (HFI) data processing procedures for the \Planck\ 2018 release. Major improvements in mapmaking have been achieved since the previous \Planck\ 2015 release, many of which were used and described already in an intermediate paper dedicated to the \Planck\ polarized data at low multipoles. These improvements enabled the first significant measurement of the reionization optical depth parameter using \Planck-HFI data. This paper presents an extensive analysis of systematic effects, including the use of end-to-end simulations to facilitate their removal and characterize the residuals. The polarized data, which presented a number of known problems in the 2015 \Planck\ release, are very significantly improved, especially the leakage from intensity to polarization. Calibration, based on the cosmic microwave background (CMB) dipole, is now extremely accurate and in the frequency range 100 to 353\,GHz reduces intensity-to-polarization leakage caused by calibration mismatch. The Solar dipole direction has been determined in the three lowest HFI frequency channels to within one arc minute, and its amplitude has an absolute uncertainty smaller than 0.35\microK, an accuracy of order $10^{-4}$. This is a major legacy from the \Planck\ HFI for future CMB experiments. The removal of bandpass leakage has been improved for the main high-frequency foregrounds by extracting the bandpass-mismatch coefficients for each detector as part of the mapmaking process; these values in turn improve the intensity maps. This is a major change in the philosophy of ``frequency maps,'' which are now computed from single detector data, all adjusted to the same average bandpass response for the main foregrounds. End-to-end simulations have been shown to reproduce very well the relative gain calibration of detectors, as well as drifts within a frequency induced by the residuals of the main systematic effect (analogue-to-digital convertor non-linearity residuals). Using these simulations, we have been able to measure and correct the small frequency calibration bias induced by this systematic effect at the $10^{-4}$ level. There is no detectable sign of a residual calibration bias between the first and second acoustic peaks in the CMB channels, at the $10^{-3}$ level.}

\maketitle

\tableofcontents

\section{Introduction}
\label{sec:introduction}
This paper, one of a series accompanying the final full release of \Planck\footnote{\Planck\ (\url{http://www.esa.int/Planck}) is a project of the European Space Agency (ESA), with instruments provided by two scientific Consortia funded by ESA member states and led by Principal Investigators from France and Italy, telescope reflectors provided through a collaboration between ESA and a scientific consortium led and funded by Denmark, and additional contributions from NASA (USA).} data products, summarizes the calibration, cleaning and other processing steps used to convert High Frequency Instrument (HFI) time-ordered information (TOI) into single-frequency maps. A companion paper \citep{planck2016-l02} similarly treats LFI data.

The raw data considered here are identical to those of the previous \Planck\ 2015 release \citep[see][hereafter ``\citehfimap'']{planck2014-a09} with one exception: we drop approximately 22 days of data taken in the final days of HFI observations because of the increasing Solar activity and some HFI end-of-life changes in the cryogenic chain operations during this period. These affected the data more significantly in the last 22 days than in any earlier period of similar length during the mission. However, for polarization studies, baseline maps at 353\,GHz are based on polarization-sensitive bolometer (PSB) observations only (for reasons explained later in this paper), although maps with spider-web bolometers (SWBs) are also made available for intensity studies.

HFI has impressive sensitivity (single-multipole power spectrum sensitivity) $C_{\ell}=1.4$ to $2.5 \times 10^{-4}\microKcarre$ at 100, 143, and 217\,GHz on the best (i.e., lowest foreground) half of the sky. We cannot yet take full advantage of this sensitivity because it requires exquisite control of systematic errors from instrumental and foreground effects, which were shown by null tests to exceed the detector noise at low multipoles. Thus, for the 2018 release, we have concentrated our efforts on improving the control of systematic effects, particularly those in the polarized data -- especially at low multipoles where they dominate -- which were not fully exploited in the 2015 release. Although this is the last full data release from the \Planck\ Collaboration, natural extensions of \sroll, some of which are demonstrated in this paper, offer the possibility of even better results from HFI data in the future. 

For the present release, full end-to-end (E2E) simulations have been developed, which include the modelling of all known instrumental systematic effects and of sky maps (CMB and foregrounds). These models are used to build realistic and full time-ordered data sets for all six HFI frequencies. These simulated data can then be propagated through the \sroll\ mapmaking process to produce frequency maps and power spectra. These simulated data have been used in this paper through the E2E simulations to characterize the mapmaking and thus the frequency maps. They are also used to produce a statistically meaningful number of simulations for likelihood analysis, taking into account that the residuals from systematic effects are, in general, non-Gaussian. This provides a powerful tool for estimating the systematic residuals in both the maps and power spectra used in \citelowell, and also used extensively in this paper and in \citet{planck2016-l05} and \citet{planck2016-l06}.

Section~\ref{sec:bulk_simulations} describes the HFI 2018 release maps and also differences with the 2015 release maps and those used in \citelowell, which were built using an early version of the \sroll\ mapmaking process. Hence we can often simply refer to the analysis of systematic effects already carried out in \citelowell. This section also assesses how representative and robust the simulations are, when compared with released maps as examined through various null tests.

Section~\ref{sec:calibration} discusses the photometric calibration, which is based on the orbital CMB dipole for the four lower frequencies; the two submillimetre channels are instead calibrated on the giant planets (as in the 2015 release). The a posteriori measurement of the dipole arising from the Solar System's motion with respect to the rest frame of the CMB (the Solar dipole) has been improved very significantly, especially for the higher HFI frequencies. The accurate determination of the Solar dipole direction and amplitude is a significant \Planck\ legacy for the calibration of present and future CMB experiments. It is used in the present work for the photometric inter-calibration of bolometers within a frequency band, and for inter-calibration between different frequency bands. It could also be used to inter-calibrate \Planck\ with other full-sky CMB experiments.

Section~\ref{sec:map_characterization} describes the E2E simulations used to determine the amplitude of systematic effects, as well as their impact at the map and power spectrum levels. The modular structure of the simulation code allows us to combine or isolate different systematic effects and to evaluate their amplitudes and residuals by comparison with noise-only TOIs. In addition, we define the version of the E2E simulations, including both the noise and the dominant systematic effects, as the ``noise'' used in the full focal-plane bulk simulations (FFP10, similar to the FFP8 simulations described in \citehfimap).

Section~\ref{sec:conclusion} gives conclusions. In order to improve the readability of this paper, in some cases only representative figures are given. Additional, complementary figures are provided in \citet{planck2016-ES}, hereafter the ``\citeES.''

\section{Data processing}
\label{sec:data_processing}
\begin{figure}[htbp!]
\includegraphics[width=\columnwidth]{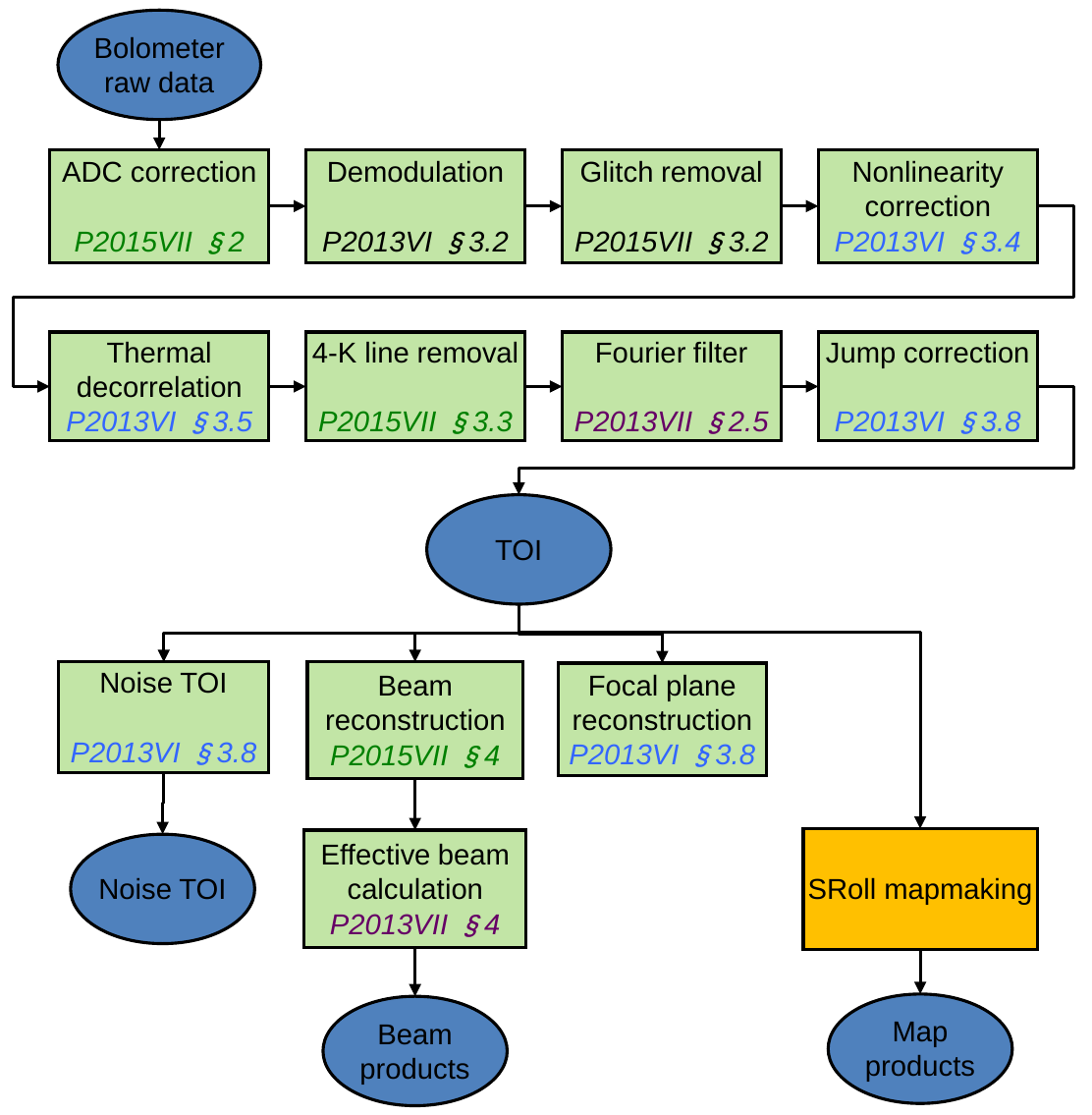}
\caption{Schematic of the HFI pipeline, referencing sections of previous papers (and this work) at each step.}
\label{fig:DPCflow} 
\end{figure}

Figure~\ref{fig:DPCflow} presents an overview of the entire HFI data processing chain, including the TOI cleaning and calibration, as well as the mapmaking.
Details are described in a series of pre-launch and post-launch papers, in particular \citet{lamarre2010}, \citet{planck2011-1.5}, \citet{planck2013-p03}, \citet{planck2013-p03c}, \citet{planck2014-a08}, \citet{planck2014-a09}, and \citet{planck2014-a10}. The schematic in Fig.~\ref{fig:DPCflow} shows the process that produces the inputs for the \sroll\ mapmaking solution. Each step is shown with a reference to the appropriate paper and section.

The HFI 2018 pipeline, up to the mapmaking step, is identical to the one used for the 2015 results and described in \citehfimap. The cleaned, calibrated TOIs used as input to the mapmaking are therefore identical to the ones used in the 2015 release and subsequent intermediate results. Improvements in the HFI 2018 maps are almost entirely due to the \sroll\ mapmaking; this removes most known systematic errors and is described in \citelowell. The HFI 2018 maps include other small changes in the mapmaking procedure that are noted below.

\subsection{TOI processing and outputs}
\label{sec:preprocessing}
\subsubsection{On-board signal processing}
\label{sec:onboard}
The HFI bolometers are current-biased, by applying a square wave voltage (of frequency $f_{\rm{mod}}$ = 90\,Hz) across a pair of load capacitors, producing a nearly square-wave current bias ($I_{\rm{bias}}$) across the bolometer. The bolometer resistance, proportional to the optical power incident on the bolometer from the sky, is then measured as a nearly square-wave voltage. The signal is amplified with a cold (50\,K) JFET source follower, and the majority of the bolometer voltage (proportional to the DC component of the sky signal) is removed by subtracting a constant-amplitude square wave $V_{\rm{bal}}$, bringing the output voltage closer to zero. A second stage of amplification in the warm electronics follows, along with a low-pass filter; then the voltage is digitized with an analogue-to-digital converter (ADC), 40 samples per half-period. Next, the 40 samples are summed to create a single science sample per half-period. The science samples are accumulated and compressed in 254-sample slices, which are passed to the spacecraft and telemetered to the ground station. On the ground, the spacecraft packets are reassembled and decompressed into a science timeline, forming for each bolometer the raw data at the input to the data processing. The compression of the data required to fit within the HFI telemetry allocation implies a small loss of accuracy. In this release, with its tighter control on other errors, the effect of compression and decompression becomes non-negligible and is discussed in Sect.~\ref{sec:Compression}.

\subsubsection{TOI processing outputs to \sroll}
\label{sec:ouputstosroll}
The first step in the data processing is to correct the TOI for the known non-linearity in the ADC. That ADC non-linearity (ADCNL) was measured during the warm phase of the mission, but not with enough accuracy to correct it at the level required for the present analysis. Next, the data samples are demodulated by subtracting a running average baseline and multiplying the digital signal by the parity of the bias voltage (alternating $+1$ and $-1$). Cosmic-ray ``glitches'' are detected and templates of the long-time-constant tails ($<3$\,s) of these glitches are fitted and subtracted (very-long-time constants, i.e., tens of seconds, not included, are discussed in Sect.~\ref{sec:xferfunction}).

A simple quadratic fit to the bolometer's intrinsic non-linear response measured on the ground is applied to the signal. A thermal template, constructed from a filtered signal of the two ``dark'' (i.e., not optically-coupled to the telescope) bolometers, is decorrelated from the time-ordered data to remove the long term drifts of the signal. Harmonics of the pickup of the \HeJT\ cooler drive current (referred to here as ``4-K lines'') are fitted and subtracted. A transfer function, based on a model with several time constants with their respective amplitudes, and with a regularizing low-pass filter, is deconvolved from the data, also in the Fourier domain. Jumps in the voltage level are detected and corrected. At this point, a cleaned TOI is available, which, as already noted has been produced using a process that is unchanged from the previous 2015 release. As described in \citelowell, TOIs are compressed per stable pointing period in the form of {\tt HEALPix} \citep{gorski2005} binned rings (HPR) which form the inputs to the mapmaking.

\subsubsection{Change in data selection}
\label{sec:1000rings}
\begin{figure}[htbp!]
\includegraphics[width=\columnwidth]{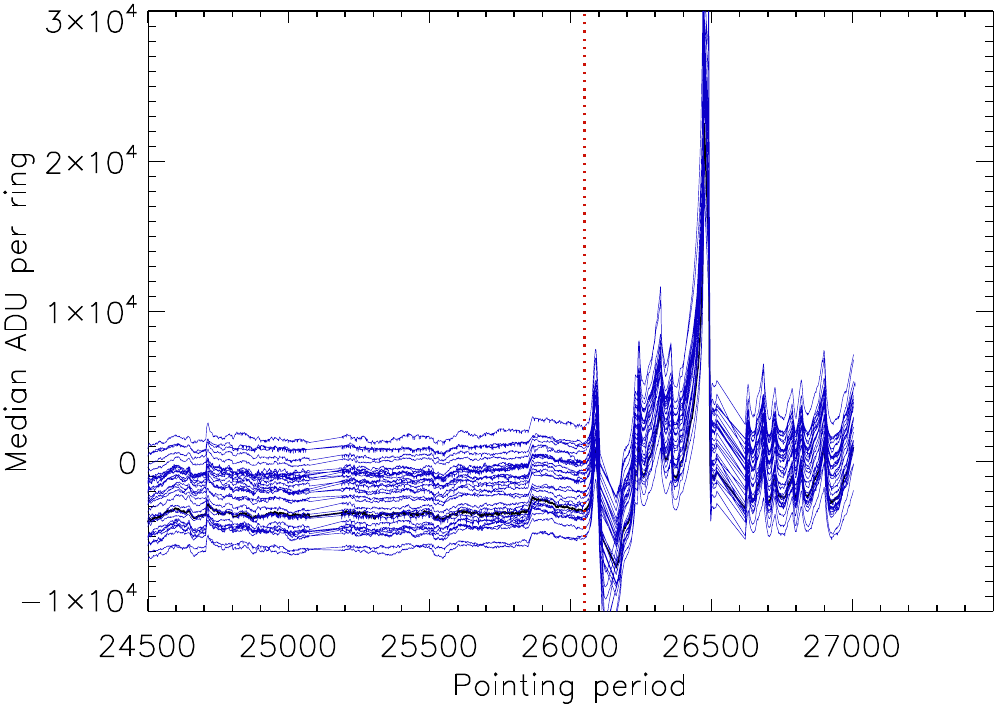}
\caption{Signal level for all HFI detectors near the end of the HFI cryogenic phase of the mission.}
\label{fig:1000rings} 
\end{figure}

A data qualification stage, which essentially remains unchanged \citep[see][]{planck2014-a08}, selects data that are the inputs to the mapmaking portion of the pipeline. In this 2018 release, we choose an earlier end point for the data we use, ending at pointing period (also called ring) 26050 instead of 27005. This cuts out the data at the very end of the HFI cryogenic phase when a larger passive $^3$He flow control was required to maintain the 100-mK temperature when the pressure in the tank became too low; this kept the 100-mK stage close to its nominal temperature at the cost of significant temperature fluctuations, inducing response drifts, associated with the long stabilization time constant of the 100-mK stage. Figure~\ref{fig:1000rings} shows the large variations of the mean signal associated with the 100-mK stage temperature unstable period.

\begin{figure}[htbp!]
\includegraphics[width=\columnwidth]{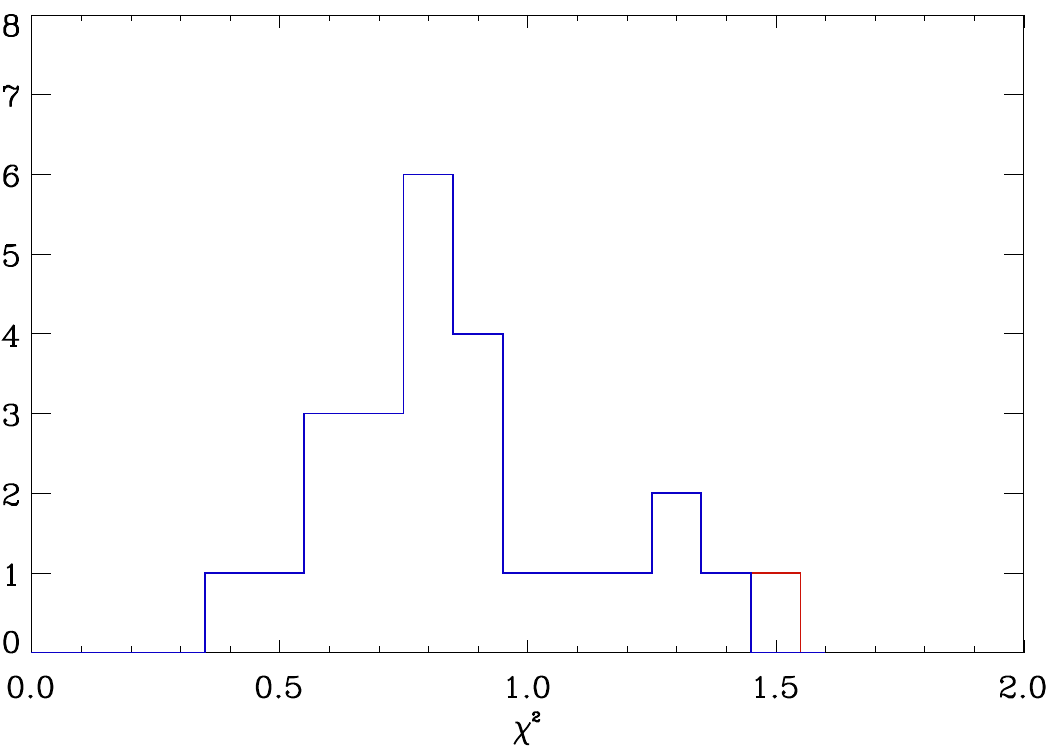}
\caption{Histogram the $\chi ^2$ distribution for the maps in which a set of 1000 consecutive pointing periods has been removed (see text). The red line shows the $\chi ^2$ for the last block of pointing periods removed.}
\label{fig:xi2br} 
\end{figure}

Although the temperature fluctuation effects are removed to first order in the processing, we demonstrate a residual effect of this unstable period. We build 26 reduced data sets of the mission (full set: 26~000 pointing periods) from which 1000 different consecutive pointing periods have been removed. We then differentiate the maps built from the full set and these reduced data sets. We compute cross-spectra ($C_{\ell}$, for $\ell=3$ to 20) of these differences at 100 and 143\,GHz and compute
\begin{eqnarray}
\chi^2 &=& \frac{1}{18} \sum_{\ell=3}^{20} \frac{C_{\ell}^2}{\sigma^2},\nonumber
\end{eqnarray}
where $\sigma$ is the variance for each $\ell$ for the 26 reduced sets. Figure~\ref{fig:xi2br} shows the distribution of this normalized $\chi ^2$.
The red line indicates this quantity for the 2018 HFI maps, where the pointing periods 26050 to 27005 have been removed, showing that those pointing periods are indeed anomalous.

\subsubsection{Noise characterization}
An estimate of the sky signal, based on redundant observations within a pointing period, is subtracted from the data to provide an estimate of the noise component of the TOI. This noise is referred to as ``TOI noise.'' The TOI noise has not changed since the 2015 release. A decomposition between white (detector photon and electronic) noise, correlated thermal noise shared by all bolometers ($\simeq 1/f^2$), and a $1/f^{\alpha}$ noise component with $\alpha \simeq 1$ adjusted to fit the data, was already carried out \citep[see for example figure~26 of][]{planck2014-a08}. After deglitching, the $1/f^{\alpha}$ component (referred to in this paper as ``the $1/f$ noise'') was confirmed to be Gaussian (figures~2 and 3 of \citelowell). Nevertheless, the $\alpha$ parameter exhibits small variations about unity.

The knee frequency is almost independent of the noise level, which goes from $1.4\,\mu{\rm K}\,{\rm s}^{1/2}$ at 143\,GHz, to $400\,\mu{\rm K}\,{\rm s}^{1/2}$ for 353\,GHz. This indicates a link between noise level and the $1/f$ component. This is discussed in Sect.~\ref{sec:undetected}. The noise above the knee frequency shows an extra component, seen as dips and bumps starting at 0.35\,Hz, and extending with regular spacing up to 3\,Hz for all frequencies and detectors, and creating a very slow rise from 3\,Hz up to 0.35\,Hz, above the white noise (see figure~1 of \citelowell). The compression algorithm uses a slice of 254 samples (see Sect.~\ref{sec:onboard}), corresponding to a frequency of about 0.35\,Hz. A simulation including compression and decompression and deglitching of added glitches accounts for this effect, see figure 26 of \cite {planck2014-a08}. The TOI noise is still modelled as white noise plus a $1/f$ component in the E2E simulations described below.

\subsubsection{TOI processing outputs to simulations and likelihood codes}

The TOI noise product is used, together with a physical model of the detector chain noise, as the noise input to the E2E simulations. This noise is adjusted with a smooth addition at the level of a few percent in order to match the noise and systematic effect residuals measured in the odd-even pointing period null test (see Appendix~\ref{sec:end2end}).

The planet-crossing data are used to reconstruct the focal-plane geometry. The relevant TOI data for each planet observation are selected and first processed to remove cosmic-ray glitches. Then scanning-beam maps are built from the planet data, accounting for the motion of the planets on the sky. The selected scanning-beam maps are passed to the effective-beam computation codes to retrieve the effective beam as a function of position on the sky, and to compute effective-beam window functions for various sky cuts and scanning strategy. This procedure is identical to the one used for the 2015 release \citep[see][]{planck2014-a08}. The overall transfer function, which is then evaluated through E2E simulations, is based on the effective beam, accounting for the scanning strategy.

\subsection{\sroll-mapmaking solution}
\label{sec:Sroll}

\subsubsection{The integrated scheme}
\label{sec:srollscheme}
To fully exploit the HFI polarization data, a better removal and control is needed for the intensity-to-polarization leakage due to calibration mismatch and bandpass mismatch than was done in the 2015 release. This implies taking advantage of the very high signal-to-noise ratio to improve knowledge of the instrument by extracting key parameters from the sky observations, instead of using the preflight, ground-based measurements. For this purpose, \sroll\ makes use of an extended destriper. Destriper methods have been used previously to remove baseline drifts from detector time streams, while making co-added maps of the data, by taking advantage of the redundancy in the scanning strategy. \sroll\ is a generalized polarization destriper, which, in addition, compares all the observations of the same sky pixel by the same detector with different polarization angles, as well as by different detectors within the same frequency band. This destriper thus fits differences between instrument parameters that minimize the difference between all polarized observations of the same sky pixel in the same frequency band. This allows a very good correction of the intensity-to-polarization leakage. \sroll\ solves consistently for:
\begin{itemize}
\item one offset for each pointing period;
\item an additional empirical transfer function to the correction already done in the TOI processing, covering the missing low-frequency parts in both the spatial and temporal domains (see Sect.~\ref{sec:xferfunction});
\item a total kinetic CMB dipole relative calibration mismatch between detectors within a frequency band;
\item a bandpass mismatch for the foreground response due to colour corrections with respect to the CMB calibration, using spatial templates of each foreground;
\item the absolute calibration from the orbital dipole which does not project on the sky using the CMB monopole temperature $T_{\rm CMB}=2.72548\,{\rm K}\pm0.57\,{\rm mK}$ from \citet{fixsen2009}.
\end{itemize}
 With all these differential measurements, the absolute value of some of the parameters is given by imposing constraints on the average of all detectors in a frequency band, specifically requiring:
 \begin{itemize}
 \item the sum of the offsets to be zero (no monopoles);
 \item the average of the additional colour corrections (for both dust and free-free emission) to be zero, thus keeping the same average as the one measured on the ground.
\end{itemize}
The use of an independent astrophysical observation of a single foreground, if its extent and quality are good enough, allows a direct determination of the response of each single detector to this foreground (for example $^{12}$CO or $^{13}$CO lines).\footnote{The contributions of other spectral lines in the HFI frequency bands are negligible with respect to the present accuracy.} We measure the accuracy of the recovery of such response parameters, as well as the reduction of the systematic effect residuals in the final maps (and their associated power spectra), through E2E simulations. The response of a given detector to a particular foreground signal, after \sroll\ correction, is forced to match the average response of all bolometers in that frequency band. This is achieved at the expense of using a spatial template for each foreground to adjust the response coefficients. The foreground template must be sufficiently orthogonal to the CMB and to the other foregrounds. In Sect.~\ref{sec:srollimplementation}, we show that when the gains converge, the leakage parameters converge towards the true value, but the quality of the template affects only the convergence speed. This improved determination of the response to foregrounds, detector by detector, could thus be used to integrate the component-separation procedure within the mapmaking process.

The effect of inaccuracies in the input templates has also been assessed through the E2E simulations. An iteration on the dust foreground template (re-injecting the foreground map obtained after a first \sroll\ iteration, followed by component separation), has been used to check that the result converges well in one iteration (see Sect.~\ref{sec:intensity_leakage}). The foreground spatial templates are used only to extract better bandpass-mismatch corrections, in order to reduce intensity-to-polarization leakage; they are not used directly in the sky-map projection algorithm, nor to remove any foregrounds, as described in the next section.

The E2E simulations show that \sroll\ has drastically reduced the intensity-to-polarization leakage in the large-scale HFI polarization data. The leakage term was previously more than one order of magnitude larger than the TOI noise, and prevented use of the HFI large-scale polarization data in the 2015 results. \sroll\ also provides a better product for component separation, which depends only on the average-band colour correction and not on the individual-detector colour corrections still affecting the polarized frequency maps and associated power spectra. Single detector maps, which were used in 2015 for different tests and for component separation, cannot be used for this 2018 release, since the differences of response of single detectors to the main (dust, free-free, and CO) foregrounds have been removed within the mapmaking.

\sroll\ thereby enables an unprecedented detection of the $EE$ reionization peak and the associated reionization parameter $\tau$ at very low multipoles in the $E$-mode power spectrum. A description of the \sroll\ algorithm has been given in \citelowell\ and is still valid. The \sroll\ algorithm scheme and equations are given again in this section, together with a small number of improvements that have been made for this HFI 2018 release. Among them is the bolometer photometric calibration scheme, which exploits the Doppler boost of the CMB created by the Earth's orbital motion (the orbital dipole) for the CMB channels (100 to 353\,GHz). It is now significantly improved with respect to \citehfimap\ by taking into account the spectral energy distribution (SED) variation of the dust foreground on large scales. The submillimetre channels (545 and 857\,GHz) are calibrated on planets, as in the previous release.

\subsubsection{\sroll\ implementation}
\label{sec:srollimplementation}
\sroll\ data model for bolometer signal $M$ is given by Eq.~\ref{eq:datamodel} where indices are:
\begin{itemize}
\item $b$ for the bolometer, up to $n_{\mathrm{bolo}}$;
\item $i$ for the stable pointing period (ring), up to $n_{\mathrm{ring}}$;
\item $k$ for the stable gain period (covering a range of pointing periods $i$);
\item $p$ for the sky pixel;
\item $h$ for bins of spin frequency harmonics (up to $n_{\mathrm{harm}}$), labelled as bin$_{h=1}$ for the first harmonic, bin$_{h=2}$ for harmonics 2 and 3, bin$_{h=3}$ for harmonics 4 to 7, and bin$_{h=4}$ for harmonics 8 to 15;
\item $f$ for the polarized foreground, up to $n_{\mathrm{comp}}$ (dust and free-free).
\end{itemize}
\begin{eqnarray}
\label{eq:datamodel}
g_{b,k}M_{b,i,p} &=& I_p + \rho_b \left[Q_p\cos(2\phi_{b,i,p}) + U_p\sin(2\phi_{b,i,p})\right] \nonumber \\
 & & + \sum_{h=1}^{n_{\mathrm{harm}}}\gamma_{b,h} V_{b,i,p,h} +
\sum_{f=1}^{n_{\mathrm{comp}}}L_{b,f} C_{b,i,p,f}\nonumber \\
 & & +\ D^{\mathrm{tot}}_{i,p} + F_{b,i,p}^{\mathrm{dip}}+ F_{b,i,p}^{\mathrm{gal}} + O_{b,i} + g_{b,k}N_{b,i,p},
\end{eqnarray}
where:
\begin{itemize}
\item $g_{b,k}$ is the absolute gain of a bolometer;
\item $M_{b,i,p}$ is the measured bolometer total signal,
\item $I_p$, $Q_p$, and $U_p$ represent the common sky maps seen by all bolometers (excluding the Solar dipole);
\item $\rho_b$ is the polarization efficiency, kept fixed at the ground measurement value;
\item $\phi_{b,i,p}$ is the detector polarization angle with respect to the north-south axis, kept fixed at the ground measurement value;
\item $V_{b,i,p,h}$ is the spatial template of the empirical transfer function added in the mapmaking;
\item $\gamma_{b,h}$ is the empirical transfer-function complex amplitude added in the mapmaking;
\item $C_{b,i,p,f}$ is the foreground-components spatial template;
\item $L_{b,f}$ is the bandpass foreground colour-correction coefficients difference with respect to the frequency bandpass average over bolometers for foreground $f$, i.e., for each foreground component, we set $\sum_{b=1}^{n_{\mathrm{bolo}}} L_{b,f} = 0$;
\item $D^{\mathrm{tot}}_{i,p}$ is the total CMB dipole signal (sum of Solar and orbital dipoles), with $D^{\mathrm{sol}}_{p}$ being the template for the Solar dipole with a fixed direction and amplitude and $D^{\mathrm{orb}}_{i,p}$ being the template of the orbital dipole with its known amplitude;
\item $F_{b,i,p}^{\mathrm{dip}}$ is the total dipole integrated over the far sidelobes (FSL);
\item $F_{b,i,p}^{\mathrm{gal}}$ is the Galactic signal integrated over the FSL;
\item $O_{b,i}$ is the offset per pointing period $i$ used to model the $1/f$ noise, and we set $\sum_{b=1}^{n_{\mathrm{bolo}}} \sum_{i=1}^{n_{\mathrm{ring}}}O_{b,i}=0$, since the \Planck\ data provide no information on the monopole;
\item $N_{b,i,p}$ is the white noise, with variance $\sigma_{b,i,p}$.
\end{itemize}
Table~\ref{tab:param} summarizes the source of the templates and coefficients used by, or solved, within \sroll.
\begin{table}[htbp!]
\newdimen\tblskip \tblskip=5pt
\caption{Source of the templates and coefficients used or solved within \sroll.}
\label{tab:param}
\vskip -4mm
\footnotesize
\openup=2pt
\setbox\tablebox=\vbox{
\newdimen\digitwidth
\setbox0=\hbox{\rm 0}
\digitwidth=\wd0
\catcode`*=\active
\def*{\kern\digitwidth}
\newdimen\signwidth
\setbox0=\hbox{+}
\signwidth=\wd0
\catcode`!=\active
\def!{\kern\signwidth}
\newdimen\pointwidth
\setbox0=\hbox{.}
\pointwidth=\wd0
\catcode`?=\active
\def?{\kern\pointwidth}
\halign{\hbox to 2.5cm{#\leaderfil}\tabskip 1em&
#\hfil\tabskip 0em\cr
\noalign{\doubleline}
\omit \hfil Quantity\hfil&\omit\hfil Source\hfil\cr
\noalign{\vskip 3pt\hrule\vskip 5pt}
$g_{b,k}$&Solved within {\tt SRoll}\cr
$M_{b,i,p}$&Input\cr
$I_p$, $Q_p$, $U_p$&Solved within {\tt SRoll}\cr
$\rho_b$&Measured on ground\cr
$\phi_{b,i,p}$&Measured on ground\cr
$V_{b,i,p,h}$&Computed from TOI\cr
$\gamma_{b,h}$&Solved within {\tt SRoll}\cr
$C_{b,i,p,f}$&\Planck\ 2015 maps smoothed at $1^\circ$\cr
$L_{b,f}$&Solved within {\tt SRoll}\cr
$D^{\mathrm{tot}}_{i,p}$&\Planck\ 2015 dipole + satellite orbit\cr
$F_{b,i,p}^{\mathrm{dip}}$, $F_{b,i,p}^{\mathrm{gal}}$&Simulated from FSL model\cr
$O_{b,i}$&Solved within {\tt SRoll}\cr
$N_{b,i,p}$& variance ($\sigma$) computed from TOI\cr
\noalign{\vskip 3pt\hrule\vskip 5pt}}}
\endPlancktable
\end{table}

Solving for gain variability necessarily involves solving a non-linear least-squares equation. {\tt SRoll} uses an iterative scheme to solve for the gains $g_{b,k,n}$. At iteration $n$, we set
\begin{equation}
g_{b, k,n} = g_{b,k} + \delta g_{b,k, n},
\label{eq:ge}
\end{equation}
where $\delta g_{b,k, n}$ is the difference between the gains $g_{b,k,n}$ and the real gain $g_{b,k}$. The goal is to iteratively fit the gain error $\delta g_{b,k, n}$, which should converge to 0. Outside the iteration, we first remove the orbital dipole and the low-amplitude foreground FSL signals, leading to a corrected measured bolometer total signal $M'_{b,i,p}$:
\begin{equation}
g_{b,k,n}~M'_{b,i,p} = g_{b,k,n}~M_{b,i,p} - F_{b,i,p}^{\mathrm{dip}}- F_{b,i,p}^{\mathrm{gal}} -D^{\mathrm{orb}}_{i,p}.
\label{eq:fsl}
\end{equation}
Using Eqs.~(\ref{eq:ge}) and (\ref{eq:fsl}), Eq.~(\ref{eq:datamodel}) becomes
\begin{eqnarray}
g_{b,k,n}~M'_{b,i,p} & = & S_{i,p} + \frac{\delta g_{b,k, n}}{g_{b,k}}~ \left(S_{i,p}+D^{\mathrm{orb}}_{i,p}\right)~ \nonumber \\
& & +\sum_{h=1}^{n_{\mathrm{harm}}}\gamma_{b,h} V_{b,i,p,h} + \sum_{f=1}^{n_{\mathrm{comp}}}(L_{b,f} + L_{f}) C_{b,i,p,f} \nonumber \\
& & +\ O_{b,i} + g_{b,k}~N_{b,i,p},
\label{eq:SimplifiedModelbis}
\end{eqnarray}
where $S_{i,p}$ is the part of the signal that projects on the sky map:
\begin{eqnarray}
S_{i,p} &=& I_p +\rho_b \left[Q_p\cos(2\phi_{i,p}) + U_p\sin(2\phi_{i,p})\right] +D^{\mathrm{sol}}_{p} \\
 &=&\widetilde{S_{i,p}} + (1+\eta_{i,p})~D^{\mathrm{tot}}_{i,p} - D^{\mathrm{orb}}_{i,p},
\end{eqnarray}
if we define $\widetilde{S_{i,p}}$ as the part of the signal $S_{i,p}$ orthogonal to $D^{\mathrm{tot}}_{i,p}$ during the period $i$. The quantity $\eta_{i,p}$ is the part of the foregrounds correlated with the total dipole.

Equation~(\ref{eq:SimplifiedModelter}) gives the difference between two unpolarized measures, ``$1$'' and ``$2$,'' of the same sky pixel $p$. Extension to polarized data is straightforward, but for readability we only write the unpolarized case here:
\begin{eqnarray}
g_{1,n}~M'_{1,p} - g_{2,n}~M'_{2,p}
&= & \left(\frac{\delta g_{1, n}}{g_{1}} \right)~ \left(\widetilde {S_{i,p}}+(1+\eta_{1,p})D^{\mathrm{tot}}_{i,p}\right) \nonumber \\
& & -\left(\frac{\delta g_{2, n}}{g_{2}}\right)~ \left(\widetilde {S_{i,p}}+(1+\eta_{2,p})D^{\mathrm{tot}}_{i,p}\right) \nonumber \\
& & +\sum_{h=1}^{n_{\mathrm{harm}}}\gamma_{1,h} V_{1,p,h} - \sum_{h=1}^{n_{\mathrm{harm}}}\gamma_{2,h} V_{2,p,h} \nonumber \\
& & +\sum_{f=1}^{n_{\mathrm{comp}}}L_{1,f} C_{1,p,f} - \sum_{f=1}^{n_{\mathrm{comp}}}L_{2,f} C_{2,p,f} \nonumber \\
& & +\ O_{1} - O_{2} + g_{1,n}~N_{1,p} - g_{2,n}~N_{2,p}.
\label{eq:SimplifiedModelter}
\end{eqnarray}
In Eq.~(\ref{eq:SimplifiedModelter}), the spatial templates $\widetilde {S_{i,p}}$ and $D^{\mathrm{tot}}_{i,p}$ are orthogonal, by definition.

A destriper minimizes the mean square of many signal differences for the same sky pixel observed either with two different detectors in the same frequency band, or several observations with the same detector. We compute the $\chi^2$ as
\begin{eqnarray}
\chi^2 &=& \sum_1 \sum_2 \frac{ \left[ \left(g_{1,n}~M'_{1,p} \right) - \left(g_{2,n}~M'_{2,p}\right) \right]^2 } {g_{1,n}^2 \sigma_1^2 +g_{2,n}^2 \sigma_2^2},
\label{eq:monchi2}
\end{eqnarray}
where $\sigma_1$ and $\sigma2$ are the noise levels associated with measurements 1 and 2.

In practice, the \sroll\ destriper minimizes this $\chi^2$ difference between one bolometer and the average of all bolometers in a frequency band. For this minimization, similarly to \citet{keihanen2004}, but with more parameters, we solve for $\Delta g_{b, k, n}$, $O_{b,i}$, $\gamma_{b,h}$, and $L_{b,f}$.

In the iteration, $\delta g_{b,k, n+1}\simeq \delta g_{b,k,n}~(1 + \eta_{b,k})$. If $|\eta|<1$, the iteration converges: $\lim\limits_{n \rightarrow \infty} {\delta g_{b,k,n+1}}= 0$, and thus Eq.~(\ref{eq:ge}) gives $\lim\limits_{n \rightarrow \infty} {g_{b,k,n}}= g_{b,k}$, which is the optimal gain implied by the combination of input parameters and templates. The terms $(\delta g_{b,k, n}/g_{b,k}) (1+\eta) D^{\mathrm{tot}}_{i,p}$ from Eq.~(\ref{eq:SimplifiedModelter}), drive the absolute calibration and inter-calibration convergence. The Solar dipole amplitude is extracted a posteriori and is not used in the mapmaking. The length of the stable calibration periods are chosen to fulfil the condition $|\eta|<1$. \citelowell\ has shown that this is possible, with a reasonable choice of such periods, even at 353\,GHz. This convergence, including the degeneracy between the determination of the gain mismatch and the determination of the bandpass mismatch leakage, which has been shown to be small, is discussed in detail in section~B.1.6 of \citelowell.

When the relative gains converge, the $O_{b,i}$, $\gamma_{b,h}$, and $L_{b,f}$ parameters converge also. This occurs even if their spatial templates are weakly correlated on the sky.

The bandpass-mismatch coefficient of one bolometer $b$ with respect to the average, for a given foreground $L_{b,f}$, are extracted by \sroll. They are used, in combination with an associated a priori template $C_{b,i,p,f}$ (which is not modified by \sroll), to remove from the HPRs the effect of different response to a foreground of each bolometer within a frequency band. This is a template correction in HPRs for bolometer $b$, computed with the single parameter $L_{b,f}$ for the whole sky and template $C_{b,i,p,f}$ for the foreground $f$. This brings all bandpass mismatch to zero and all detectors to the same colour correction as the frequency average for the foreground $f$. Nevertheless, it does depend on the accuracy of the a priori template chosen, these being the \Planck\ 2015 foreground maps. In Sect.~\ref{sec:intensity_leakage}, we estimate the error induced by the inaccuracy of the dust template by iterating on the 353-GHz maps, taking the one coming from a component-separation procedure with the frequency maps generated at the previous iteration. Using the E2E simulations, we show that the residuals measured by the ${\rm input}-{\rm output}$ difference decrease with iterations (see Fig.~\ref{fig:dustiter2}), being smaller than the noise at the first iteration by two orders of magnitude, and showing fast convergence of \sroll\ at the second iteration with a further reduction of the residuals.

Thus, for component-separation methods, the colour correction of the frequency map for each foreground should be taken as the straight (non-noise-weighted) average of the ground-based bandpasses for all bolometers at that frequency. \textit{The component-separation schemes must not adjust bandpass mismatch between detectors of the same frequency for dust, CO, and free-free emission components}. Such adjustments can be done for the synchrotron component in the HFI frequencies (which is too weak to be extracted by \sroll) within the uncertainties of the ground measurement, as discussed in Sect.~\ref{sec:caveats}. The residuals of this systematic effect are simulated and discussed in Sect.~\ref{sec:intensity_leakage}.

The CMB total dipole dominates only for the 100- to 353-GHz bands, so we use a smoothed sky at 545 and 857\,GHz, where the dust emission provides the inter-calibration of detectors inside each frequency band. The absolute calibration is provided by the planet model, as in the 2015 release. Finally, we check a posteriori, using the Solar dipole in the 545-GHz data, that the planet calibration is within $(0.2\pm0.5)\,\%$ of the CMB calibration.

\sroll\ then projects the TOIs to pixel maps using the parameters extracted in the destriping procedure, with noise weighting. Badly conditioned pixels, for which the polarization pointing matrix cannot be computed, are defined as unseen {\tt HEALPix} pixels. As detailed in Table~\ref{tab:unseenpixels}, in the full-mission frequency maps, the number of these unseen pixels is only one pixel at 100\,GHz, and none at the other frequencies; however, they appear in significant numbers in the null-test maps (the worse case is for 217-GHz hm2, with 20\,\% missing pixels), as explained in more detail in Sect.~\ref{sec:highmultipoles}.
\begin{table}[htbp!]
\newdimen\tblskip \tblskip=5pt
\caption{Number of unseen pixels per frequency band and data sets, specifically for full-mission, half-mission and ring splits (out of $5\times10^7$ pixels at \Nside=2048.)}
\label{tab:unseenpixels}
\vskip -6mm
\footnotesize
\setbox\tablebox=\vbox{
\newdimen\digitwidth
\setbox0=\hbox{\rm 0}
\digitwidth=\wd0
\catcode`*=\active
\def*{\kern\digitwidth}
\newdimen\signwidth
\setbox0=\hbox{+}
\signwidth=\wd0
\catcode`!=\active
\def!{\kern\signwidth}
\newdimen\pointwidth
\setbox0=\hbox{.}
\pointwidth=\wd0
\catcode`?=\active
\def?{\kern\pointwidth}
\halign{\tabskip 0pt\hbox to 2.85 cm{#\leaderfil}\tabskip 1em&
\hfil#\hfil&
\hfil#\hfil&
\hfil#\hfil&
\hfil#\hfil&
\hfil#\hfil\tabskip 0em\cr
\noalign{\doubleline}
\omit\hfil Frequency [GHz]\hfil&Full&hm1&hm2&Odd&Even\cr
\noalign{\vskip 3pt\hrule\vskip 5pt}
\noalign{\vskip 2pt}
100&1&*2\,528&*50\,917&245\,567&250\,204\cr
143&0&*6\,531&*64\,307&**5\,967&**4\,609\cr
217&0&13\,358&115\,439&*23\,002&*20\,655\cr
353 PSB only&0&*1\,402&*39\,793&**3\,095&**2\,477\cr
353&0&*1\,334&*39\,309&**3\,093&**2\,479\cr
545&0&*1\,187&**2\,450&***\,**1&***\,**9\cr
857&0&*4\,106&**1\,998&***\,**2&***\,**1\cr
\noalign{\vskip 3pt\hrule\vskip 5pt}}}
\endPlancktable
\end{table}

The solution for the map parameters is built with the sky partly masked. This masking has two goals: firstly, to avoid regions of the sky with time-variable emission; and secondly to avoid regions with a strong Galactic signal gradient. The brightest point sources (e.g., strong variable radio sources) are thus removed. The Galactic plane and molecular cloud cores are also masked, due to strong signal gradients. Nevertheless, a relatively large sky coverage is needed to properly solve for the bandpass mismatch extracted from the Galactic signal. The fraction of the sky used is $f_{\mathrm{sky}} = 86.2\,\%$ at 100\,GHz, $f_{\mathrm{sky}} = 85.6\,\%$ at 143\,GHz, $f_{\mathrm{sky}} = 84.6\,\%$ at 217\,GHz, and $f_{\mathrm{sky}} = 86.2\,\%$ at 353\,GHz. Figure~B.4 of \citelowell\ shows the masks used in the solution. 
The mismatch coefficients are estimated on a masked sky. Of course, the bandpass-mismatch coefficients, used in the mapmaking process, are detector parameters, and are constant, and applied on the full sky.

\subsubsection{Approximations in the pipeline}
\label{sec:approx}
We now describe the potential corrections identified in the pipeline, but not implemented because they would induce only very small corrections.

HPRs contain residuals due to imperfect removal of time-variable signals. \sroll\ begins by binning the data for each detector, from each pointing period, into HPRs at $\Nside\,{=}\,2048$. In addition to the detector HPR data, we produce templates of the signals that do not project on the sky: Galactic emission seen through the far sidelobes; the zodiacal dust emission; and the orbital dipole (including a higher-order quadrupole component). These templates are used to remove those small signals from the signal HPRs, which thus depend on the quality of the FSL and zodiacal models. Because of the asymmetry of the FSL with respect to the scanning direction and the zodiacal cloud asymmetry, FSL and zodiacal signals do not project on the sky in the same way for odd and even surveys. This gives a very good test for the quality of this removal and it is fully used in the destriper process.

Aberration of the beam direction is not included in the removed dipole. The kinetic dipole is the sum of the Doppler effect and the change in direction of the incoming light, the aberration. The auto- and cross-spectra of the difference maps, built with and without accounting for the aberration, show high multipole residuals induced by the striping of the maps affected by the small discontinuities in the apparent time variation of the gain. The levels are negligible (less than $10^{-3}\microKcarre$ in $TT$ and $10^{-5}\microKcarre$ in $EE$ and $BB$).

Polarization efficiency $\rho$ and angle cannot be extracted directly in \sroll\ because they are degenerate with each other. They induce leakage from $T$ to $E$- and $B$-modes and from $E$- to $B$-modes. \citet{rosset2010} reported the ground measurement values of the polarization angles, measured with a conservative estimate of $1\deg$ error. It has been shown, using $TB$ and $EB$ cross-spectra, that the frequency-average polarization angles are known to better than 0\pdeg5 (appendix A of \citelowell). The same paper also reports, in table B.1, the ground measurements of the polarization efficiency deviations from unity (expected for a perfect PSB) of up to 17\,\%, with statistical errors of 0.1 to $0.2\,\%$; the systematic errors, expected to be larger, are difficult to assess but probably not much better than 1\,\%. These polarization angles and efficiencies are used in \sroll. Testing these parameters within \sroll\ is discussed in Sect.~\ref{sec:polareff}, which shows that these effects are very significant for the SWBs.

The previously known temporal transfer function for the CMB channels cannot be extracted by \sroll\ and is kept as corrected in the TOI processing. Nevertheless additional very long time constants were found to shift the dipole and thus to contribute to the calibration errors in the 2013 release. In the 2015 release, the dipole shifts were taken into account (see \citehfimap). Such very long time constants are not detected in the scanning beams and associated temporal transfer functions. Instead these longer ones have been detected through the shift in time of the bolometer responses to a temperature step in the proportional-integral-differential (PID) control power of the 100-mK plate. The average time constant was 20\,seconds. These very long time constants, which cannot be detected in scanning beams, are the main contributors to the residuals corrected by the empirical transfer function at the HPR level in \sroll, between spin harmonics $h=2$ and 15, described in Sect.~\ref{sec:xferfunction}. The residuals from uncorrected spin harmonics $h=1$ affect differently the calibration in odd and even surveys, while the uncorrected spin harmonics $h >15$ generate striping. This is also discussed in Sect.~\ref{sec:zebra}.

Although second-order terms for kinetic boost are negligible for the CMB channels, we removed in an open-loop the Solar velocity induced quadrupole. The frequency-dependent cross-term between the Solar and orbital velocity is the other second-order correction term included in \sroll.

\subsection{Improvements with respect to the previous data release}
\label{sec:improvements}
Comparison between HFI 2015 and 2018 maps and associated power spectra are discussed in Sect.~\ref{sec:powerspectra}. In this section we provide a detailed discussion of the main specific differences in the treatment of the data.

\subsubsection{PSB$_a$ versus PSB$_b$ calibration differences}
\begin{figure}[htbp!]
\includegraphics[width=\columnwidth]{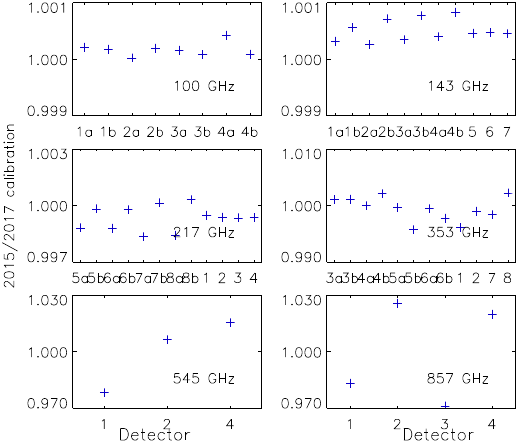}
\caption{Ratio between the HFI 2015 and 2018 absolute calibrations. The differences observed between PSB$_a$ and PSB$_b$ detectors come mostly from the 2015 calibration.}
\label{fig:AversusBCalibration} 
\end{figure}

Figure~\ref{fig:AversusBCalibration} shows the difference between the HFI 2015 and 2018 absolute calibrations based on single-bolometer maps (see Sect.~\ref{sec:monobolomaps}).
At 143 and 217\,GHz, there is an obvious PSB$_a$ versus PSB$_b$ pattern, PSB$_a$ detectors being systematically lower than PSB$_b$ ones. As the rms of the bolometer inter-calibration for the CMB channels in \sroll\ is better than $10^{-5}$ (see Fig.~\ref{fig:interbol} and figure~13 of \citelowell), the differences, apparent in Fig.~\ref{fig:AversusBCalibration}, are dominated by the errors in the 2015 inter-calibration. This can also be seen directly in figure~1 of \citehfimap, which shows this pattern. We conclude that polarization was not properly modelled in the 2015 HFI mapmaking, inducing these residuals in the 2015 single-bolometer calibration. The figure also illustrates the improvement on the calibration in the 2018 release. A detailed discussion of the quality of detector calibration is presented in Sect.~\ref{sec:calibration}.

\subsubsection{Intensity-to-polarization leakage}
In the 2015 release, the intensity-to-polarization leakage, due to calibration mismatch and bandpass mismatch, was removed at 353\,GHz, following the mapmaking step, by a global fit solution using template maps of the leakage terms (these template maps were also released). As shown in section~3.1 of \citelowell\, the intensity-to-polarization leakage dominates at very low multipoles; however, it was also shown that the ADCNL dipole distortion systematic effect was strong at very low multipoles. The degeneracy between these two systematic effects severely limited the accuracy of this template-fitting procedure. {\tt SRoll} accurately extracts these intensity-to-polarization leakage effects, as shown by using the E2E simulations (see Sect.~\ref{sec:intensity_leakage}) and thus, in the present 2018 release, the leakage coefficients are now solved within \sroll.

\section{Map products}
\label{sec:bulk_simulations}

To make a ``global'' assessment of the noise and systematics and of improvements in their residuals after applying corrections, we first compare the maps from the 2015 and 2018 releases. Our characterization of the data also relies on null-test maps, and power spectra for the 2018 release. We analyse these in reference to the suite of E2E simulations of null tests, which were also used (as further described in Sect.~\ref{sec:map_characterization}) to characterize the levels of residuals from each separate systematic effect. While some systematic residuals remain in the maps at some level, all are smaller than the known celestial signals, and also smaller than the noise. Cross-spectra between frequency maps, averaged over a multipole range, are used in likelihood codes to test cosmological models; however, this requires knowledge of the data at a much lower level than the pixel TOI noise in the maps or in single multipoles in the power spectra. We therefore construct tests sensitive to such small signals.

We have identified a number of null tests in which correlated noise or signal modifications appear:
\begin{itemize}
\item the half-mission (``hm'') null test, which was extensively used in the previous release, is one of the best such tests, since it is sensitive to the time evolution of instrumental effects over the 2.5 years of the HFI mission (e.g., the ADCNL effect);
\item the survey null-test splits the data between those for the odd surveys
(S1+S3) and those for the even surveys (S2+S4), and is sensitive to asymmetries due to time constants and beam asymmetries;
\item the detector-set null test (hereafter ``detset'' null test) separates the four PSB detectors, at 100, 143, 217, and 353\,GHz, into two detector sets and make two maps out of these, then the difference of detset maps can be used as a very useful test of systematics arising from the detector chain specificities;
\item a formerly much used null test was to split the data into half rings (comparing the first and second half of each stable pointing period), but the noise in this case is affected by the glitch removal algorithm, in which we mask the same parts of the ring in the two halves of the pointing period;
\item we introduce the ``ring'' null test, using the difference of two pointing periods that just follow each other, one odd and the other even pointing periods (not to be confused with the odd-even survey differences).
\end{itemize}

Two of these tests (those for half missions and detsets) are tests from which we expect very small differences, whereas the other ones are not built with an exactly equivalent observation strategy: either scanning in opposite directions (for the survey null test) or using a different scanning strategy (for the ring null test). In the following, we compare the null tests and make a recommendation on their use.

\subsection{Frequency maps}
\subsubsection{2018 frequency maps}
\label{sec:freqmaps}

The main 2018 products from the HFI observations are the Stokes $I$, $Q$, and $U$ maps at 100, 143, 217, and 353\,GHz, and the intensity maps at 545 and 857\,GHz.

The \Planck\ 2015 Solar dipole is removed from those 2018 maps to be consistent with LFI maps and to facilitate comparison with the previous 2015 ones. The best Solar dipole determination from HFI 2018 data (see Sect.~\ref{sec:solardipole}) shows a small shift in direction of about 1\arcm, but a 1.8\microK\ lower amplitude (corresponding to a relative correction of $5\times 10^{-4}$). Removal of the 2015 Solar dipole thus leaves a small but non-negligible dipole residual in the HFI 2018 maps. To correct for this, and adjust maps to the best photometric calibration, users of the HFI 2018 maps should:
\begin{itemize}
\item put back into the maps the \Planck\ 2015 Solar dipole $(d,l,b) = (3.3645 \pm 0.0020\,{\rm mK}, \text{264\pdeg00} \pm \text{0\pdeg03}, \text{48\pdeg24} \pm \text{0\pdeg02)}$ (see \citehfimap);
\item include the calibration bias (column~E of Table~\ref{tab:ratios}), i.e., multiply by 1 minus the calibration bias;
\item remove the HFI 2018 Solar dipole.
\end{itemize}

The monopole of the 2018 HFI maps has been defined as in the previous 2015 release.

\begin{figure*}[htbp!]
\includegraphics[width=0.32\textwidth]{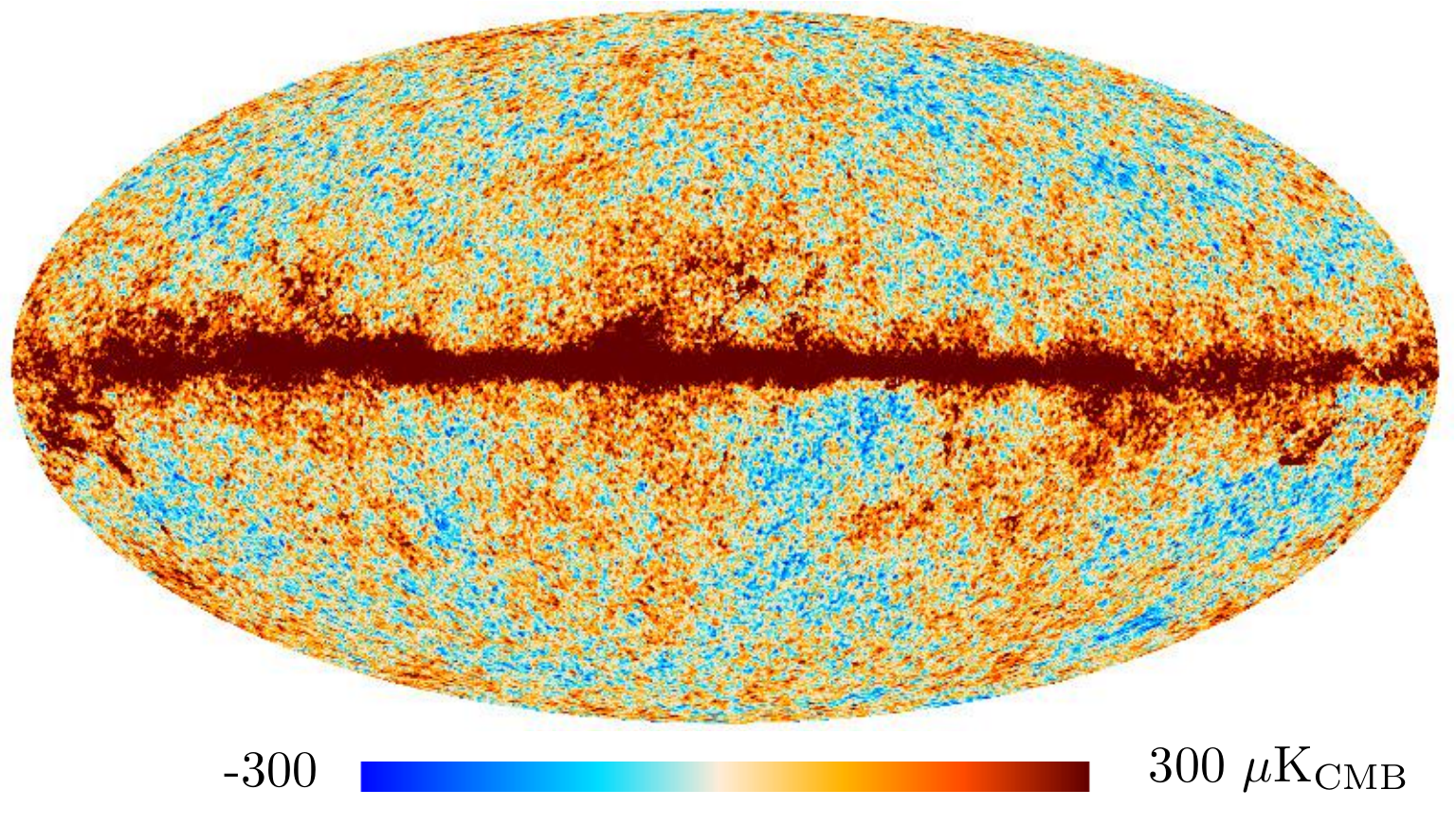}
\includegraphics[width=0.32\textwidth]{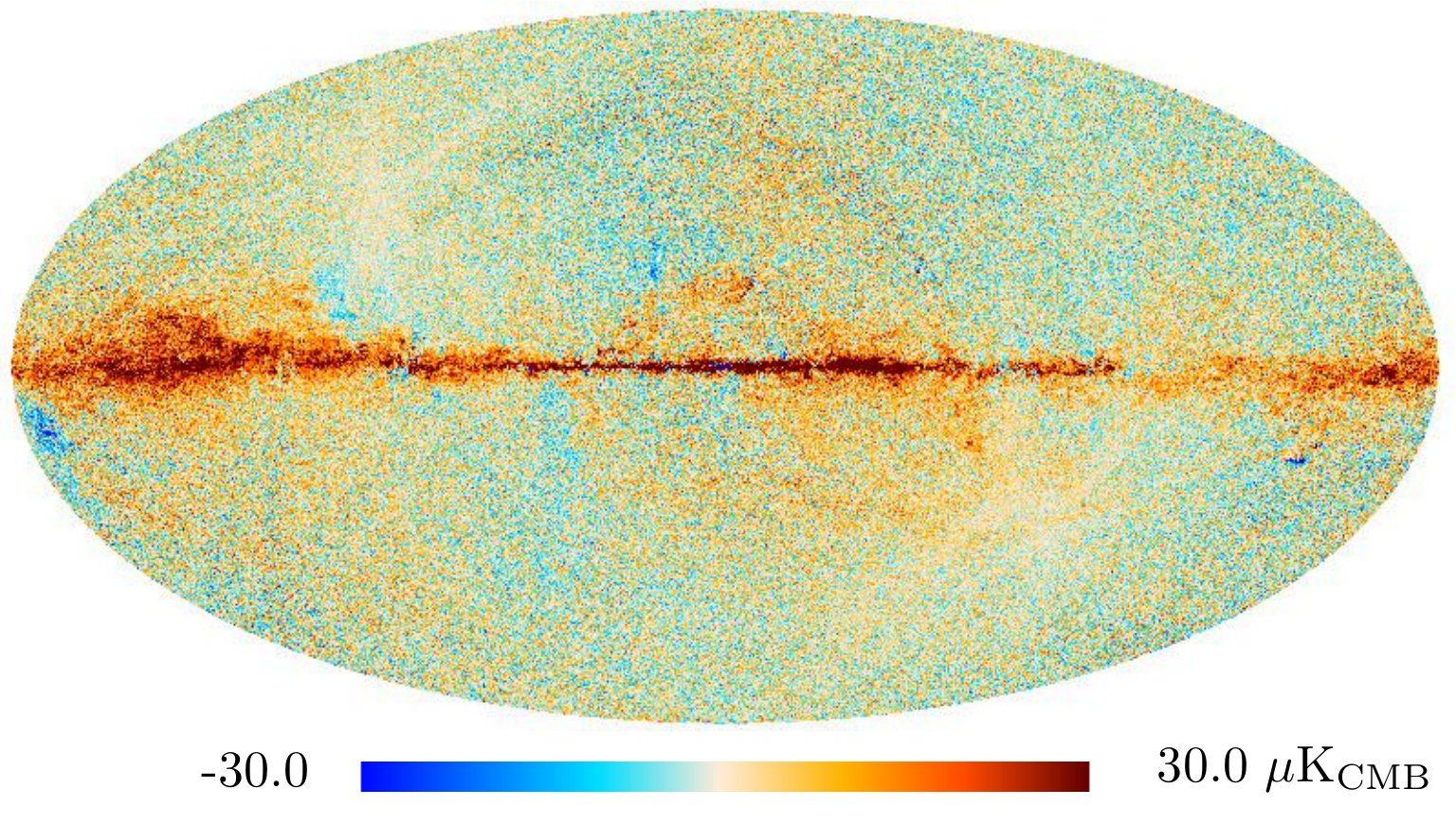}
\includegraphics[width=0.32\textwidth]{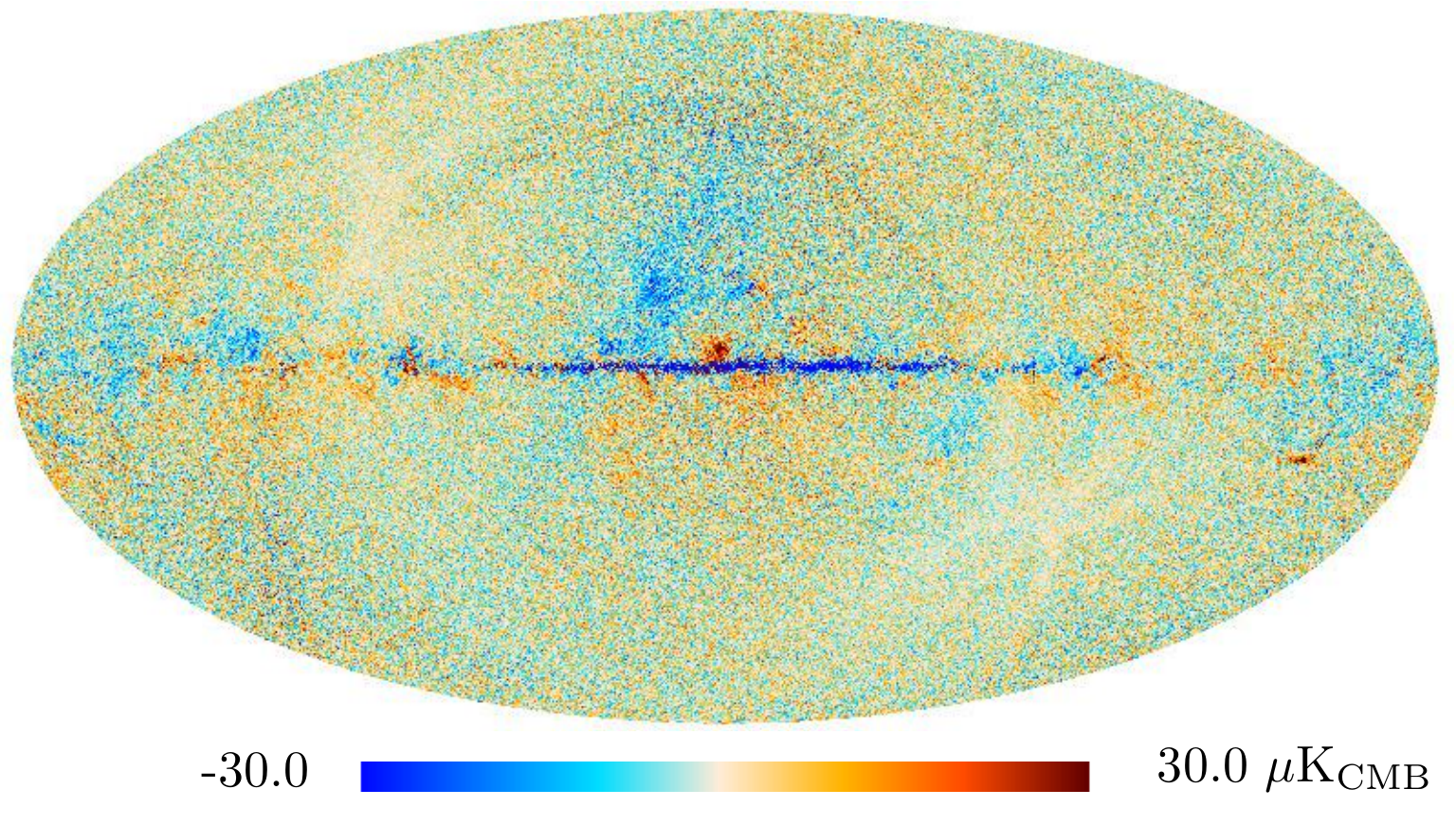}\\
\includegraphics[width=0.32\textwidth]{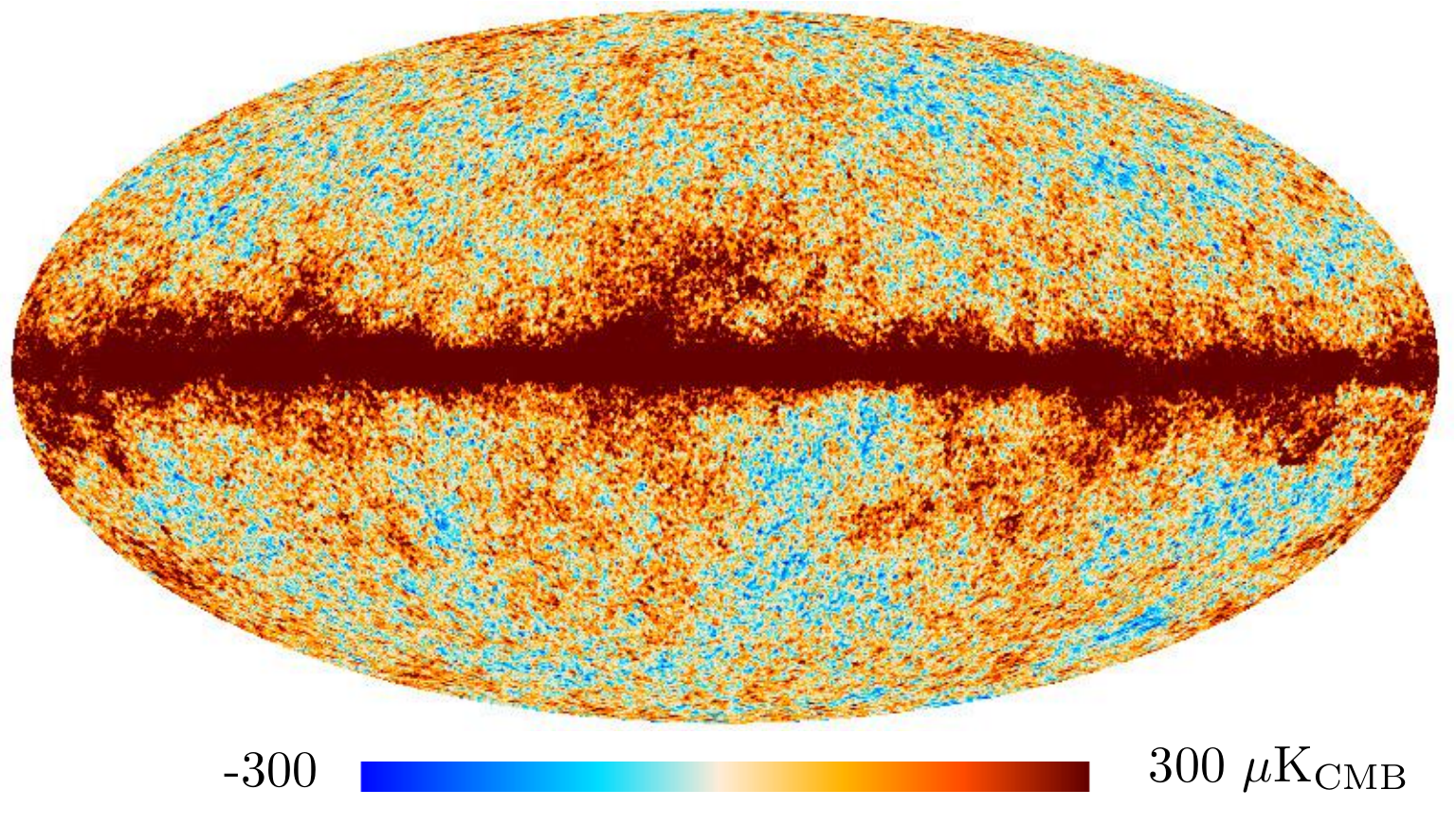}
\includegraphics[width=0.32\textwidth]{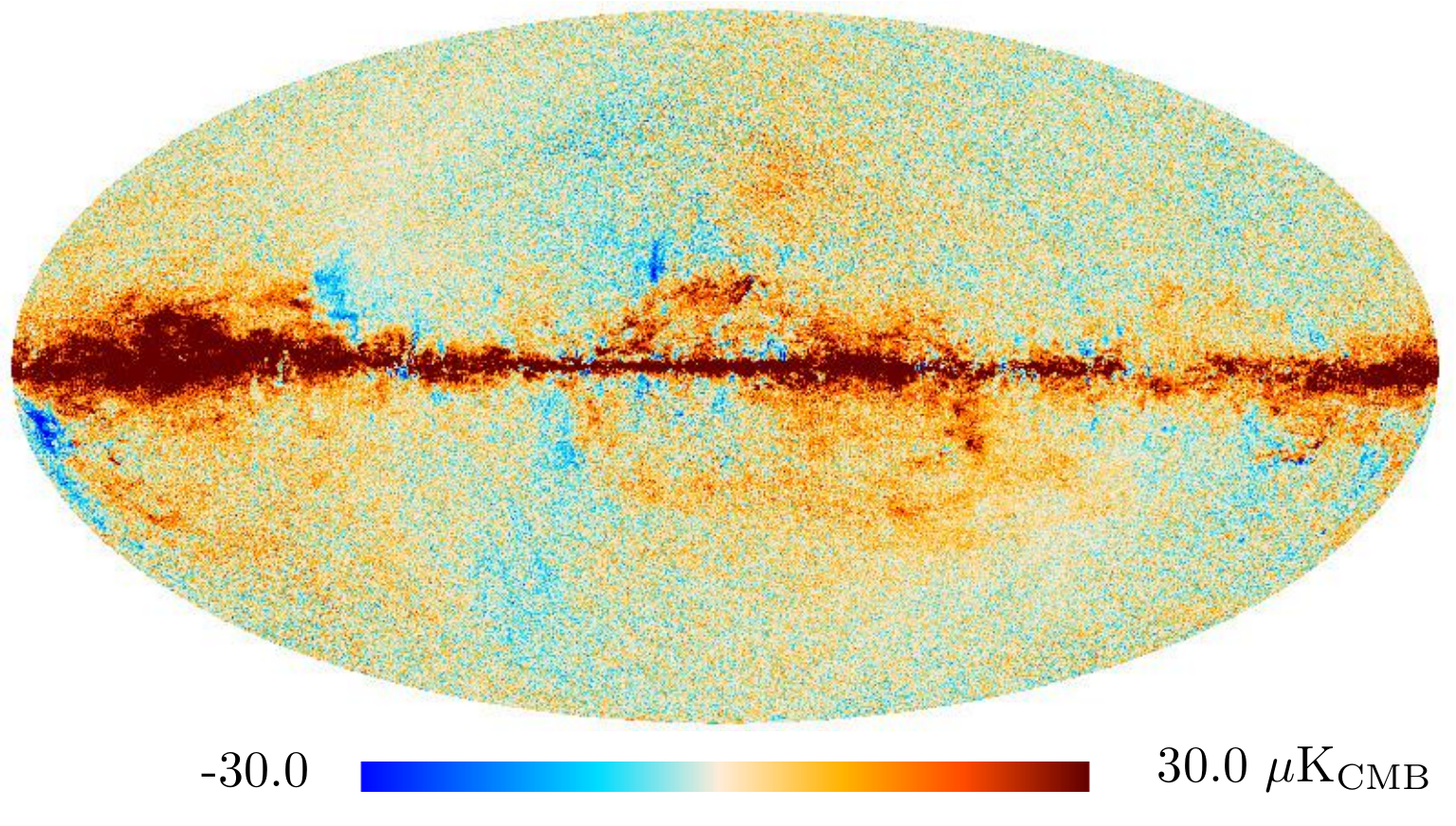}
\includegraphics[width=0.32\textwidth]{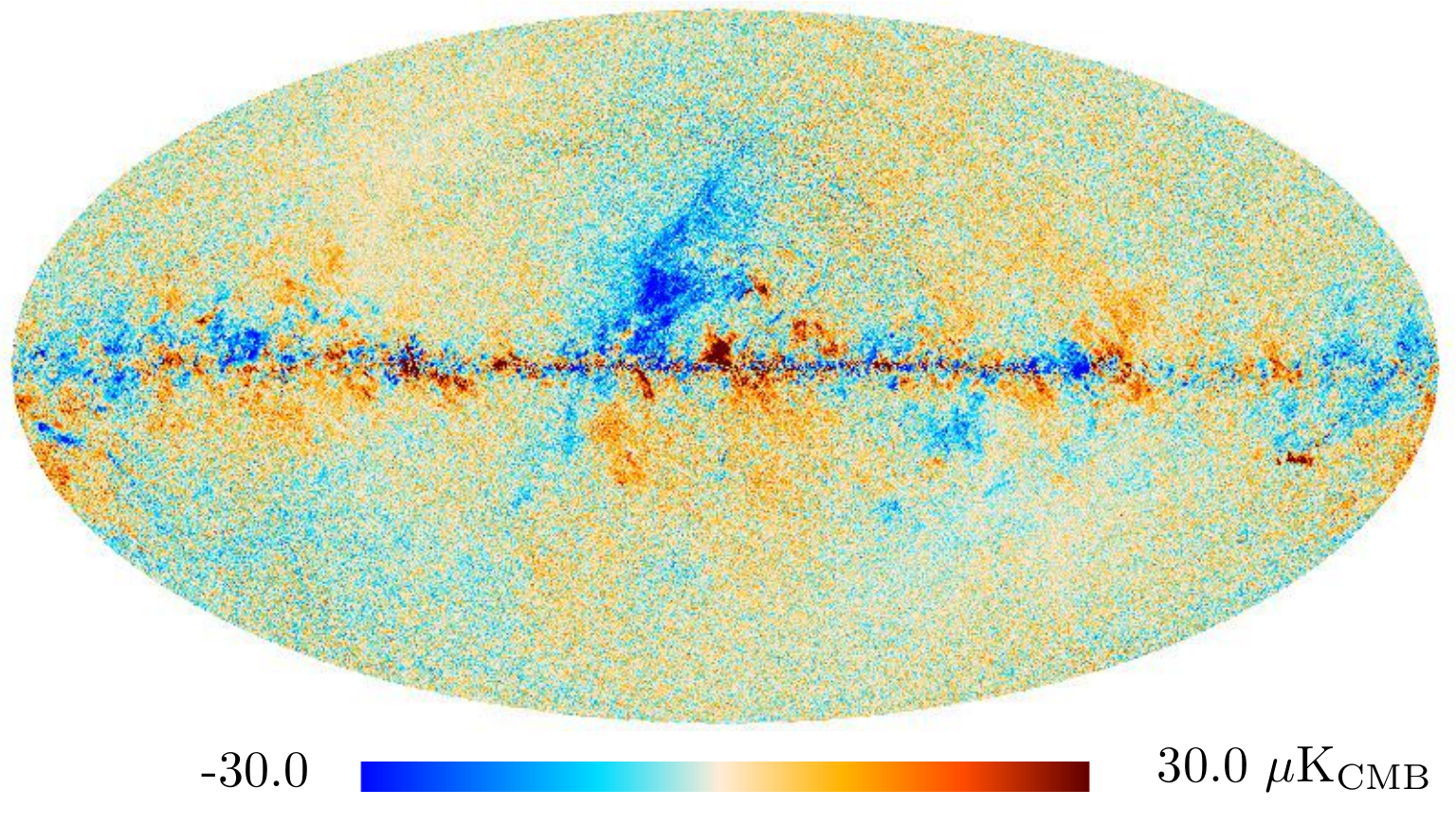}\\
\includegraphics[width=0.32\textwidth]{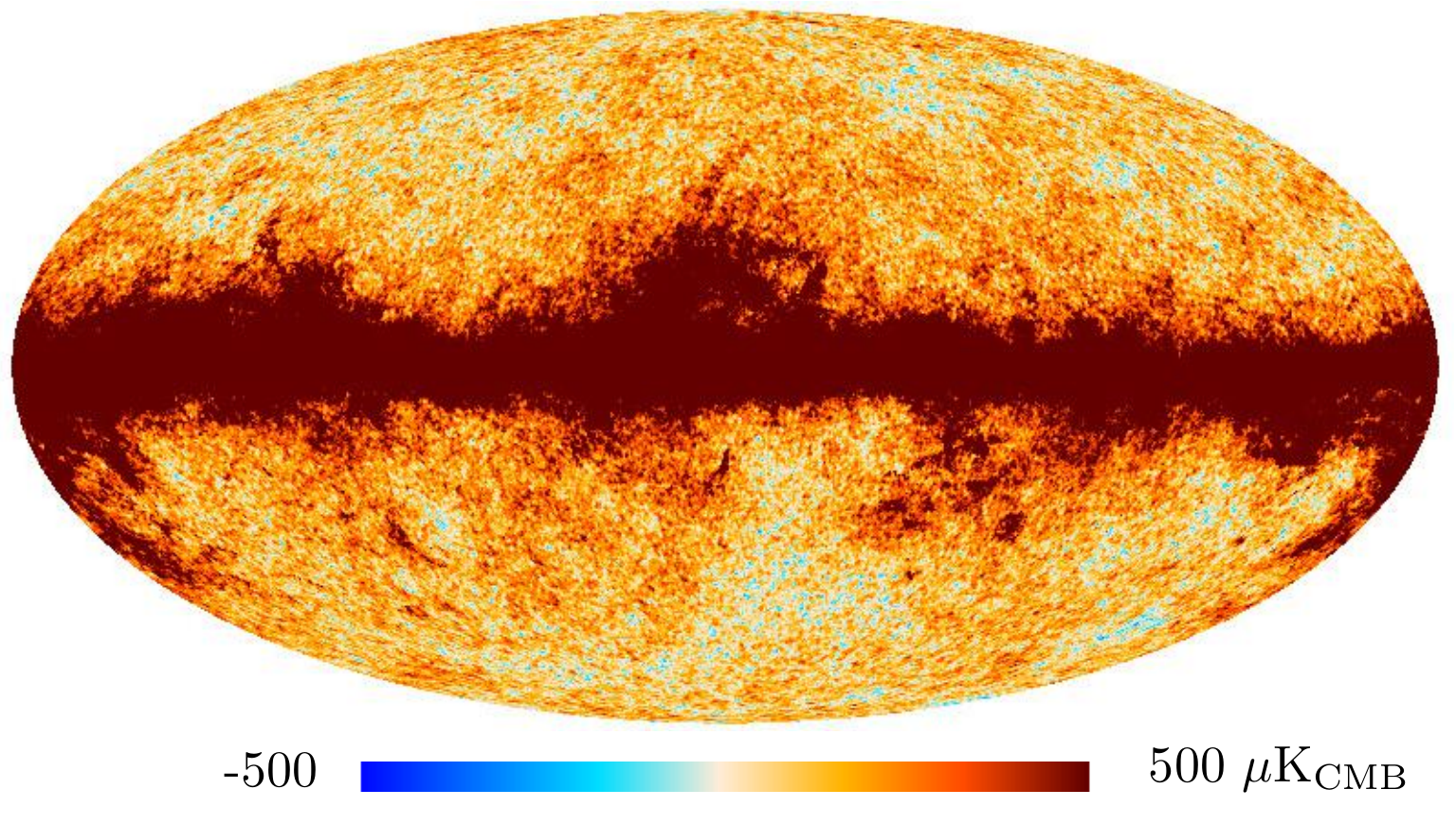}
\includegraphics[width=0.32\textwidth]{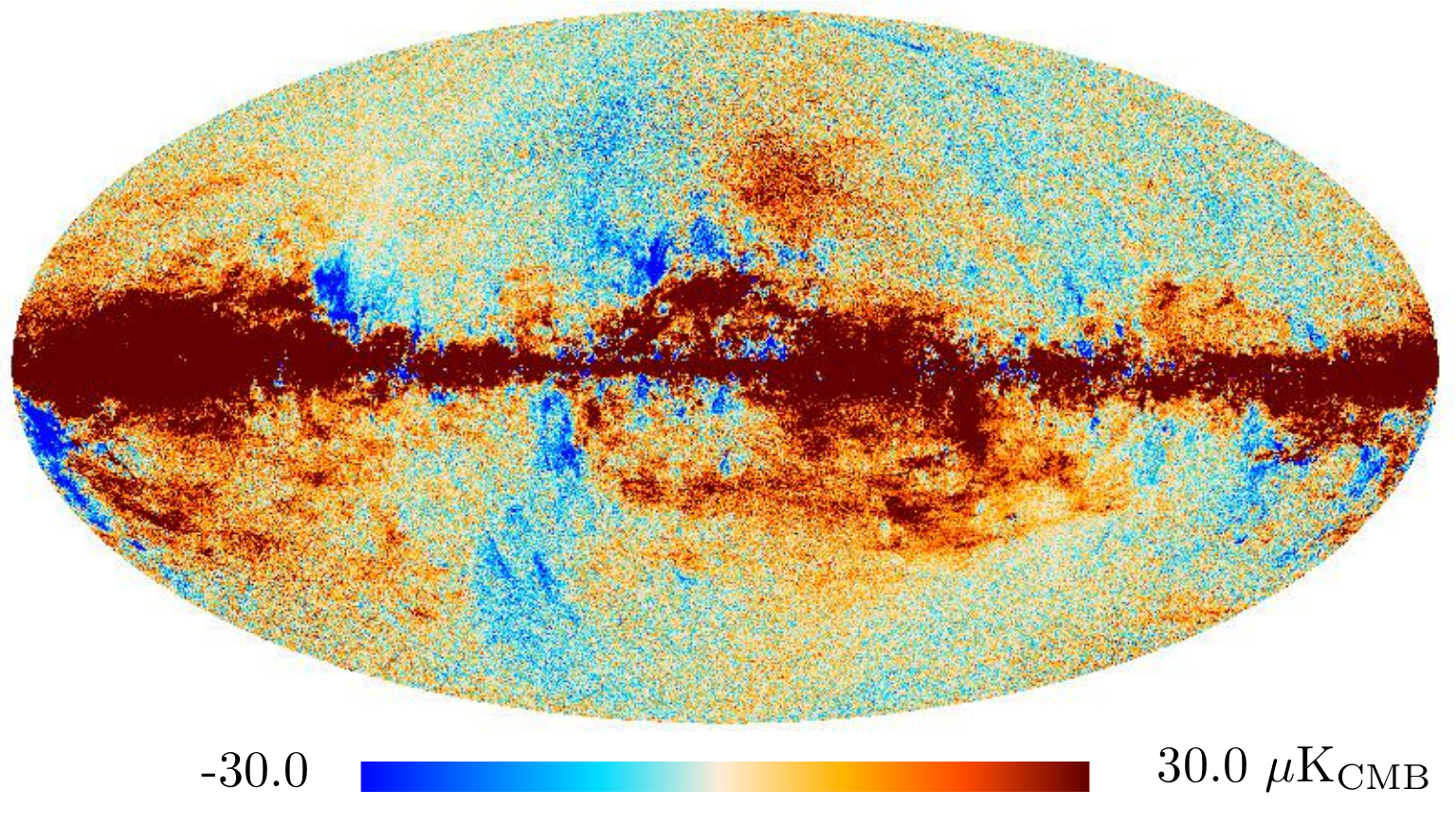}
\includegraphics[width=0.32\textwidth]{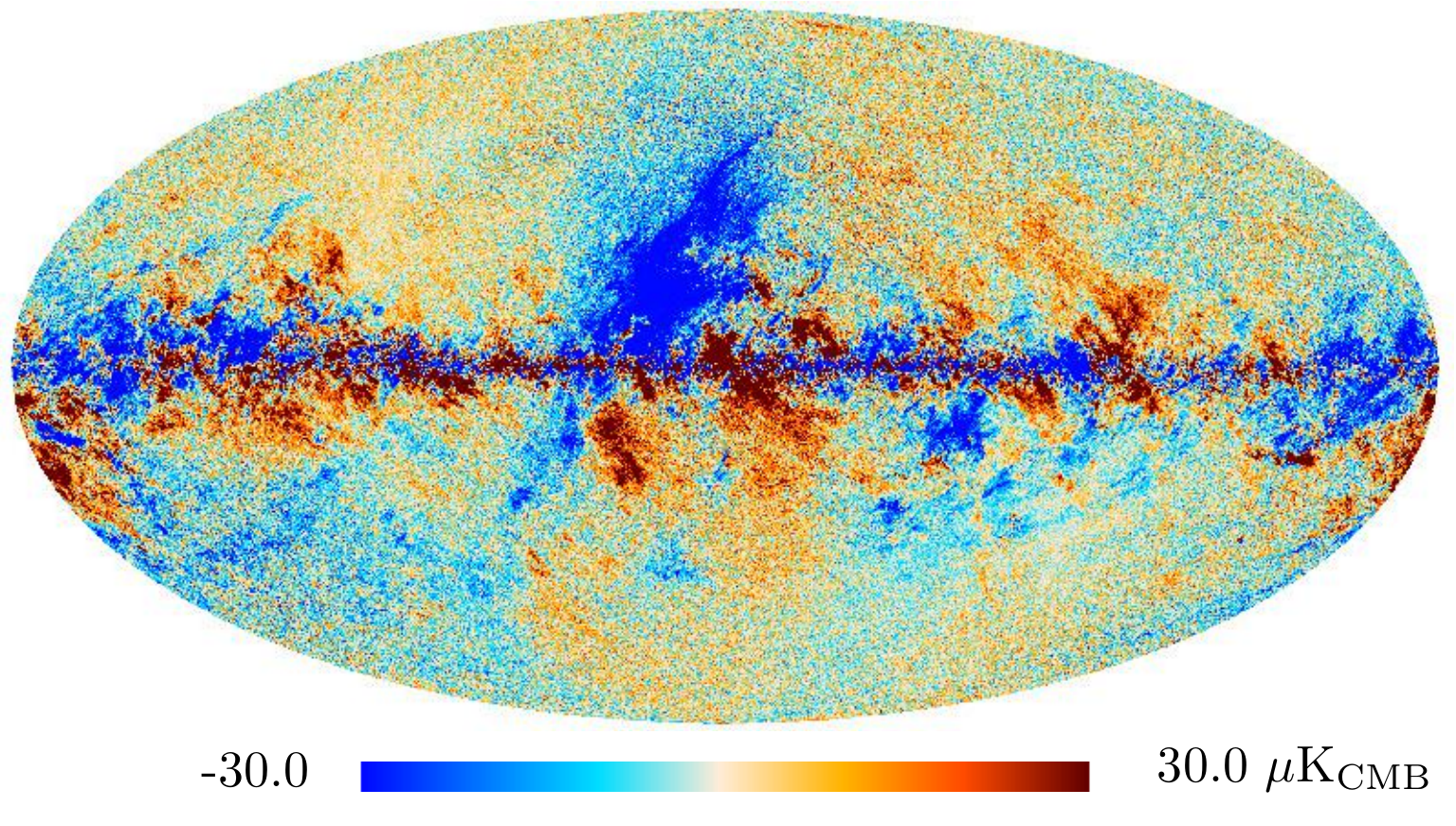}\\
\includegraphics[width=0.32\textwidth]{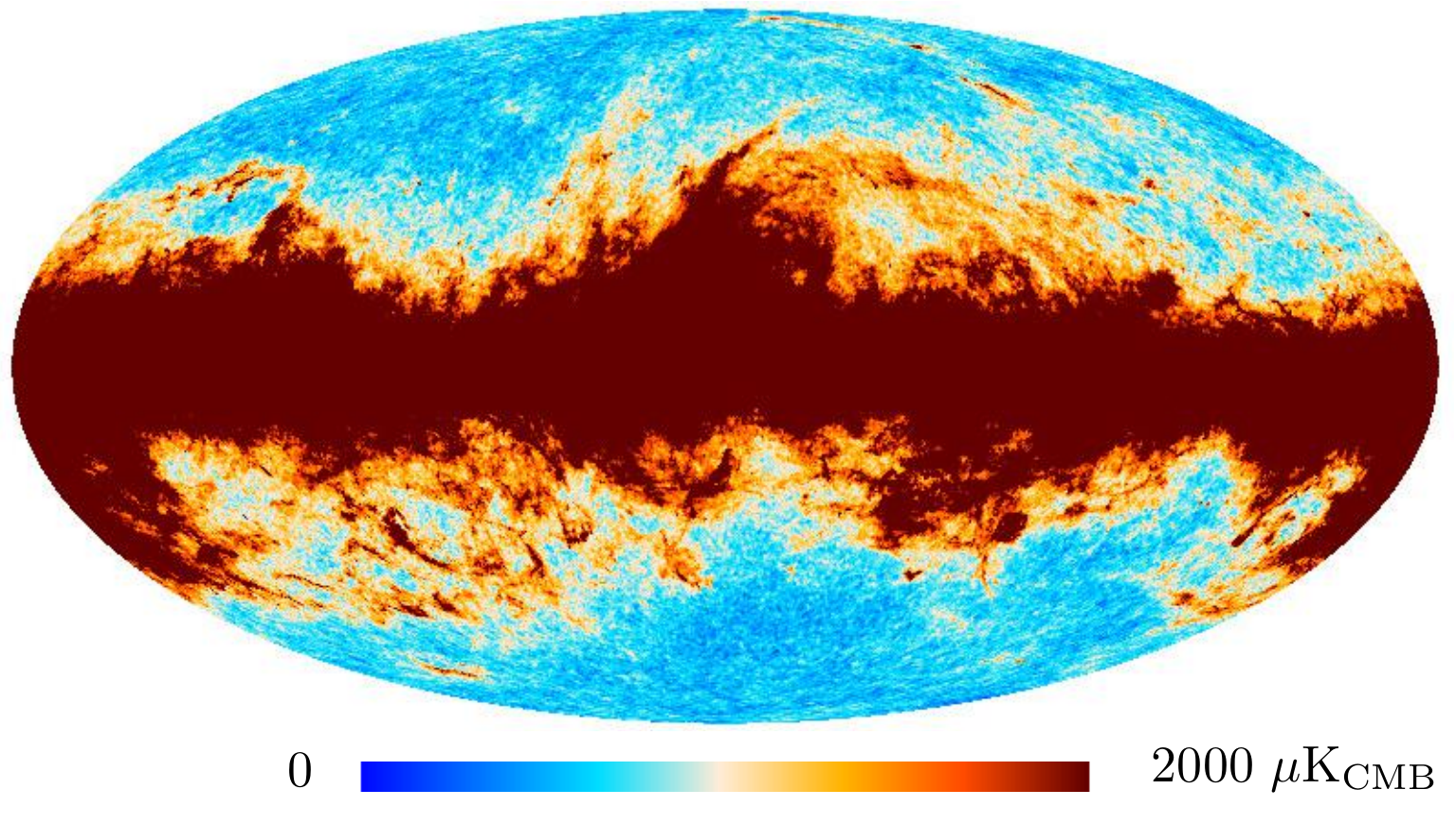}
\includegraphics[width=0.32\textwidth]{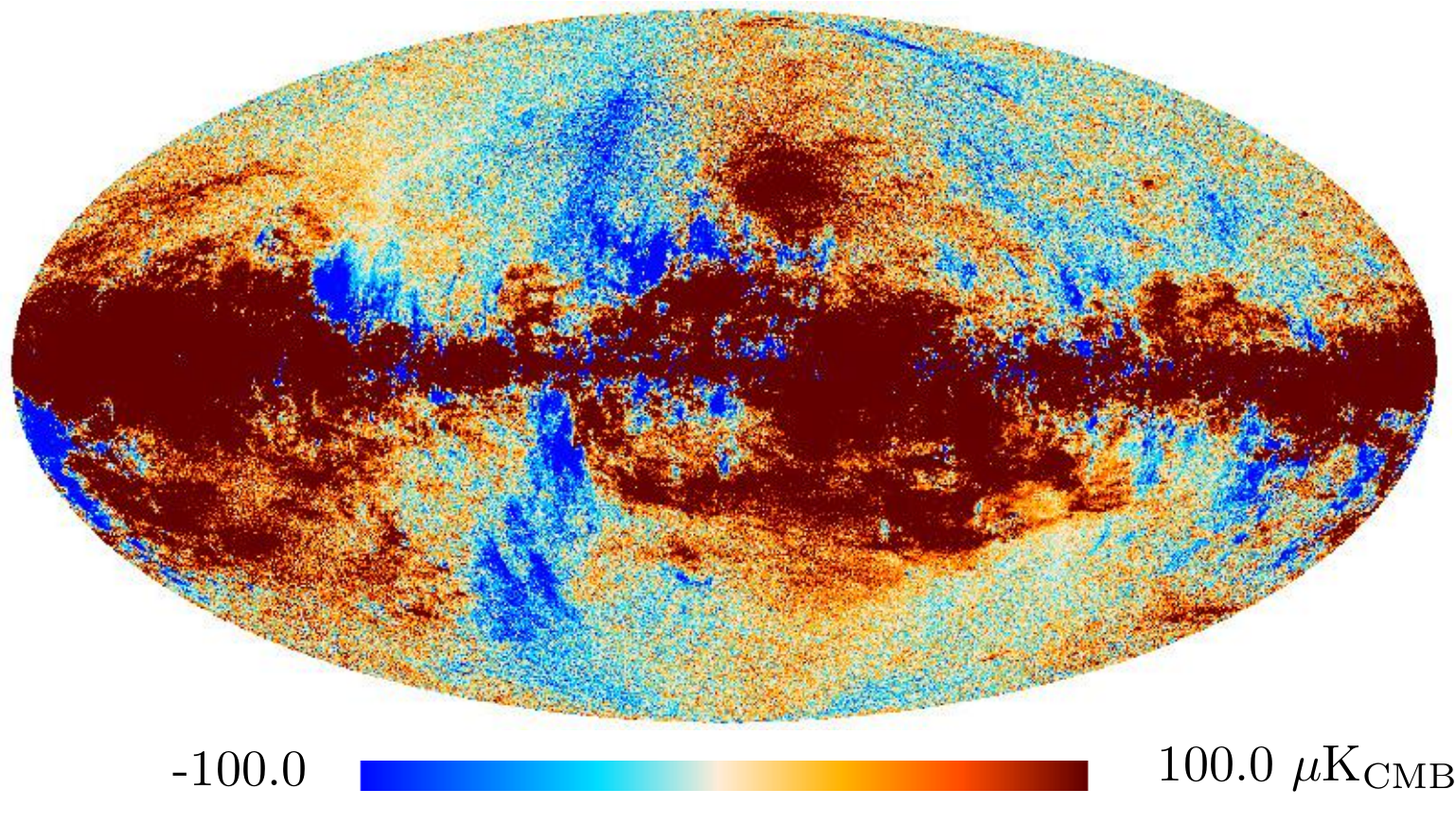}
\includegraphics[width=0.32\textwidth]{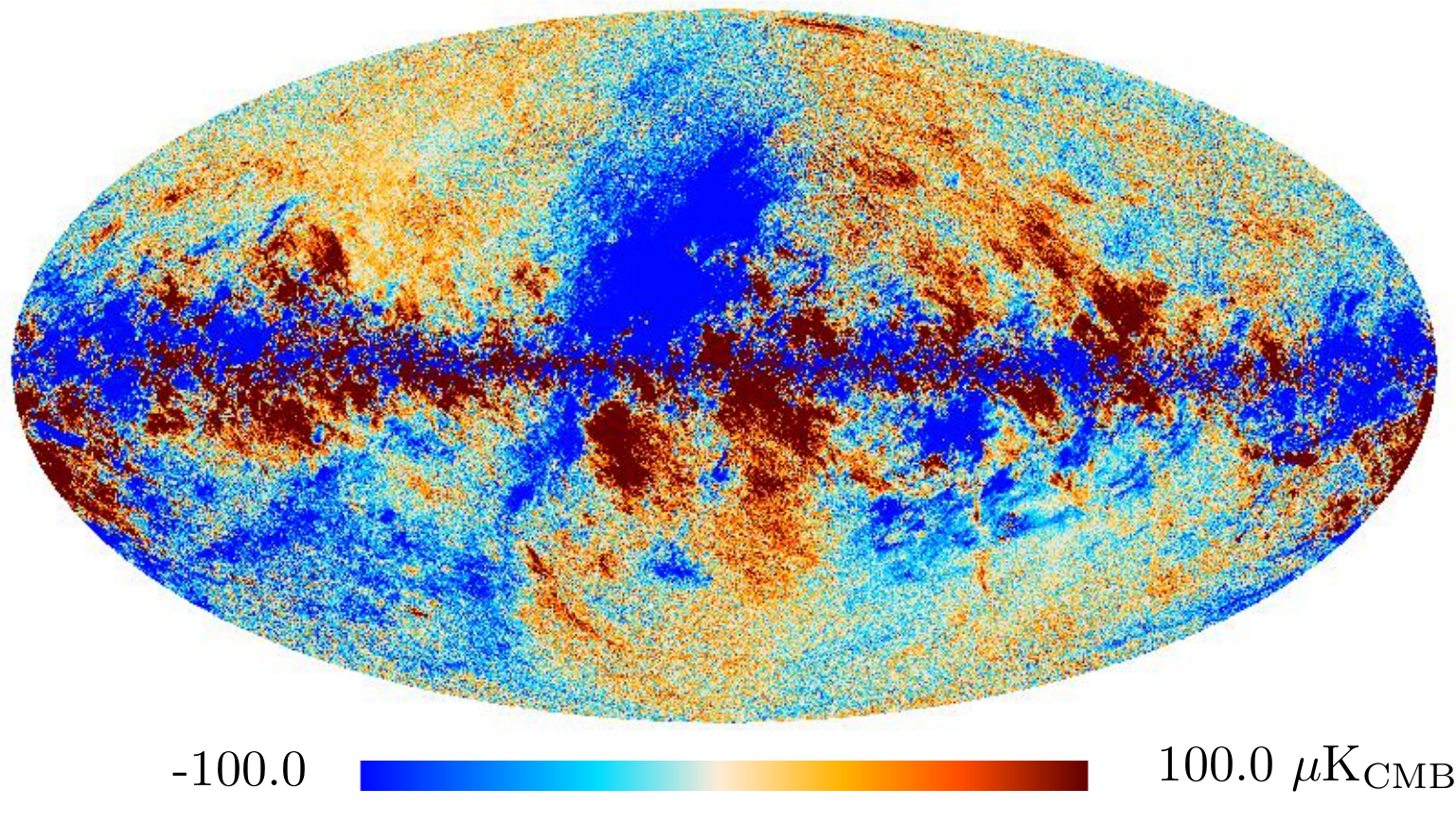}\\
\includegraphics[width=0.32\textwidth]{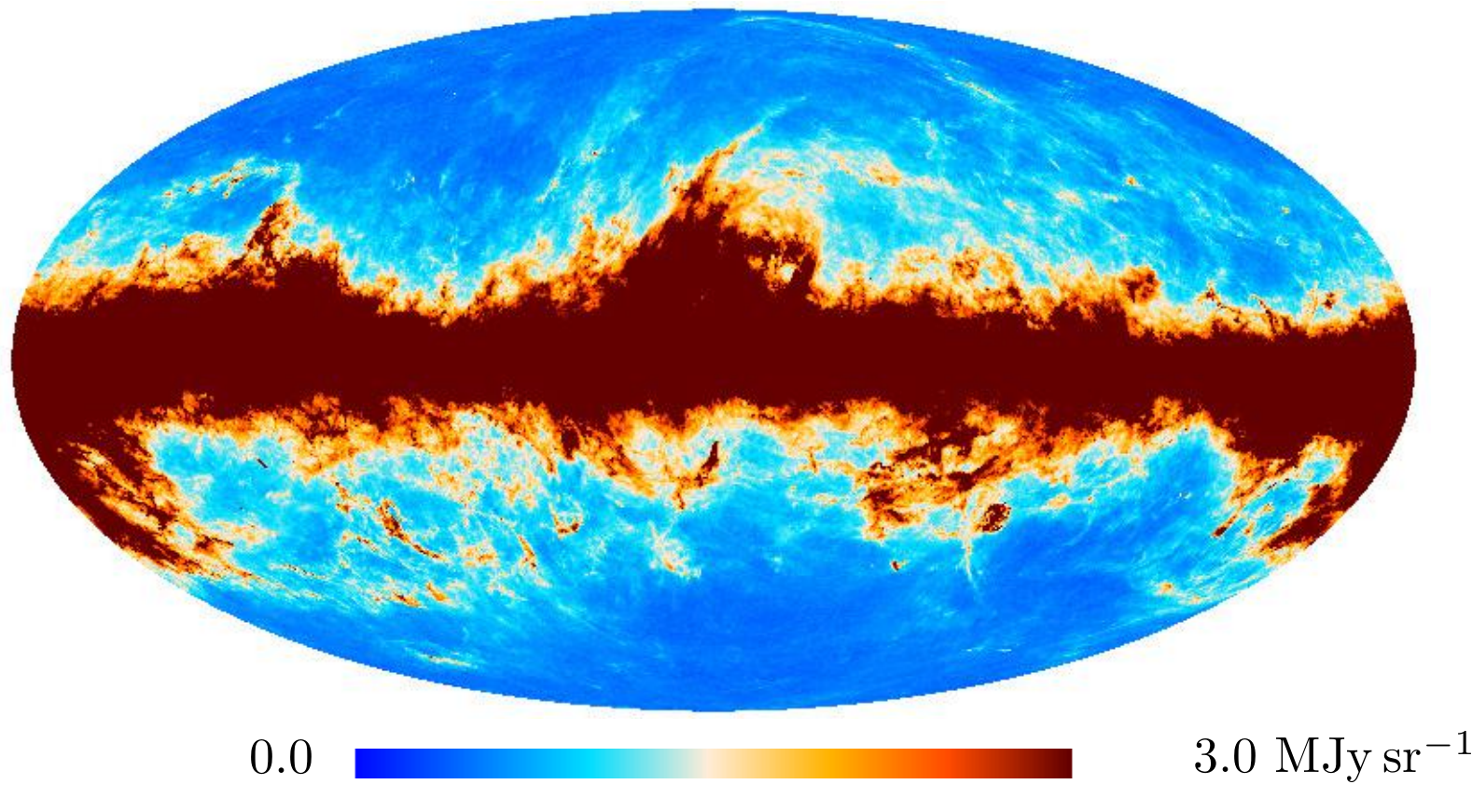}\\
\includegraphics[width=0.32\textwidth]{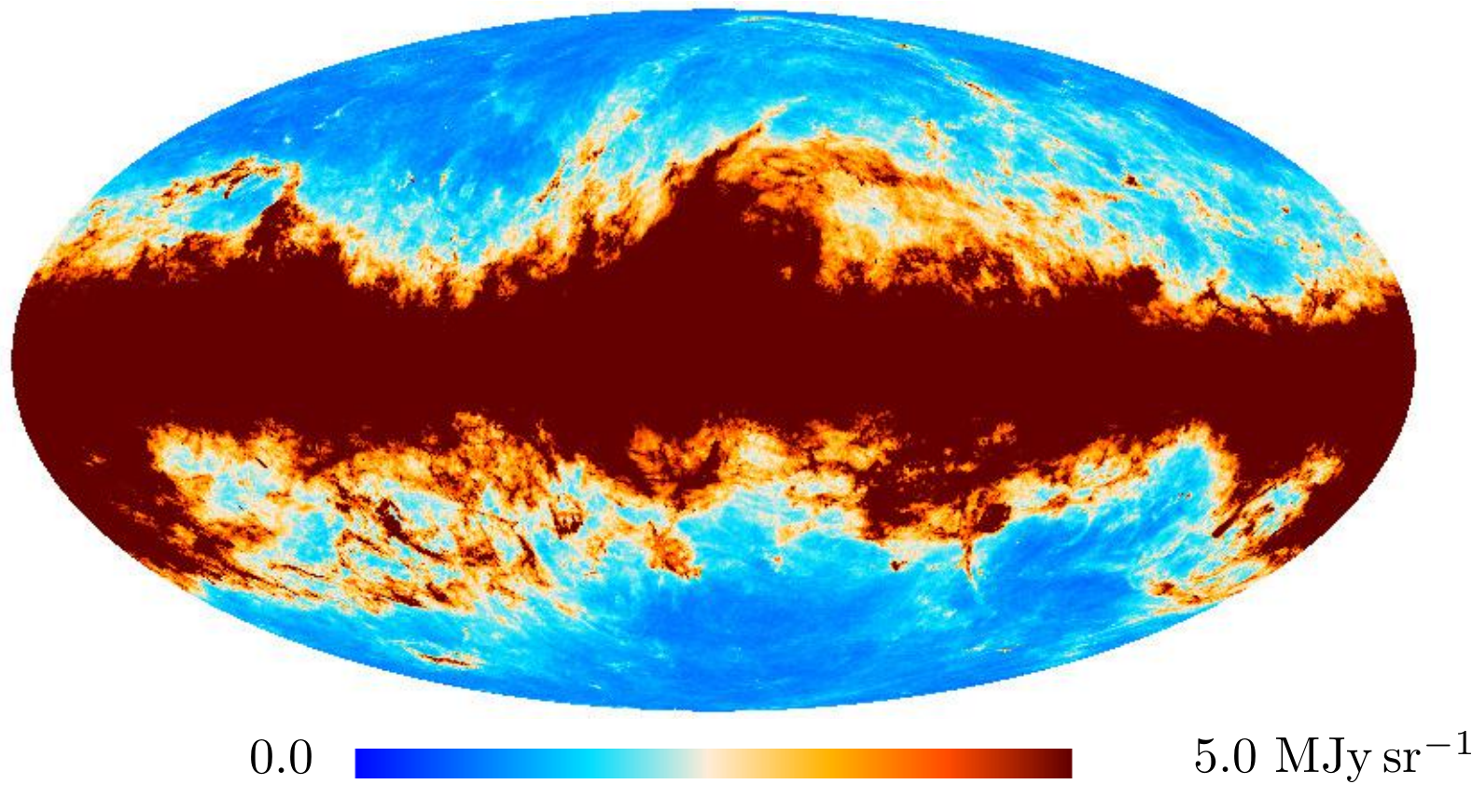}
\caption{\Planck-HFI Solar dipole-removed maps at 100 to 857\,GHz (in rows), for Stokes $I$, $Q$,
and $U$ (in columns).}
\label{fig:maps_all} 
\end{figure*}

Figure~\ref{fig:maps_all} shows the 2018 $I$, $Q$, and $U$ maps for 100 to 353\,GHz, and the $I$ maps for 545 and 857\,GHz.

\subsubsection{Beams}

The scanning beam is defined to be the beam measured from the response to a point source of the full optical and electronic system, after filtering. The effective beam at the map level is the overall angular response to the sky in a map pixel, which results from the combination of the scanning beam, the scanning strategy, and the mapmaking. For the 2015 release, a self-consistent separation of residual time-transfer effects and the optical response was performed to build the scanning beams using planet observations \citep[as described in appendix~B of][]{planck2014-a08}. These have not been updated for this 2018 release.

Effective beams for frequency maps are built with the 10\arcs\ resolution scanning beams, taking into account the scanning strategy, detector weighting, and sky area. As in \citep{planck2013-p03c}, {\tt FEBeCoP} was used to compute the 100\arcm-cut-off effective beams for each pixel at \Nside=2048, incorporating all the dependencies just listed.

Mean values of the effective beam properties, averaged across the entire sky, are given in Table~\ref{tab:mapcharacteristics}. These are identical to those provided in table~3 of \citet{planck2014-a08}, since the only change to the input TOI information was the omission of the last 1000 pointing periods (see Sect.~\ref{sec:1000rings}), which affects only a small fraction of the sky. For the remainder of the sky, the effective beams are fully identical to those employed in 2015. Associated beam window functions are discussed in Sect.~\ref{sec:summaryTF}.

\subsubsection{Specific maps for testing purposes}
\label{sec:monobolomaps}

For testing purposes, we deliver the maps used in both the half-mission and ring null tests. We also build single-bolometer maps, taking the signal of one bolometer and, using for this bolometer the model obtained from the \sroll\ global solution, we remove the polarized part and all or (depending of the specific test purpose) a subset of the systematic effects modelled in the mapmaking process. This map thus contains the intensity signal and the sum of all, or part of, systematics residuals for this bolometer. This means that some of these single-bolometer maps only contain part of the sky signal and thus cannot be used for component separation.

We employ a spatial template of a given foreground to solve for the bandpass mismatch. Recall that this mismatch arises because the detectors are calibrated on the CMB dipole, and foregrounds have different spectra. Since we rely on spatial templates, we can separate only those foregrounds that have sufficiently different spatial distributions.

\begin{figure}[htbp!]
\includegraphics[width=0.48\columnwidth]{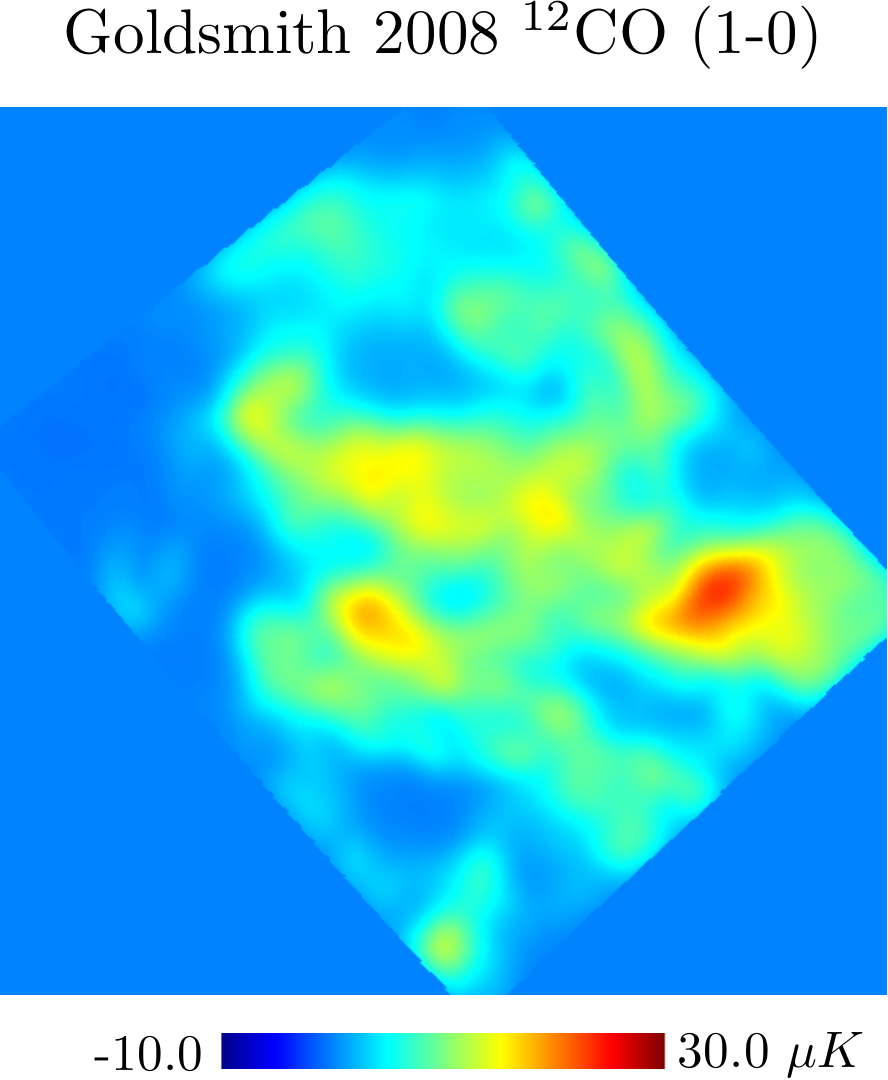}
\includegraphics[width=0.48\columnwidth]{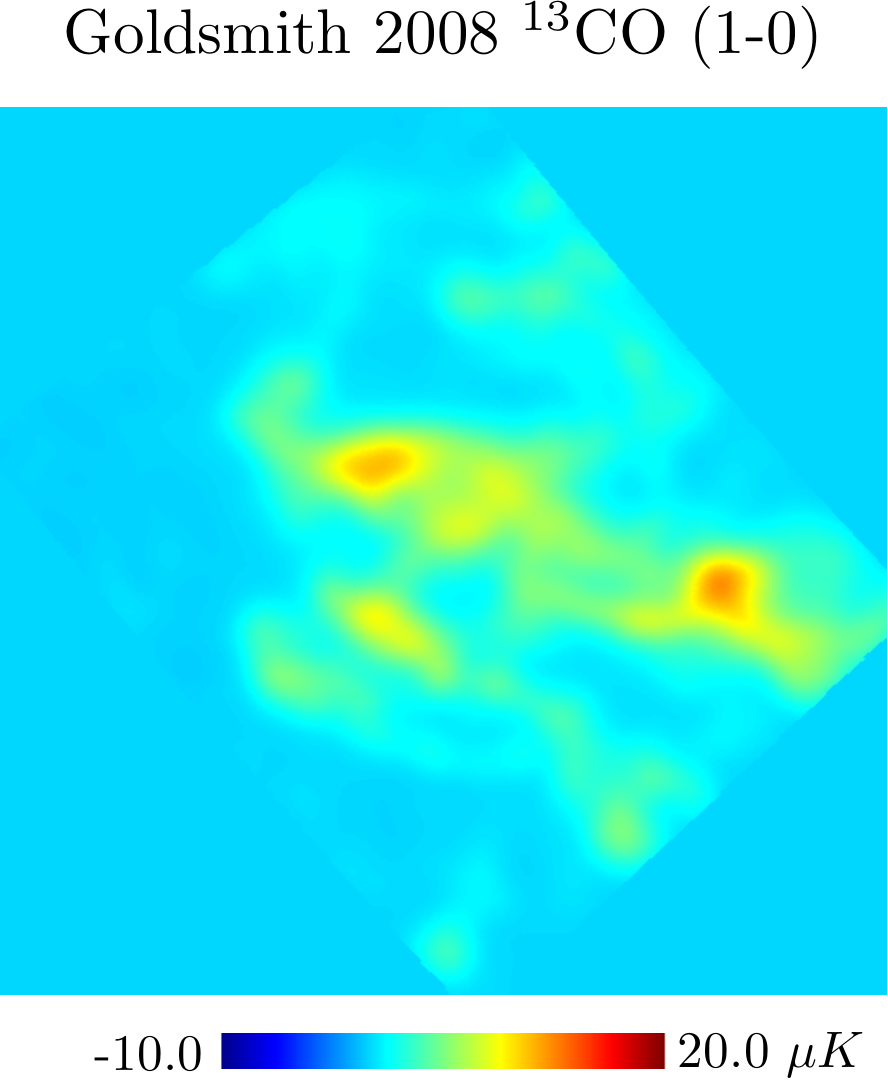}\\
\\
\includegraphics[width=0.48\columnwidth]{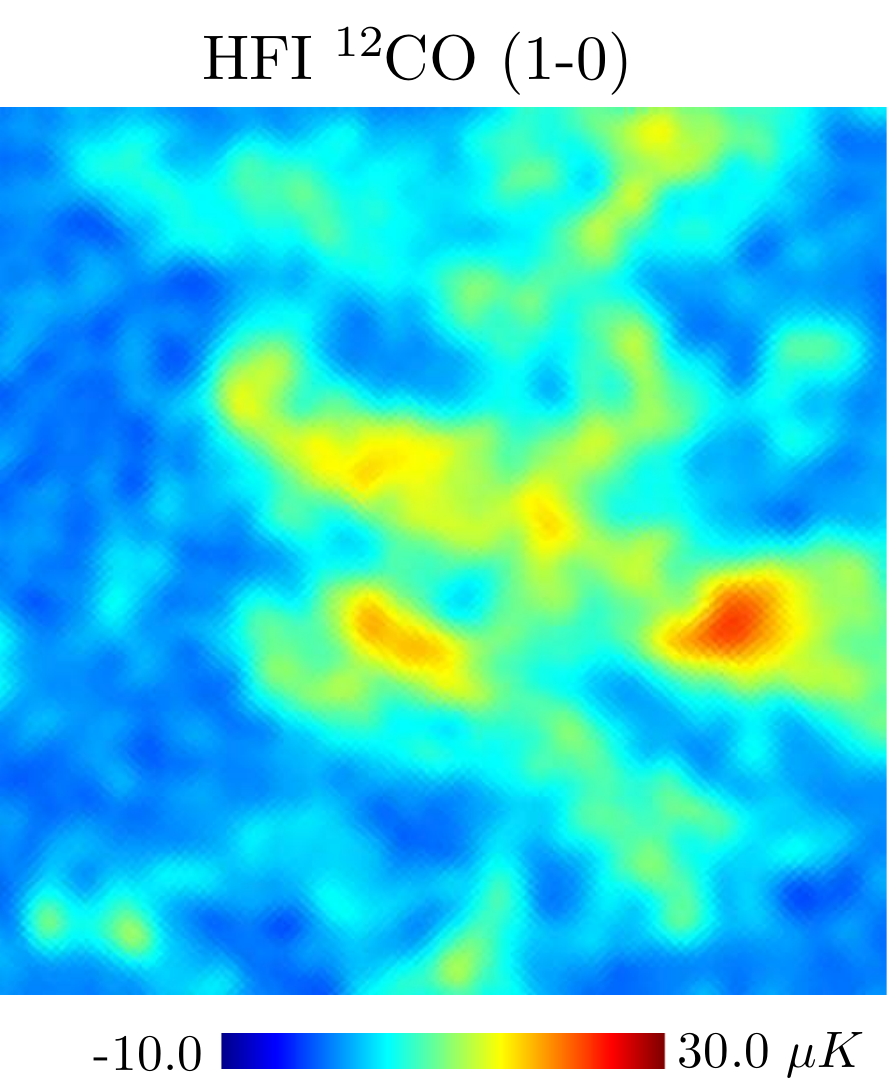}
\includegraphics[width=0.48\columnwidth]{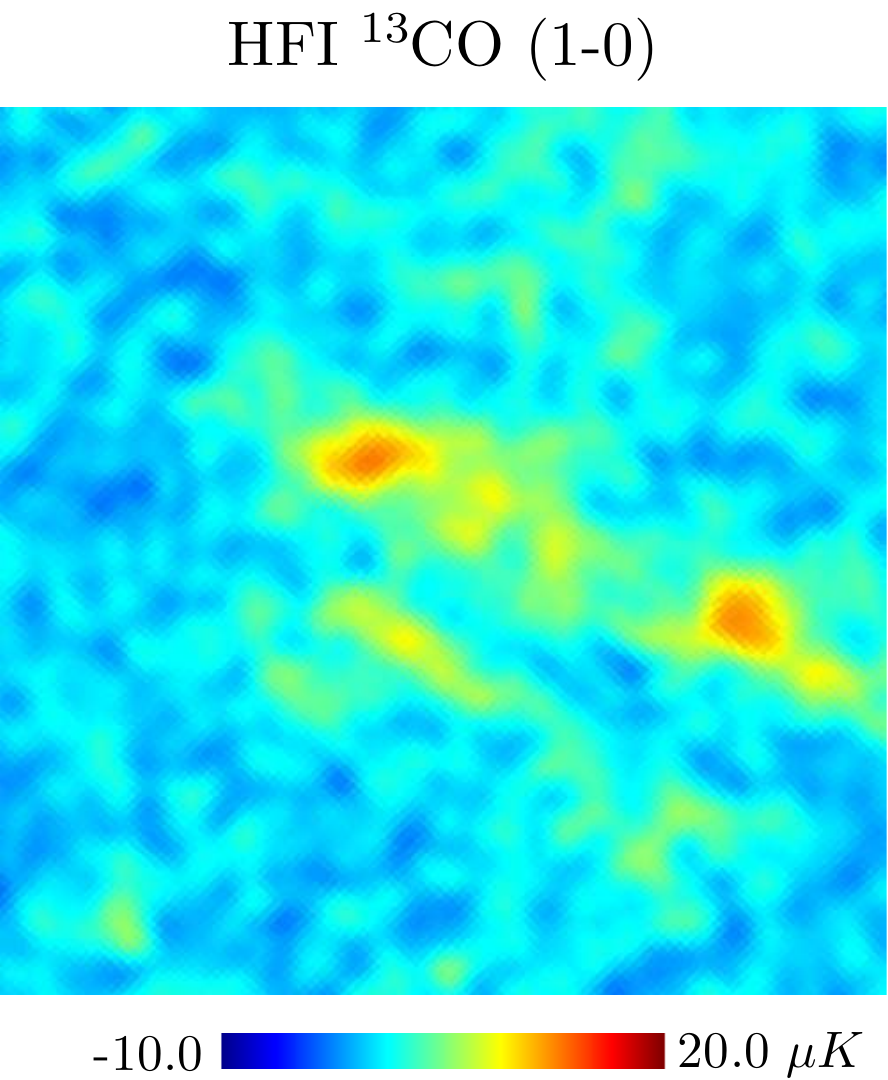}
\caption{{\it Top panel}: two ground-based radio astronomy maps, centred on (173\pdeg0, $-$16\pdeg0) in Galactic coordinates, at $^{12}$CO and $^{13}$CO in the Taurus region. {\it Bottom panel}: HFI $^{12}$CO and $^{13}$CO maps extracted by \sroll.}
\label{fig:cmpCOmaps} 
\end{figure}

\begin{figure}[htbp!]
\includegraphics[width=\columnwidth]{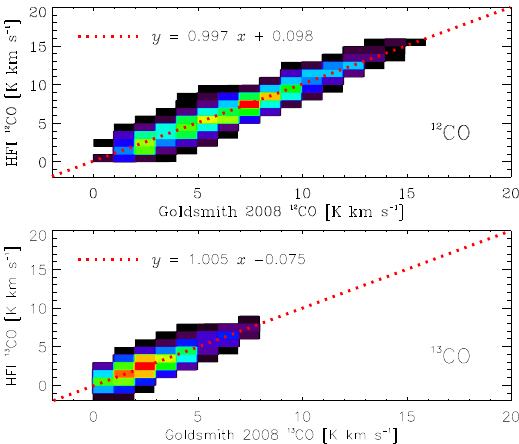}
\caption{Correlation between the radio astronomy $^{12}$CO and $^{13}$CO maps and the HFI ones shown in Fig.~\ref{fig:cmpCOmaps}. The plots are constructed from bins of CO $J$=1$\rightarrow$0 line intensity of 1\,K km s$^{-1}$ and show the colour-coded histogram of the distribution of points in each bin.}
\label{fig:COTaurusCorrelation} 
\end{figure}

For the CO line emission foreground at 100\,GHz, \sroll\ removes only one CO component, with a template based on the \Planck\ 2015 {\tt Commander} component-separation maps. As a test of the \sroll\ destriper capabilities, we attempted at 100\,GHz to extract the response differences to the two isotopologue (i.e., $^{12}$CO and $^{13}$CO) lines within its global minimization, using as templates millimetre spectroscopy maps of 64\,deg$^2$ in the Taurus region for these two lines \citep{Goldsmith2008}. The extracted parameters found from a small fraction ($1.5\times10^{-3}$) of the sky are then used in \sroll\ to build all-sky maps of the two isotopologues. We can compare how well those two CO template maps have been reconstructed and this is shown in Fig.~\ref{fig:cmpCOmaps}.
The absolute calibration is performed using the average bandpass measured on the ground and then colour corrected (using coefficients from the \citeES). The HFI maps have recovered well the radio astronomy maps, including the small differences between isotopologues. More quantitatively, the accuracy of the solved $^{12}$CO and $^{13}$CO response coefficients is evaluated via the correlation plots of the input and ouput maps, as shown in Fig.~\ref{fig:COTaurusCorrelation}.
The excellent correlation over the different brightness ranges of the two isotopologues demonstrates that the \sroll\ method can separate accurately foregrounds even if they do not show very different spatial distributions.

\begin{figure}[htbp!]
{\includegraphics[width=\columnwidth]{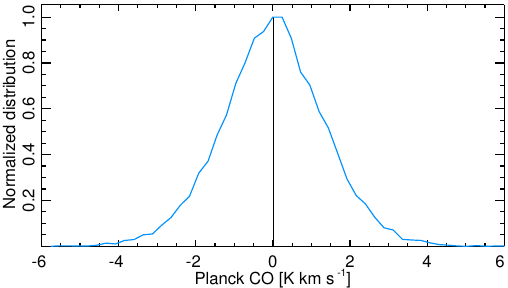}}
\caption{Histogram of the $^{12}$CO intensity for the lines of sight where CO was not detected in the high-Galactic-latitude survey.}
\label{fig:COvalidation} 
\end{figure}

Another test at high Galactic latitudes, where there is no large fully sampled map of CO, can be carried, out as was done in \citet{planck2013-p03a}. We compare the CO $J$=1$\rightarrow$0 detection at high Galactic latitude from the 15\,000 lines of sight of \citet{Hartmann1998} and \citet{Magnani2000}. As expected, there is a correlation for the 1\,\% of the lines of sight where CO was detected \citep[see figure~17 of][]{planck2013-p03a}. Figure~\ref{fig:COvalidation} shows that, for all the other lines of sight where CO was not detected by \Planck, the distribution of \Planck\ signals is centred on zero and the FWHM is 3\,K\,km\,s$^{-1}$, to be compared with a FWHM of 5\,K\,km\,s$^{-1}$ in the previous release.
This shows that we could directly take out the CO from each frequency map. Furthermore, we should be able to extend the method to the CO $J$=2$\rightarrow$1 and $J$=3$\rightarrow$2 lines of the two isotopologues if limited areas have been mapped to high enough accuracy to provide a good template.

\subsubsection{Caveats on the usage of the frequency maps}
\label{sec:caveats}

Some imperfections have shown up in the tests of the HFI 2018 maps that were previously hidden by higher-level systematics in the 2015 data. These lead to guidelines for the proper use of the HFI 2018 data.

\paragraph{Monopoles}
~~\\
Monopoles, which cannot be extracted from \Planck\ data alone, are adjusted at each frequency (as was done in the previous 2015 release). For component separation, this provides maps that can be used directly in combination with other tracers.

First, the monopole is consistent with an intensity of the dust foreground at high Galactic latitudes proportional to the column density of the ISM traced by the 21-cm emission at low column densities ($N_{\rm H}<2 \times 10^{20}$\,cm$^{-2}$), neglecting the dust emission in the ionized component. Second, a CIB monopole coming from a galaxy evolution model \citep{2011A&A...529A...4B} is added. Third, a zodiacal emission zero level (monopole) has also been added; this is taken to be representative of the high ecliptic latitude emission in the regions where the interstellar zero level was set. The values of these additions are given in Table~\ref{tab:mapcharacteristics} and serve to give HFI frequency maps colour ratios that are compatible with foregrounds, although this requires us to introduce astrophysical observations and models that are not constrained by the HFI data. The colour ratios measured on the absolute values of the \Planck\ maps for the lowest interstellar column densities become significantly dependent on the CIB, interstellar, and zodial monopoles uncertainties (also reported in Table~\ref{tab:mapcharacteristics}).

Scientific analysis that depends on the value of the monopoles in the HFI maps needs to be adjusted to each specific problem. In particular, high latitude polarization fractions calculated using frequency maps would be hard to interpret. This should instead be done after component separation, since the values would be affected by the uncertainties on the monopoles, which are not derived from the \Planck\ data themselves and are not the same for different components.

\begin{figure}[htbp!]
\includegraphics[width=\columnwidth]{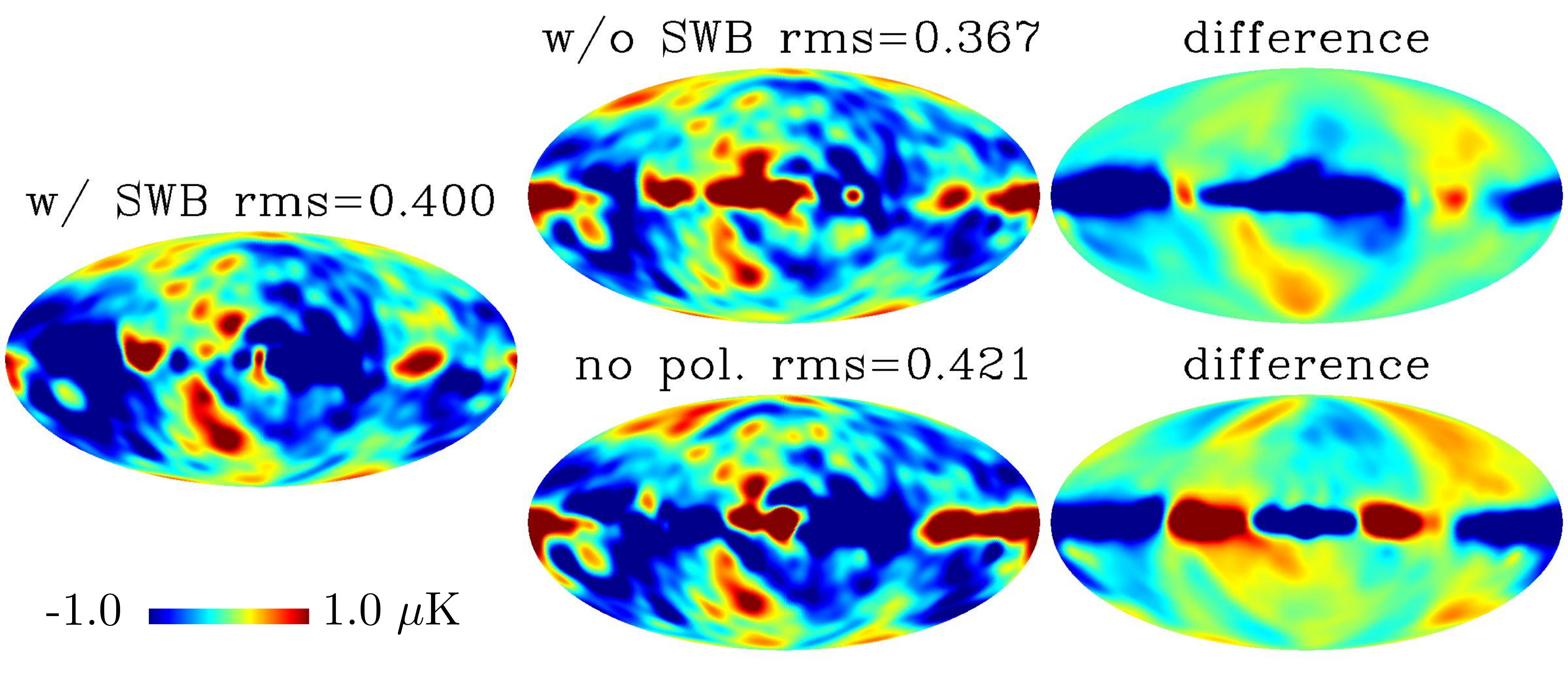}
\caption{217-GHz $Q$ maps cleaned using the 353-GHz map built with the SWBs (left), without the SWBs (top centre), and with the SWBs but with a polarization efficiency taken to be zero (bottom centre). The differences (right two maps) leave quadrupole terms of amplitude smaller than 1\microK.} 
\label{fig:swb353effect} 
\end{figure}

\paragraph{Drawbacks when including SWBs}
~~\\
Recall that all maps have been generated using the polarization efficiencies that were measured on the ground \citep{rosset2010}. These polarization efficiencies for the SWB detectors are in the range of 1.5 to 8\,\%. The polarized maps at 353\,GHz have been produced without using the SWB bolometers, since there were indications of problems mostly at large scales in the differences between the intensity maps with and without SWBs, with a level of around $10\microK$. This has been confirmed in 217-GHz maps cleaned of dust using a 353-GHz map as a template. Three such maps were made in Stokes $Q$ using 353-GHz maps, both with and without SWBs, as well as with SWBs but with a polarization efficiency taken to be zero in \sroll\ (as a worst case). The results are shown in Fig.~\ref{fig:swb353effect}.

Removing the SWBs improves the residuals (lower rms) and leaves quadrupole residuals at high Galactic latitude, at a level of only about $0.5\microK$ at 217\,GHz. Ignoring the polarization efficiency in \sroll\ (lower centre plot) increases the rms, demonstrating that the SWB efficiencies are measured with an uncertainty not much smaller than the value itself (this is also demonstrated by simulations in Fig.~\ref{fig:CrossPol}). We will show later (Fig.~\ref{fig:fig17}) that including SWBs in the data used to build the 353-GHz polarization maps induces a very large intensity-to-polarization effect, this leakage dominating all systematics at very low multipoles. \textit{Thus, for polarization studies, the 2018 353-GHz maps built without the SWBs should be used.}

We nevertheless also deliver 2018 intensity maps using both PSBs and SWBs. The maps including SWBs present a higher signal-to-noise ratio, which is important at high multipoles, where the maps are not significantly affected by the systematic effect investigated above (which dominates only at very low multipoles, especially the quadrupole). The quadrupole residual is a small (a few tenths of a \microK), but not negligible residual, and thus care should be exercised when using CMB channel maps. The same discrepancy between PSBs and SWBs is likely to also be present at 143 and 217\,GHz, but cannot be measured because of the lower polarized fraction of the CMB compared to dust. There is no reason why the polarization efficiencies of SWBs should be better than that at 353\,GHz, and the inclusion of the SWB data at these frequencies could affect the average polarization efficiency by 1--2\,\%.

\paragraph{Colour correction and component separation}
~~\\
The general destriper extracts the single-detector colour-correction mismatch for the three main HFI foregrounds (free-free, dust, and CO) SED responses, and adjusts the signal to the one that would have been obtained with the unweighted frequency average response (the noise was negligible in the ground measurements). This implies that any component-separation procedure using the HFI 2018 frequency maps has to use, for these three foreground components, the unweighted average colour correction for different foregrounds at a given frequency. For other foregrounds, the single-bolometer colour corrections are still different from the average, and the same as in 2015. For convenience in component separation, the relevant colour-correction factors for this 2018 release (using a straight average) are extracted from the 2015 data (see the \citeES) and gathered in Table~\ref{tab:colorcorrection}. Further adjustment of single detector response, as was done in \citet{planck2014-a12}, should not be applied.
\begin{table*}[htbp!] 
\newdimen\tblskip \tblskip=5pt
\caption{Foreground colour-correction coefficients extracted from the \citeES, expressed for dust and free-free as effective frequencies. The dust colour corrections are for modified blackbody SEDs with: $T\,{=}\,18$\,K and $\beta\,{=}\,1.6$; $T\,{=}\,17$\,K and $\beta\,{=}\,1.5$; and $T\,{=}\,21$\,K and $\beta\,{=}\,1.48$. Numbers have been rounded up to take into account systematic effects that by far dominate the statistical uncertainties.}
\label{tab:colorcorrection}
\vskip -3mm
\footnotesize
\setbox\tablebox=\vbox{
\newdimen\digitwidth
\setbox0=\hbox{\rm 0}
\digitwidth=\wd0
\catcode`*=\active
\def*{\kern\digitwidth}
\newdimen\signwidth
\setbox0=\hbox{+}
\signwidth=\wd0
\catcode`!=\active
\def!{\kern\signwidth}
\newdimen\pointwidth
\setbox0=\hbox{.}
\pointwidth=\wd0
\catcode`?=\active
\def?{\kern\pointwidth}
\halign{\hbox to 2.5 cm{#\leaderfil}\tabskip 1.5em&
\hfil#\hfil\tabskip 0.5em&
\hfil#\hfil\tabskip 0.5em&
\hfil#\hfil\tabskip 1.5em&
\hfil#\hfil\tabskip 0.5em&
\hfil#\hfil\tabskip 1.5em&
\hfil#\hfil\tabskip 1.5em&
\hfil#\hfil\tabskip 0pt\cr
\noalign{\doubleline}
\omit&&&&\multispan2\hfil CO\hfil\cr
\noalign{\vskip -5pt}
\omit&\multispan3& \multispan2\hrulefill\cr
\noalign{\vskip 3pt}
\omit\hfil\sc Frequency\hfil&\multispan3\hfil\sc Dust\hfil& $F_{^{12}{\rm CO}}$& $F_{^{13}{\rm CO}}$& \sc Free-free&\sc Unit conversion\cr
\noalign{\vskip -5pt}
\omit&\multispan3\hrulefill\cr
\noalign{\vskip -3pt}
\noalign{\vskip 3pt}
\omit\hfil [GHz]\hfil& $\nu_{\rm dust1}$& $\nu_{\rm dust2}$& $\nu_{\rm dust3}$& [$\mu{\rm K}_{\rm CMB}\,/({\rm K}_{\rm RJ}\,{\rm km}\,{\rm s}^{-1})$]& [$\mu{\rm K}_{\rm CMB}\, / ({\rm K}_{\rm RJ}\,{\rm km}\,{\rm s}^{-1})$]& (spectral index 0)&[${\rm MJy}\,{\rm sr}^{-1}\,{\rm K}_{\rm CMB}^{-1}$]\cr
\noalign{\vskip 3pt\hrule\vskip 5pt}
\noalign{\vskip 2pt}
100& 104.6& 104.7& 104.6& 14.78& 15.55& 101.307& 244.1\cr
143& 147.2& 147.4& 147.3& $4.7\times10^{-4}$& $1.8\times10^{-5}$& 142.709& 371.7\cr
217& 227.6& 227.8& 227.7& 45.85& 35.37& 221.914& 483.7\cr
353& 369.2& 369.6& 369.5& 175.1& 117.1& 361.289& 287.5\cr
\noalign{\vskip 3pt\hrule\vskip 5pt}}}
\endPlancktable
\end{table*}
These colour corrections are identical for all pixels on the sky. This was not the case in previous releases, where each pixel was dependent on noise and hit counts, thereby complicating the component separation. Single-bolometer maps, intended, for example, to achieve absolute calibration on the Solar dipole, thus cannot be used for polarized component separation. This can only be done by using the full \sroll\ model.

\paragraph{Sub-pixel effects in the foreground-template inputs to \sroll}
~~\\
Among the 2018 HFI map improvements over the 2015 release, a critical one is the correction by \sroll\ of the bandpass mismatch leakage of intensity to polarization. As described in Sect.~\ref{sec:Sroll}, the correction of bandpass mismatch in the HPRs requires a spatial foreground template. The CO template is taken from the 2015 \Planck\ CO {\tt Commander} map to extract the leakage coefficients. The very same template is used to remove the intensity-to-polarization leakage. To avoid introducing a significant amount of correlated noise in the maps at high latitude, we choose to limit the template resolution to $\Nside\,{=}\,128$. As a consequence of this lower resolution, very strong gradients in CO emission are not well represented and affect the leakage correction. Artefacts of the $\Nside\,{=}\,128$ gridding appear in the frequency maps, especially at 100 GHz, in regions where there are sharp gradients in CO emission. These occur in star-forming regions and dense molecular clouds, so the delivered frequency maps are not suitable for cosmological or astrophysical analysis in regions such as the Galactic centre, regions tangent to the molecular ring in the Galactic disc, and the central regions of the Orion and Rho Ophiuchi molecular clouds. As noted, these artefacts from CO gradients are largest at 100\,GHz.

These effects are well reproduced qualitatively in the E2E simulations. While the simulations of this effect have a very similar sky pattern to the artefacts seen in the data, the simulations cannot be used directly to correct the data since the foreground used in the simulations is not identical to the foreground sky.


We show, in the \citeES, from simulations, maps of the relative level of this artefact with respect to both noise and full intensity in the $\Nside\,{=}\,128$ pixels. The maps and information supplied there will allow users to construct specialized masks adapted to their specific needs. As noted in the \citeES, Galactic science investigations using HFI data from these regions of strong CO emission should properly start from specific maps to be built from the HPRs with the bandpass mismatch correction at full resolution. For polarization maps, at all frequencies (including 100\,GHz), only 0.04\,\% or less of the sky is affected by this bias at a level equal to 7\,\% of the noise. For intensity maps, the bias is reduced by a factor of 3. If it is necessary to reduce the bias to 1\,\% of the noise, only 2\,\% of the sky needs to be excluded. Users can also employ information in the \citeES\ to set limits on the ratio of the absolute value of this bias relative to the intensity in a given sky region, and hence to construct appropriate masks. For polarization, these approaches are comparable.

Dust also contributes to map noise, but with an order of magnitude smaller amplitude. The 2018 maps are optimized for diffuse emission, and detailed studies of these very bright regions require specific mapmaking procedures.

\subsubsection{Calibrated HPRs}
\label{sec:HPRs}

In addition to the HFI frequency maps, we also produce the HPRs used to project the calibrated data into the 2018 maps. These HPRs are available, together with the various (bandpass and dipole) corrections in the PLA, and are described in the \citeES.

\subsection{Comparison with previous HFI frequency maps}

\begin{figure}[htbp!]
\includegraphics[width=\columnwidth]{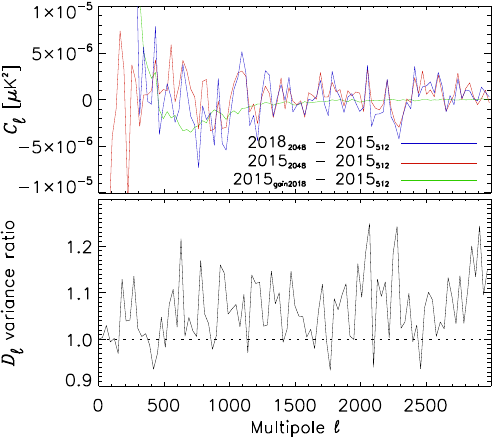}
\caption{{\it Top\/}: the blue curve shows the difference between 143hm1$\times$143hm2 cross-spectra between the 2015 (destriped at $\Nside\,{=}\,512$) and 2018 (destriped at $\Nside\,{=}\,2048$) data. The red curve shows the difference between the 2015 solution destripped at $\Nside\,{=}\,512$ and at $\Nside\,{=}\,2048$. The green curve shows the improvement brought to the 2015 data by the use of the better 2018 gain solution keeping the destriping at $\Nside\,{=}\,512$. {\it Bottom\/}: associated level of improvement of the variance ratio between the destriped 2015 data at $\Nside\,{=}\,512$ and $\Nside\,{=}\,2048$.}
\label{fig:diffdx11rd12} 
\end{figure}

In the 2018 release, the destriping solution is obtained using {\tt HEALPix} $\Nside\,{=}\,2048$ maps, where earlier versions of the HFI maps used $\Nside\,{=}\,512$. Figure~\ref{fig:diffdx11rd12} (top panel) shows that this accounts for most of the improvement between 2015 and 2018 data releases at high multipoles ($\ell\,{\ga}\,1000$, blue curve). Indeed, this improvement is reproduced using $\Nside\,{=}\,2048$ for the 2015 solution (the red curve follows the blue one for multipoles larger than 1000). The green curve shows, using the 2015 data release, the difference brought by introducing, in the 2015 data, the improved 2018 gain solution. The better gain solution accounts for the sharp rise of the blue curve at $\ell\,{<}\,400$, which was not explained by the increase of \Nside.
 
Figure~\ref{fig:diffdx11rd12} (bottom panel) shows the variance improvement as a function of multipoles induced by increasing the \Nside\ used in the destriping procedure. This shows that 2015 data had a significant noise excess at 143\,GHz of order 7\,\% for $\ell >100$, due to the use of the lower resolution for the destriping.

The improvements between the 2015 and the 2018 releases have been driven by the need to improve the HFI polarization maps on large scales, through a better correction of systematic effects (as discussed above), but also by the intensity-to-polarization leakage due to bandpass mismatch. In 2015, these corrections were not performed for the delivered maps and instead an a posteriori template fitting procedure was applied (see appendix~A of \citehfimap), but shown to be partially degenerate with other systematic effects (\citelowell). The global correction map for leakage at 353\,GHz, which was nevertheless made available with the 2015 data release, were the associated dust correction maps that can be found in figure~19 of \citehfimap. This correction was overestimated, due to the degeneracy with the ADCNL systematic effect.

The correction of the leakage was first carried out in \citelowell, which reduced it enough to allow the measurement of the reionization optical depth $\tau$ from the $EE$ reionization peak at $\ell<10$. Section~\ref{sec:polareff} describes this correction and the further improvements, including the measurements of the polarization efficiency from the sky data.

\begin{figure}[htbp!]
\includegraphics[width=0.32\columnwidth]{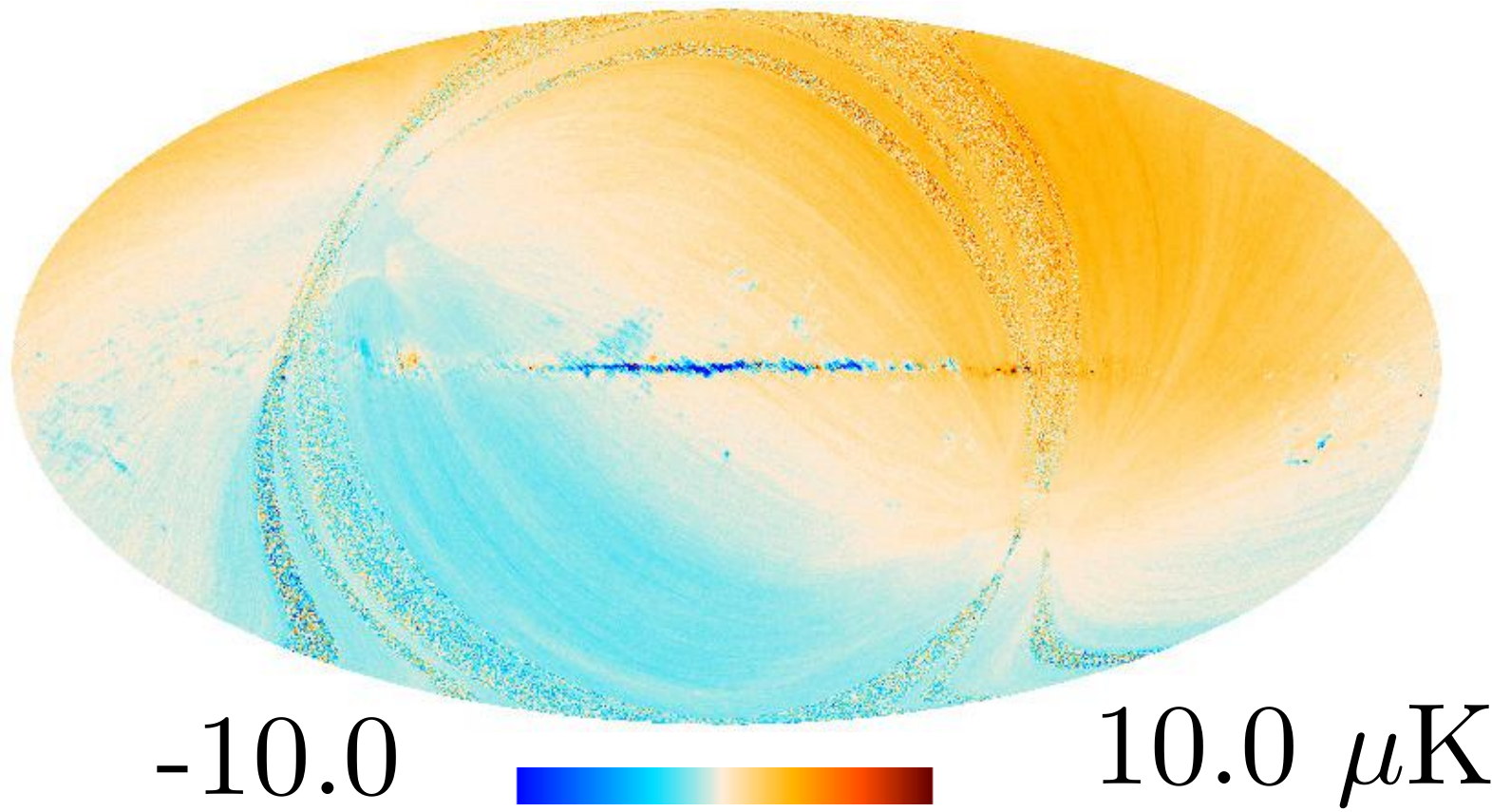}
\includegraphics[width=0.32\columnwidth]{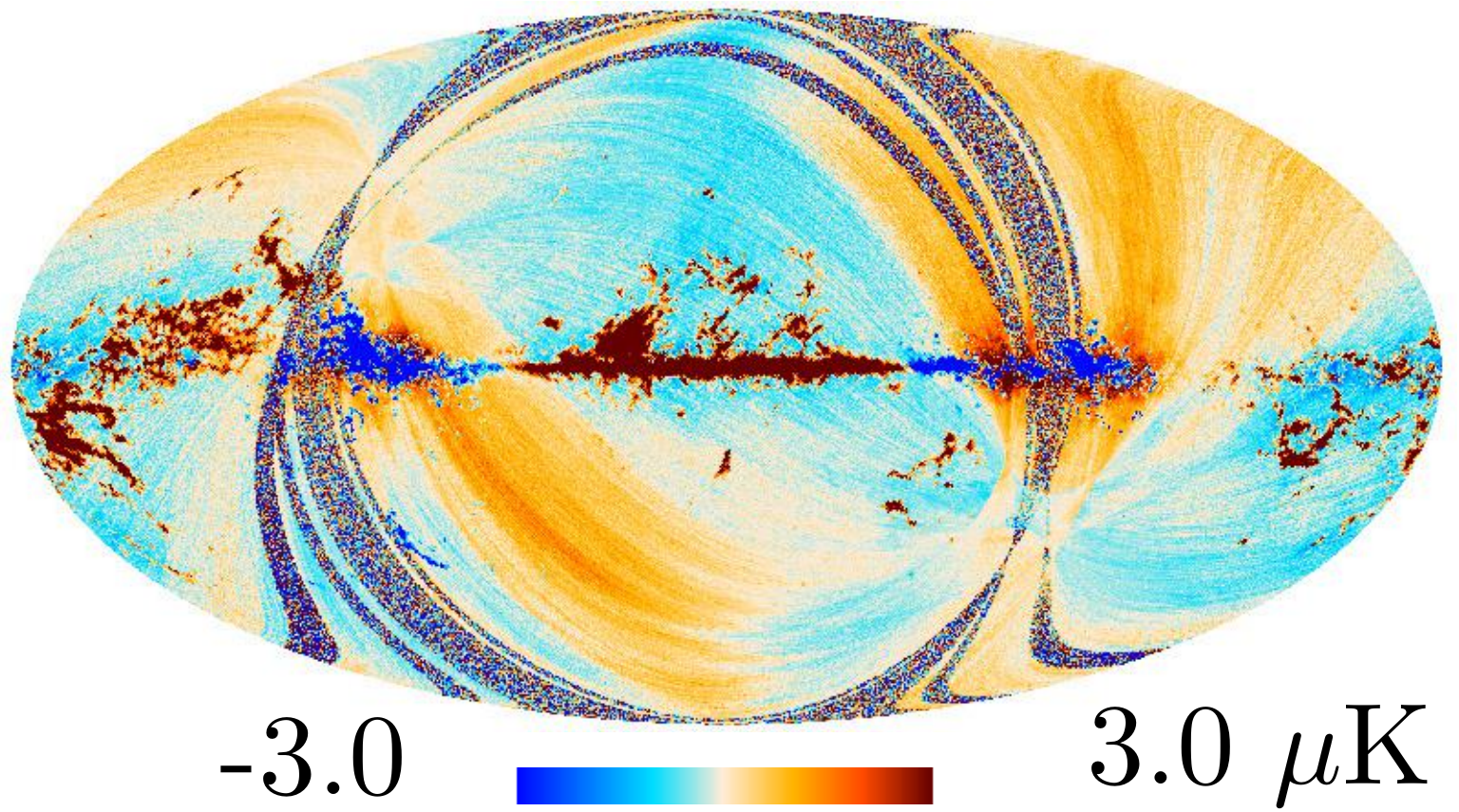}
\includegraphics[width=0.32\columnwidth]{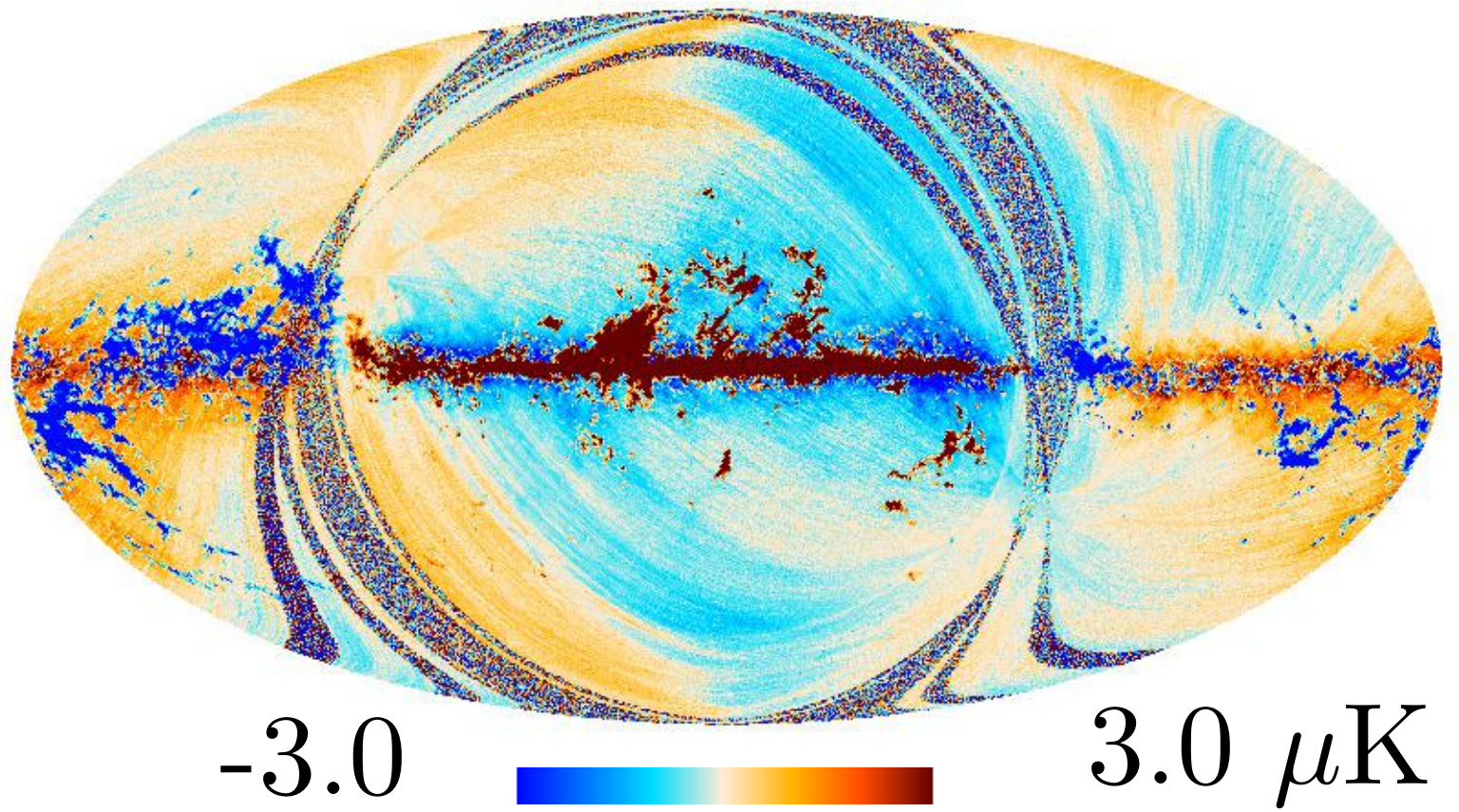}\\
\includegraphics[width=0.32\columnwidth]{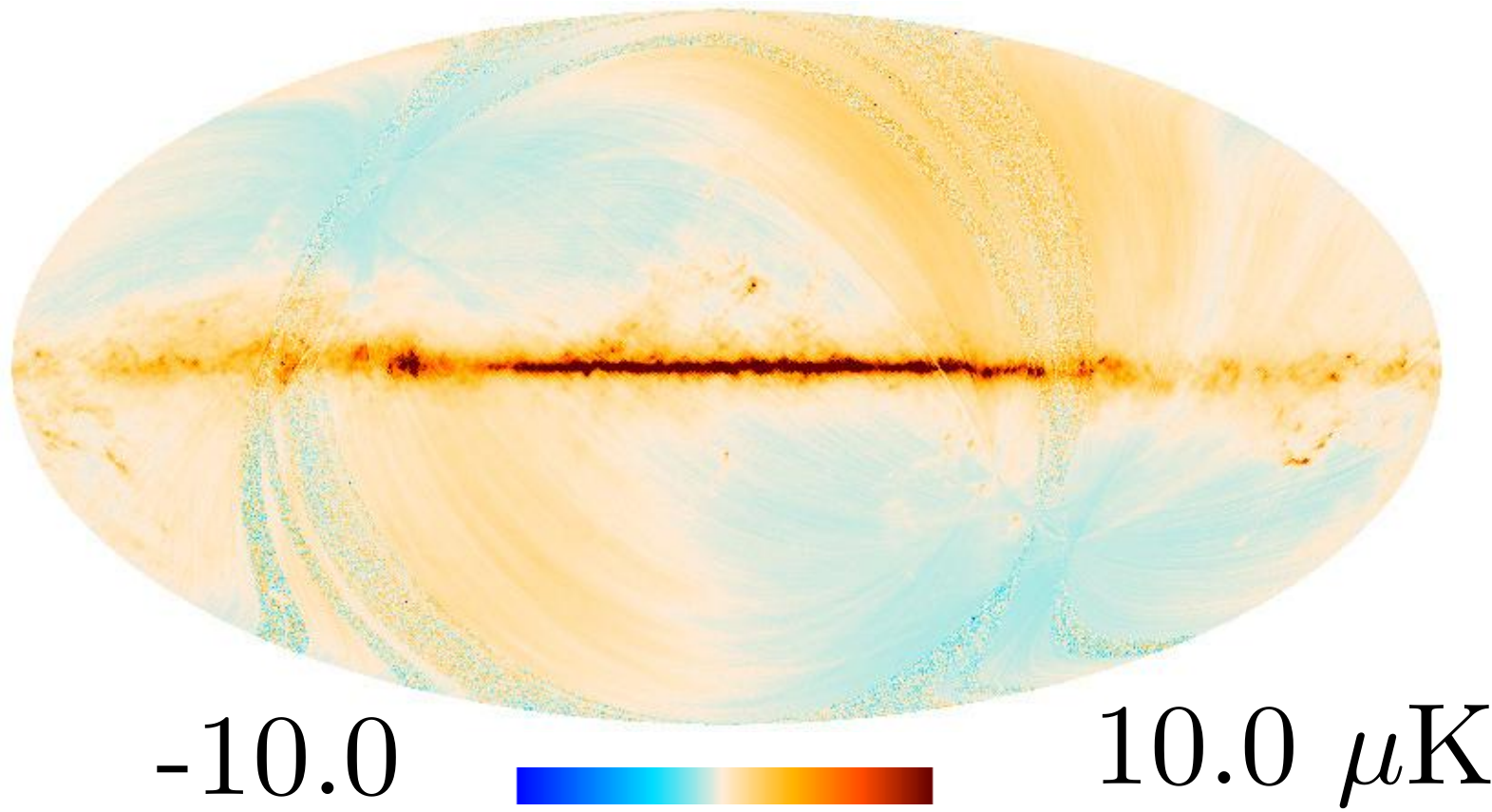}
\includegraphics[width=0.32\columnwidth]{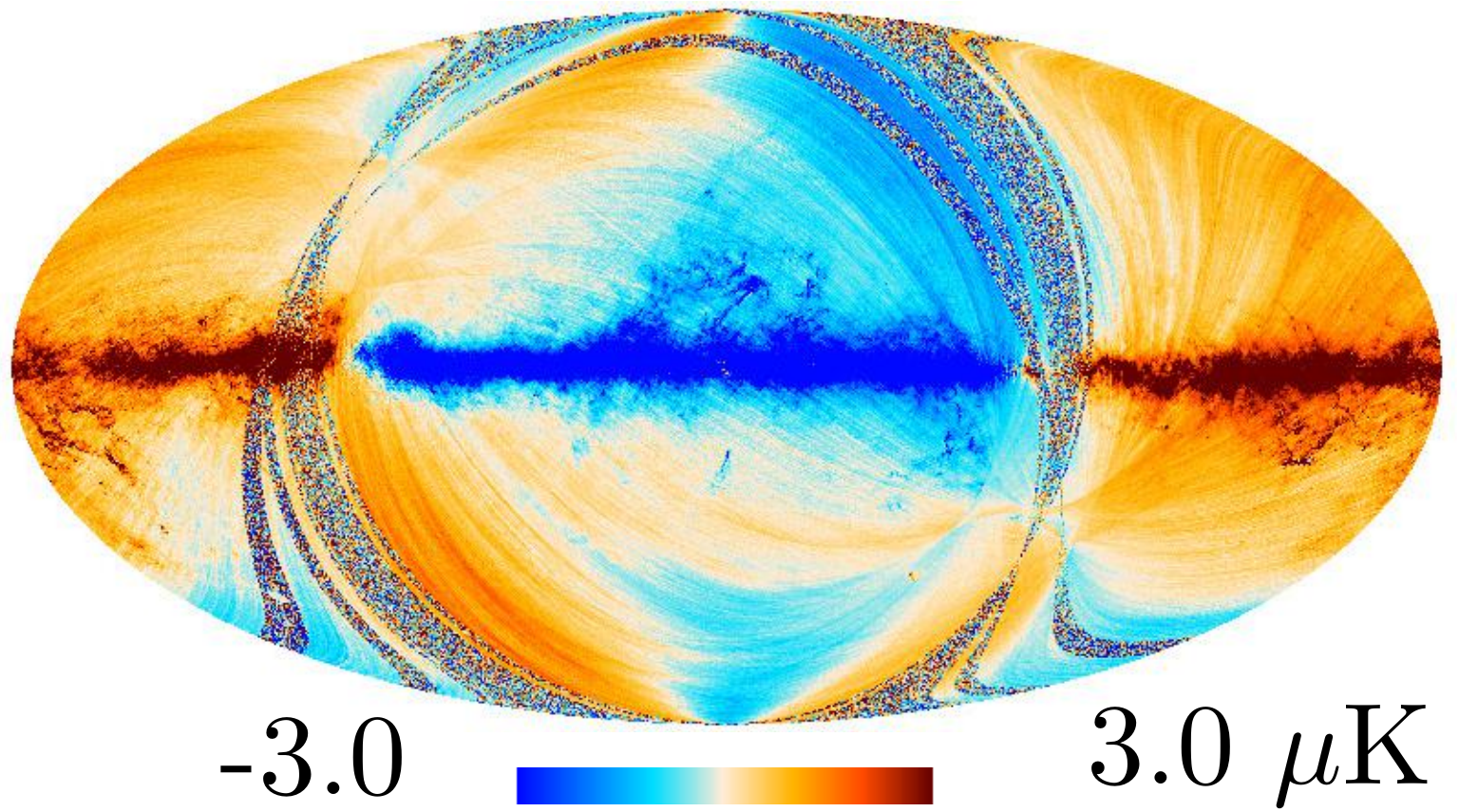}
\includegraphics[width=0.32\columnwidth]{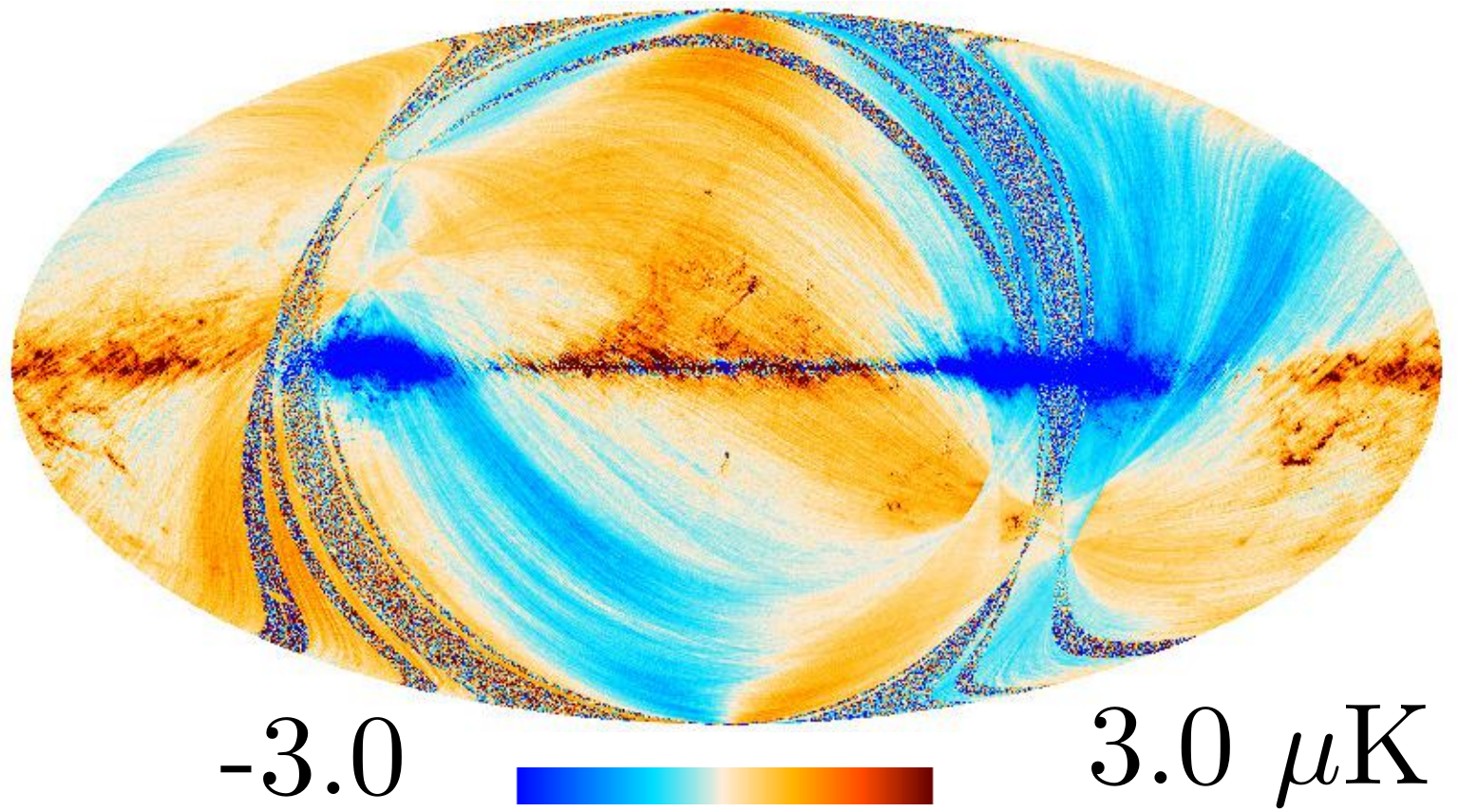}\\
\includegraphics[width=0.32\columnwidth]{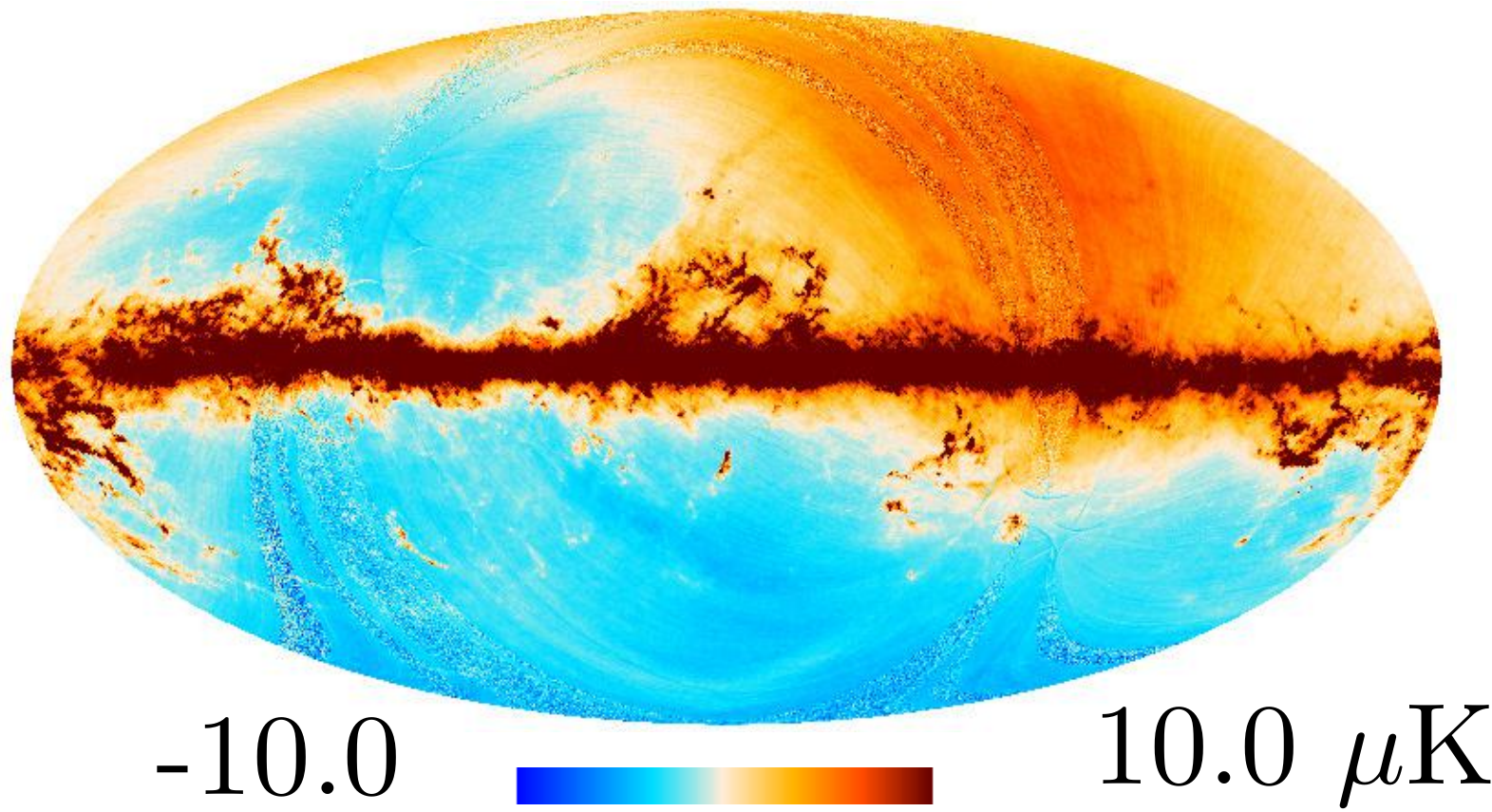}
\includegraphics[width=0.32\columnwidth]{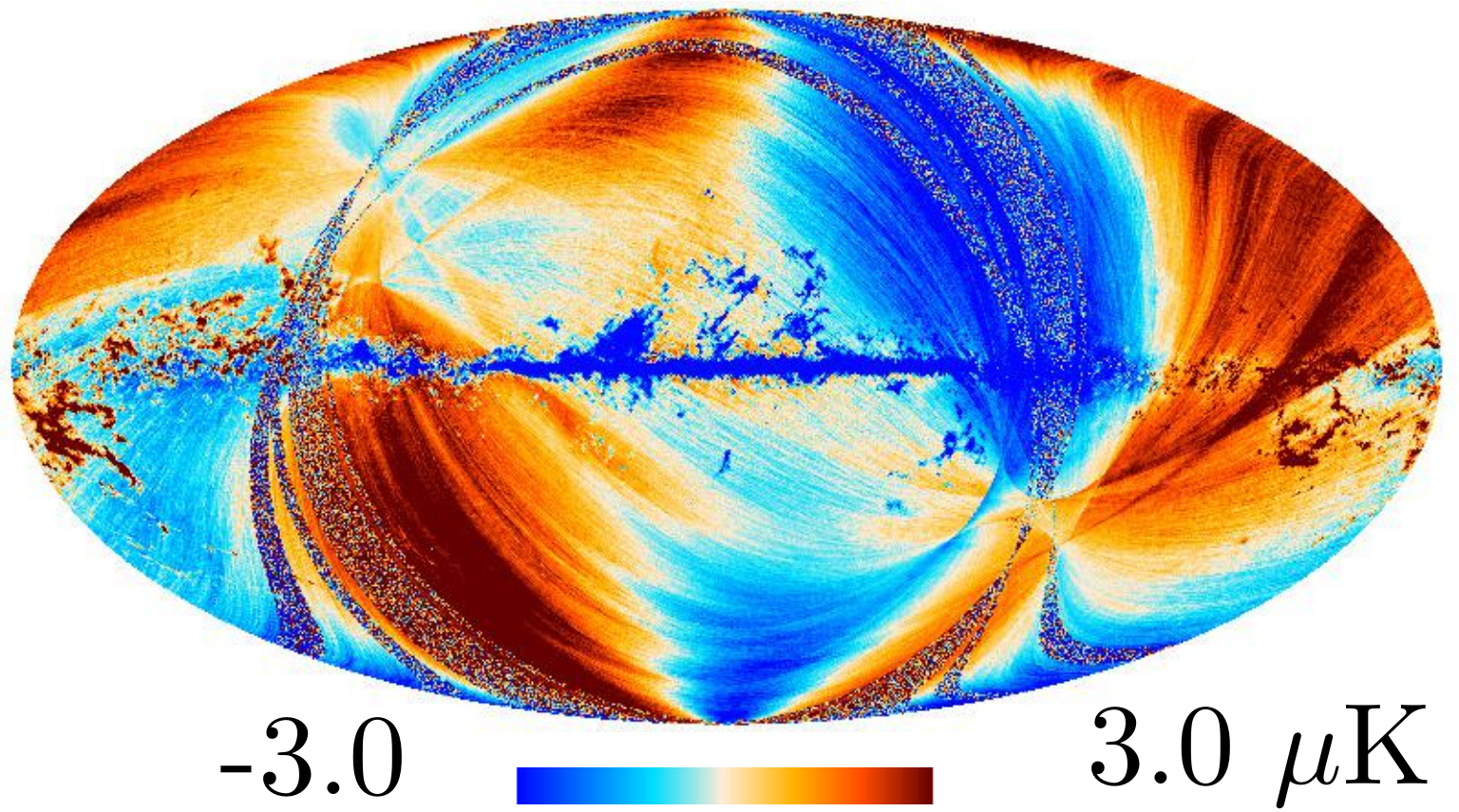}
\includegraphics[width=0.32\columnwidth]{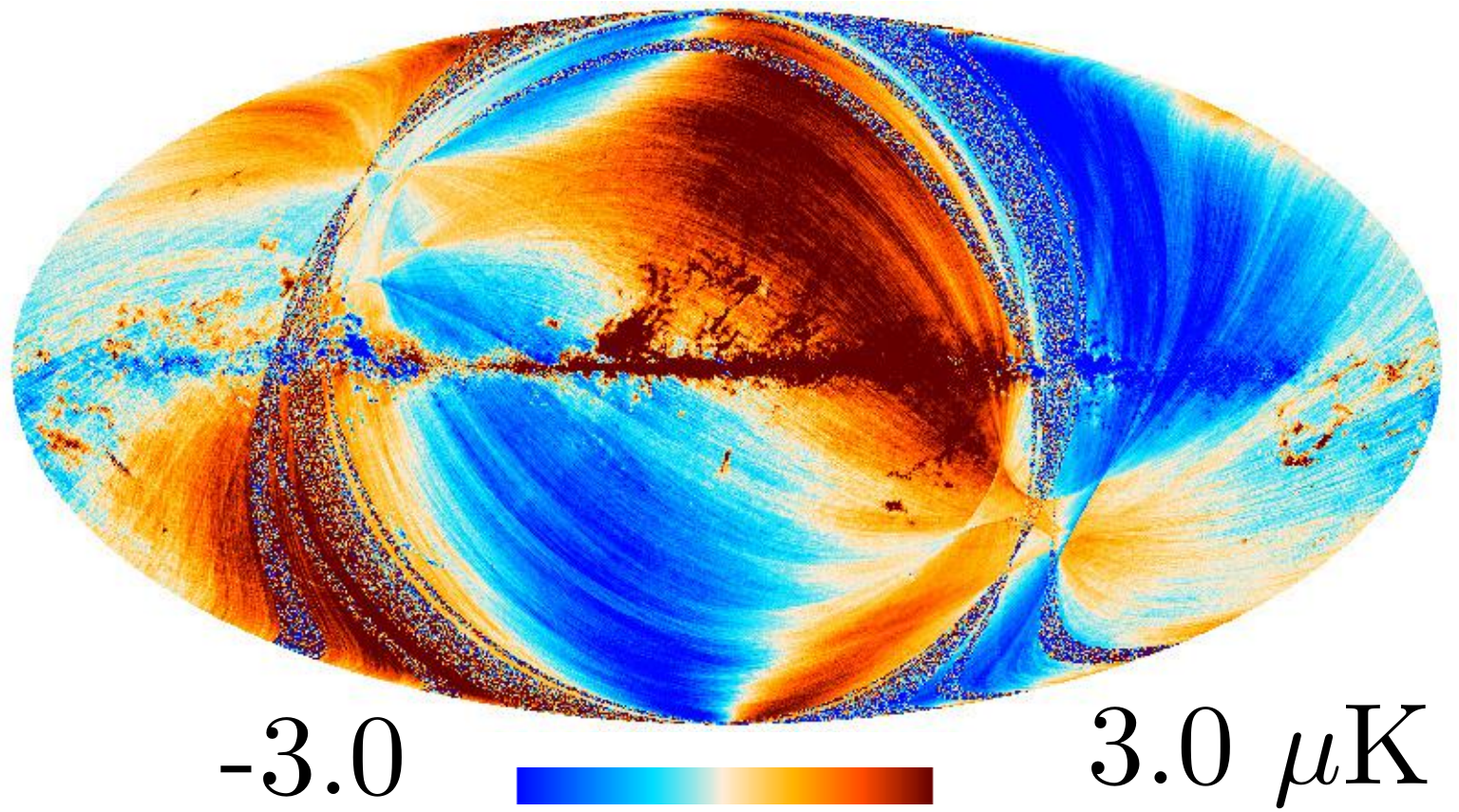}\\
\includegraphics[width=0.32\columnwidth]{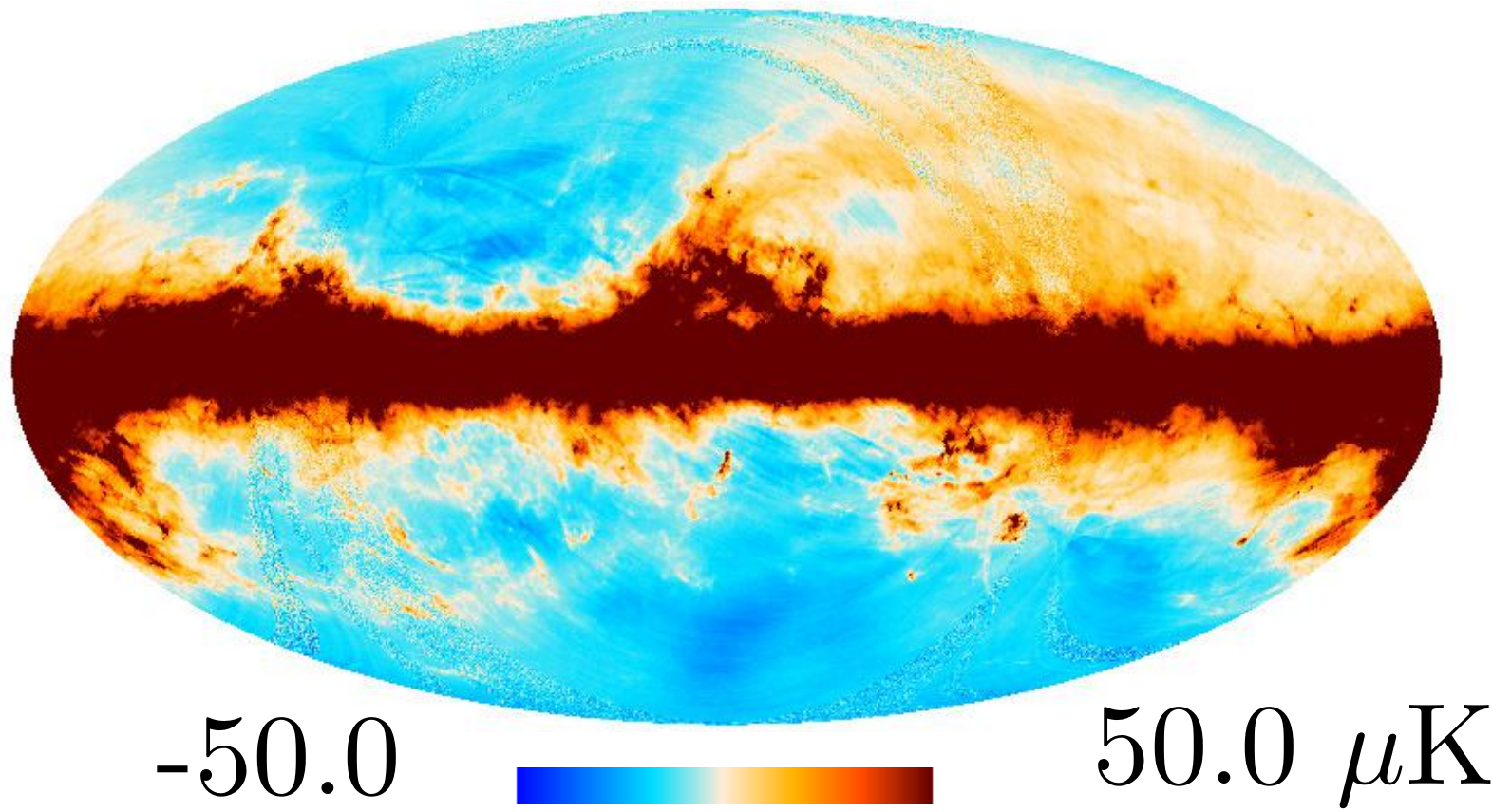}
\includegraphics[width=0.32\columnwidth]{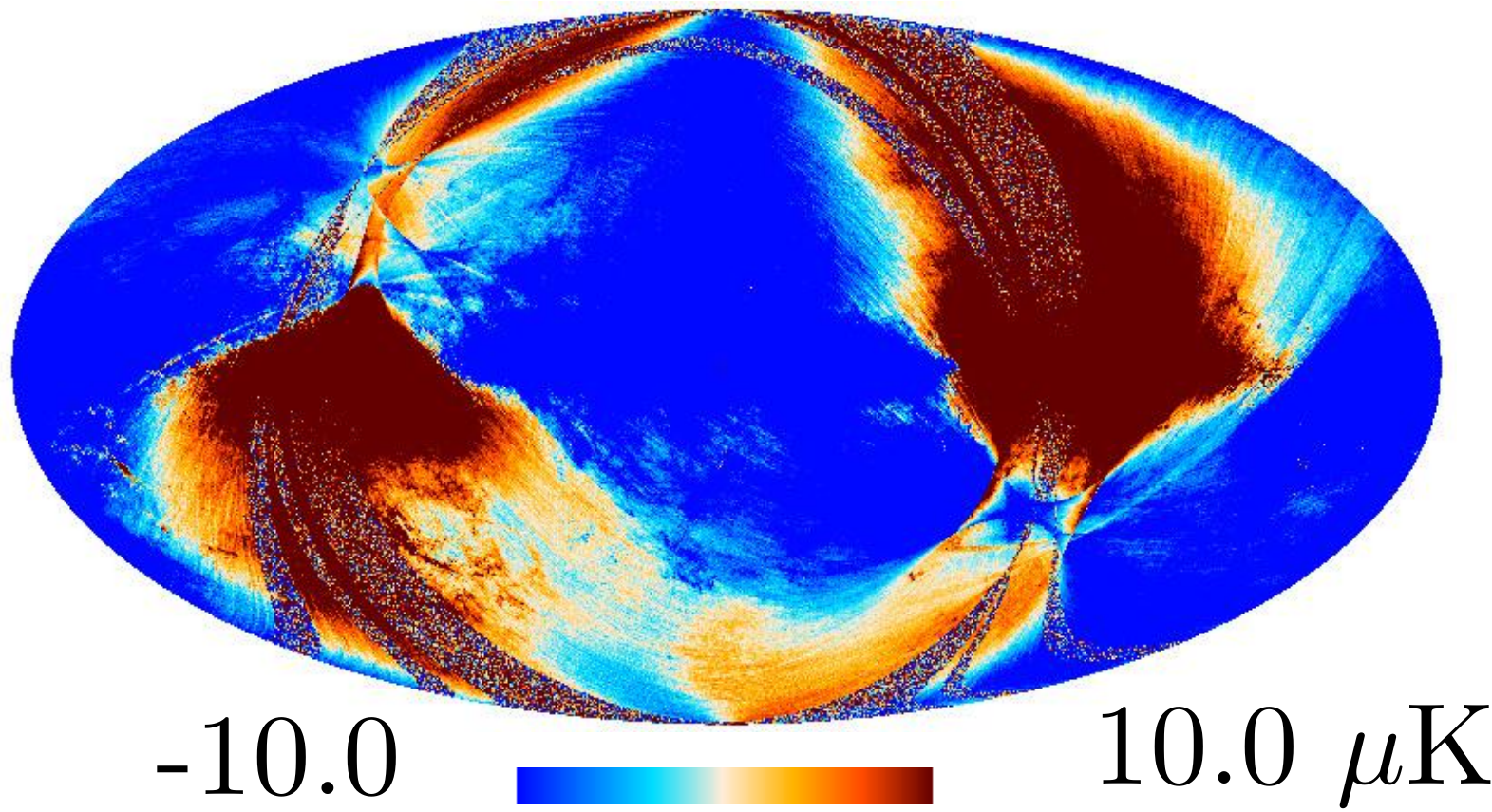}
\includegraphics[width=0.32\columnwidth]{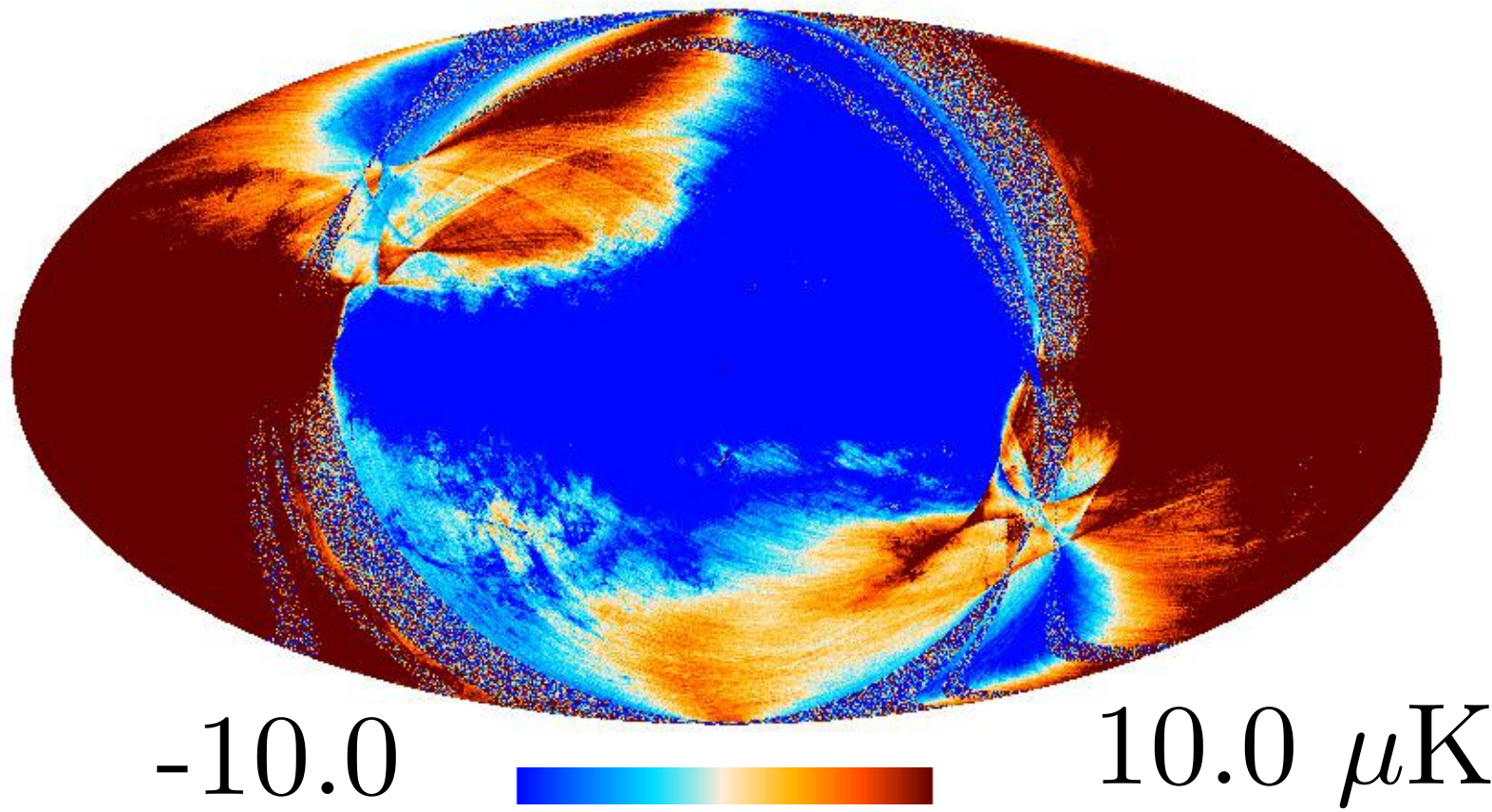}\\
\includegraphics[width=0.32\columnwidth]{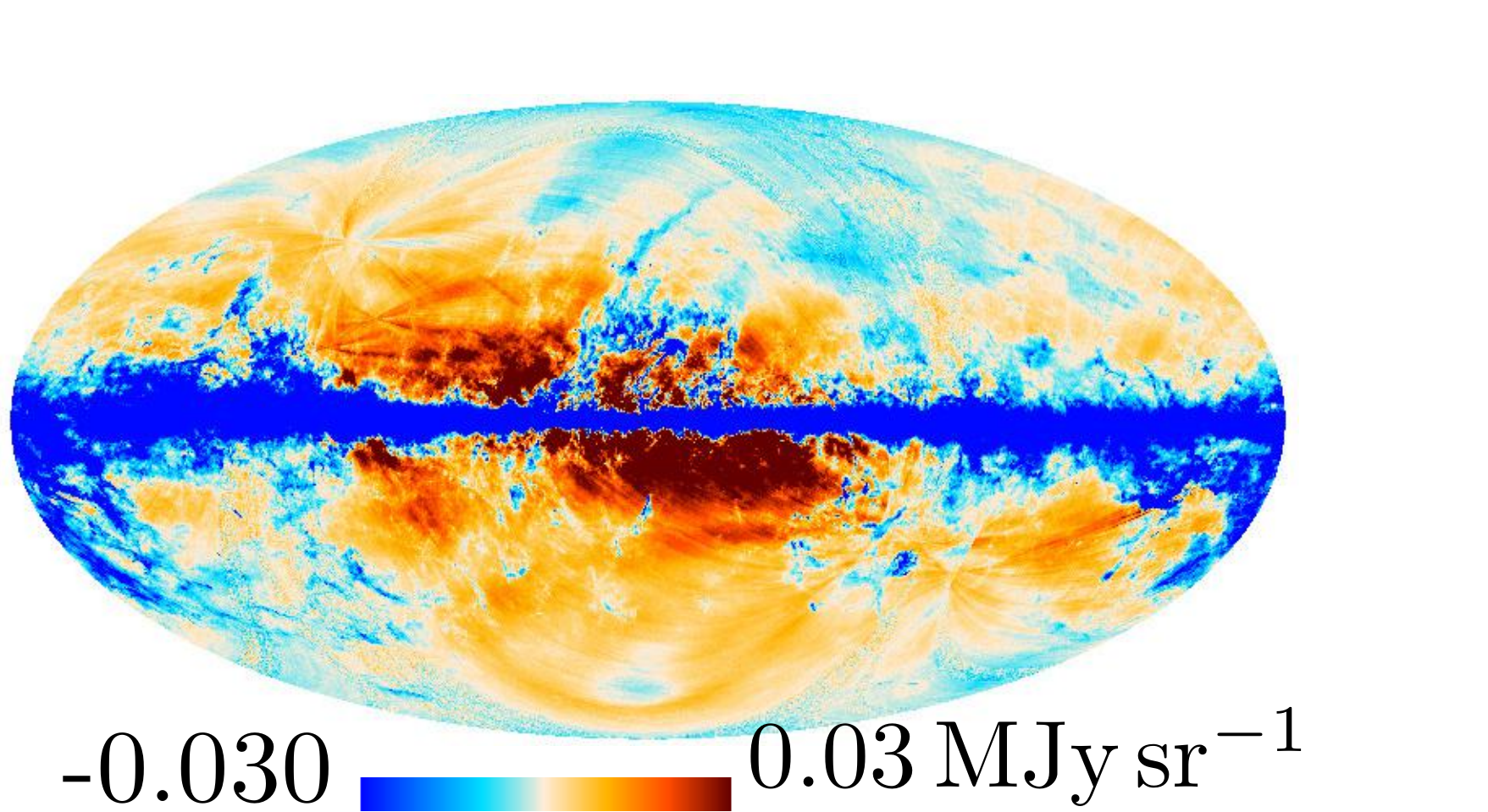}\\
\includegraphics[width=0.32\columnwidth]{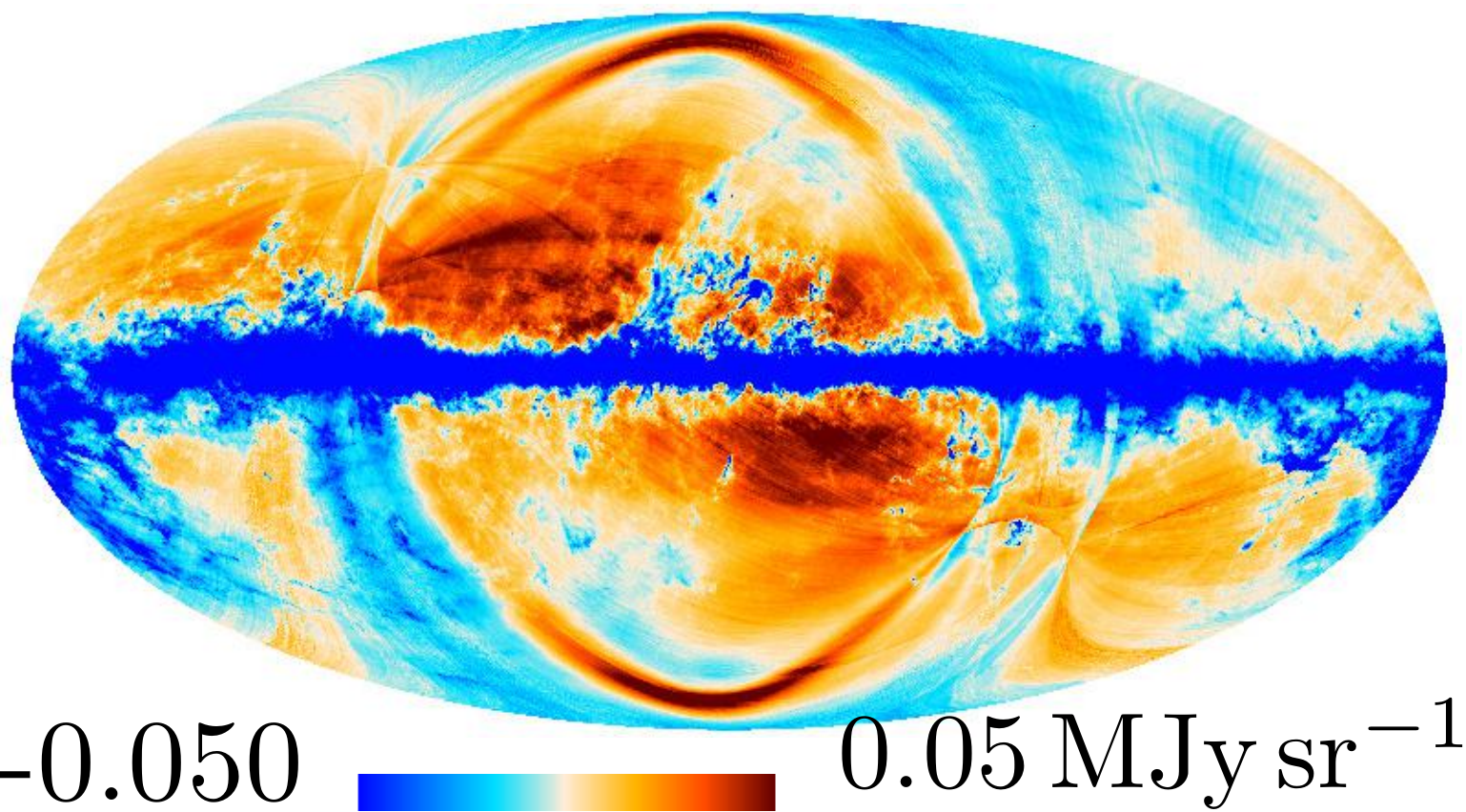}
\caption{Difference between the HFI 2015 and 2018 maps. Frequencies (100 to 857\,GHz) are in rows, while Stokes parameters ($I$, $Q$, and $U$) are in columns.} 
\label{fig:diffDX11vsRC4}
\end{figure}

Differences between the HFI 2018 maps and the 2015 ones are shown in Fig.~\ref{fig:diffDX11vsRC4}, specifically plotting the ($2015-2018$) difference maps in $I$, $Q$, and $U$.
Of course, these differences do not directly show evidence for a reduction of the systematics level in 2018. It is only after discussion based on simulations of the improvement mentioned above (and further work presented in Sects.~\ref{sec:oddeven} and \ref{sec:powerspectra}), that we can demonstrate that the differences are mostly due to a decrease of systematic residuals in the 2018 release.

In regions of strong Galactic signal (the Galactic ridge and molecular clouds above and below the Galactic disc), we can use the behaviour of the differences between 100, 143, and 217\,GHz maps to disentangle the contribution to the bandpass leakage due to dust increasing monotonically over these frequencies from the leakage due to CO lines decreasing from 100 to 143\,GHz, where there is no CO line. The 143-GHz map is smoother then the 100-GHz one outside the Galactic disc, indicative of a dominance of dust in the more diffuse ISM, with only a patchy distribution of CO seen only at 100 GHz (Sect.~\ref{sec:bpeffect}). Finally, we recommend that users of the 2018 data, mask CO in the high latitude sky for high sensitivity cosmology studies, using the 2015 \Planck\ CO maps.

The \sroll\ mapmaking has also been used for the first time on the submillimetre channels at 545 and 857\,GHz. The difference between the 2015 and 2018 releases at these frequencies are also shown in Fig.~\ref{fig:diffDX11vsRC4}. The difference map at 857\,GHz shows clearly an FSL signature at 857\,GHz, at a level of 2--$5\times 10^{-2}\,{\rm MJy}\,{\rm sr}^{-1}$. This is expected, since the FSL contributions were not removed in the 2015 maps.

The zodiacal cloud and bands were removed using the same model in this release as in 2015, but improving the fit of the emissivities, as discussed in Sect.~\ref{sec:zodi}. Nevertheless those improvements are too small to be seen, even at 857\,GHz. The large-scale features seen in the difference maps are due instead to improved control of systematics (ADCNL, bandpass, and calibration).

In summary, the differences between the 2015 and 2018 maps all show the improvements expected in the new maps from better correction of systematic effects.

\subsubsection{Survey null tests on the data}
\label{sec:oddeven}

The odd-even survey differences are used implicitly in \sroll\ to detect systematics sensitive to either scan direction, or different orientations of the beam, between two different measurements of the same sky pixel by a given bolometer. These systematics are thus mostly removed. Survey null tests remain a very sensitive tool for investigating the residual systematic effects that are mostly cancelled in the 2018 maps when averaging odd and even surveys in full- or half-mission maps.

\begin{figure*}[htbp!]
\begin{center}
\begin{tabular}{cccc}
	\textbf{2015}&I&	Q&U\\
	\raisebox{25pt}{\parbox{.03\textwidth}{100}}&
	\includegraphics [width=0.18\textwidth]{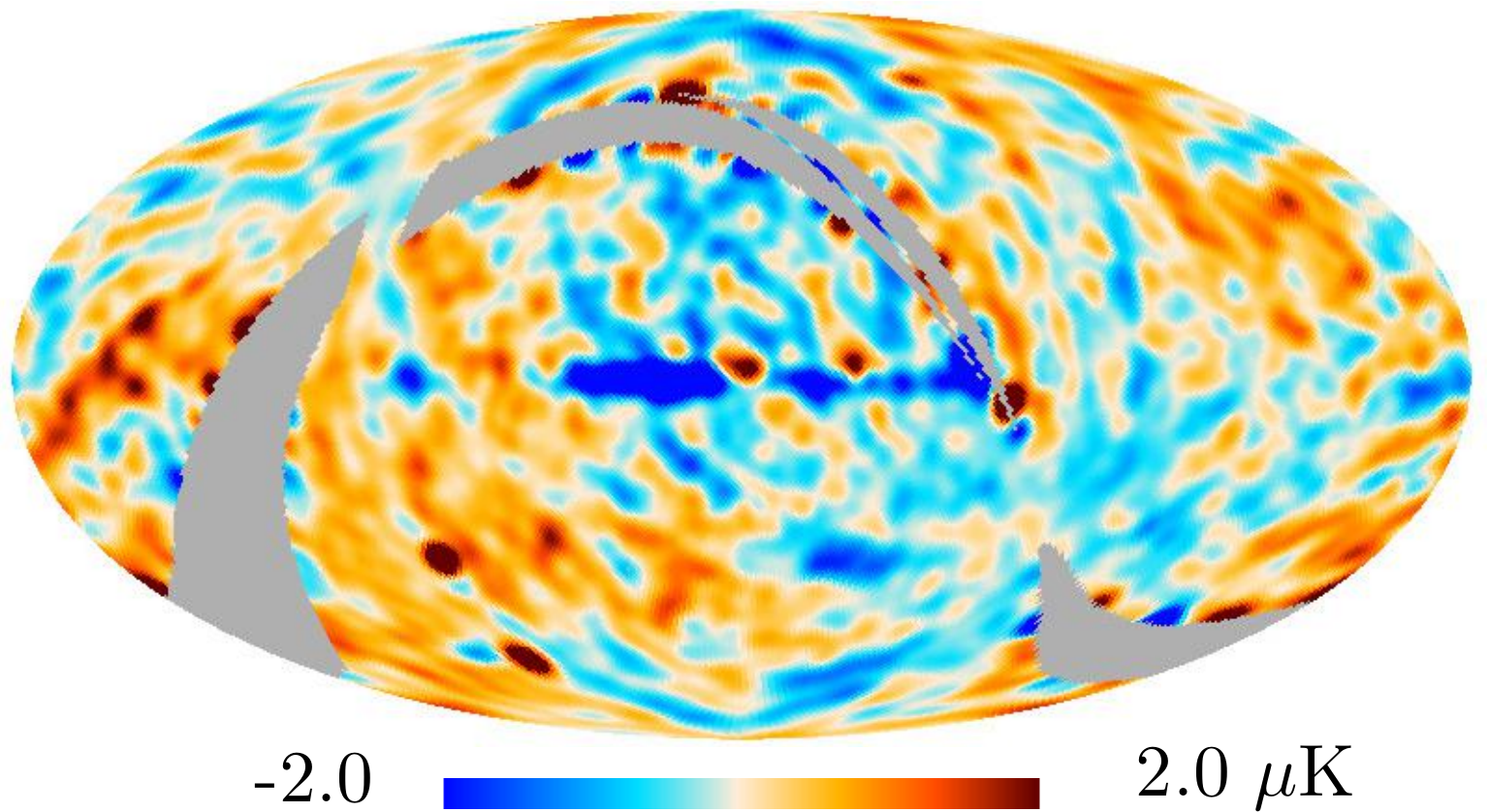}&
	\includegraphics [width=0.18\textwidth]{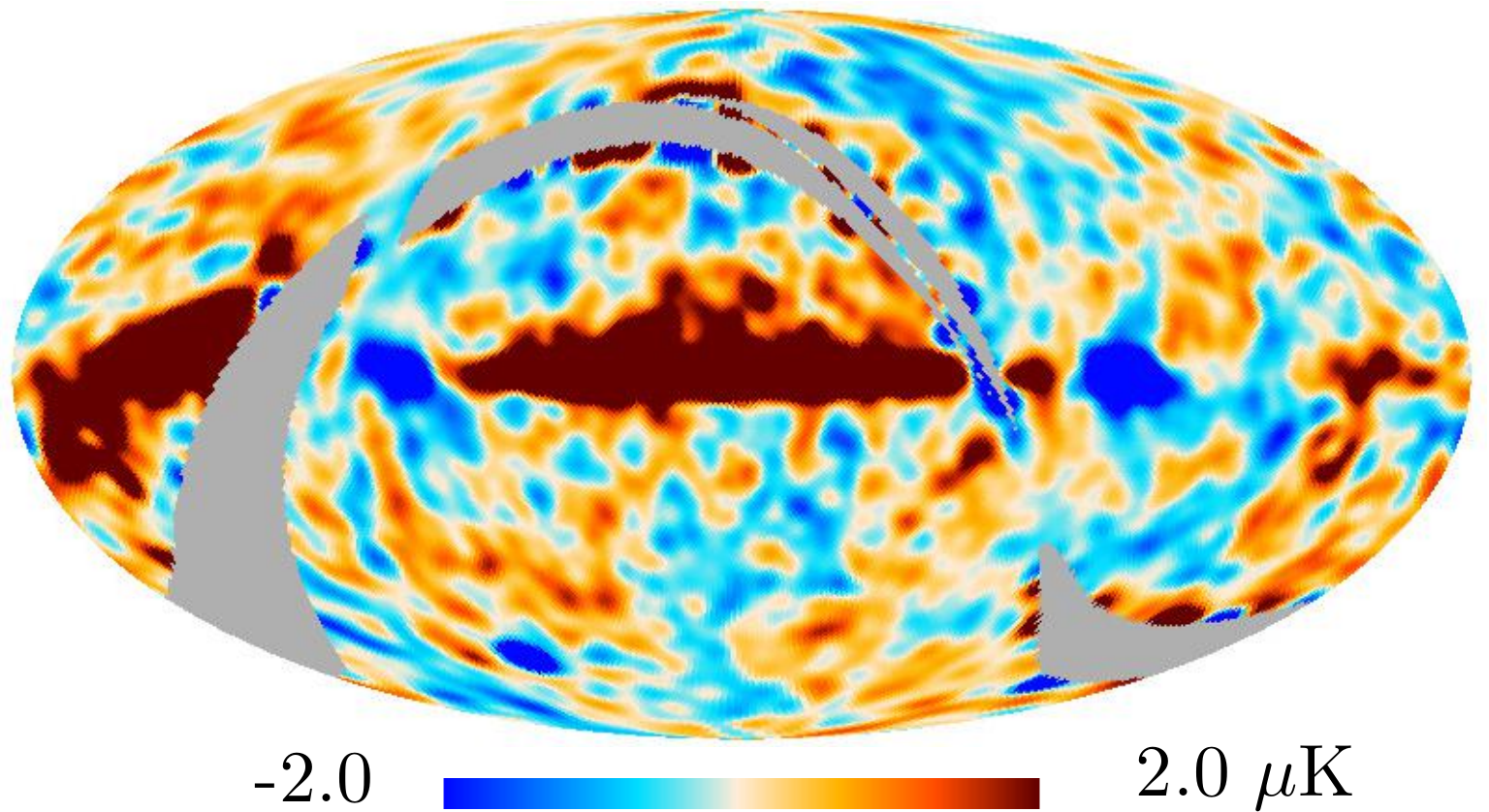}&
	\includegraphics [width=0.18\textwidth]{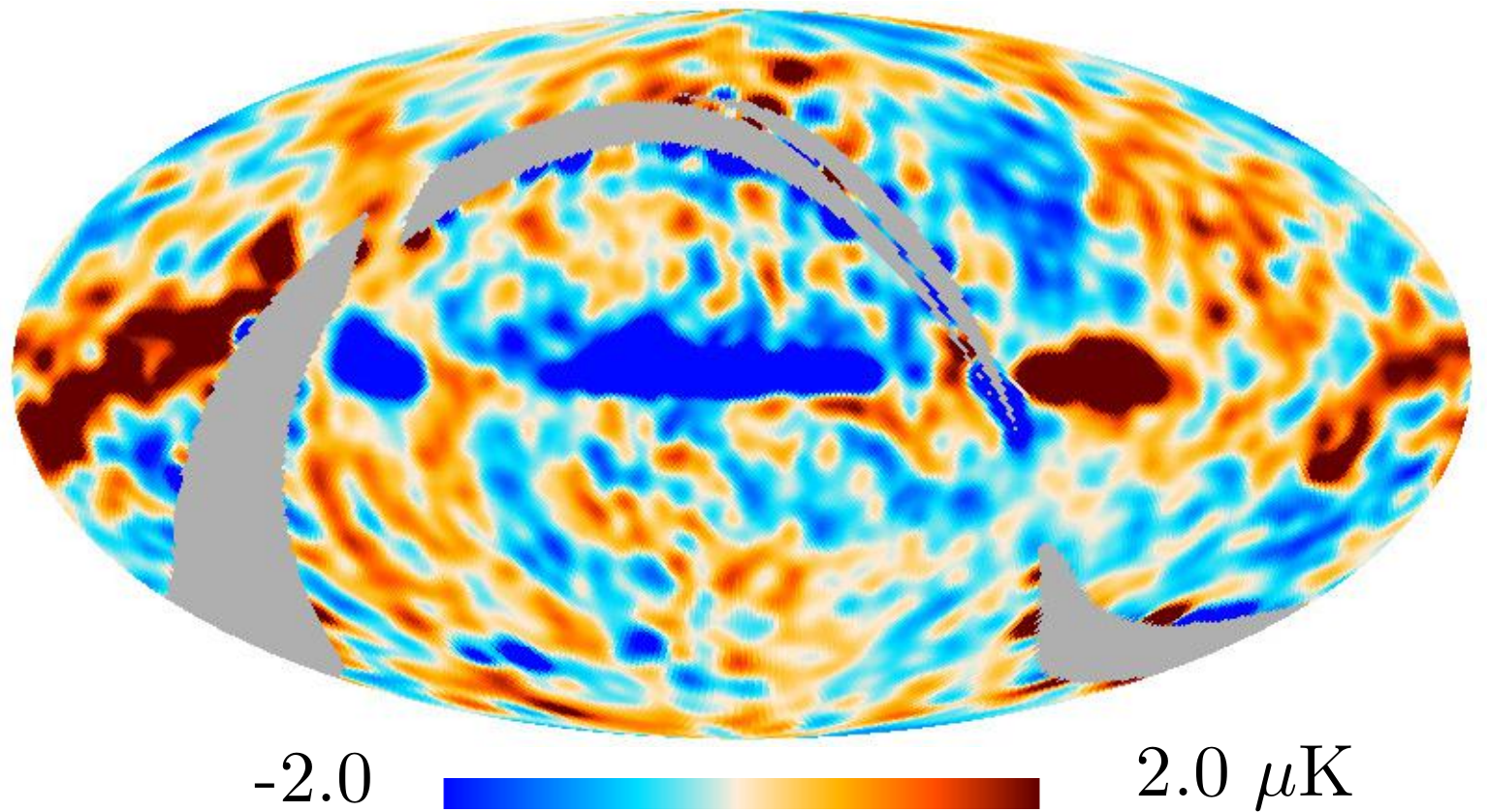}\\
	\raisebox{25pt}{\parbox{.03\textwidth}{143}}&
	\includegraphics [width=0.18\textwidth]{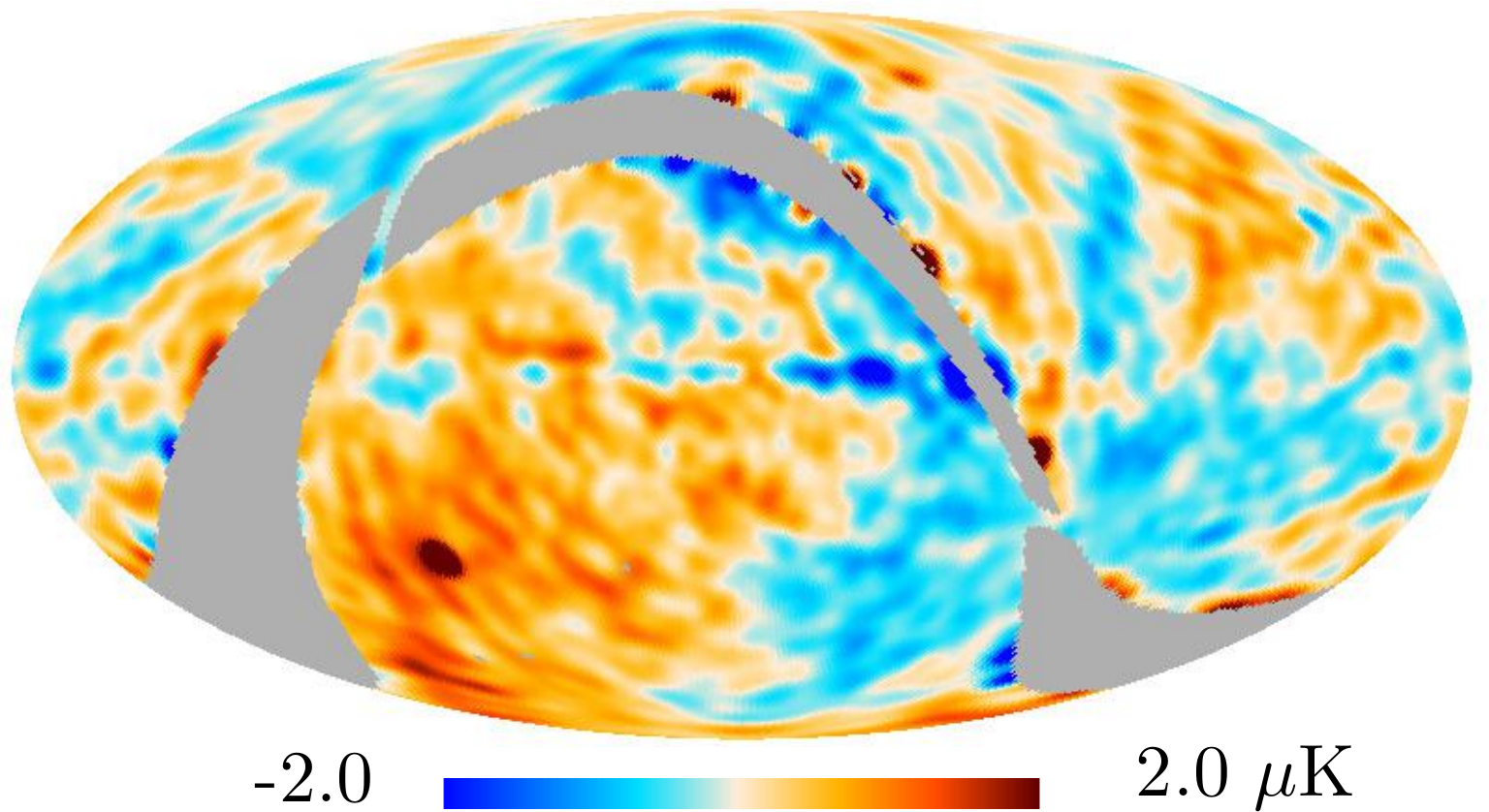}&
	\includegraphics [width=0.18\textwidth]{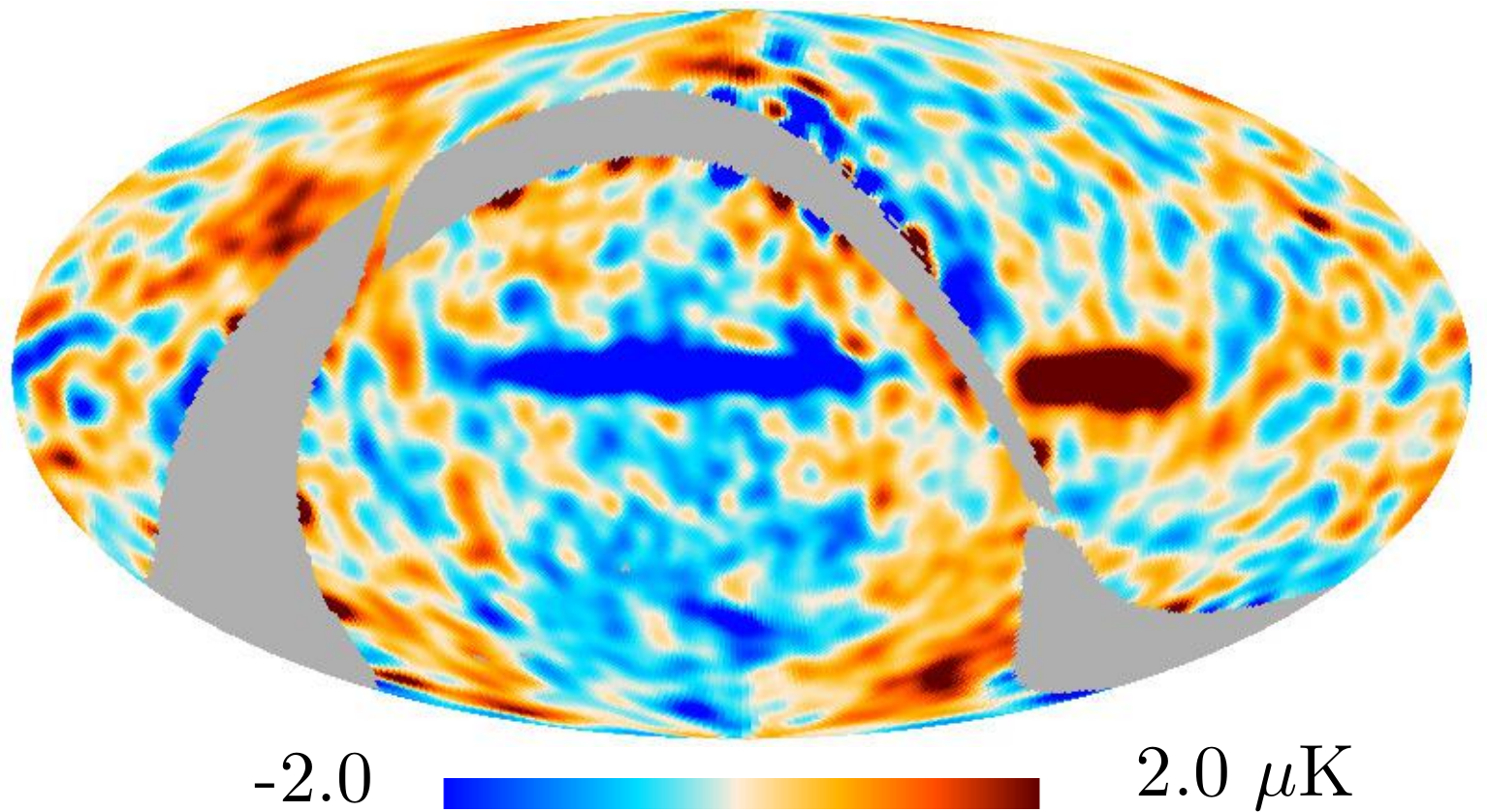}&
	\includegraphics [width=0.18\textwidth]{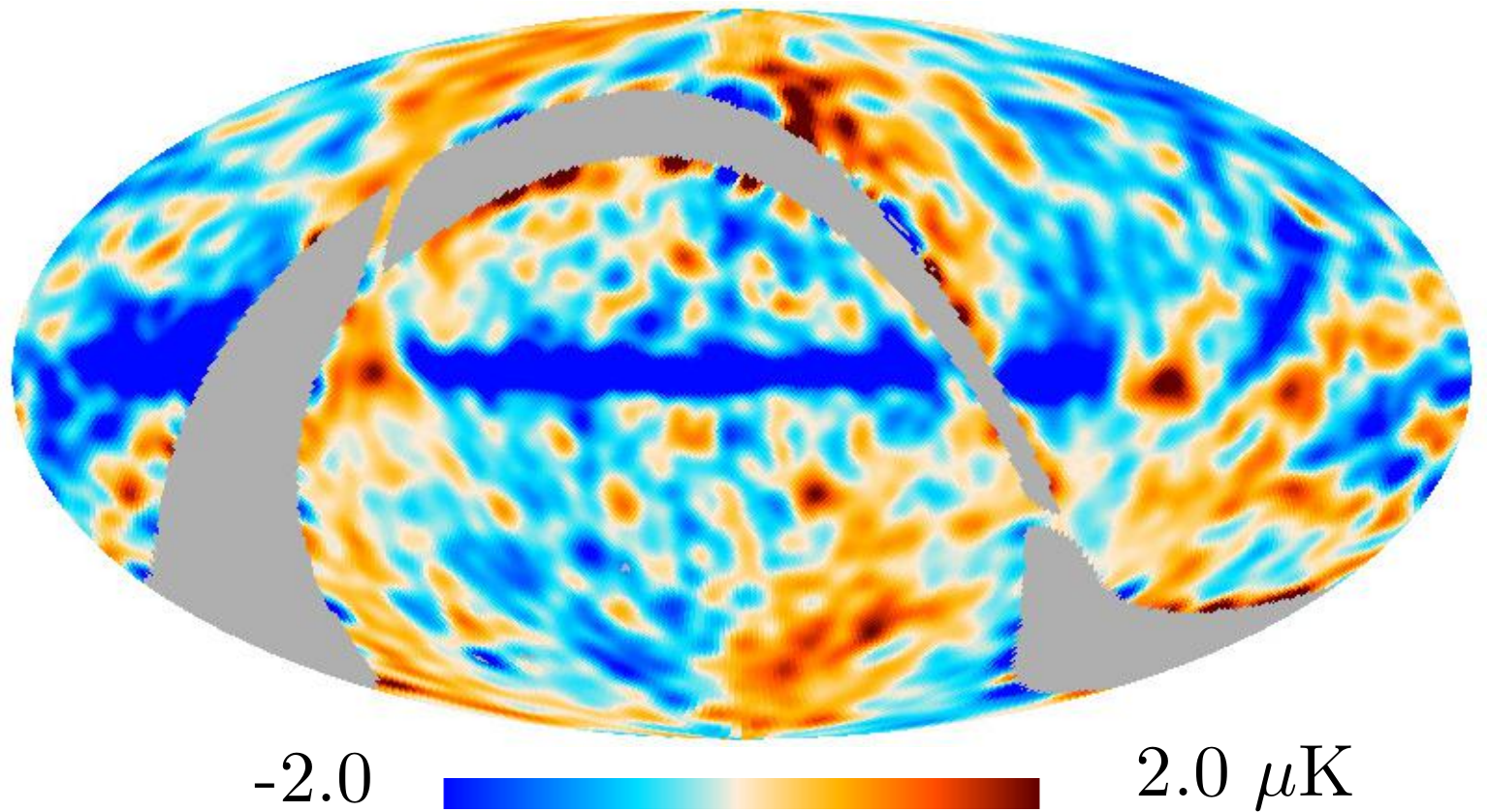}\\
	\raisebox{25pt}{\parbox{.03\textwidth}{217}}&
	\includegraphics [width=0.18\textwidth]{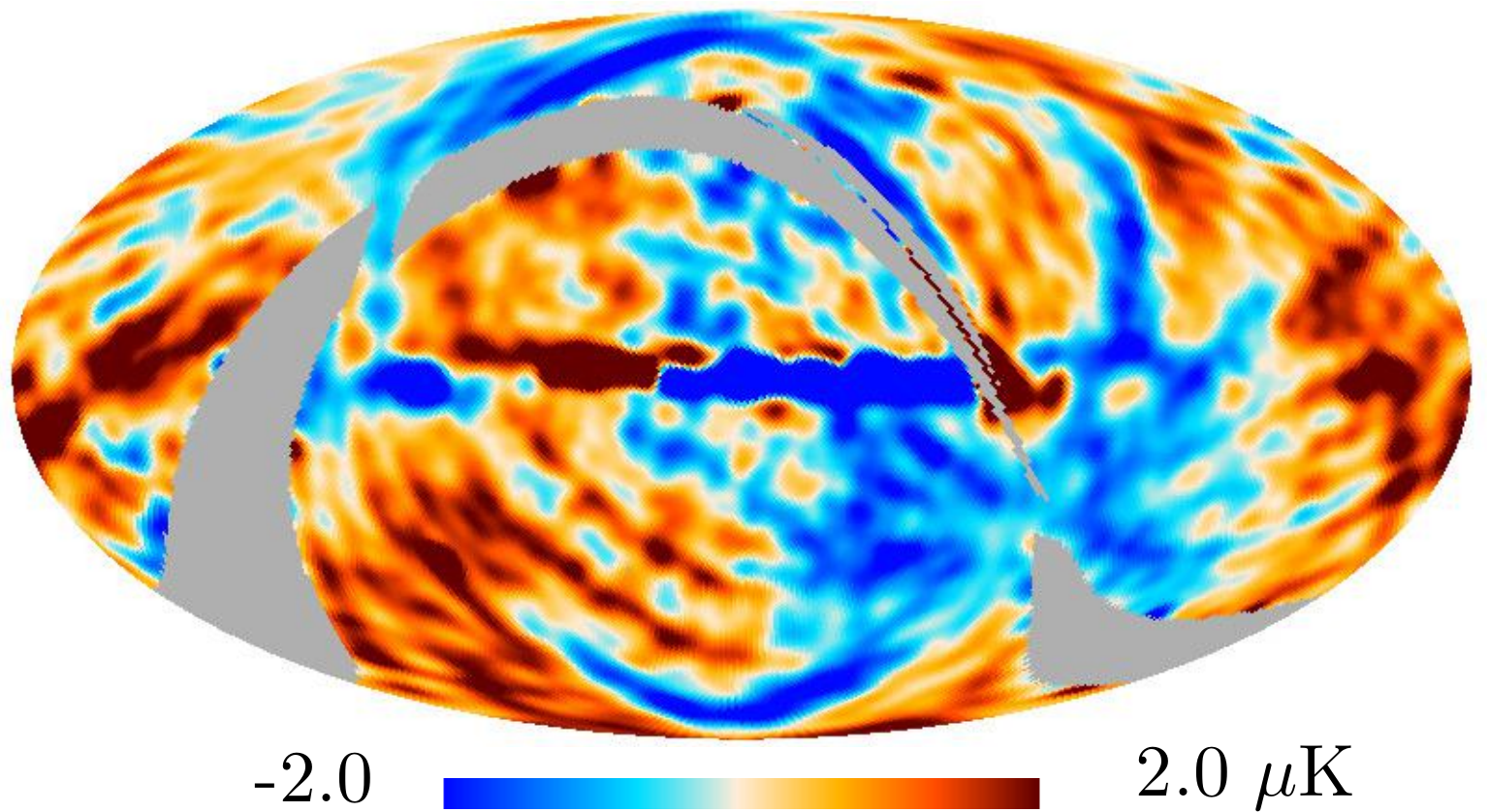}&
	\includegraphics [width=0.18\textwidth]{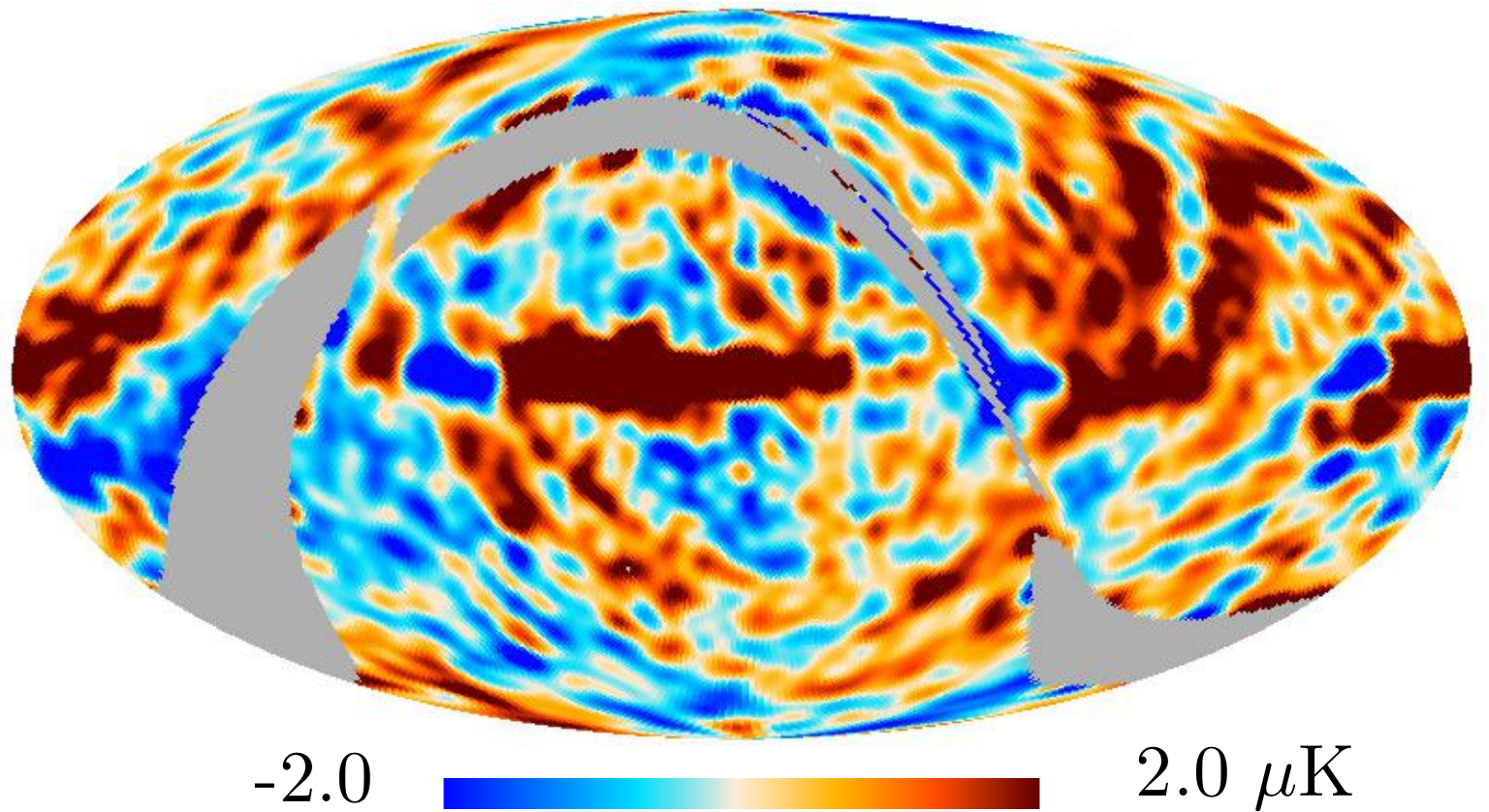}&
	\includegraphics [width=0.18\textwidth]{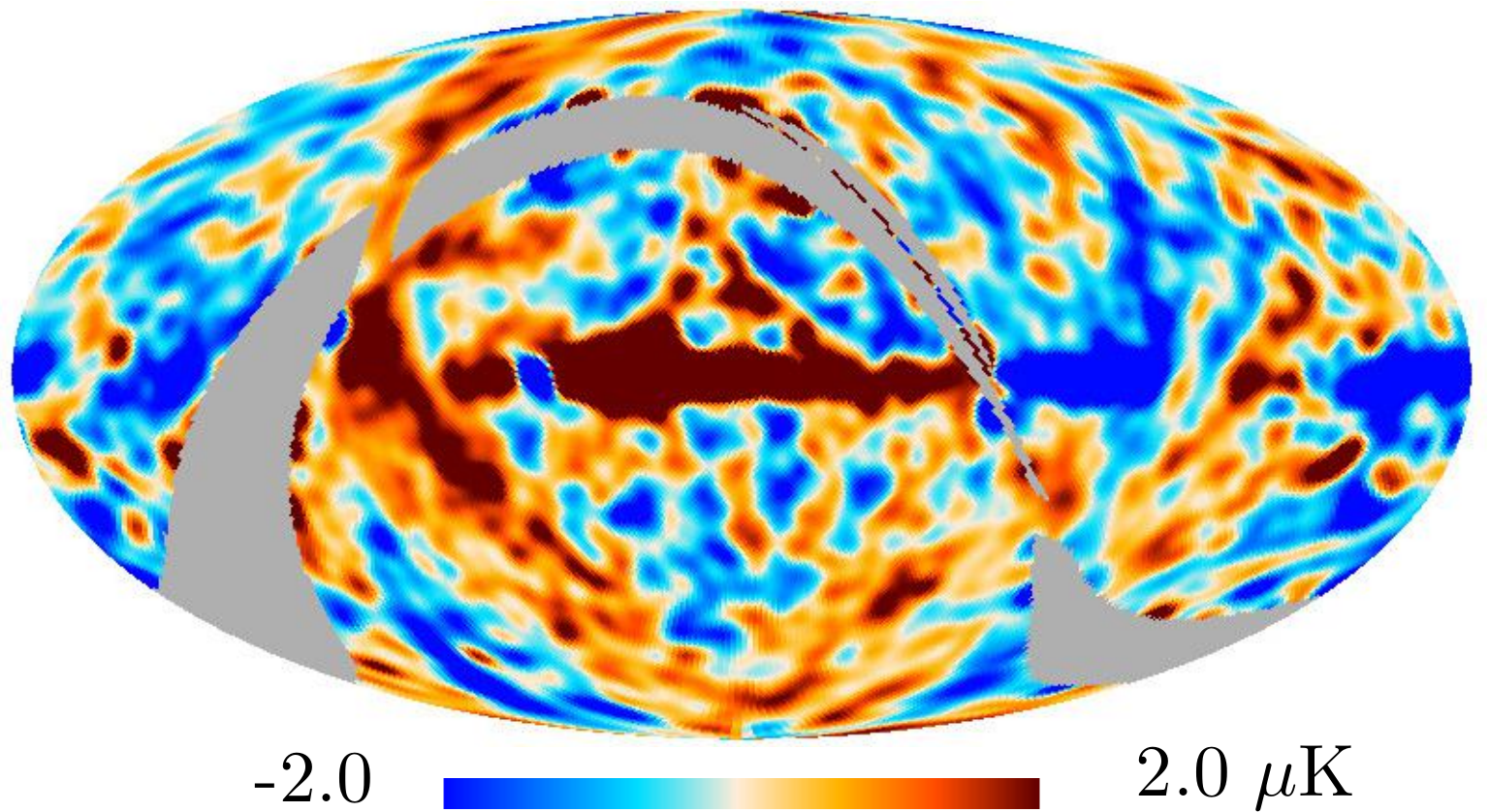}\\
	\raisebox{25pt}{\parbox{.03\textwidth}{353}}&
	\includegraphics [width=0.18\textwidth]{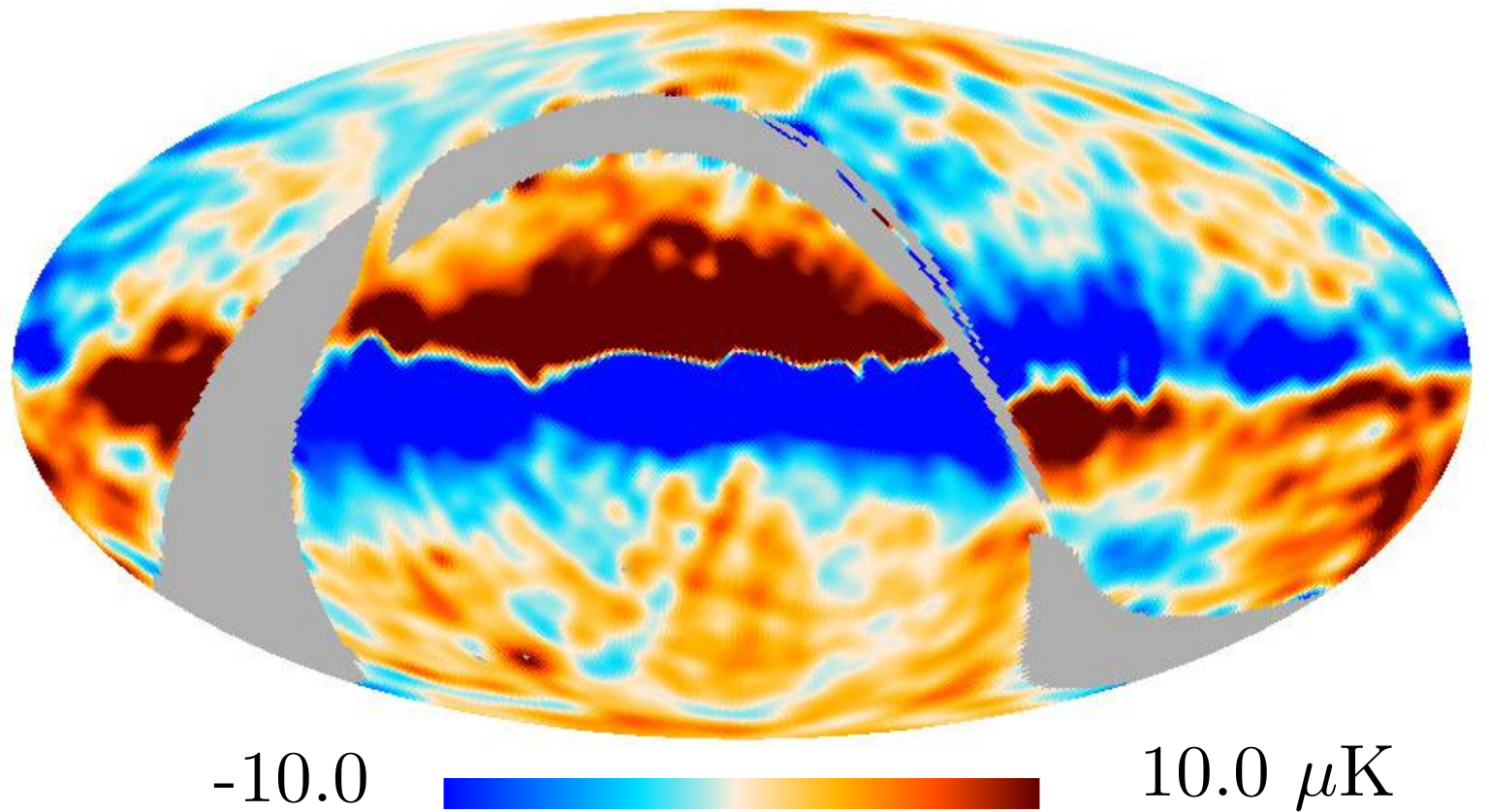}&
	\includegraphics [width=0.18\textwidth]{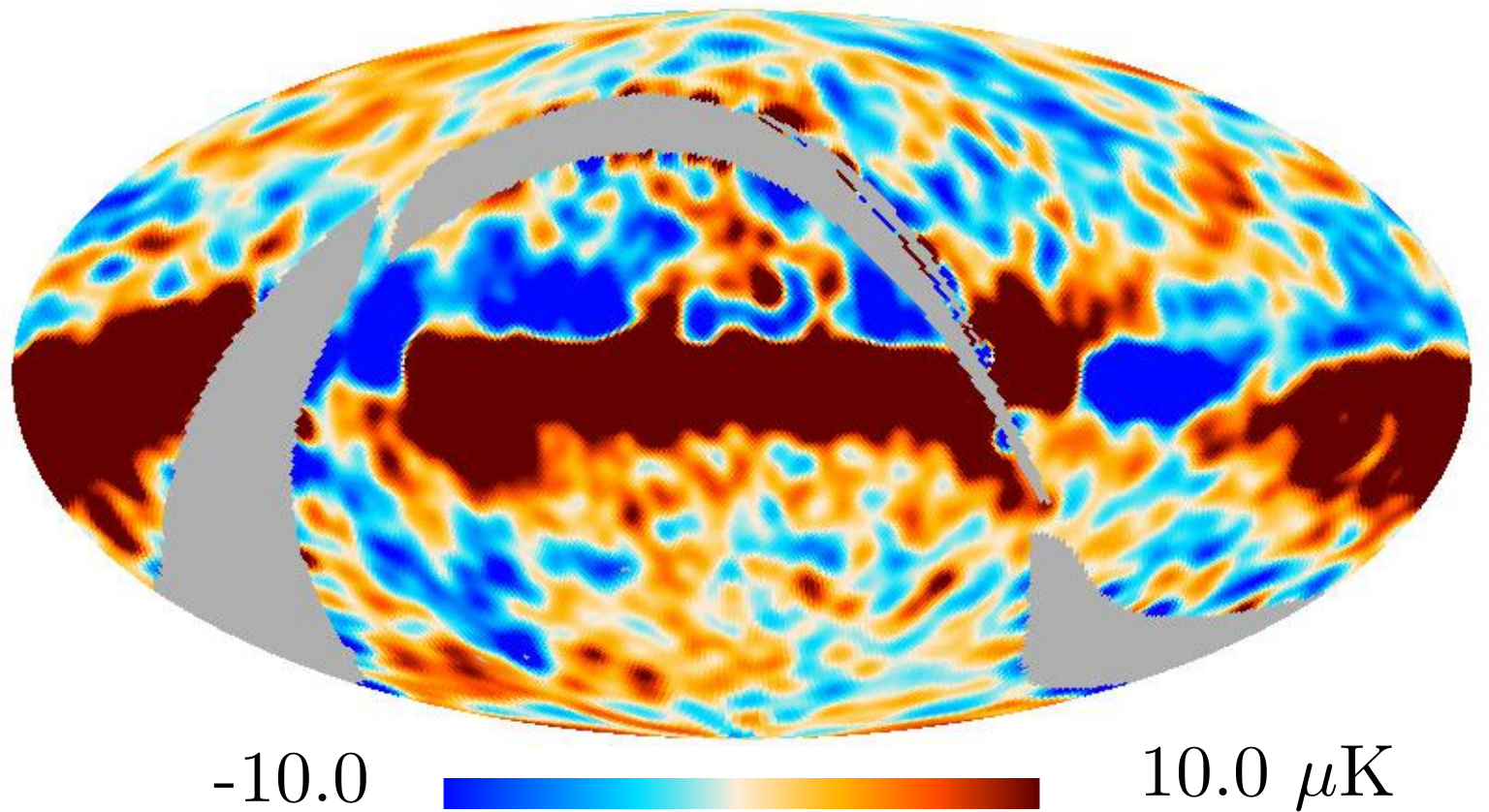}&
	\includegraphics [width=0.18\textwidth]{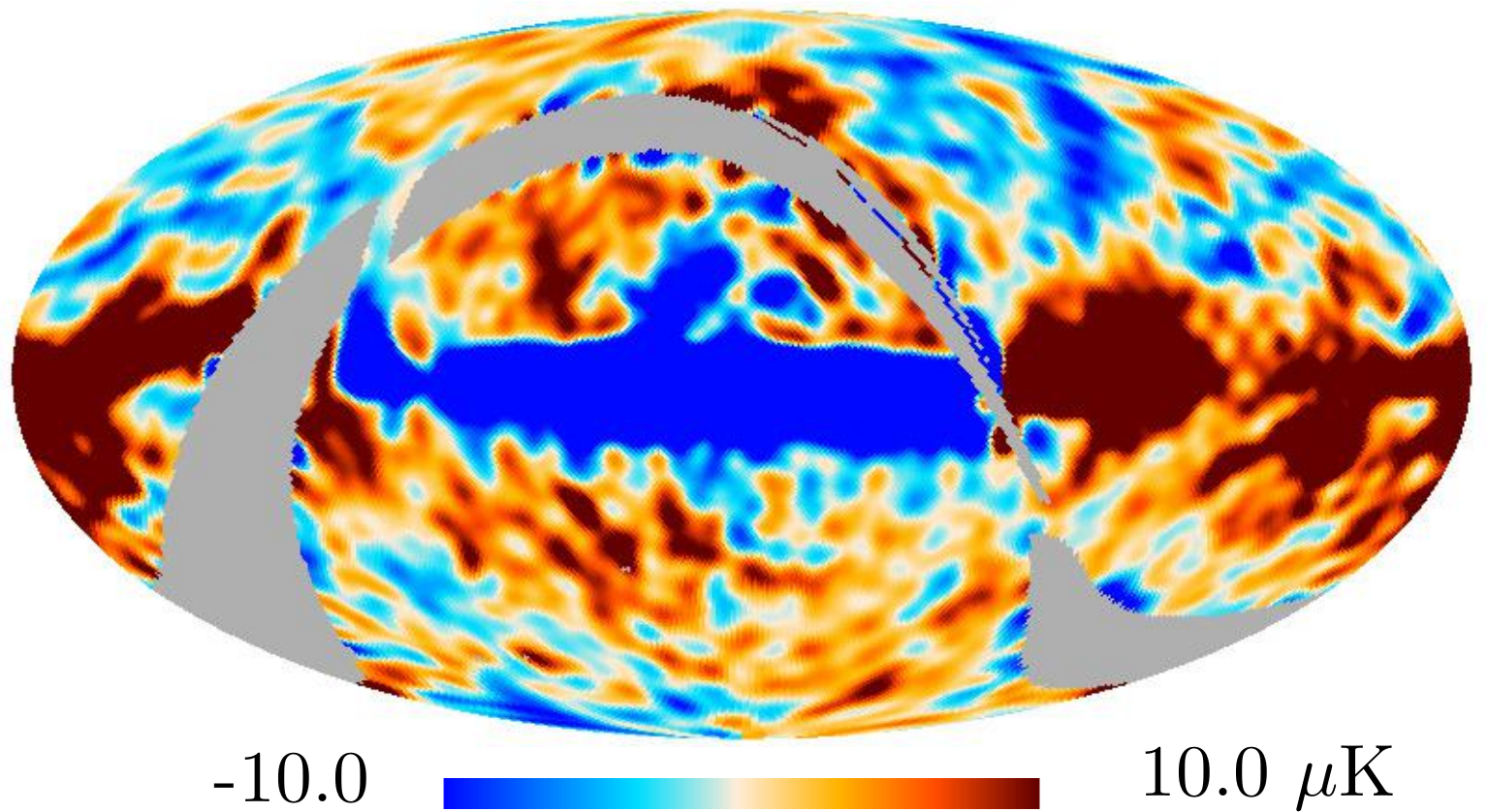}\\
	\raisebox{25pt}{\parbox{.03\textwidth}{545}}&
	\includegraphics [width=0.18\textwidth]{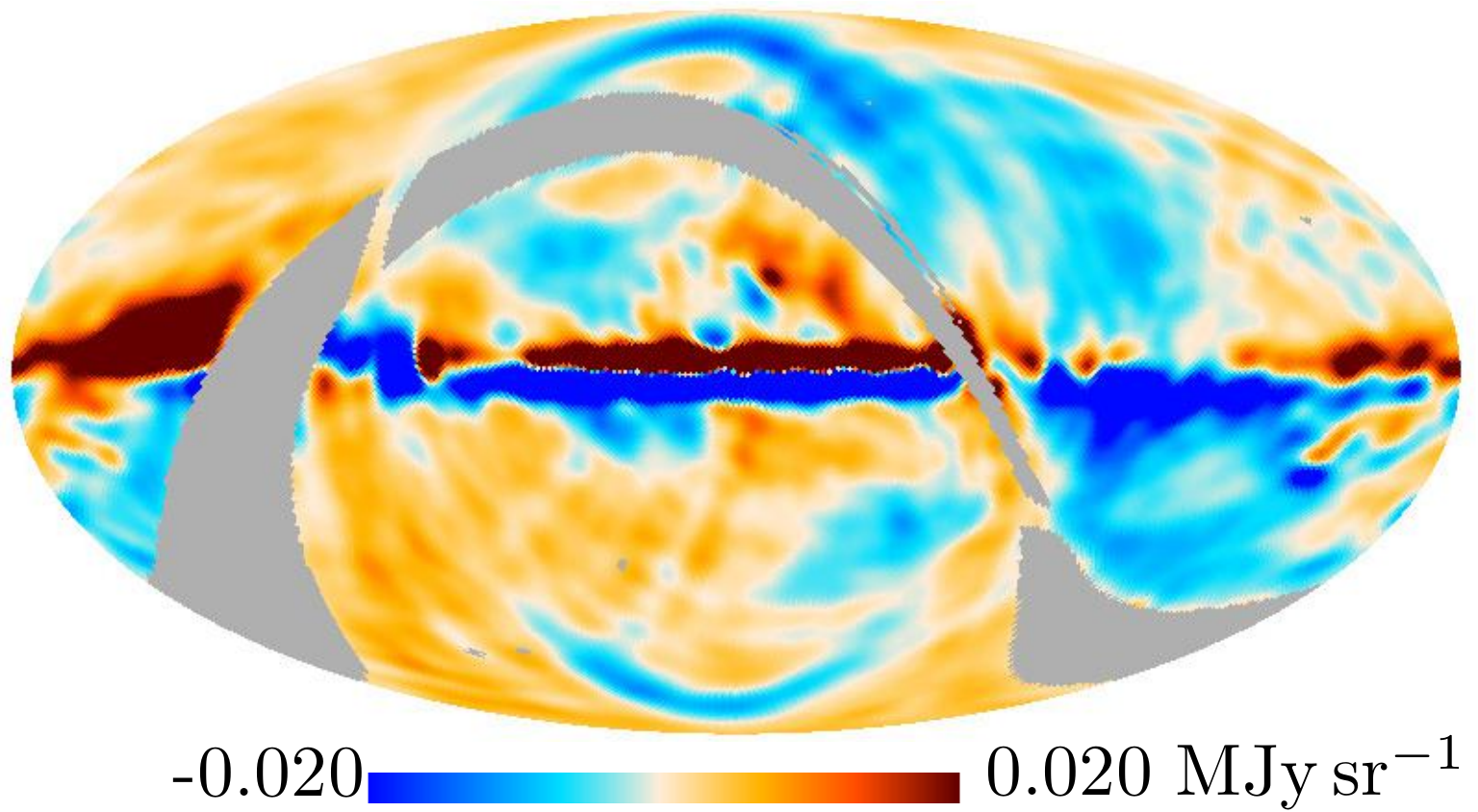} \\
	\raisebox{25pt}{\parbox{.03\textwidth}{857}}&
	\includegraphics [width=0.18\textwidth]{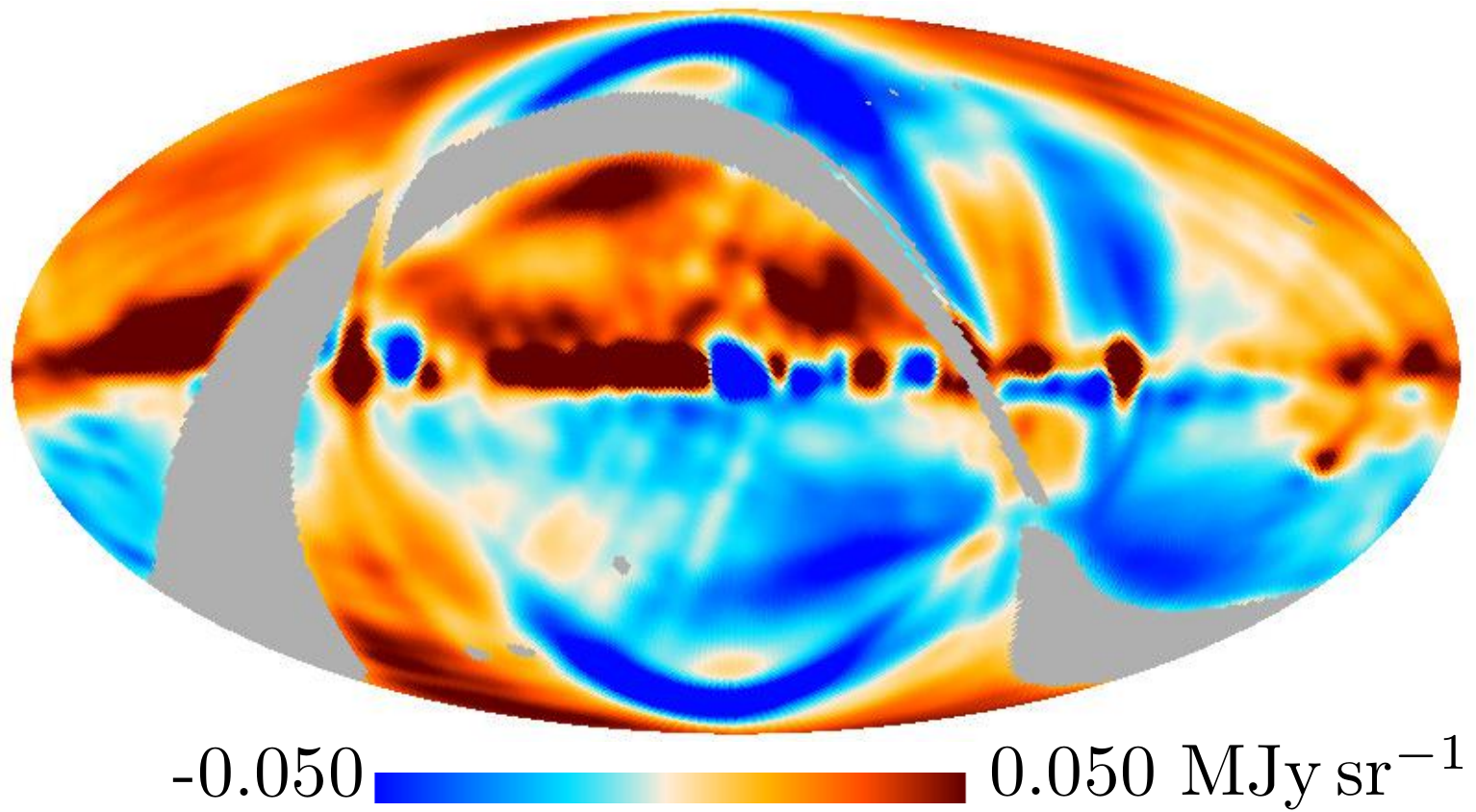}&
	\\
	\\
	\textbf{2018}&I&	Q&U\\
	\raisebox{25pt}{\parbox{.03\textwidth}{100}}&
	\includegraphics [width=0.18\textwidth]{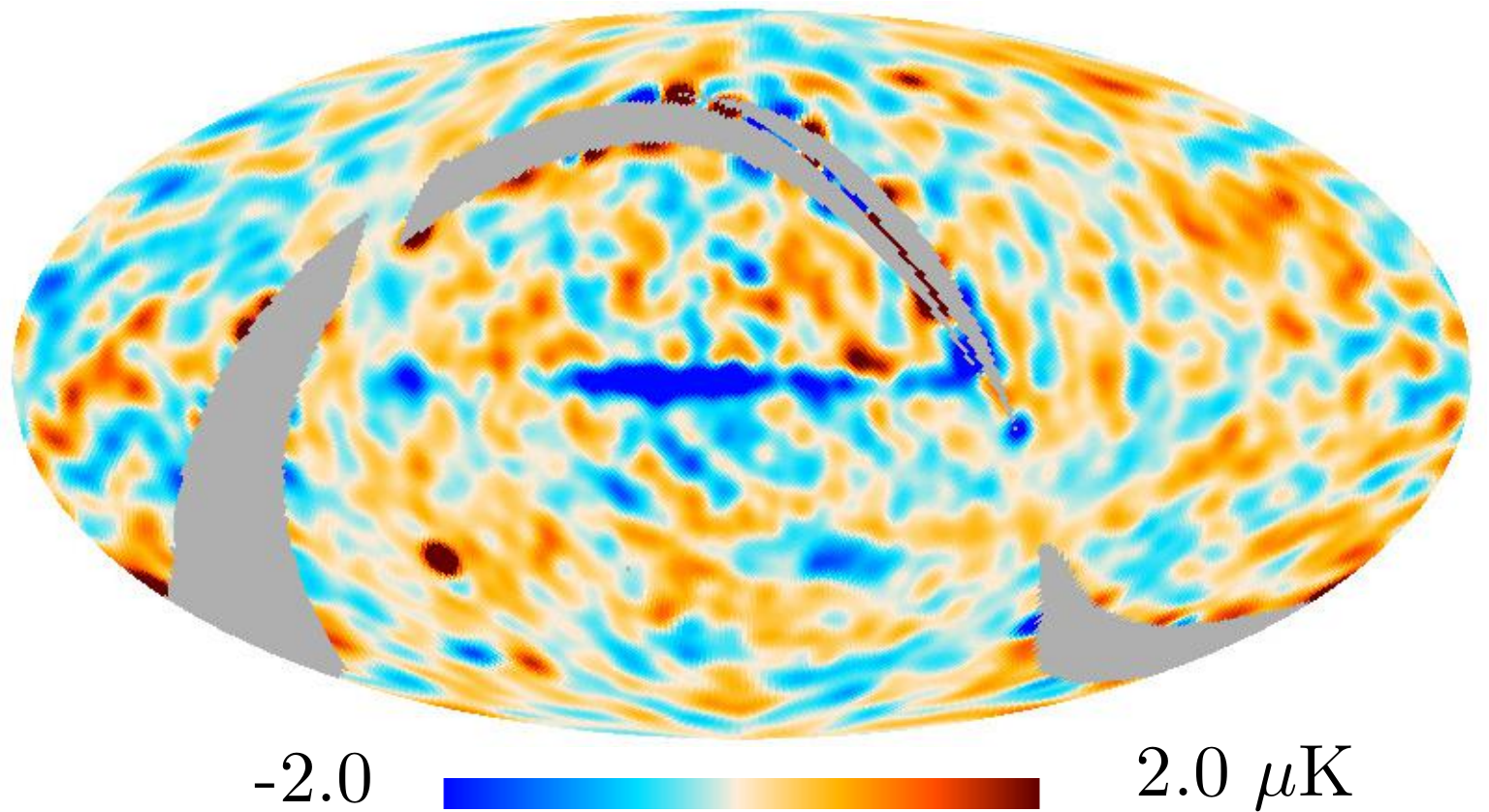}&
	\includegraphics [width=0.18\textwidth]{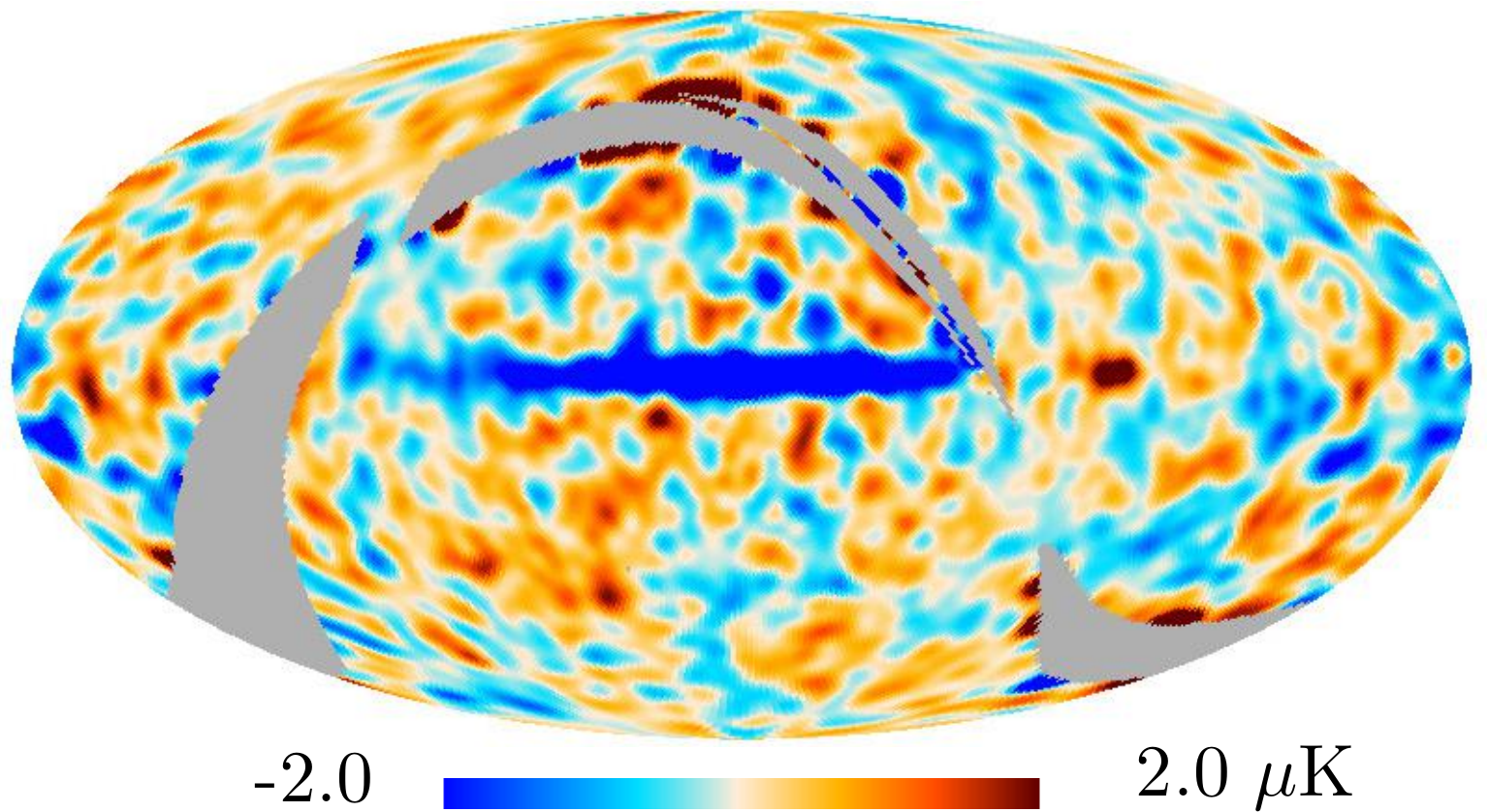}&
	\includegraphics [width=0.18\textwidth]{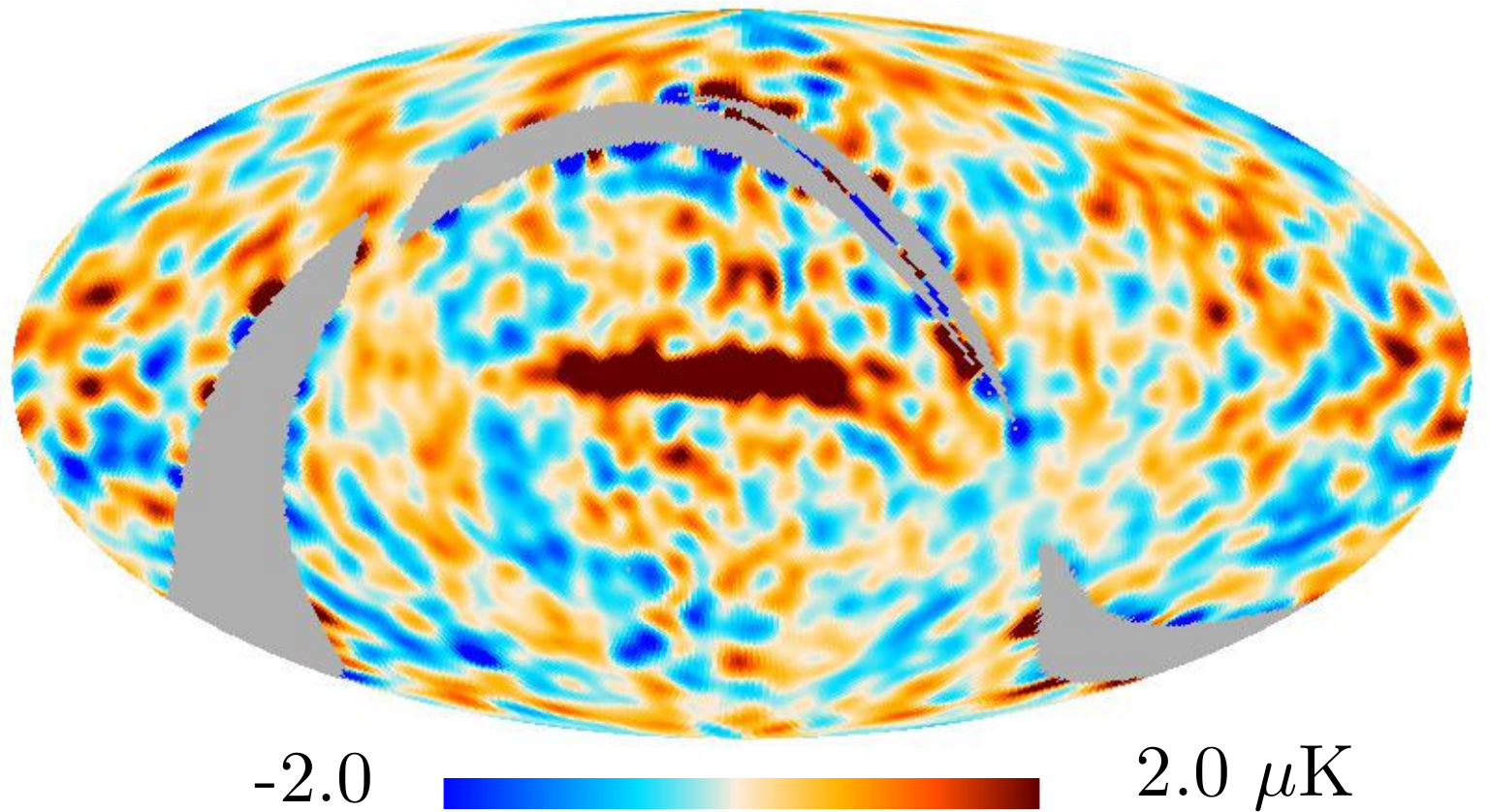}\\
	\raisebox{25pt}{\parbox{.03\textwidth}{143}}&
	\includegraphics [width=0.18\textwidth]{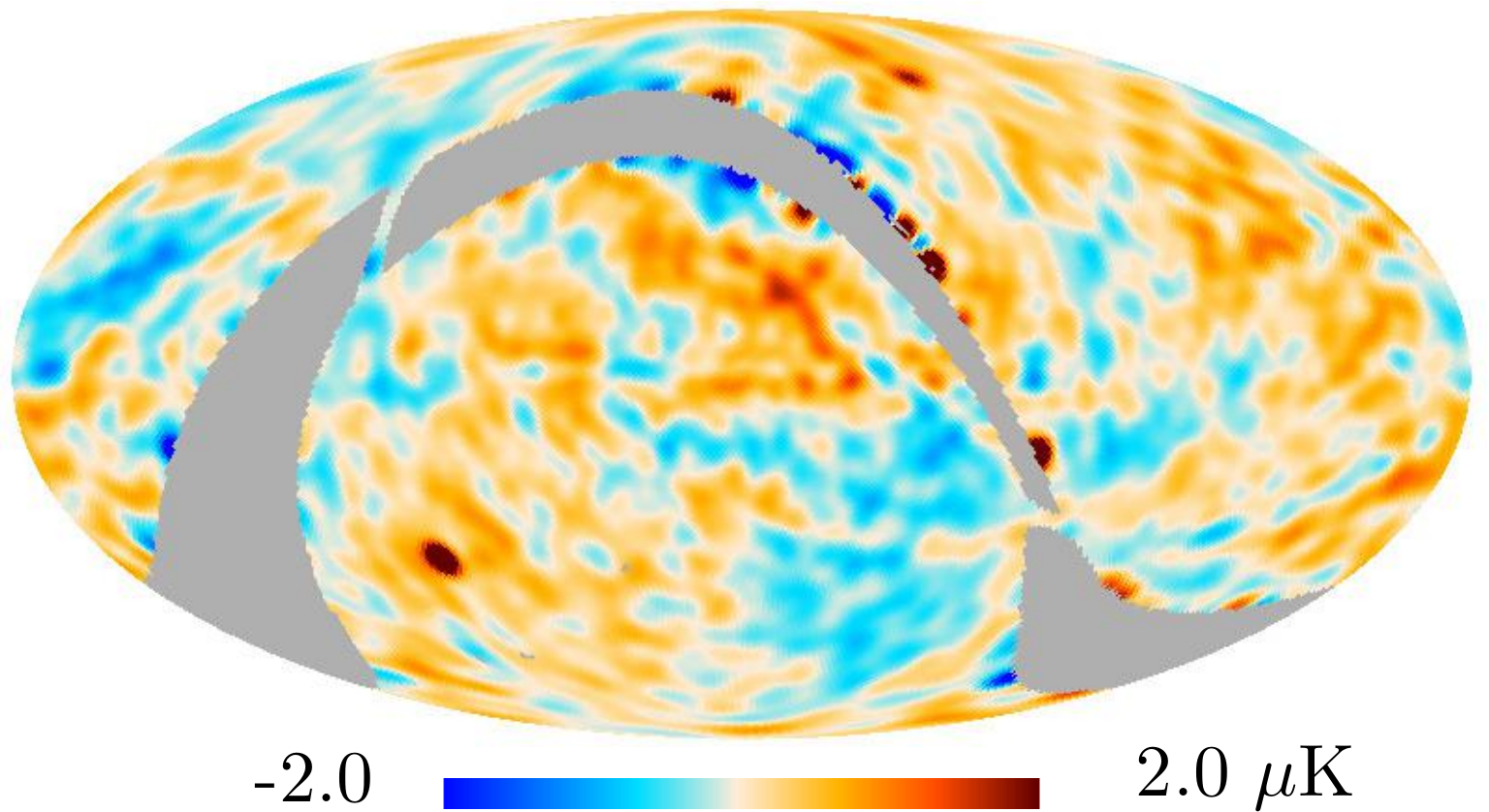}&
	\includegraphics [width=0.18\textwidth]{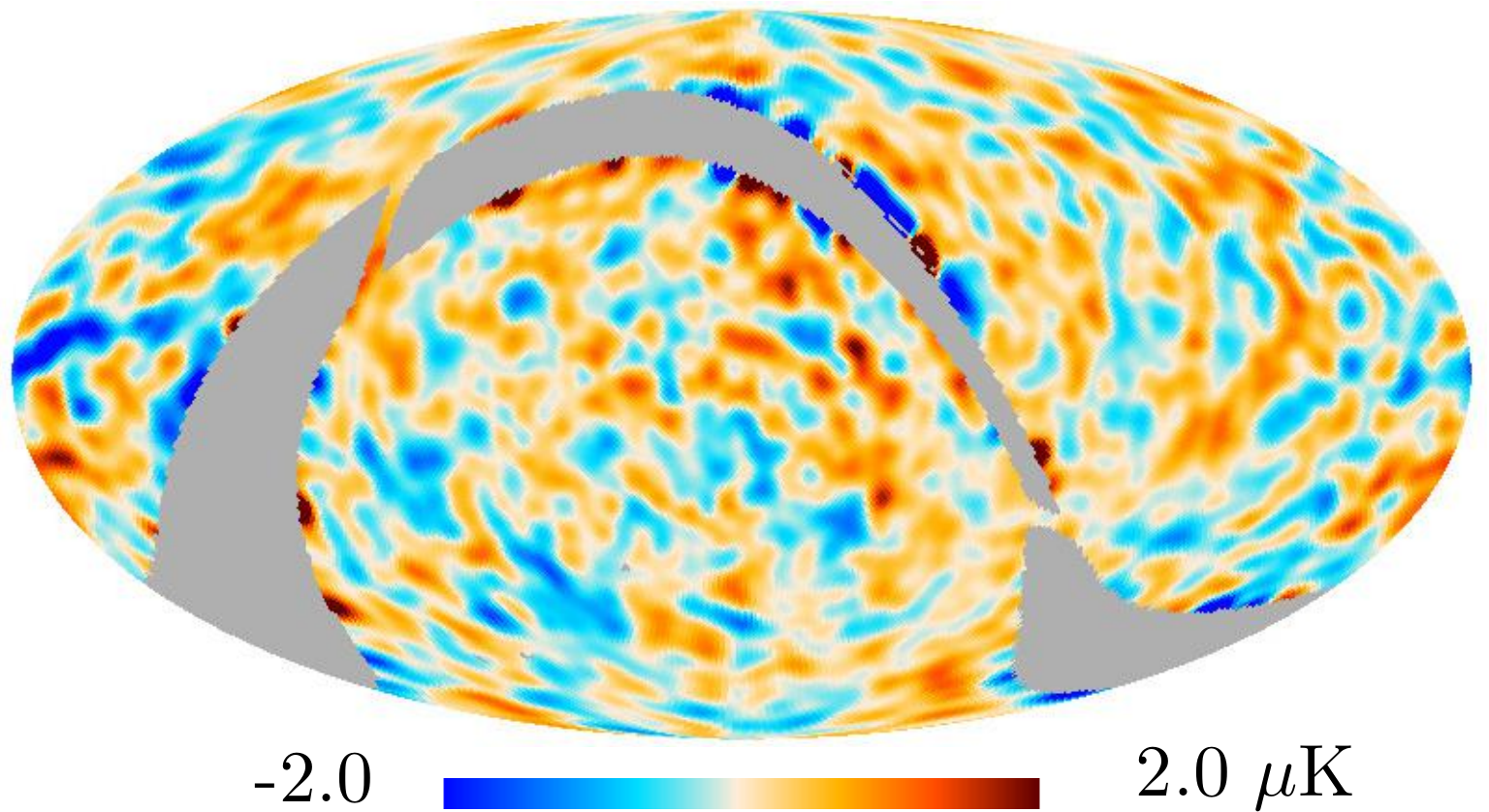}&
	\includegraphics [width=0.18\textwidth]{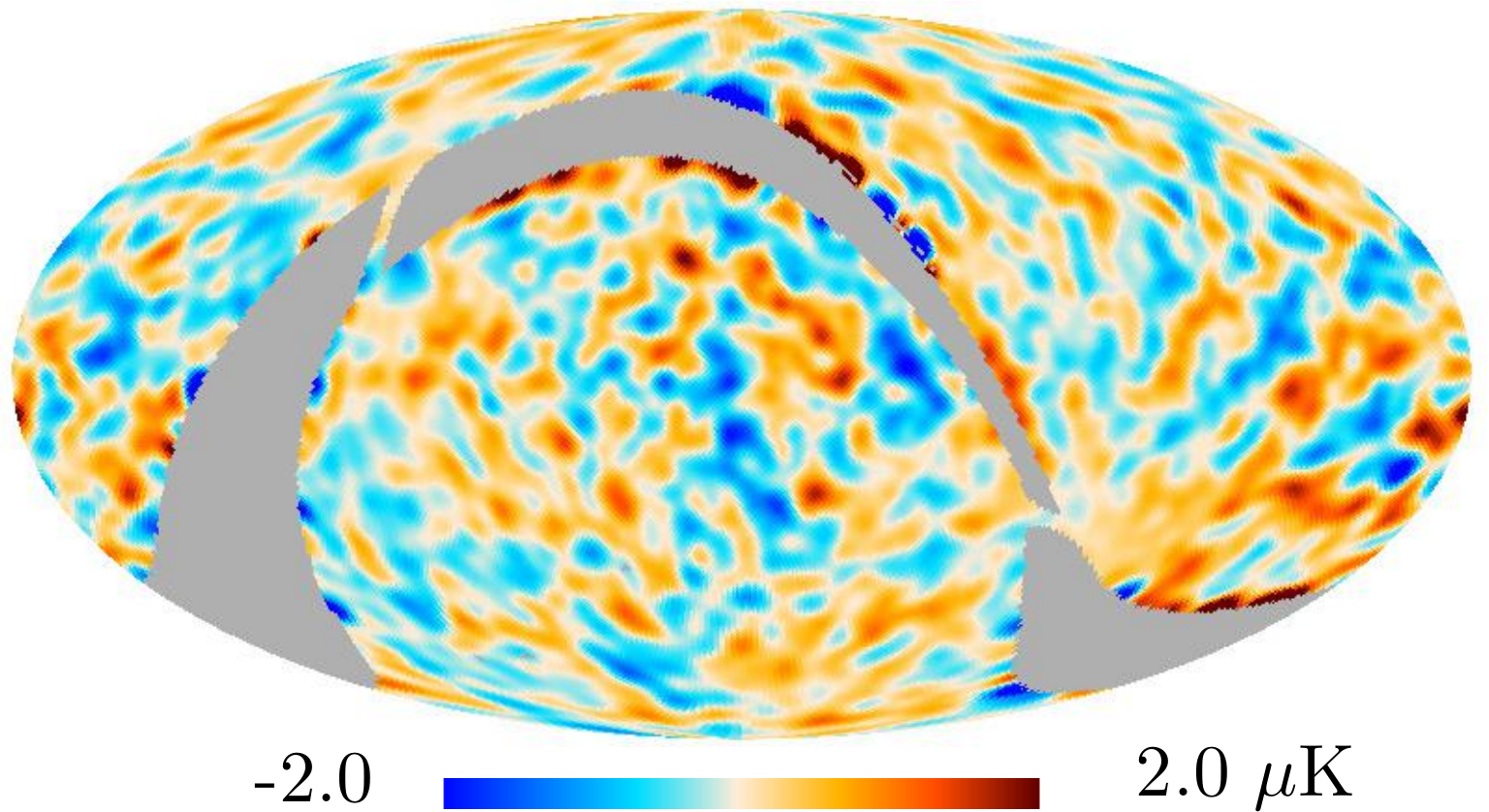}\\
	\raisebox{25pt}{\parbox{.03\textwidth}{217}}&
	\includegraphics [width=0.18\textwidth]{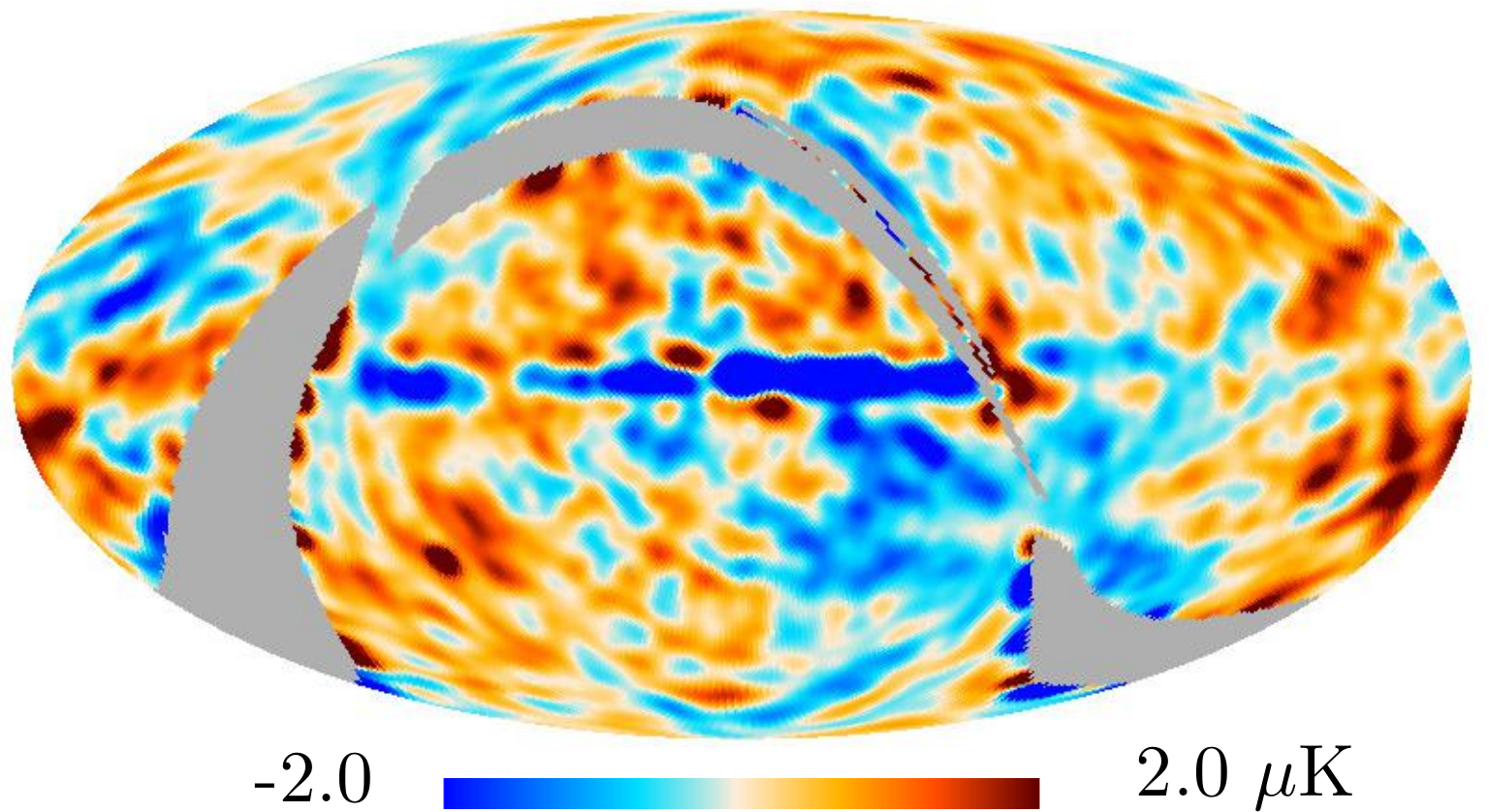}&
	\includegraphics [width=0.18\textwidth]{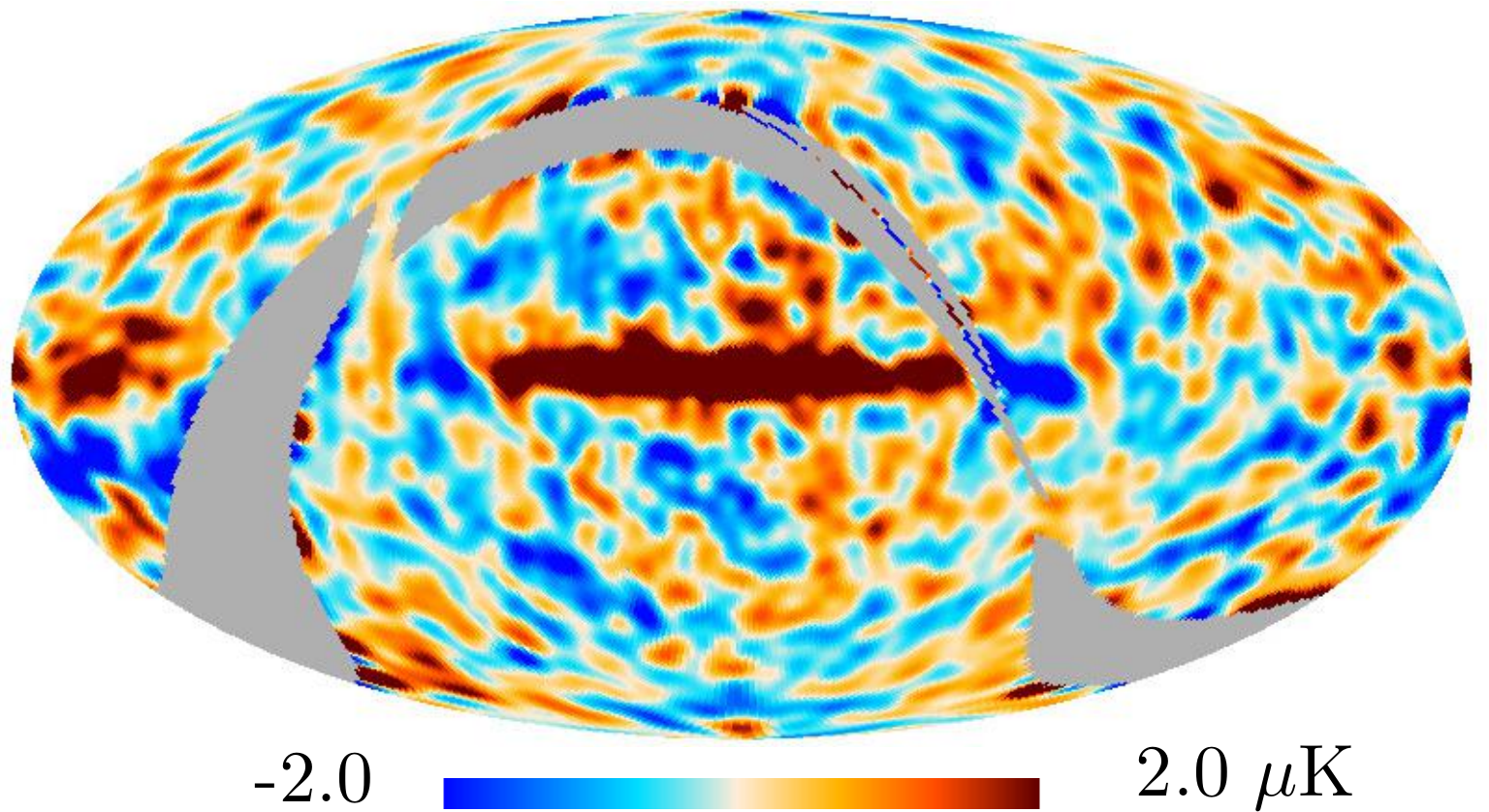}&
	\includegraphics [width=0.18\textwidth]{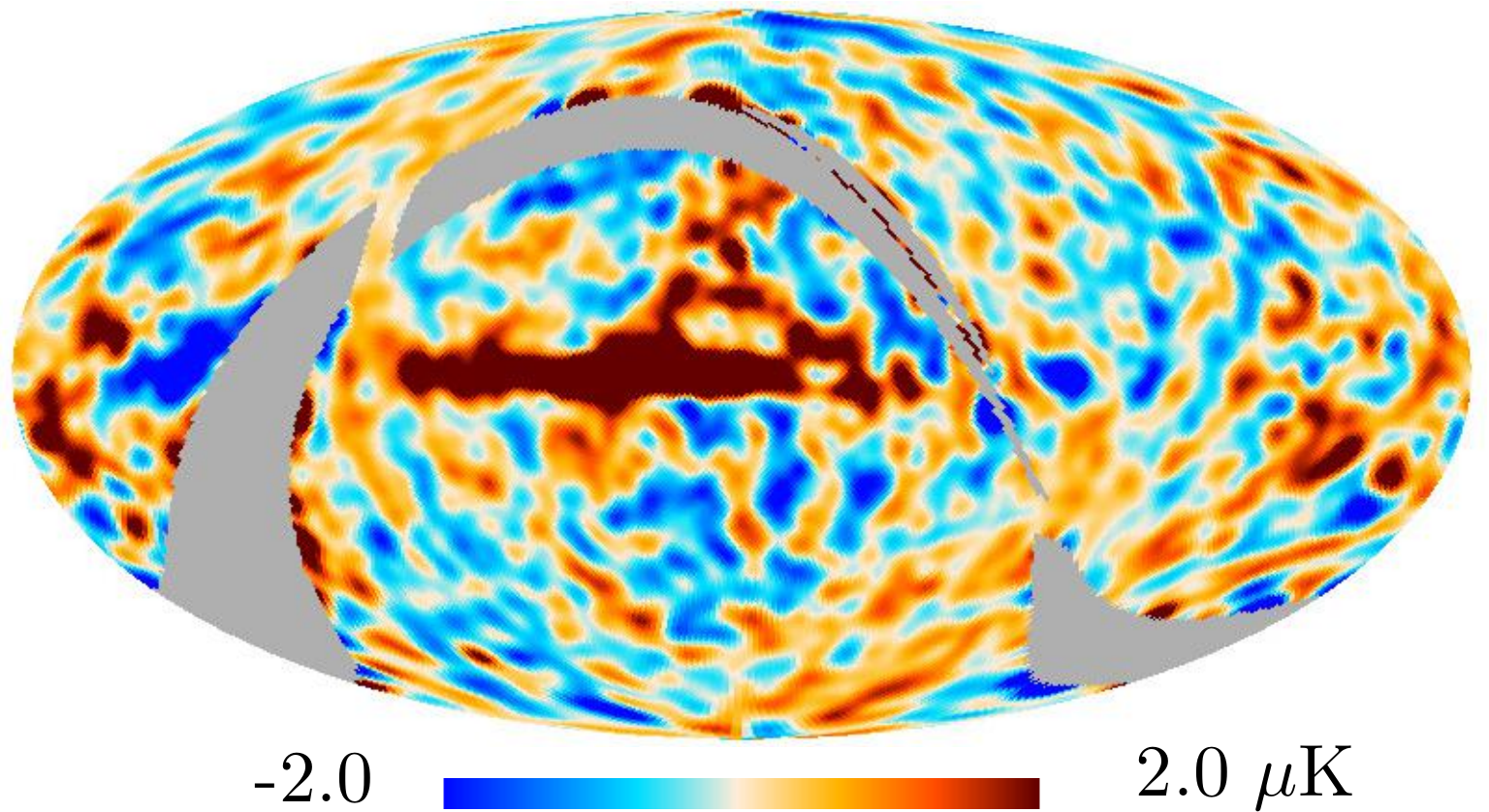}\\
	\raisebox{25pt}{\parbox{.03\textwidth}{353}}&
	\includegraphics [width=0.18\textwidth]{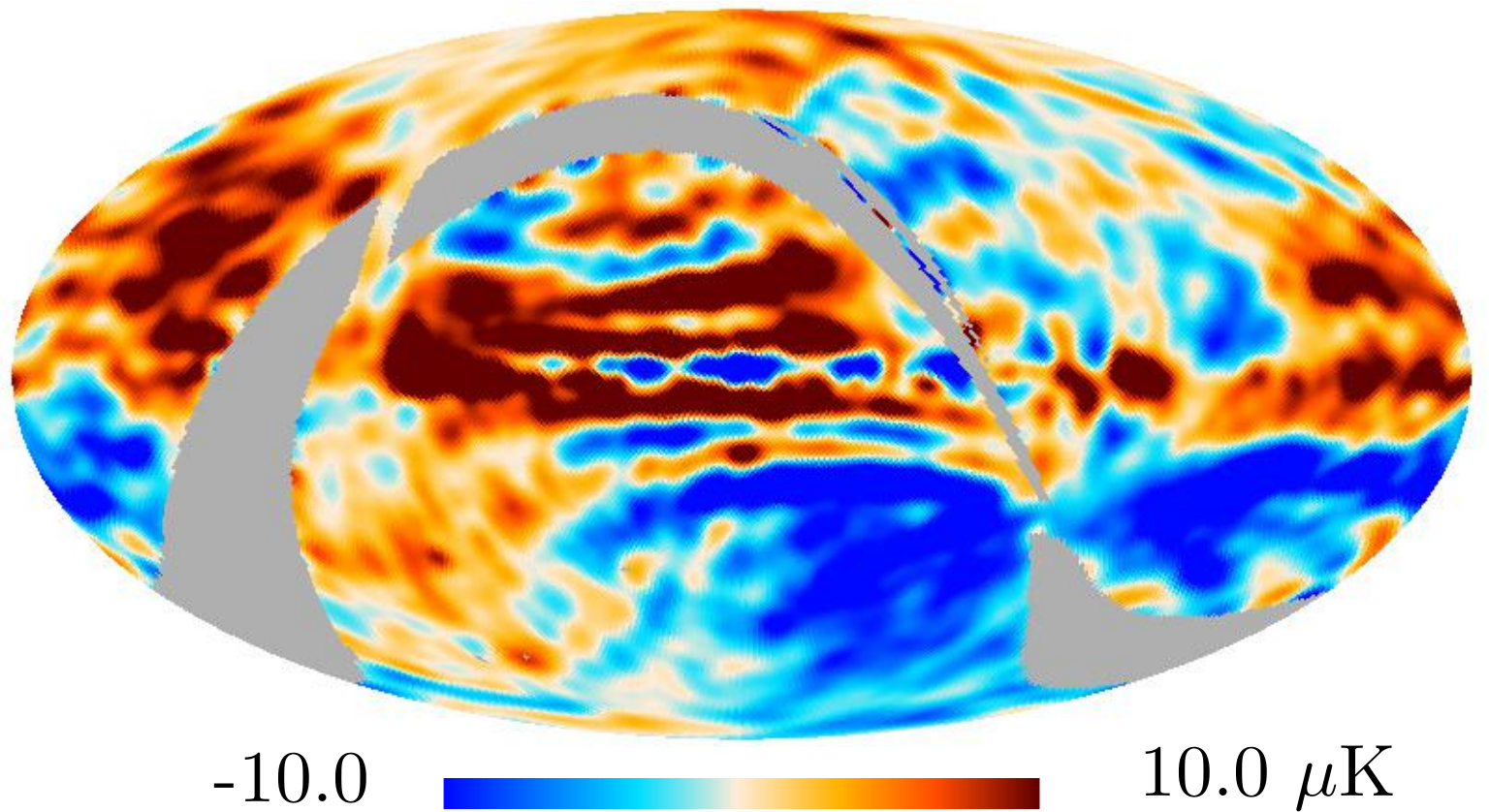}&
	\includegraphics [width=0.18\textwidth]{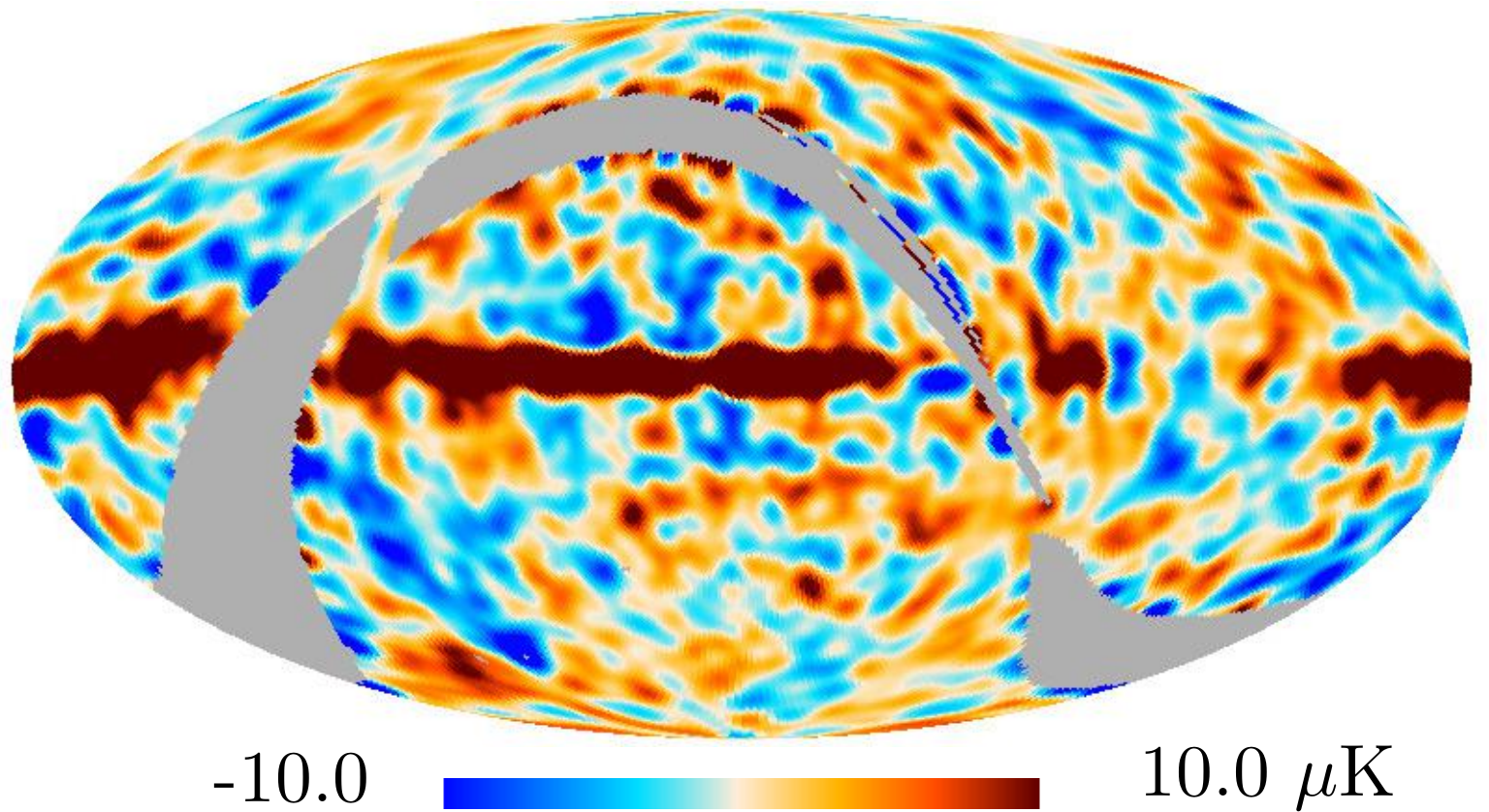}&
	\includegraphics [width=0.18\textwidth]{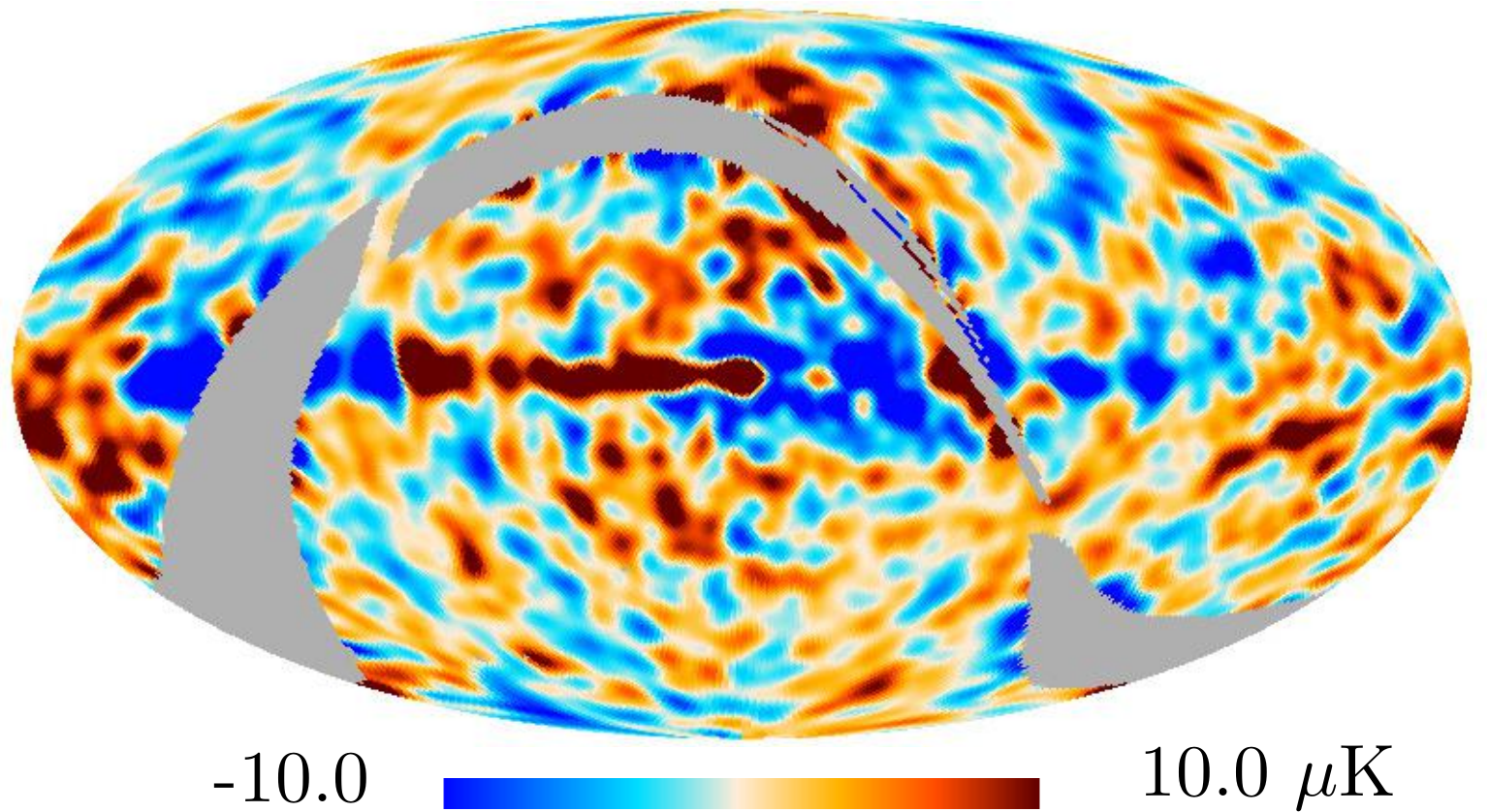}\\
	\raisebox{25pt}{\parbox{.03\textwidth}{545}}&
	\includegraphics [width=0.18\textwidth]{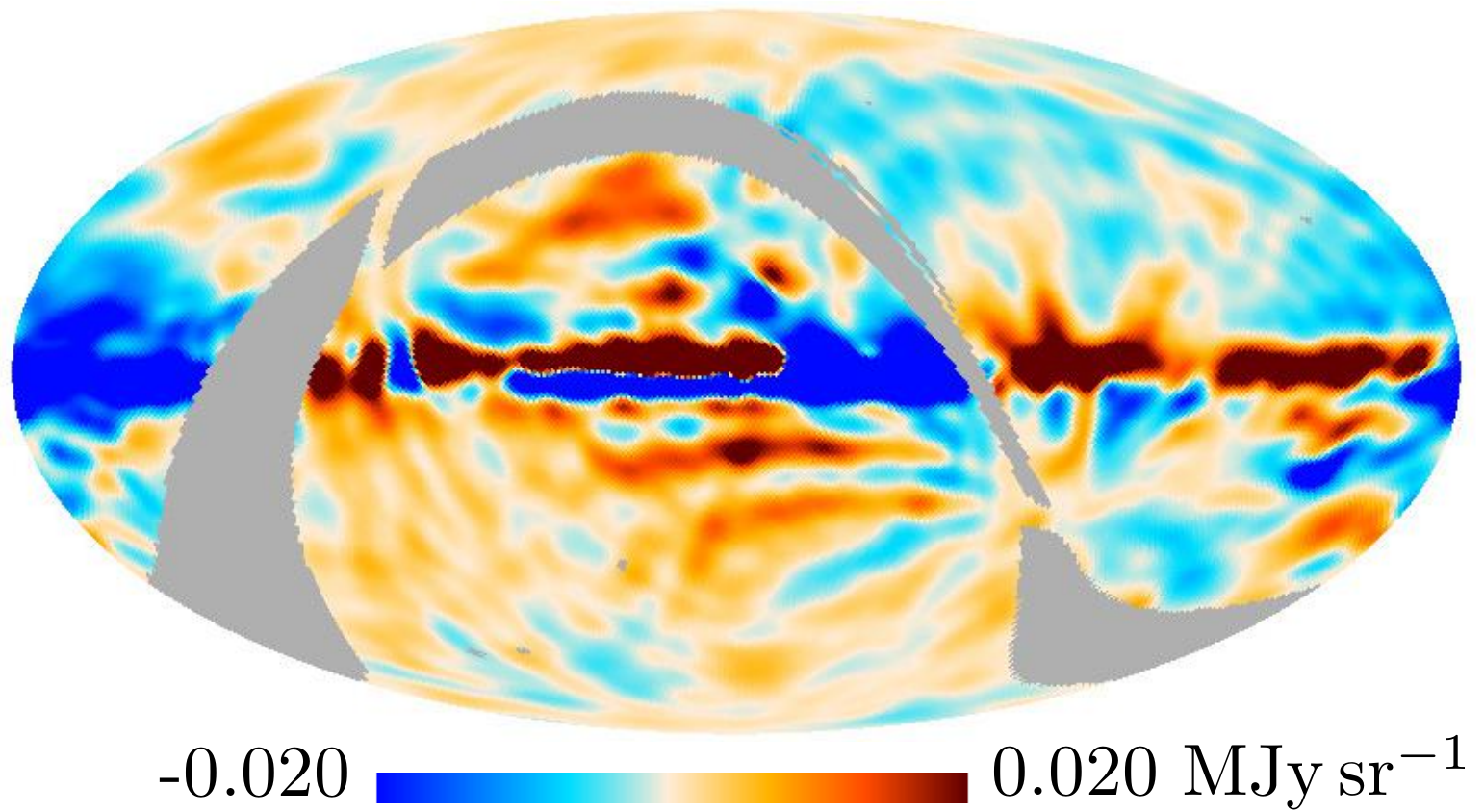}\\
	\raisebox{25pt}{\parbox{.03\textwidth}{857}}&
	\includegraphics [width=0.18\textwidth]{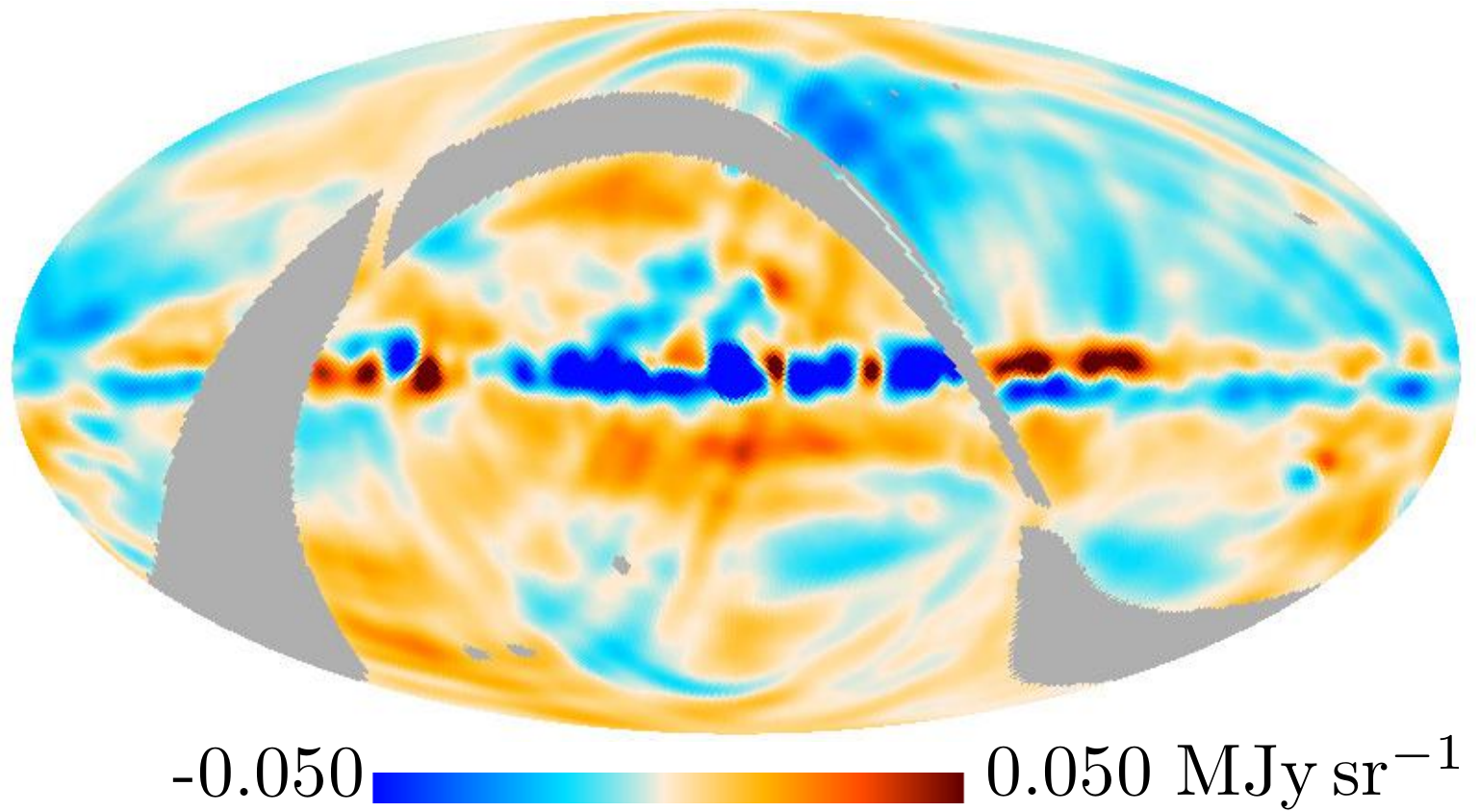}&
\end{tabular}
\end{center}
\caption{Survey differences of the full-mission ($({\rm S1}+{\rm S3})-({\rm S2}+{\rm S4})$) maps in $I$, $Q$, and $U$ (in columns) for the 2015 data (top panel) and the 2018 data (bottom panel). There are large improvements at CMB frequencies in the residuals for the 2018 data compared to those in the 2015 data. Bands are visible in the 353-GHz and 545-GHz intensity maps, due to the incomplete correction of the transfer function.}
\label{fig:s1234} 
\end{figure*}

For the 2013 data release, figure 10 of \citet{planck2013-pip88} presented the survey difference ${\rm S1} -{\rm S2}$, showing weak residuals of zodiacal bands at a level of 1--$2\times 10^{-2}{\rm MJy}\,{\rm sr}^{-1}$ at 857\,GHz and a negligible level at 545\,GHz. Survey differences of the full mission ($({\rm S1}+{\rm S3})-({\rm S2}+{\rm S4})$) for the 2015 and 2018 data are shown in Fig.~\ref{fig:s1234}, in which a 5\deg\ low-pass filter has been applied in order to reveal the 1-$\mu$K systematic residuals for CMB frequencies.
\renewcommand{\arraystretch}{0.5}
\renewcommand{\arraystretch}{1.}
At frequencies of 100 to 353\,GHz, comparisons of 2015 and 2018 5\deg\ smoothed maps show a dramatic reduction of the residuals for $I$, $Q$, and $U$, although systematic effects are still seen in the Galactic plane. At 143\,GHz, the large-scale residuals seen in the 2015 maps have disappeared, even in the Galactic ridge. At 100 and 217\,GHz, if one excludes the Galactic ridge, the same is true at high latitude, leaving noise-like residuals with peak-to-peak amplitudes at the 1-$\mu$K level, but with the main systematic effects not much below the noise on the largest scales, as will be shown in Sect.~\ref{sec:map_characterization}. This is the reason why such a survey null test is not used for the detection of systematic residuals in the full-mission 2018 maps, which, at these frequencies, will be shown to be dominated by other systematic effects.

In intensity at frequencies higher than 217\,GHz, Fig.~\ref{fig:s1234} shows a decrease, from the 2015 to the 2018 data, of the strong residuals, changing sign across the Galactic plane. Nevertheless, the 2018 data show the emergence of a new systematic effect not apparent in the 2015 data, the so-called ``zebra'' band striping (so-named to distinguish them from striping along the scan direction), more or less parallel to the Galactic plane. Their origin is explained in Sect.~\ref{sec:xferfunction}, resulting from only partial correction of the transfer function.

These maps still marginally show the narrow zodiacal bands at 857\,GHz at a weak level in the un-smoothed maps (not displayed). Figure~\ref{fig:s1234} shows that the 2015 data contain signatures at 545 and 857\,GHz that are typical of FSL at high Galactic latitudes; the central Galactic disc region aligned with long features of the FSL can be seen. These are significantly reduced and barely visible in the 2018 submillimetre maps.

\begin{figure}[htbp!]
\includegraphics[width=\columnwidth]{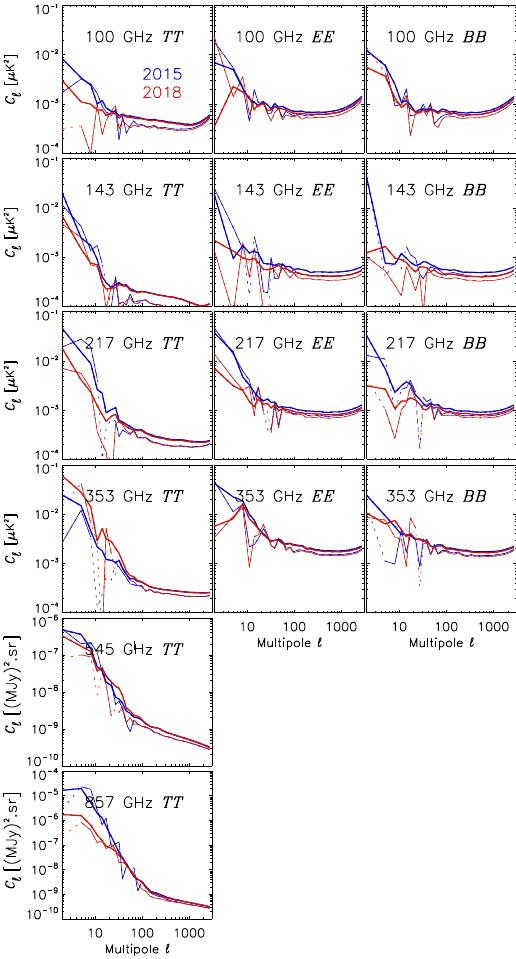}
\caption{Power spectra of the maps shown in Fig.~\ref{fig:s1234} not corrected for the sky fraction. Here blue is for the 2015 data and red for 2018. Thick and thin lines are for sky fractions of 43\,\% and 80\,\%, respectively. Dashed lines indicate negative values.}
\label{fig:diffdx11dx12} 
\end{figure}

We quantify the impact of this systematic effect on the power spectra in Fig.~\ref{fig:diffdx11dx12}, showing the power spectra\footnote{Throughout this paper, we denote by $C_{\ell}$ and $D_{\ell}$ ($\equiv\ell(\ell+1)C_\ell/2\pi$) the deconvolved spectra and pseudo-power spectra (using {\tt Spice}), respectively.} associated with the 2015 and 2018 maps, for two sky fractions, 43\,\% and 80\,\%.
The difference between the 43 and $80\,\%$ results for the white noise at $\ell\,{>}\,100$ is mostly accounted for by the different sky area (which was not corrected for). From 100 to 217\,GHz, the 2018 spectra at low multipoles are all well below the 2015 levels. This is not true at 353\,GHz for the $TT$ spectra, for the reasons mentioned above. The $EE$ and $BB$ spectra are still at the $10^{-2}\microKcarre$ level, due to the very long time constant transfer function not being corrected well enough. The zebra bands are seen as peaks in the $EE$ and $BB$ power spectra around $\ell\,{=}\,8$ and 20 at 353\,GHz. The power spectra at 545 and 857\,GHz show a big rise over the noise in the 2015 data, which is much reduced in the 2018 data and which, to first order, does not depend on the sky fraction. We have also tested through simulations that this procedure does not introduce significant artefacts.

\subsubsection{Power spectra null tests on the data}
\label{sec:powerspectra}

\begin{figure*}[htbp!]
\includegraphics[width=\textwidth]{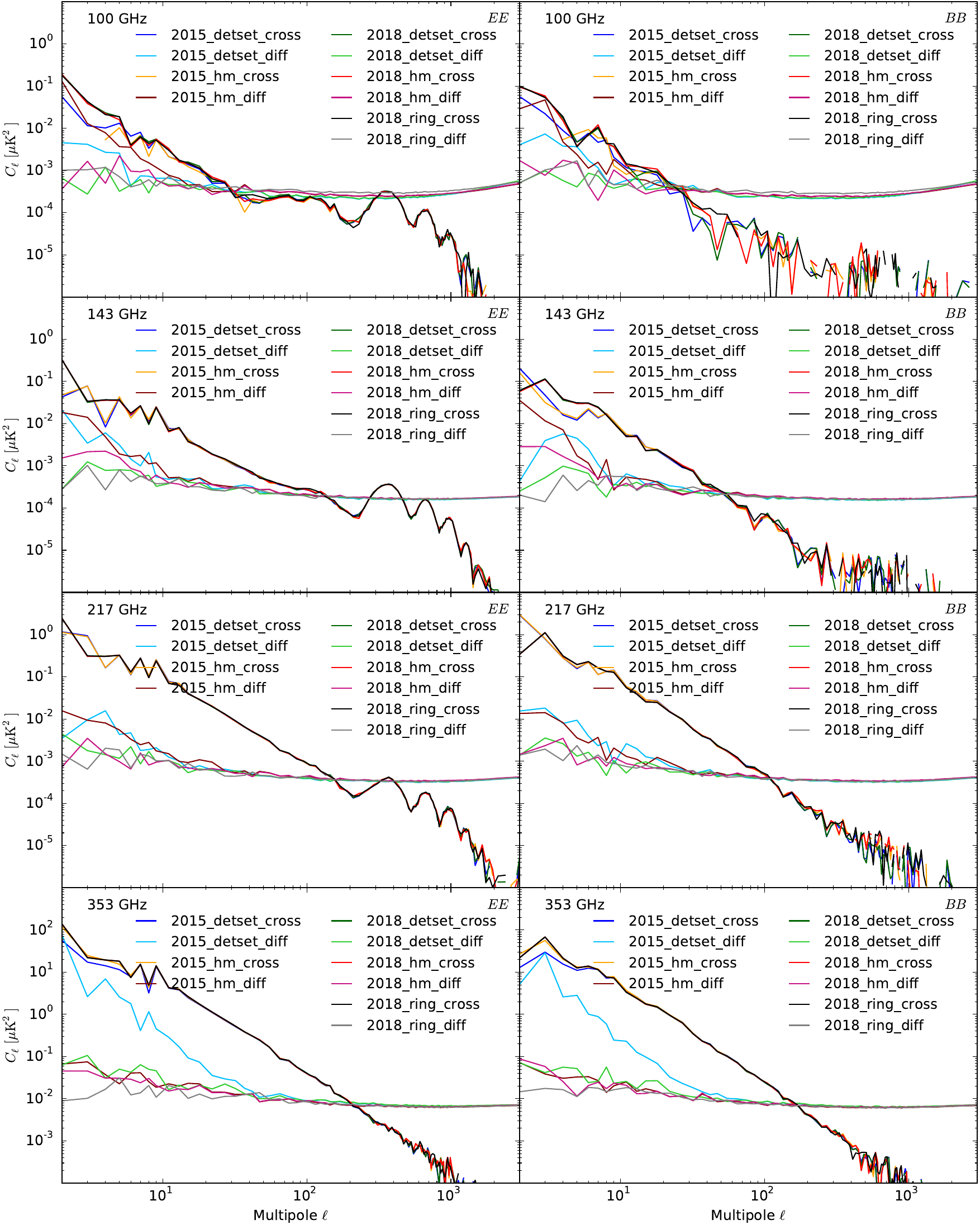}
\caption{$EE$ and $BB$ spectra of the 2015 and 2018 detset, half-mission, and rings (for 2018 only) maps at 100, 143, 217, and 353\,GHz. The full-mission auto-spectra of the difference maps, corrected for sky fraction, and the cross-spectra between the maps are shown. The sky fraction used here is 43\,\%. The binning is : $\delta\ell=1$ for $2\leq\ell<30$; $\delta\ell=5$ for $30\leq\ell<50$; $\delta\ell=10$ for $50\leq\ell<160$; $\delta\ell=20$ for $160\leq\ell<1000$; and $\delta\ell=100$ for $\ell>1000$. Fig.~\ref{fig:JKspectradiff} shows an enlargement of part of these spectra.}
\label{fig:JKspectraDX11vsRC4} 
\end{figure*}

Figure~\ref{fig:JKspectraDX11vsRC4} uses a suite of maps built from half split-data sets, namely detsets, half missions, and rings. It shows $EE$ and $BB$ power spectra of differences and cross-spectra of such maps for the 2015 and 2018 data. This gives another sensitive and quantitative estimate of the level of improvements in 2018 over the 2015 release.

The splits used in the 2015 release were detsets and half-mission sets. For the 2018 data release, we add the ring sets, replacing the half-ring ones used in 2015, which introduced correlated noise. These are sensitive to systematics that are stable in time (for example, mismatch in intensity-to-polarization leakage or scanning strategy). Conversely, the half-mission split is mostly sensitive to long-time drifting systematic effects, like the ADCNL effect, and are insensitive to the scanning strategy. The 2018 detset split (green lines) is very much improved from 2015 (blue lines), through the use of \sroll, which very accurately extracts from the data inter-calibration and bandpass-mismatch coefficients. This brings the detset differences at $\ell\,{>}\,30$ below those of the other null tests. At $\ell\,{<}\,30$, the improvement between 2015 and 2018 is striking.

The cross-spectra show the sky signal up to a very high multipole limit, where the chance correlations of the noise starts to hide the signal. The spectra of the differences show the noise plus systematic residuals, including differences in distortions of the sky signal between the two halves. The power spectra of the differences are normalized for the full data set, but not corrected for the sky fraction used (43\,\%).

The ring null-test results (grey lines) are close to those of the detset and also to the FFP8 TOI noise one, as shown in figure~18 of \citelowell. At all frequencies and in both $EE$ and $BB$, the three 2018 null tests are within a factor 2--5 of the white noise extrapolation, even at very low multipoles. The half-mission null tests show a higher level (2--$3 \times 10^{-3}\microKcarre$) at very low multipoles, explained by the ADCNL residual analysis (Sect.~\ref{sec:ADC}).

At the lowest multipoles, the two 2015 difference spectra (detset and half-mission), were higher than the white noise by factors of 10--30 at CMB frequencies. At 353\,GHz, in 2015, the very low multipole detset null test was a factor of 100 higher than the white noise and a factor of more than 30 higher than the 2018 one. In the half-mission null test, the improvement is only a factor of 2--4. The opposite behaviour is seen at the lower frequencies, i.e., the detset null test is higher than the half-mission one. The 217-GHz spectra show an intermediate behaviour.

\begin{figure}[htbp!]
\includegraphics[width=\columnwidth]{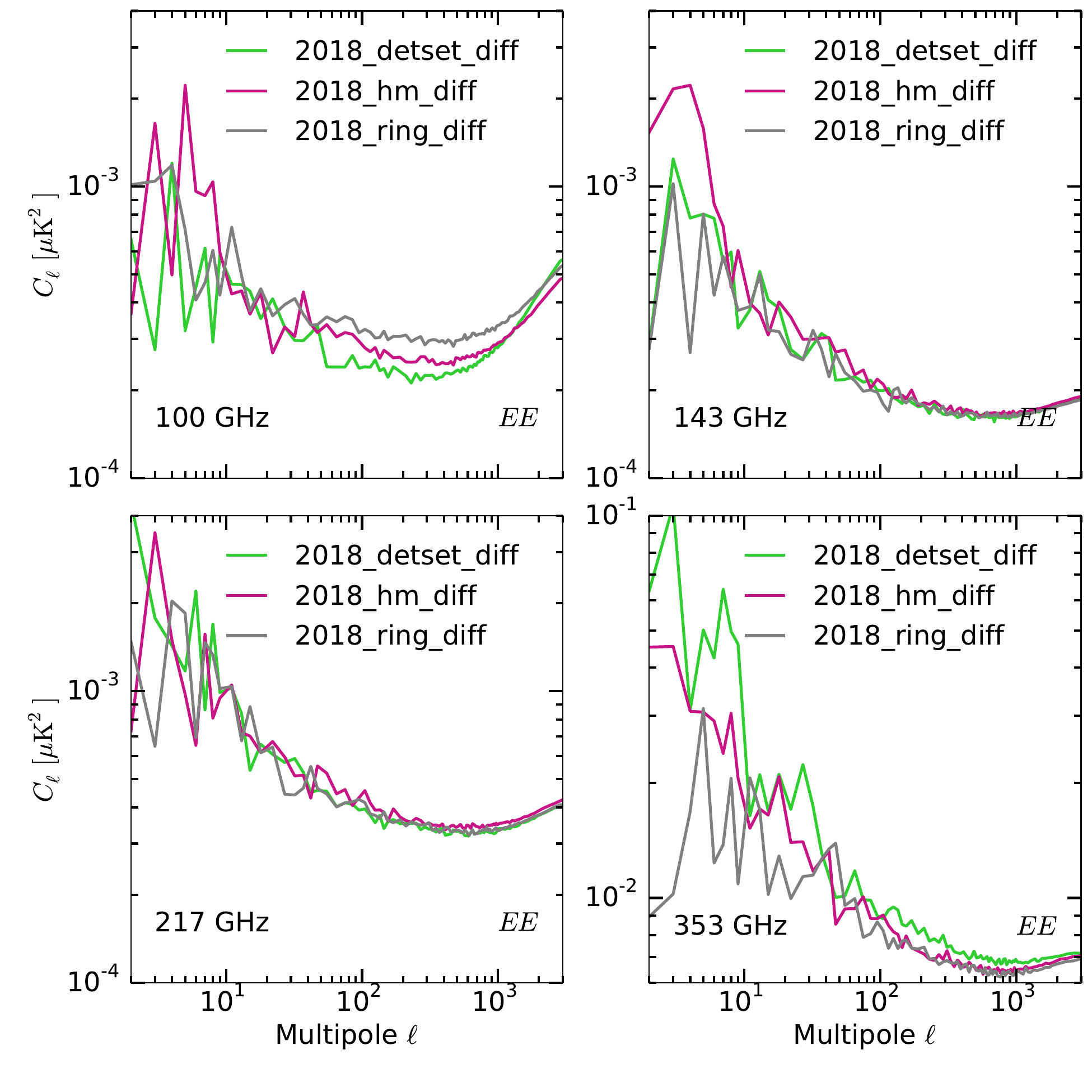}
\caption{Enlargement of Fig.~\ref{fig:JKspectraDX11vsRC4} restricted to $EE$ auto-spectra of the 2018 detset, half-mission, and ring difference maps at 100, 143, 217, and 353\,GHz on 43\,\% of the sky.}
\label{fig:JKspectradiff} 
\end{figure}

Figure~\ref{fig:JKspectradiff} shows an enlargement of the 2018 $EE$ auto-spectra of detset, half-mission, and ring map differences.
At 100 and 143\,GHz, the half-mission differences rise above the two others at low multipoles where the ADCNL dominates. Above $\ell\,{=}\,10$, at all frequencies, the half-mission differences also show a slightly steeper decrease of power with multipoles than the two other null tests. At 100\,GHz the ring null test is the highest at high multipoles. Both detset and half-mission tests show correlated systematic effects at very significant levels. At 100\,GHz, due to the different polarization angle distribution, pixels badly configured for solving the polarization are more numerous, since the ring split leaves a lower redundancy of observations per pixel and leads to an excess noise (as discussed in the next section). At 353\,GHz, the detset spectrum dominates at all multipoles.

{\it{The difference between these three null tests gives an approximate measure of how the noise plus the systematic residuals change, depending on what systematic effects a given null test is sensitive to.}} The estimates of systematic effects, presented in Sect.~\ref{sec:map_characterization}, give the best quantitative indications on this matching of null tests and systematic effects.

\subsection{Statistical analysis of noise and systematics residuals}

The E2E simulations contain the best statistical representation of the knowledge of the uncertainties after processing. A tool often used to assess these uncertainties is the ``probability to exceed'' (PTE), which assumes that a null test removes the signal and gives the total noise map, or spectrum, with no signal bias and Gaussian statistics. Meaningful PTE values on null tests could be constructed before likelihood analysis, in the days when CMB experiments were dominated by detector noise, and at HFI frequencies where the CMB dominates over the foregrounds (e.g., temperature maps near the peak of the CMB). When high sensitivity polarization data are dominated by systematic effects and foregrounds, this cannot be done meaningfully, since correlated systematic effects, or foregrounds between the two halves, makes the PTE not informative enough. This explains why our three null tests give significantly different power spectra for the noise plus systematic residuals (Fig.~\ref{fig:JKspectradiff}).

We will discuss in this section the extent to which our E2E simulations are statistically representative, and do this separately for low and high multipoles. For low multipoles, we directly compare the data with noise plus the signal's statistical distribution from the simulations for the two null tests, and do this multipole by multipole. For high multipoles, we bin every ten multipoles and compare the simulated noise plus systematics from null tests in the data with the same quantity for the simulations, including their dispersion.

The results show that the simulations do well at reproducing the differences between the two null tests observed in the data and that the use of a single null test for the likelihood analysis will not do as well.

\subsubsection{Low multipoles}

\begin{figure*}[htbp!]
\includegraphics[width=\textwidth]{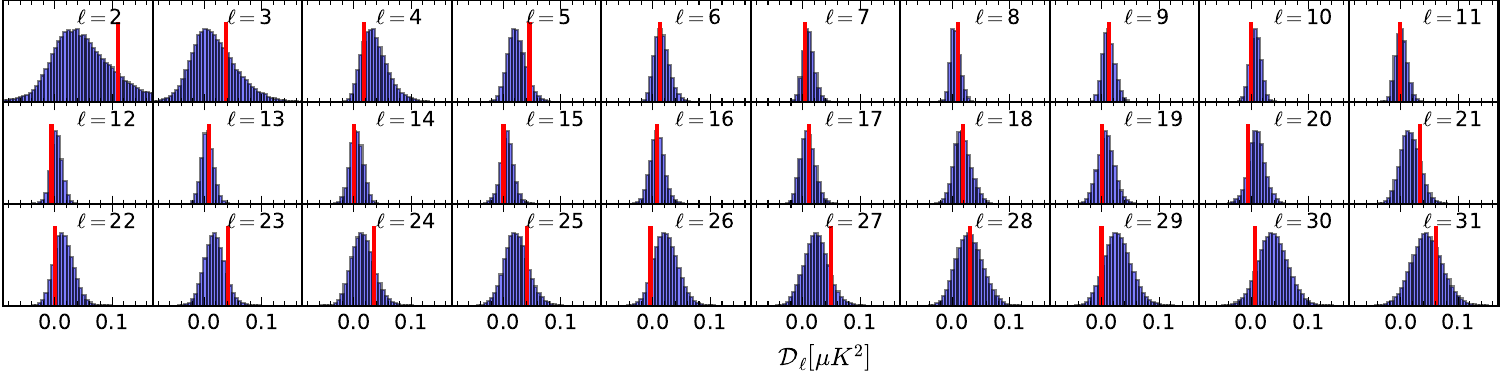}
\includegraphics[width=\textwidth]{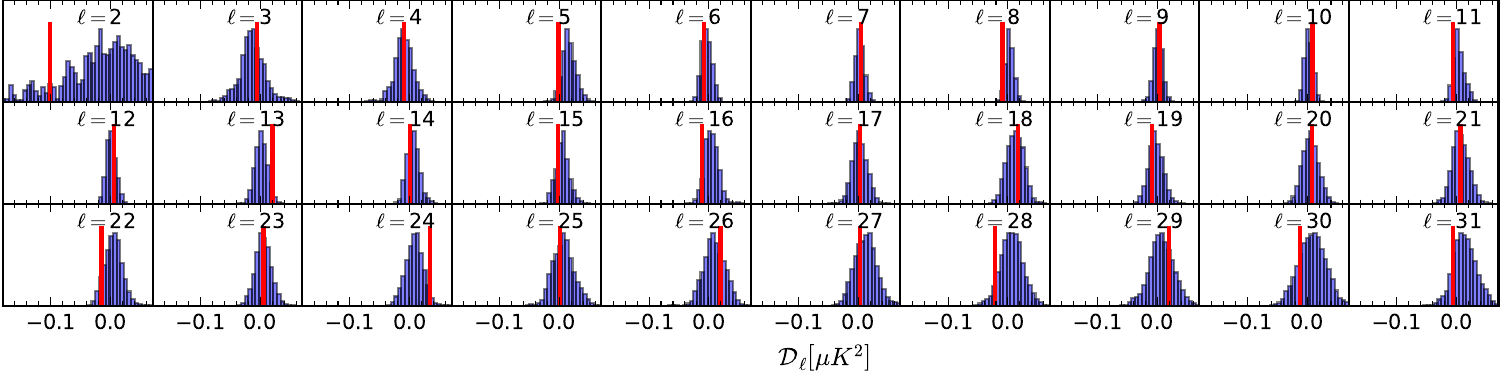}
\includegraphics[width=\textwidth]{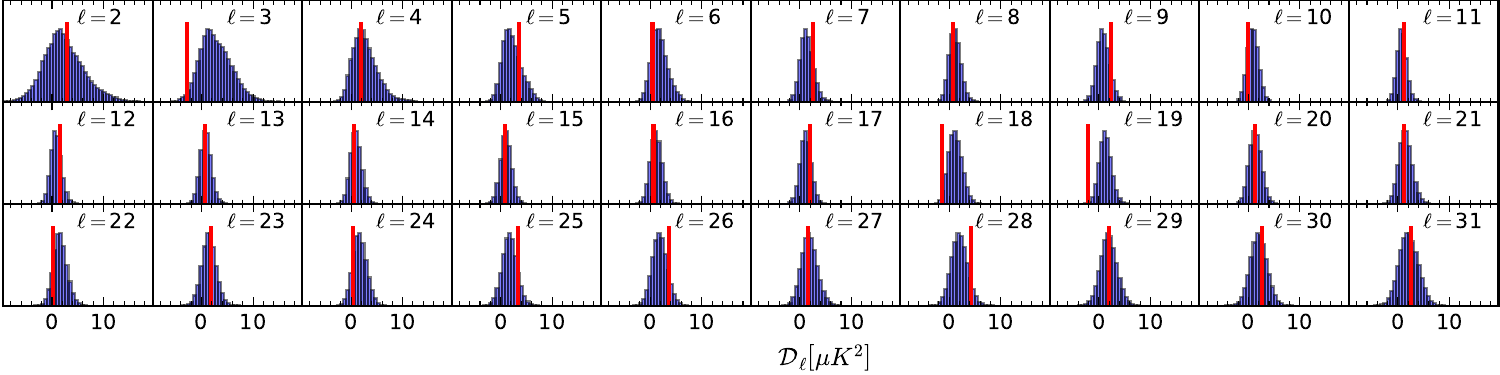}
\includegraphics[width=\textwidth]{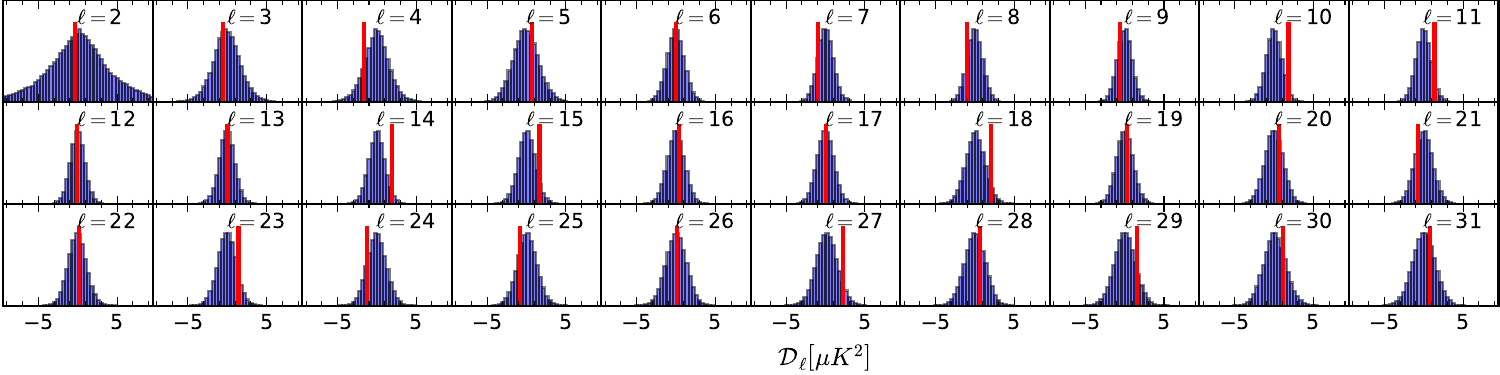}
\includegraphics[width=\textwidth]{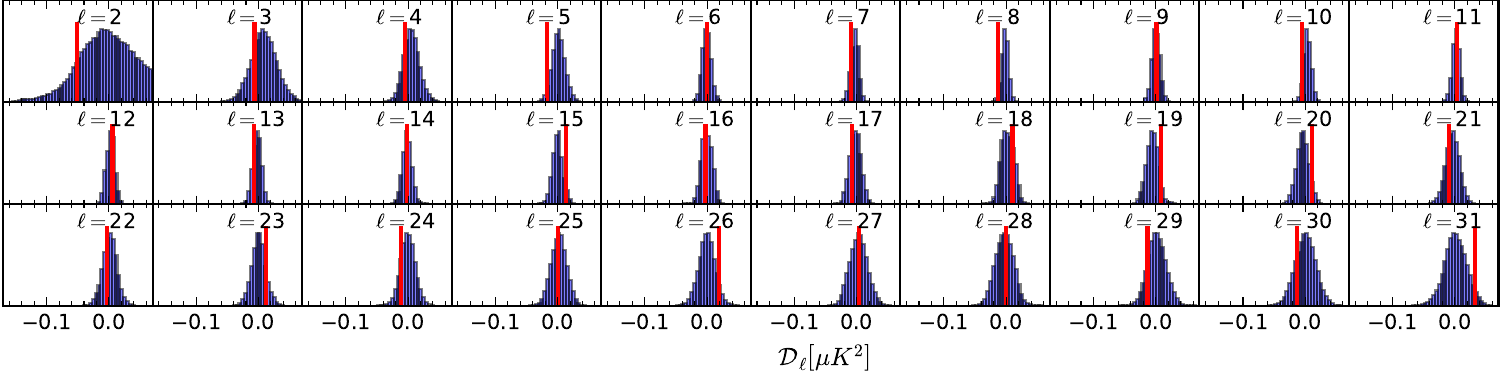}
\caption{Probability distribution for each multipole ($\ell=2$--31) of the 100$\times$143 cross-spectra, from the top panels to the bottom ones, showing $EE$, $BB$, $TE$, $TB$, and $EB$. Red vertical lines show the values for the data. The associated PTEs are given in Table~\ref{tab:PTE}.}
\label{fig:pteLuca} 
\end{figure*}

In Fig.~\ref{fig:pteLuca}, for the 100$\times$143 cross-spectra, we show the empirical distributions for each multipole up to $\ell\,{=}\,31$ for $EE$, $BB$, $TE$, $TB$, and $EB$ spectra of 300 noise plus systematic simulations, combined with 1000 signal realizations.\footnote{The signal realizations are extracted from a CMB fiducial model with the following cosmological parameters: $10^{9}A_{\rm s}=2.1$; $\tau=0.05$; and $r=0$.} The non-Gaussian character of the distribution for the lowest multipoles is very clear, especially in the case of no (or very low) signal.
\begin{table}[htbp!]
\newdimen\tblskip \tblskip=5pt
\caption{Probabilities to exceed (in percentage), related to Fig.~\ref{fig:pteLuca}.}
\label{tab:PTE}
\vskip -3mm
\footnotesize
\setbox\tablebox=\vbox{
\newdimen\digitwidth
\setbox0=\hbox{\rm 0}
\digitwidth=\wd0
\catcode`*=\active
\def*{\kern\digitwidth}
\newdimen\signwidth
\setbox0=\hbox{+}
\signwidth=\wd0
\catcode`!=\active
\def!{\kern\signwidth}
\newdimen\pointwidth
\setbox0=\hbox{.}
\pointwidth=\wd0
\catcode`?=\active
\def?{\kern\pointwidth}
\halign{\hbox to 2.0cm{#\leaderfil}\tabskip 2em&
\hfil#\hfil\tabskip 1.5em&
\hfil#\hfil&
\hfil#\hfil&
\hfil#\hfil&
\hfil#\hfil\tabskip 0em\cr
\noalign{\doubleline}
\omit\hfil Multipole\hfil& $EE$& $BB$& $TE$& $TB$& $EB$\cr
\noalign{\vskip 3pt\hrule\vskip 5pt}
*2& 51.5& 16.5& 92.2& 65.6& 12.6\cr
*3& 66.1& 54.3& *7.0& 74.7& 42.0\cr
*4& 39.3& 73.1& 78.9& 33.0& 50.1\cr
*5& 15.9& 17.5& 54.6& 46.5& *8.0\cr
*6& 87.9& 79.0& 42.2& 99.9& 85.5\cr
*7& 40.1& 96.7& 67.6& 26.4& 30.3\cr
*8& 47.6& *7.9& 75.4& 37.8& *2.7\cr
*9& 84.6& 69.0& 32.4& 60.3& 83.6\cr
10& 34.2& 35.2& 44.9& *5.3& 10.7\cr
11& 69.6& 38.8& 82.6& 27.4& 89.1\cr
12& 36.5& 58.2& 55.3& 89.4& 50.8\cr
13& 90.4& 10.3& 84.3& 82.3& 37.2\cr
14& 46.7& 80.4& 59.5& 12.5& 83.6\cr
15& 57.1& 50.3& 89.7& 30.9& *9.1\cr
16& 99.2& 23.6& 67.9& 79.7& 76.7\cr
17& 84.7& 50.9& 53.5& 90.5& 55.0\cr
18& 93.8& 72.7& *1.6& 14.6& 70.3\cr
19& 49.5& 76.9& *1.2& 91.3& 14.0\cr
20& 16.3& 64.5& 91.9& 55.8& *5.4\cr
21& 34.0& 97.0& 86.7& 57.2& 69.7\cr
22& 51.2& 13.4& 31.8& 83.5& 97.6\cr
23& 57.2& 74.6& 99.0& 23.1& 22.6\cr
24& 33.5& 16.0& 41.1& 13.0& 40.2\cr
25& 49.0& 92.2& 22.7& 56.5& 94.7\cr
26& 46.7& 90.0& 20.5& 50.0& *7.8\cr
27& 28.1& 43.0& 57.7& *9.4& 74.2\cr
28& 84.0& 11.1& *9.4& 54.0& 26.7\cr
29& 16.1& 44.1& 94.9& 46.3& 25.3\cr
30& 39.7& 57.9& 56.7& 63.1& 56.5\cr
31& 70.8& 36.1& 78.6& 79.4& *6.9\cr
32& 27.9& 20.9& 40.2& 16.8& 73.0\cr
\noalign{\vskip 3pt\hrule\vskip 5pt}}}
\endPlancktable
\end{table}
We can compare these distributions with the power observed in the data. Simulations and data are processed with the following procedure:
\begin{itemize}
\item in temperature, use the {\tt SMICA} CMB solution for the data and pure CMB realizations for the simulations;
\item in polarization, use template-fitting component separation incorporating 30\,GHz and 353\,GHz maps both for data and simulations;
\item adopt quadratic maximum likelihood (QML) power-spectrum estimation using 94\,\% of the sky in temperature and 50\,\% of the sky in polarization.
\end{itemize}
In each panel of Fig.~\ref{fig:pteLuca}, we compute the two-tailed PTE value (reported in Table~\ref{tab:PTE}). This is obtained by integrating left and right over the distribution, starting from the mean until we reach the data value and computing the corresponding PTE. The PTEs are almost everywhere within $2\,\sigma$ probabilities, with only a few not significant outliers between $2$ and $3$ $\sigma$ at intermediate multipoles (i.e., $\ell=18,19$ in $TE$ and $\ell=8$ in $EB$), showing that, at large scales, the data are well described by the E2E simulations.

The low multipole likelihood \citep{planck2016-l05} uses the full statistics of the 300 simulations to derive the likelihood of the cosmological parameters that are sensitive to polarized low multipoles in $EE$, $BB$, and $TE$, namely $\tau$, $A_{\rm s}$, and $r$ ($TB$ and $EB$ are also useful to test that they are compatible with zero cosmological signal). For other cosmology studies (e.g., isotropy and statistics or $B$ modes from lensing), specific tests need to be adopted, using statistics based on the full set of simulations.

\subsubsection{High multipoles}
\label{sec:highmultipoles}

\begin{figure*}[htbp!]
\includegraphics[width=\textwidth]{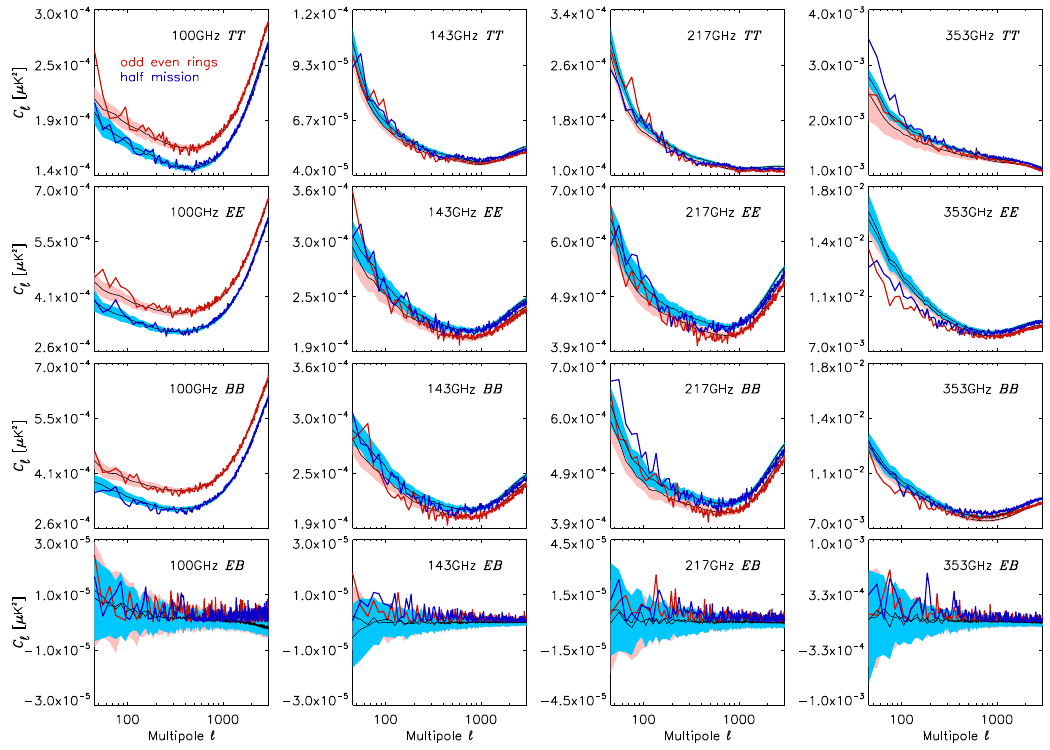}
\caption{Noise and systematic residuals in $TT$, $EE$, $BB$, and $EB$ spectra, for difference maps of the ring (red) and half-mission (blue) null tests binned by $\Delta \ell =10$. Spectra are given for the full mission and corrected for the 70\,\% sky fraction. Data spectra are represented by thick lines, and the averages of simulations by thin black lines. For the simulations, we show the 16\,\% and 84\,\% quantiles of the distribution with the same colours. The linear $y$-axis scales are adapted to show the full ranges on each panel.}
\label{fig:newpte} 
\end{figure*}
At high multipoles, we compare $TT$, $EE$, $BB$, and $EB$ cross-spectra of data and 100 E2E simulations\footnote{We ran 300 end-to-end simulations for the final analysis and likelihood, but for some tests, a smaller number gives accurate enough results and only the first 100 iterations have been used.} in Fig.~\ref{fig:newpte} using 70\,\% of the sky. In temperature and polarization, for $\ell>100$ at 100, 143, and 217\,GHz, and for $\ell>300$ at 353\,GHz, the data are well within the range of the 16\,\% and 84\,\% quantiles of the simulation distribution. In polarization, we see significant correlated excursions outside of this range for all frequencies in $EE$ and $BB$. For the CMB channels, these are smaller than $5\times10^{-5}$\microKcarre. The quasi-white noise levels (measured from $\ell=200$ to 2000) are higher by 20\,\% than the corresponding spectra in Fig.~\ref{fig:JKspectradiff}, although these spectra have been corrected to first order for sky fraction. The differences are due to the large-scale distribution of noise induced by the scanning strategy. 

\begin{figure}[htbp!]
\includegraphics[width=\columnwidth]{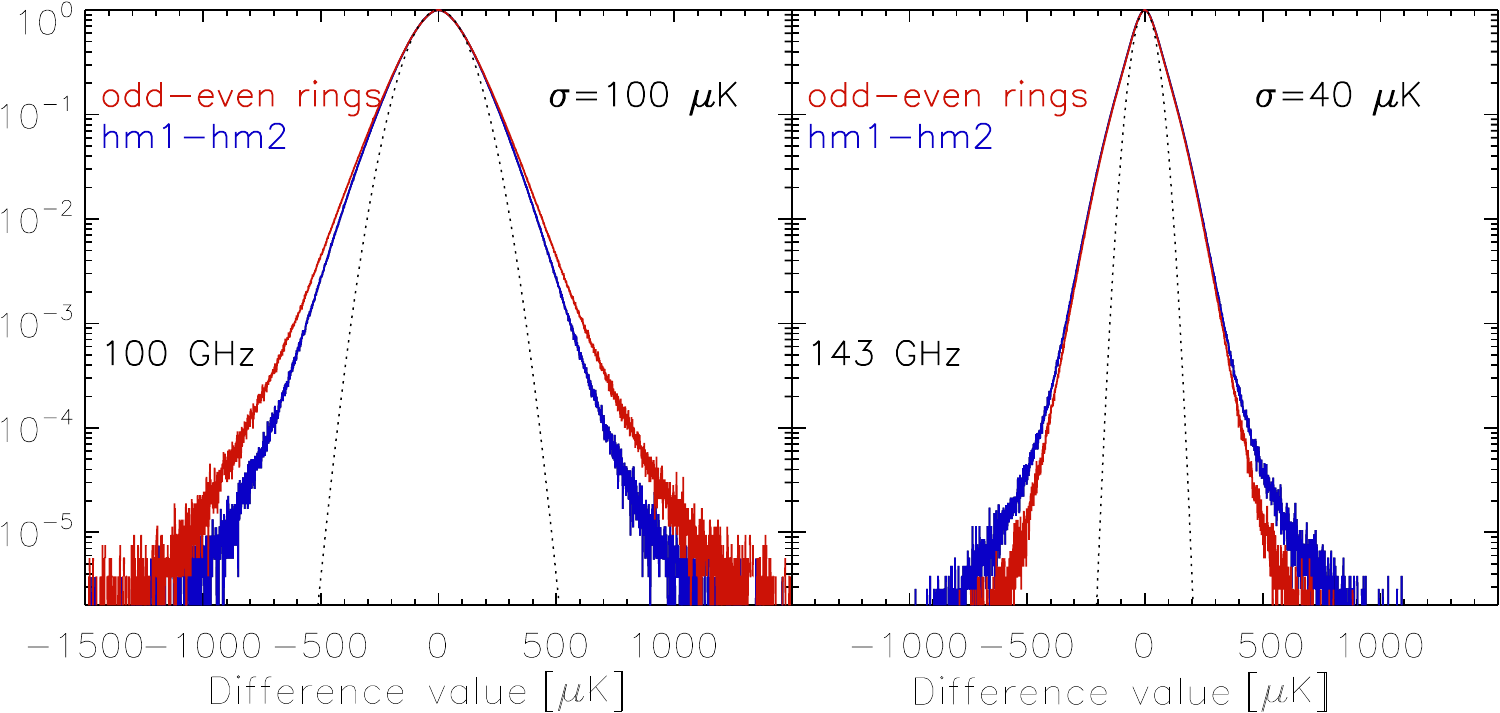}
\caption{Histogram of the rings and half-mission difference 100 and 143-GHz polarization maps at $N_{\rm side}\,{=}\,2048$, showing the different behaviour in the wings of the two null tests.}
\label{fig:ploestat1} 
\end{figure}

Figure~\ref{fig:ploestat1} shows the statistical distributions of pixels in the difference maps built with the two null tests.
These distributions show a common large kurtosis compared to Gaussian distributions (in dashed lines) expected from the scanning strategy and its associated distribution of the noise in the map. Nevertheless, in the wings, the two null tests show different behaviour; the kurtosis is higher for the ring null test at 100\,GHz and for the half-mission null test at 143\,GHz, reflecting the differences in noise level seen in Fig.~\ref{fig:newpte}. Furthermore, in all 100\,GHz spectra, there is a very significant difference between the two null tests, which is well modelled by the simulations.

\begin{figure}[htbp!]
\begin{center}
\includegraphics[width=0.48\columnwidth]{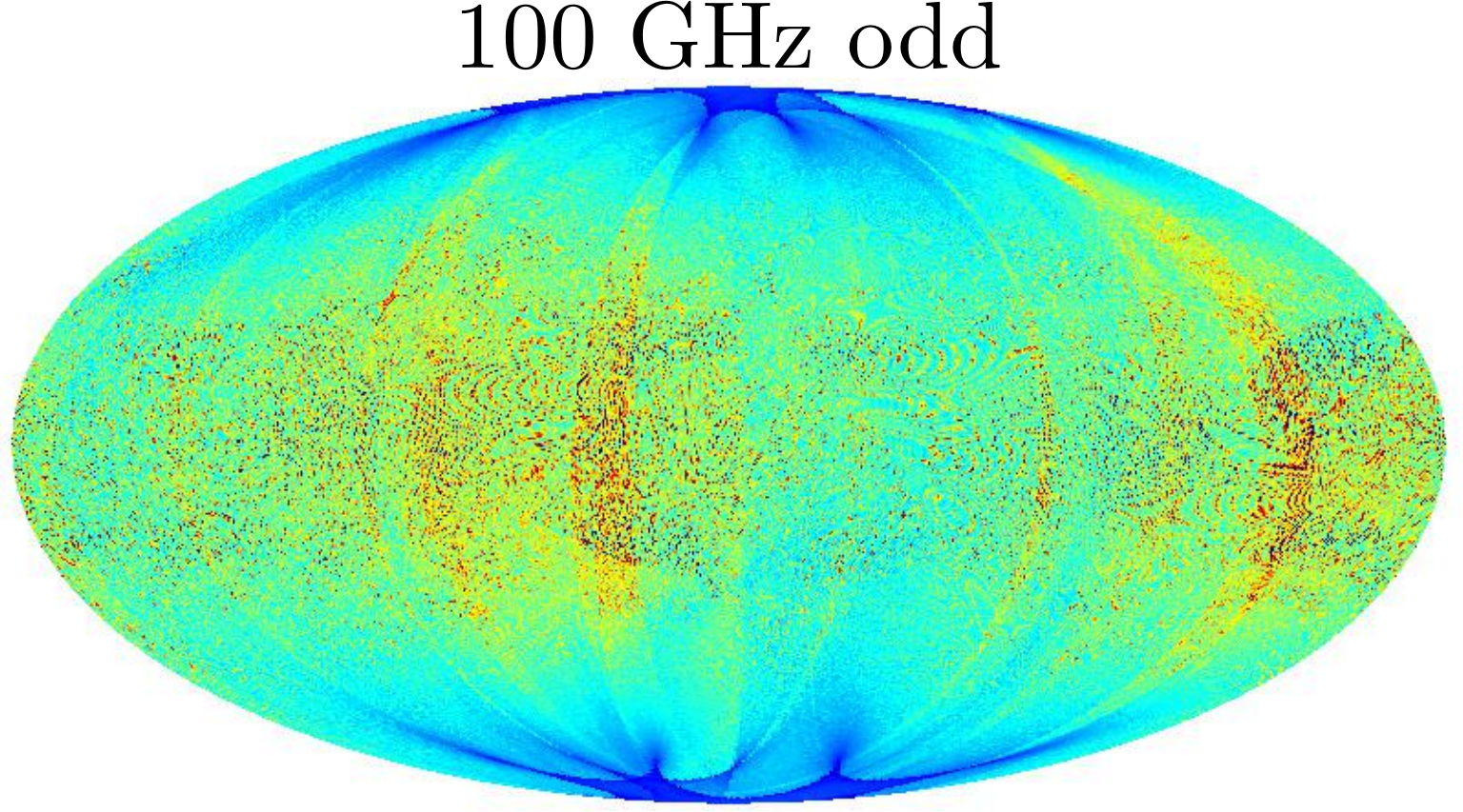}
\includegraphics[width=0.48\columnwidth]{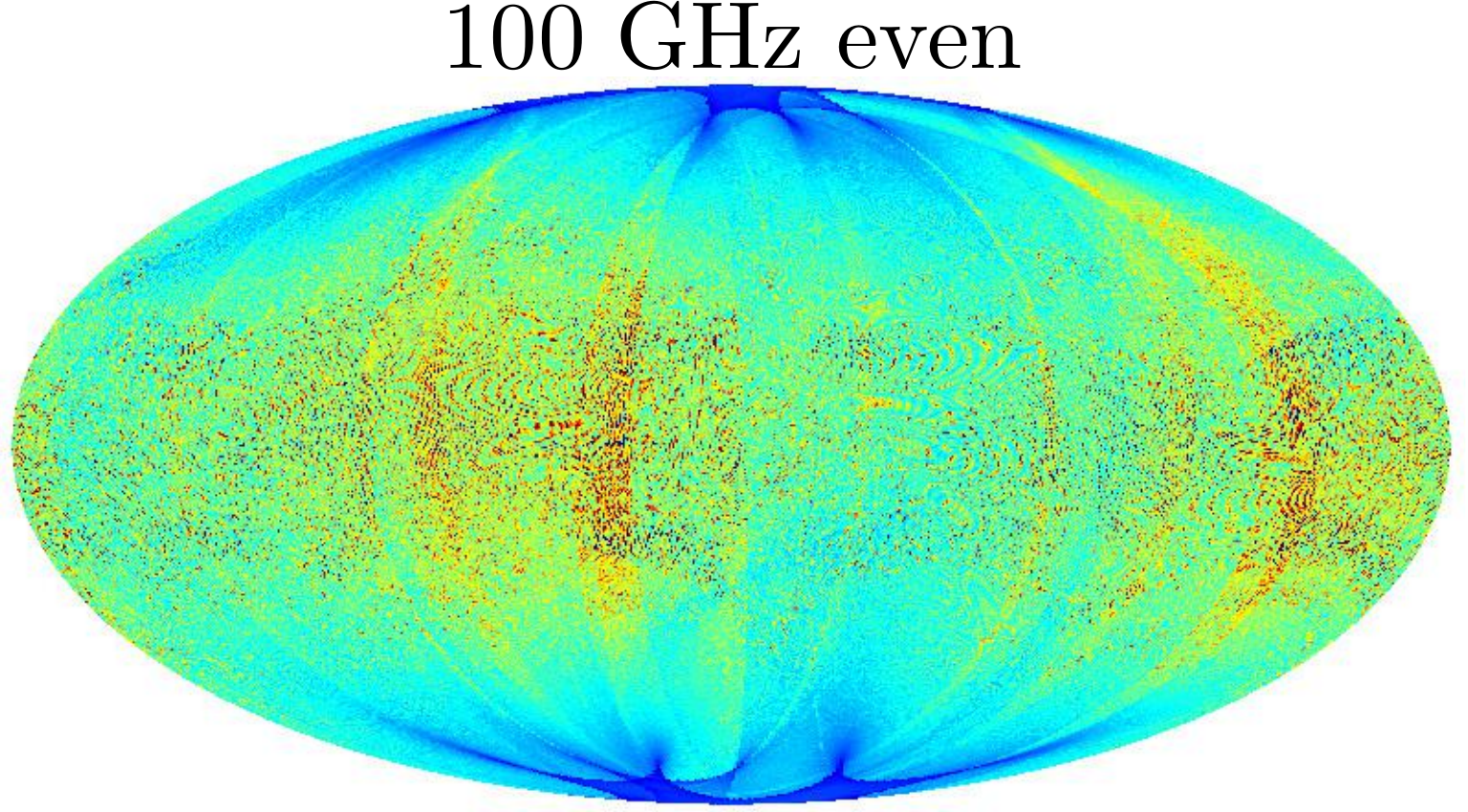}\\
\includegraphics[width=0.48\columnwidth]{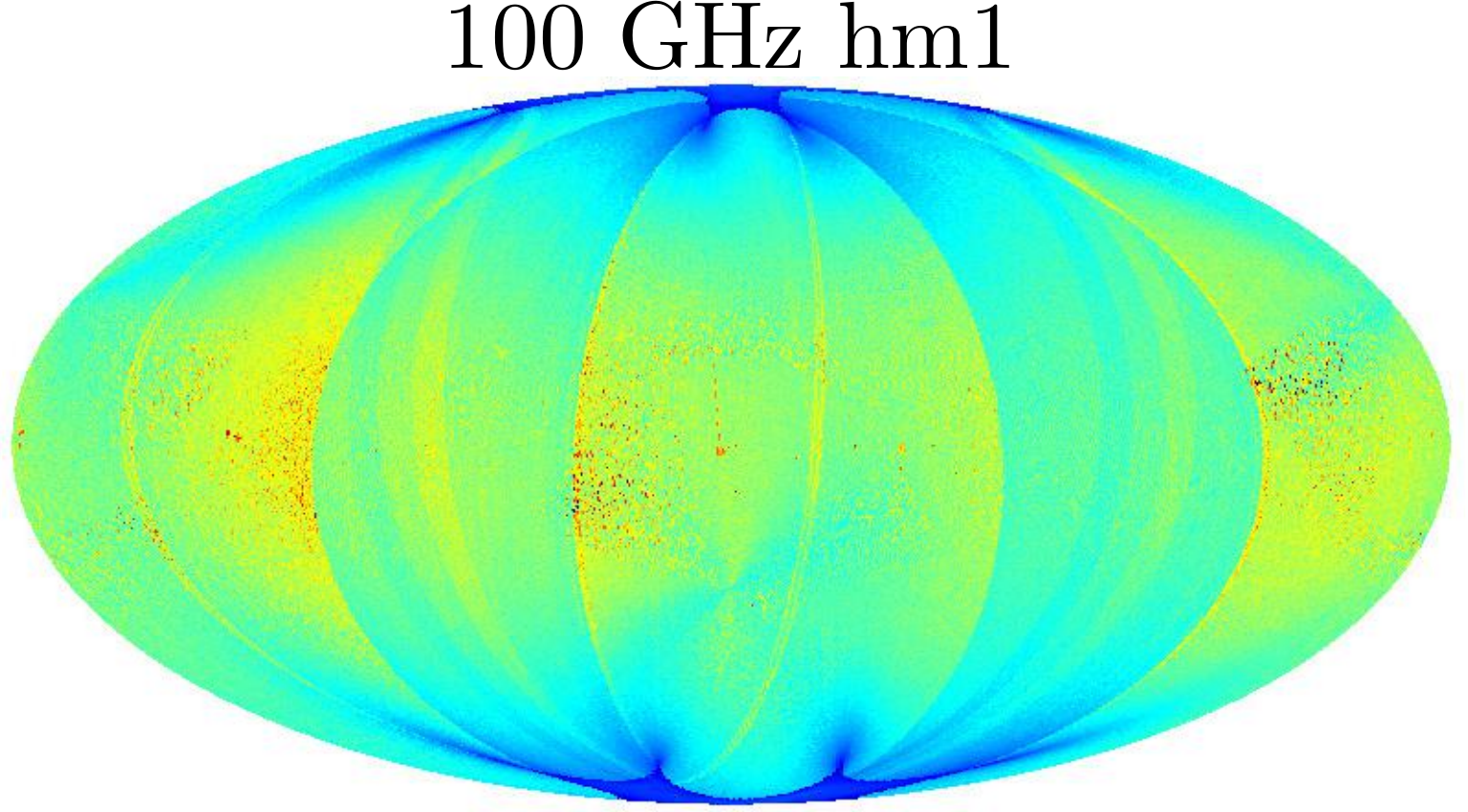}
\includegraphics[width=0.48\columnwidth]{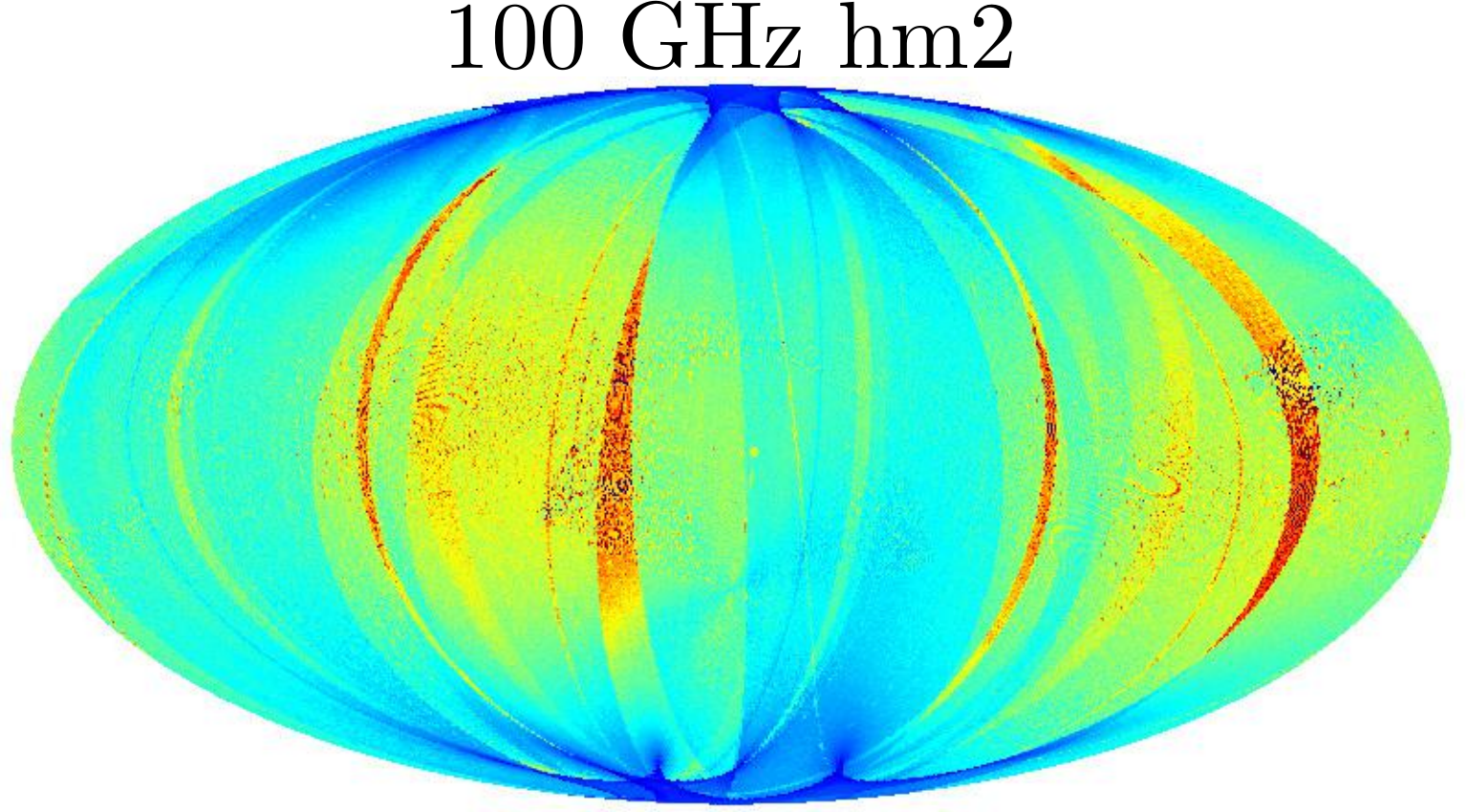}\\
\includegraphics[width=0.48\columnwidth]{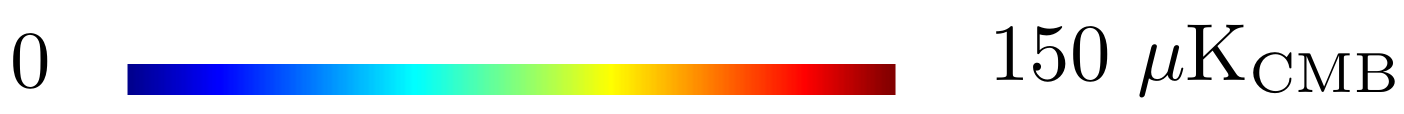}\\
\includegraphics[width=0.48\columnwidth]{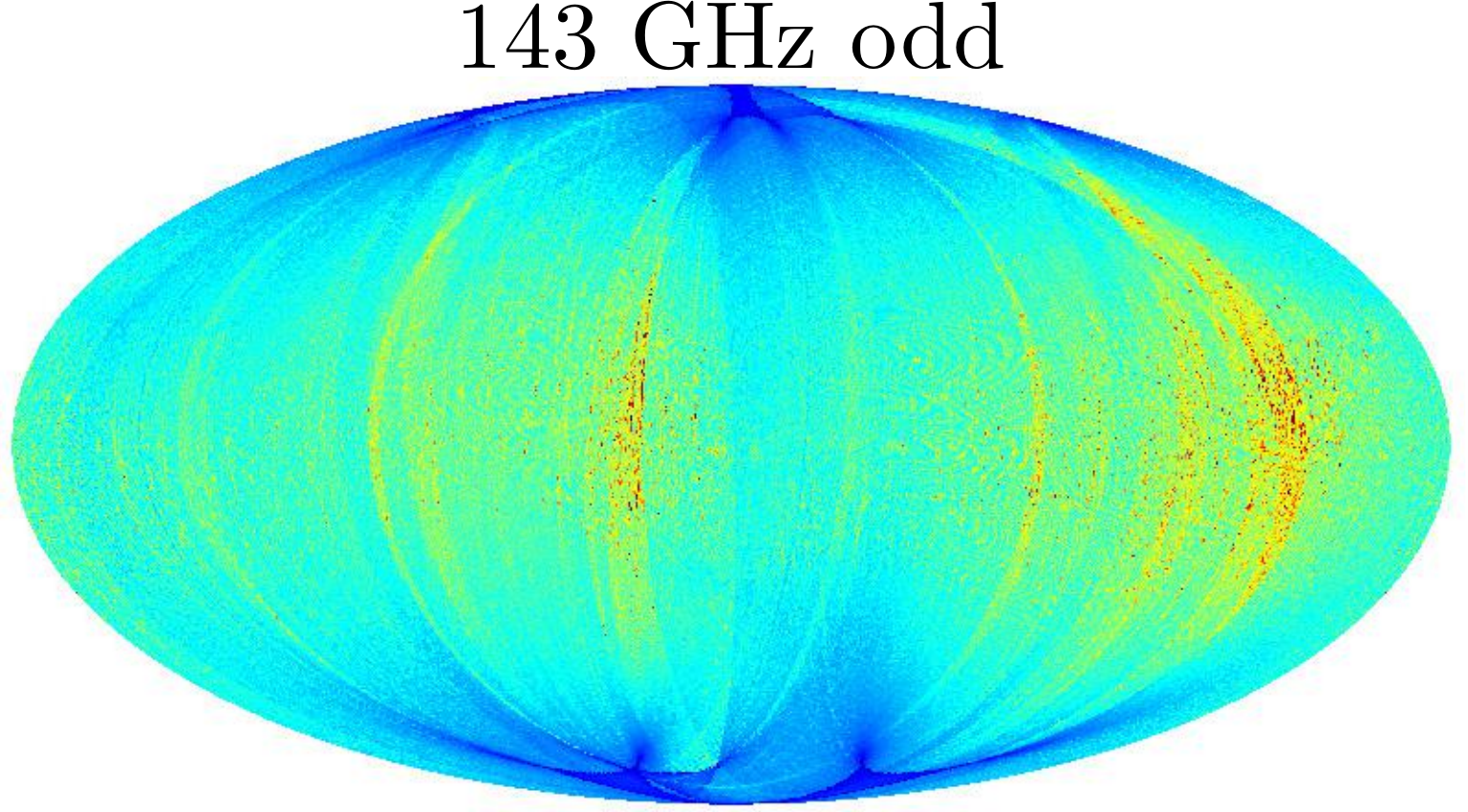}
\includegraphics[width=0.48\columnwidth]{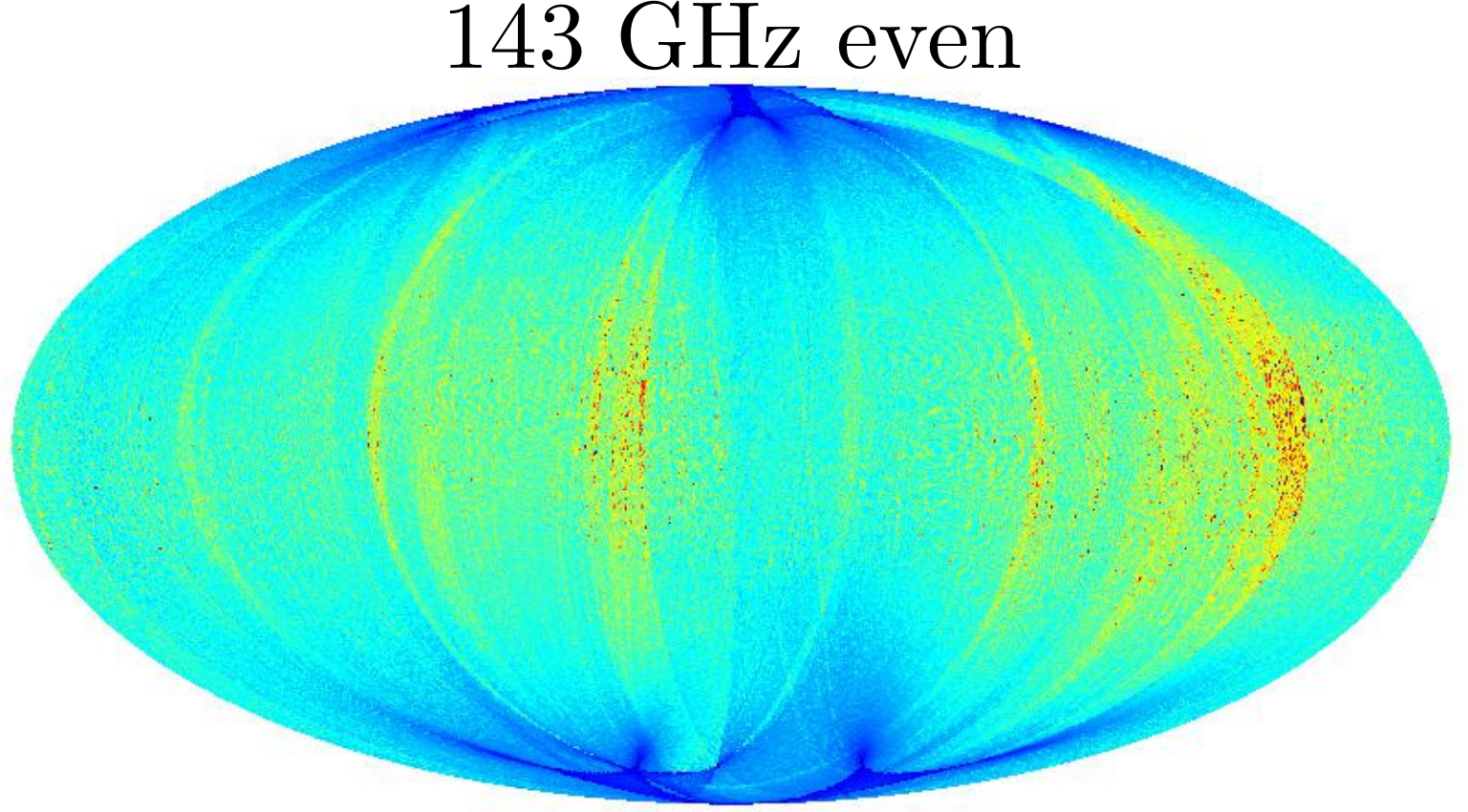}\\
\includegraphics[width=0.48\columnwidth]{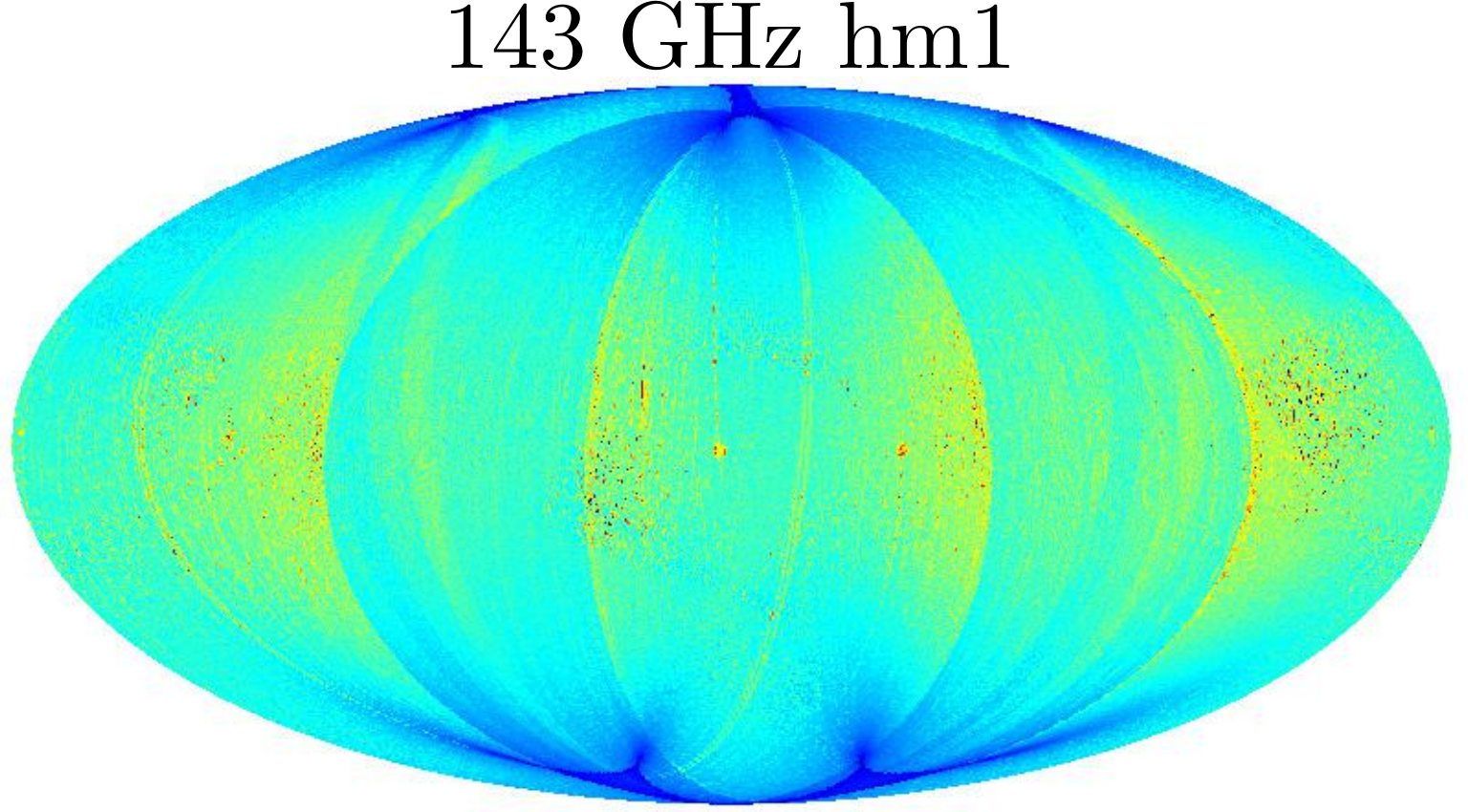}
\includegraphics[width=0.48\columnwidth]{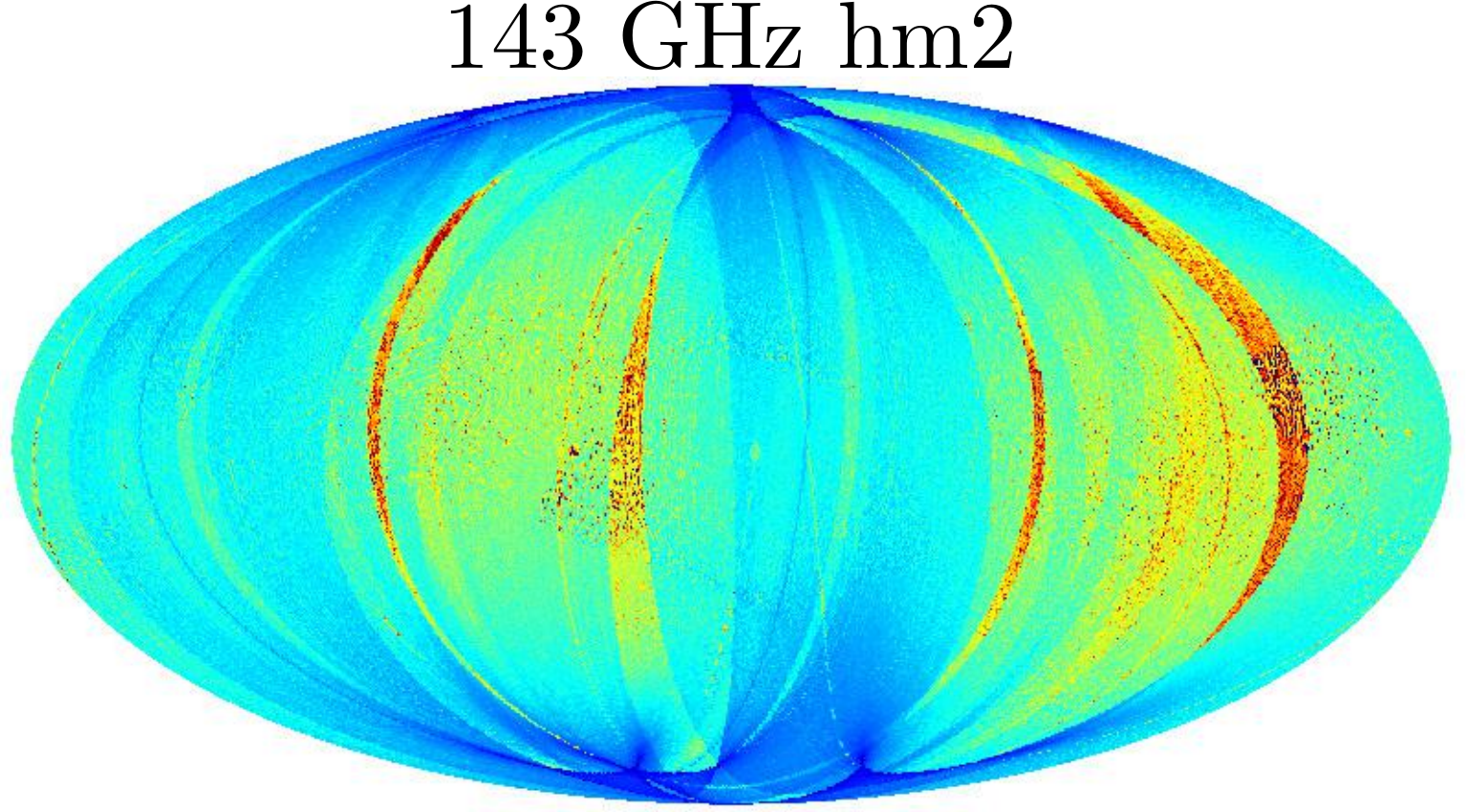}
\includegraphics[width=0.48\columnwidth]{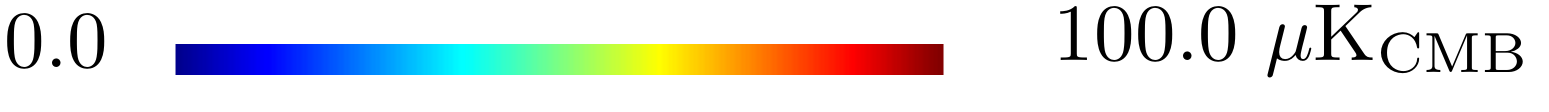}
\end{center}
\caption{100- and 143-GHz polarized maps obtained with white noise input only, showing the noise variance in ecliptic coordinates for the odd rings, for the even rings, and for the hm1 and hm2 data sets.}
\label{fig:ploestat2} 
\end{figure}

This difference is explained by simulations in Fig.~\ref{fig:ploestat2}. It shows that, when starting only from white Gaussian noise, we see an increase of the noise variance of the polarization at low ecliptic latitude, induced by the pointing matrix when the redundancy of the polarization angles per pixel is too small.
At 100\,GHz, the increase of the white noise in the TOI is reinforced by the difference of the distribution of pixels seen by the two sets of PSBs, which are not aligned on the same scanning circle. Thus in the odd (or equivalently in the even) data set, at low ecliptic latitude, each pixel is seen mostly by only one PSB set, decreasing the redundancy in angles. For the half-mission sets, this effect is much reduced because the redundancy of the rings stays the same. For the other frequencies, where the four PSBs are nearly aligned on the same circle, the effect is not seen.

The differences induced by the number of pointing periods removed in the two half-mission data sets induce differences in the angle redundancy at low ecliptic latitudes. This is the source of the asymmetry between the two half missions, both at 100 and 143\,GHz.

\subsubsection{Summary}

The above sections have shown that, when possible, scientific analysis should be performed in comparison with the simulated data and not rely on a single null test to describe the statistical behaviour. The main null tests used to characterize the full maps are:
\begin{itemize}
\item the detset null test;
\item the half-mission null test;
\item the ring null test.
\end{itemize}

Table~\ref{tab:JKsystes} gives qualitative estimates of the main systematic effect residuals for the three main null tests (detset, ring, and half-mission), with the symbols meaning: ``0,'' negligible; ``+,'' contribute significantly; and ``++,'' dominant. This shows that choosing which null-test split is more appropriate depends on the scientific objective one has in mind. The differences between these null tests provide only a first estimate of the noise, plus a set of systematic residuals to which that particular null test is sensitive. The estimates performed in Sect.~\ref{sec:map_characterization} provide the best indications on the matching of null tests to systematic effects and have been used in Table~\ref{tab:JKsystes}.

\begin{table}[htbp!]
\newdimen\tblskip \tblskip=5pt
\caption{Qualitative estimate of systematic effects best detected in different data splits.
Different systematic effects are shown in rows and half data splits in columns. A ``0'' sign indicates no effect, while ``+'' signifies an effect and ``++'' a strong effect, of a particular systematic on a particular null test.}
\label{tab:JKsystes}
\vskip -6mm
\footnotesize
\setbox\tablebox=\vbox{
\newdimen\digitwidth
\setbox0=\hbox{\rm 0}
\digitwidth=\wd0
\catcode`*=\active
\def*{\kern\digitwidth}
\newdimen\signwidth
\setbox0=\hbox{+}
\signwidth=\wd0
\catcode`!=\active
\def!{\kern\signwidth}
\newdimen\pointwidth
\setbox0=\hbox{.}
\pointwidth=\wd0
\catcode`?=\active
\def?{\kern\pointwidth}
\halign{\hbox to 3.75 cm{#\leaderfil}\tabskip 1em&
\hfil#\hfil\tabskip 1em&
\hfil#\hfil\tabskip 0.8em&
\hfil#\hfil\tabskip 0.5em&
\hfil#\hfil\tabskip 0em\cr
\noalign{\doubleline}
\omit&&&& Half\cr
\omit\hfil Systematic\hfil& Frequency& Detset& Ring& mission\cr
\noalign{\vskip 3pt\hrule\vskip 5pt}
Bandpass&100& 0& 0& 0\cr
\omit& 143, 217& +& +& 0\cr
\omit& 353& ++& +& 0\cr
\noalign{\vskip 8pt}
Polarization efficiency& 100& 0& 0& 0\cr
\omit& 143, 217& +& +& 0\cr
\omit& 353& ++& +& 0\cr
\noalign{\vskip 8pt}
ADCNL& 100& +& 0& +\cr
\omit& 143, 217& +& 0& +\cr
\omit& 353& 0& 0& +\cr
\noalign{\vskip 8pt}
Badly-conditioned& 100& 0& ++& 0\cr
\omit\hglue 1em pixels\hfil& 143, 217& +& 0& ++\cr
\omit& 353& 0& 0& ++\cr
\noalign{\vskip 8pt}
Transfer function&100& 0& 0& 0\cr
\omit& 143, 217& +& 0& 0\cr
\omit& 353& ++& 0& +\cr
\noalign{\vskip 8pt}
Calibration& 100& 0& 0& 0\cr
\omit& 143, 217& 0& 0& 0\cr
\omit& 353& +& 0& 0\cr
\noalign{\vskip 3pt\hrule\vskip 5pt}
}}
\endPlancktable
\end{table}

The non-Gaussian distribution of the systematics and foreground residuals will affect the optimal likelihood scheme used. We thus recommend to users looking for very small cosmological effects to perform their scientific analysis using the 300 E2E simulation maps in order to directly check the dispersion of their results.

\section{Calibration and dipoles}
\label{sec:calibration}

\subsection{Absolute primary photometric calibration}
Thanks to the better removal of some large-scale systematics, the HFI 2018 maps at 100, 143, 217, and 353\,GHz include a more accurate overall photometric calibration, based upon a more accurate measurement of the dipole induced by the Earth's velocity in the Solar system. This dipole modulation (denoted the ``orbital dipole'') has a quasi-constant amplitude of $271\microK$ over the mission. The orbital dipole modulation is predictable to very high precision and does not project onto the sky because the scanned rings change orientation as the spin axis sweeps across the sky, remaining always anti-Solar. Thus a given ring is scanned with an orbital velocity of the opposite sign 6~months later, and the orbital dipole signal averages to zero over one year.

As was done for the 2015 data release, the 545- and 857-GHz, channels are calibrated using comparisons of HFI measurements of Uranus and Neptune with the ESA~2 model of Uranus and the ESA~3 model of Neptune (\citehfimap).

\subsection{Updated Solar dipole determination}
\label{sec:solardipole}
\subsubsection{Method}

The kinetic dipole induced by the motion of the Solar system with respect to the CMB rest frame (denoted the ``Solar dipole'') has an amplitude 12 times larger than the orbital dipole (3.36\,mK) and remains constant in the sky maps. For the high precision calibration required in this release, we must consider second-order terms in $\beta$ ($= v/c$) and give up the first-order approximate relation between $\delta T$ and $\delta B_{\nu}$ for the dipole. These second-order terms induce a cross-term between the Solar and orbital dipoles, which makes this separation less straightforward. This cross-term is taken into account in the 2018 release, but we retain the usual linear relation between $\delta T$ and $\delta B_{\nu}$ for other anisotropies. Kinetic dipoles associated with the Solar system motion with respect to the CIB and with CMB distortions are negligible at the accuracy considered in this paper.

The noise-limited sensitivity achieved through the destriper for calibrating detectors within the same frequency band on the orbital dipole is excellent (signal to noise nearly $10^{5}$ for CMB channels). Nevertheless, the accuracy is still limited by foreground removal and systematic effects, such as uncertainties in phase shifts in the dipole due to long time constants. After reaching a stage in which the foreground residuals have been shown to be negligible, the amplitude of the Solar dipole provides an excellent a posteriori check on the relative calibration between different HFI detectors, between different frequencies, and also for comparisons of \Planck's calibration with other space-based CMB experiments. We estimate the Solar dipole amplitude and direction in the best HFI CMB frequency bands. This Solar dipole can then be used in a much broader context of inter-calibration of other instruments at higher frequencies, as demonstrated by the detection of the Solar dipole at 545\,GHz.

To extract the Solar dipole, we start by removing the CMB anisotropies from the frequency maps, over 95\,\% of the sky, using the four \Planck\ 2015 component-separation methods ({\tt SMICA}, {\tt Commander}, {\tt NILC}, and {\tt SEVEM}). This is necessary to limit spurious dipoles induced from the low multipole anisotropies when different sky masks are used for the dipole extraction, as a test of the quality of the dust emission removal. This induced dipole term is fully degenerate with the kinetic dipole, which is much larger than the CMB low multipole anisotropies. The intrinsic dipole in the CMB anisotropies is not measurable and is set to zero. The four component-separation methods use different procedures, as described in \citet{planck2016-l04}, and will leave different dipole residuals after CMB anisotropy removal. One might suspect that the extraction of the CMB anisotropies might not be very accurate in the narrow bright Galactic disc, and could leave some residual of the Galactic centre-anticentre dipole term in the Galactic plane. In that case, the zero dipole terms on the whole sky would induce a spurious high latitude dipole, compensating for the residual one from low Galactic latitudes.

The next step is to construct a base model of the Galactic dust emission, based on the correlation of CMB maps using the 857-GHz map. We model each HFI map $I_\nu$ as
\begin{equation}
\label{eq:model857}
I_\nu = q(\nu)\, I_{857} + D_{\rm res} + C,
\end{equation}
where $C$ is the CMB anisotropy map and $q(\nu)$ is the projection coefficient from 857\,GHz to frequency $\nu$. In the above equation $I_{857}$ is the HFI 857-GHz map and the use of 857\,GHz as a reference for the Galactic dust emission is guided by the fact that the CMB Solar dipole is of negligible amplitude at this frequency; it is only 0.0076\,MJy\,sr$^{-1}$, and furthermore it is mostly removed in the mapmaking process. The CIB dipole, which has not been removed, is of the same order.

In bright regions of the sky, the residual term $D_{\rm res}$ also contains extra Galactic emission, for instance free-free and CO emission, which is not (or only partially) correlated with $I_{857}$. Finally $D_{\rm res}$ also includes any residual from potential errors in the removed Solar dipole in the mapmaking process. If the assumed Solar dipole direction and/or the calibration of the channel is slightly incorrect, there will be a residual dipole in $D_{\rm res}$. The true Solar dipole can then be characterized by adding back the removed dipole to $D_{\rm res}$. The dust is removed to first order from other frequency maps using the 857-GHz map, with a single SED $q(\nu)$ taken from \citet{planck2014-XXII}.

When this was done in \citehfimap\ and \citelowell, the Solar dipole amplitudes found for the different frequencies and Galactic masks showed a drift with frequency from 100 to 545\,GHz, as well as changes with the Galactic sky masks (sky fraction from 30 to 60\,\%) for 143 to 545\,GHz. These variations, reported in table~2 of \citelowell, were indicative of a problem with the dust removal at all frequencies except 100\,GHz.

\begin{figure}[htbp!]
\includegraphics[width=0.24\textwidth]{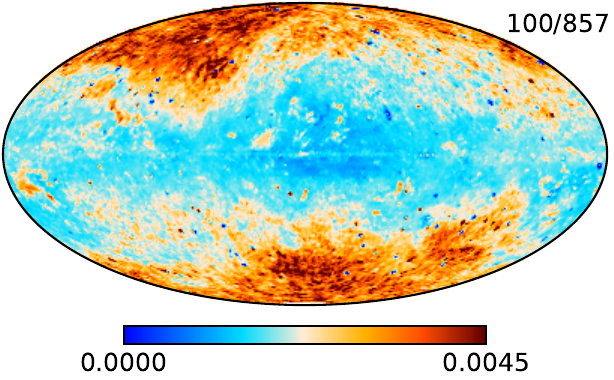}
\includegraphics[width=0.24\textwidth]{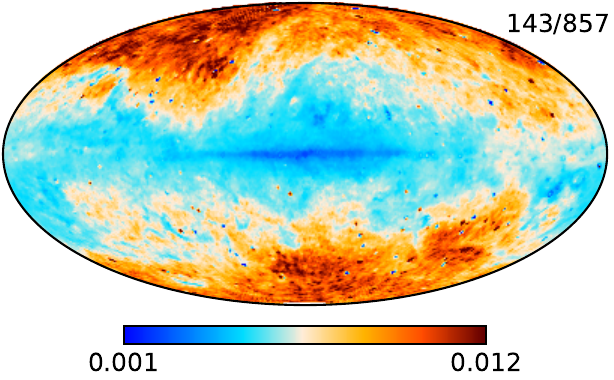}\\
\includegraphics[width=0.24\textwidth]{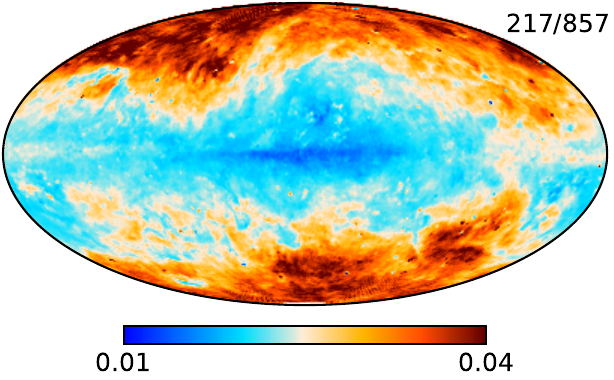}
\includegraphics[width=0.24\textwidth]{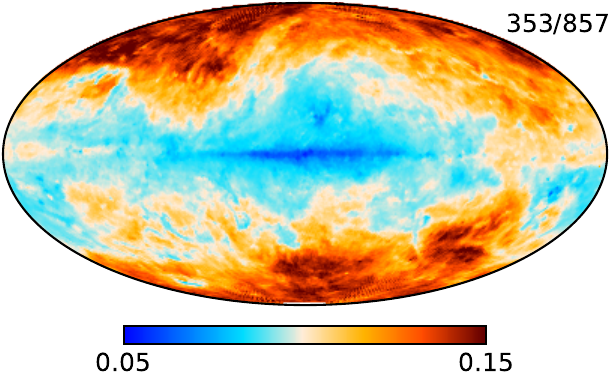}\\
\includegraphics[width=0.24\textwidth]{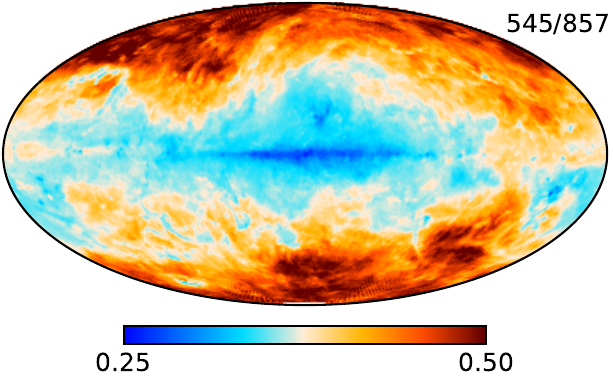}
\caption{Ratio between maps at other frequencies and the 857\,GHz map. At 100 and 143\,GHz, low-frequency foregrounds (synchrotron, free-free) have been removed before taking the ratio. The CMB anisotropies have been removed for the 100 to 353\,GHz maps, as described in the text. These ratios are thus indicative of the Galactic dust SED variations.}
\label{fig:ratiomaps} 
\end{figure}

We have now refined the procedure for the extraction of the Solar dipole. The improvements from \citelowell\ are: (i) removing the small leakage to the dipole from very low-multipole CMB anisotropies caused by the sky cut; and (ii) improving the dust removal. In the dust removal at each frequency above 100\,GHz, we add a correction to account for SED variations on the sky as shown by the ratios of frequency maps, after removal of the CMB anisotropies, as well as synchrotron and free-free, as seen in Fig.~\ref{fig:ratiomaps}. These ratios are representative of the dust SED variation on large scales, especially in Galactic latitude. We note the similarity of the patterns in the figure.

An empirical correction map containing only the two lowest multipoles ($\ell\,{=}\,1$ and 2) is used to correct the variations of the dust SED, and the small CMB anisotropy residual dipole. These two multipoles have the largest effect on the dipole with different Galactic masks. The $a_{lm}$s of the SED correction maps are fitted by imposing the condition that the resulting Solar dipole direction at 545, 353, 217, and 143\,GHz minimizes the difference of its direction from the 100-GHz one, known to be almost unaffected by the sky fraction. Four sky fractions are used in this minimization, namely 36, 44, 52, and 60\,\%. At each frequency, removing the drift of the dipole direction with sky fraction is what drives the determination of the dust SED correction, and CMB anisotropy dipole residuals.

The method takes the 100-GHz Solar dipole direction as a reference. The uncertainties attached to this assumption induce small systematic effects on the extracted Solar dipoles for the other frequencies.

Checks of the validity of this procedure are that the fit for a common dipole direction should lead to the removal of the drift of the Solar dipole amplitudes when increasing the frequency and changing the sky mask fraction.

\begin{figure}[htbp!]
\includegraphics[width=0.48\columnwidth]{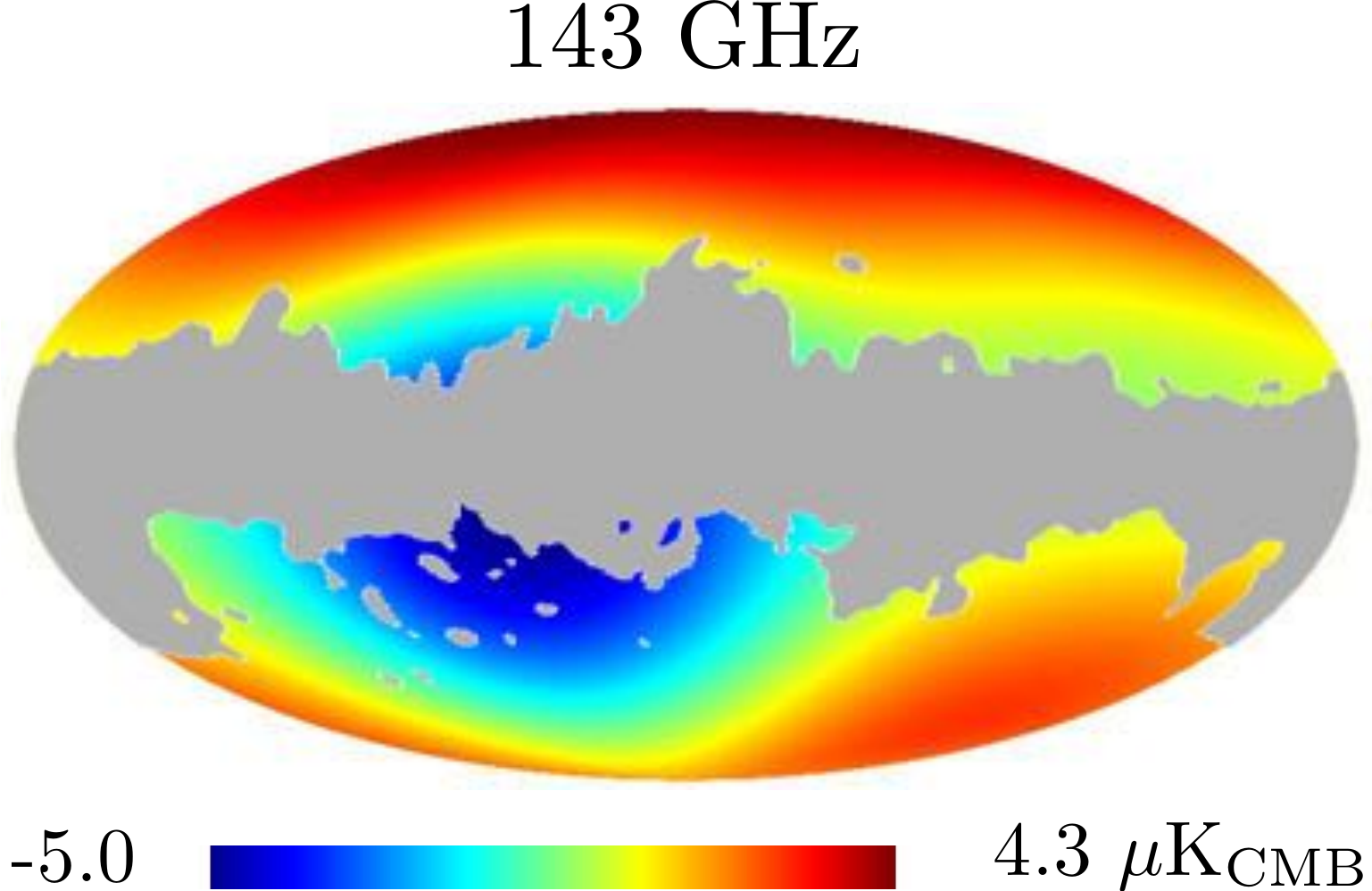}
\includegraphics[width=0.48\columnwidth]{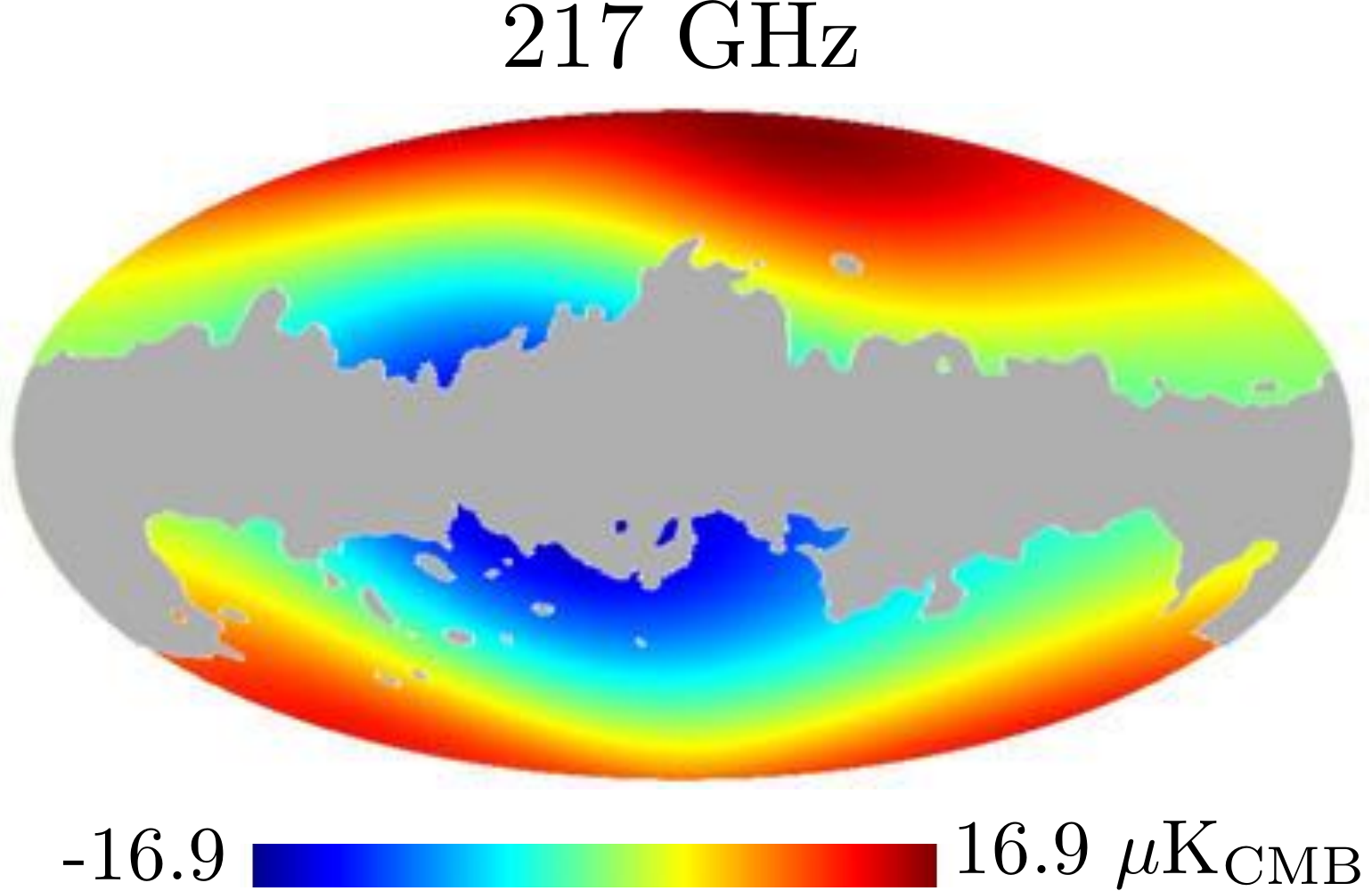}\\
\\
\includegraphics[width=0.48\columnwidth]{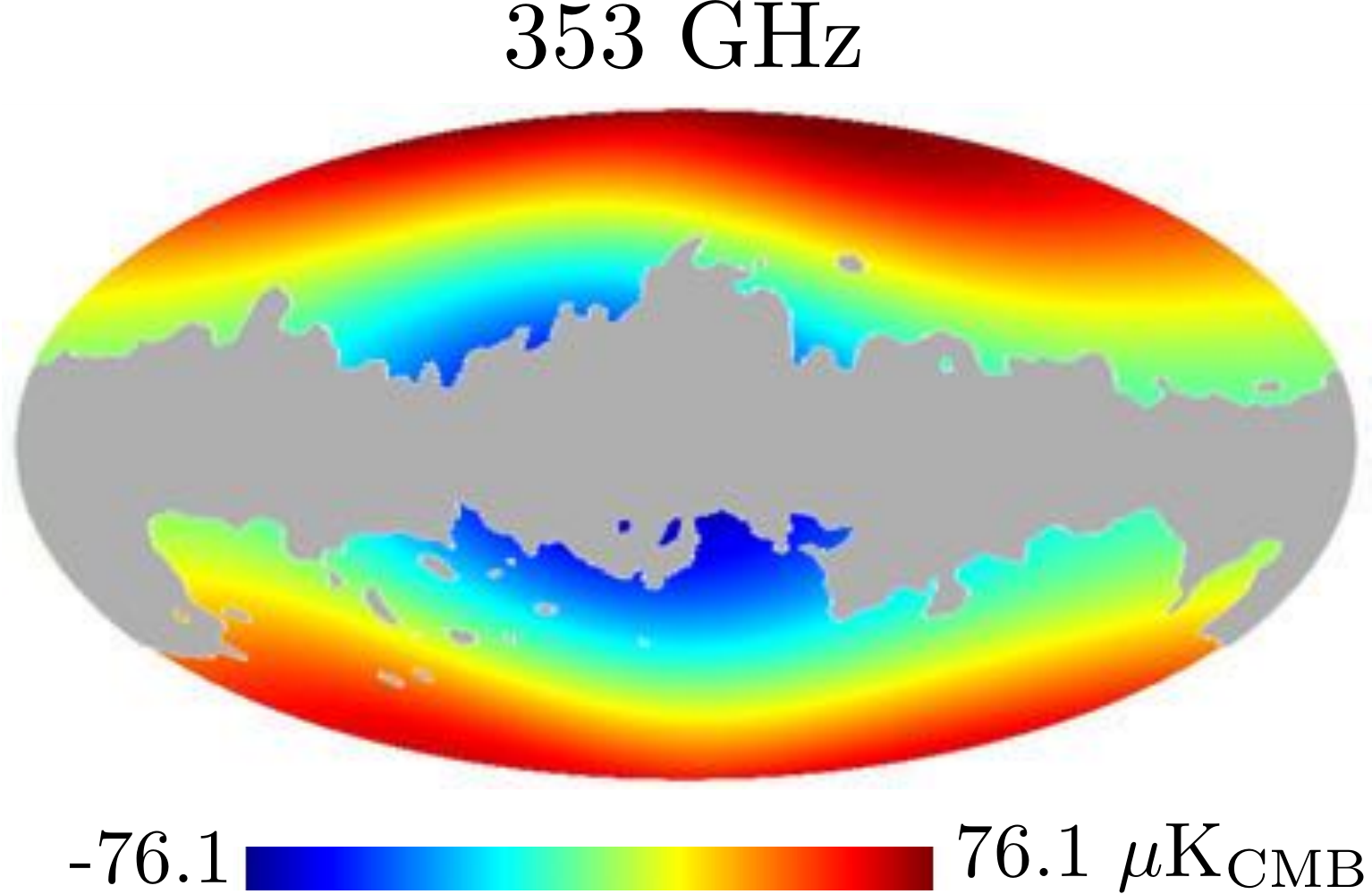}
\includegraphics[width=0.48\columnwidth]{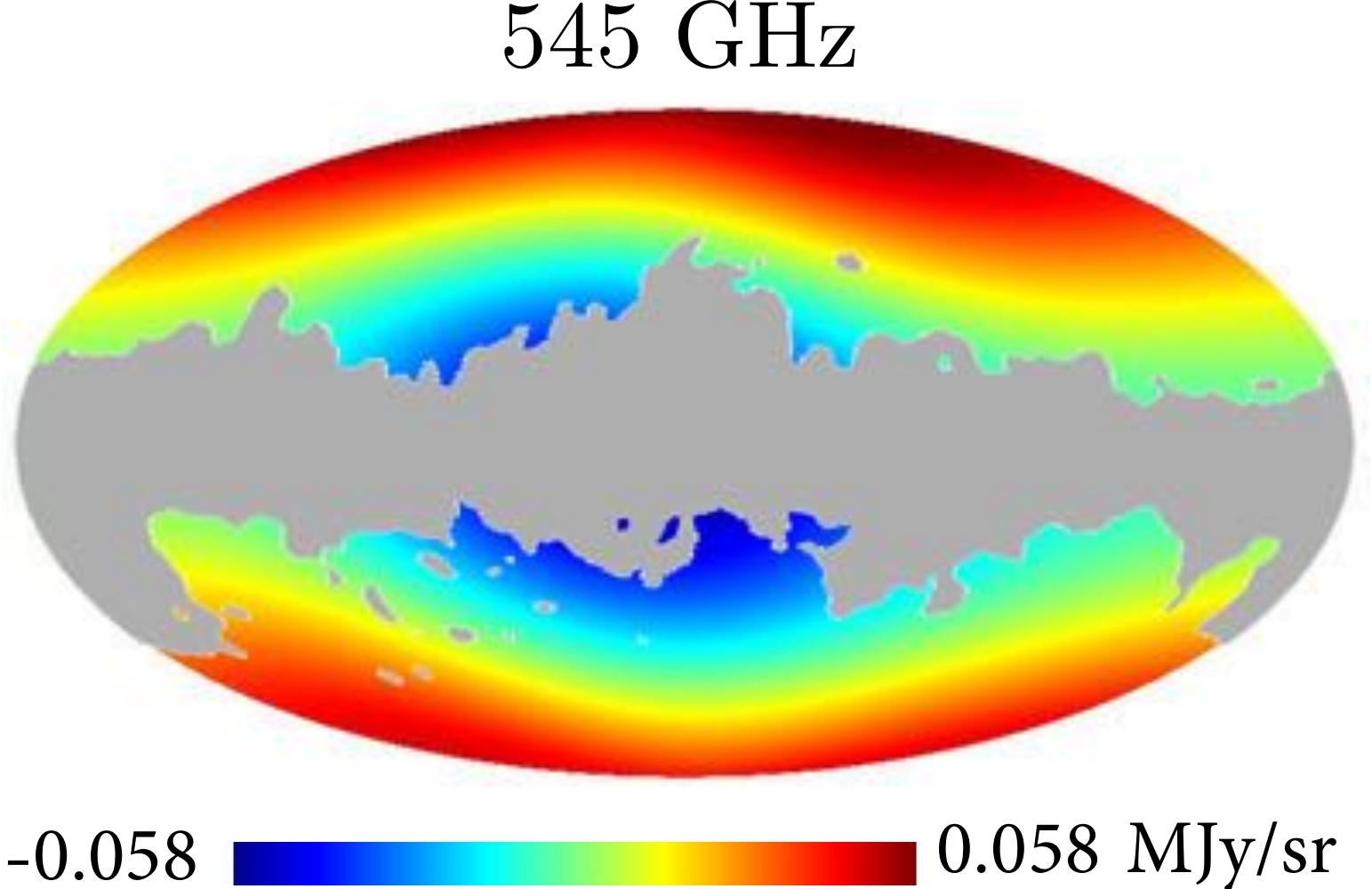}\\
\caption{Maps at $\Nside\,{=}\,32$ for $f_{\rm sky}=60$\,\% of the ad hoc empirical correction at the four frequencies for which this correction has been extracted in \sroll.}
\label{fig:dipresidual} 
\end{figure}

Another validity check is that the ad hoc dust correction maps (independently fitted for each spherical harmonic) show comparable distributions on the sky at all frequencies. If successful, these should, and do remove the previous evidence seen in \citelowell, which pointed to a need for a better dust removal process. The empirical correction maps are shown in Fig.~\ref{fig:dipresidual} and are indeed very similar for 545, 353, 217, and 143\,GHz, fulfiling the second check mentioned above.
The dominant term is a quadrupole in Galactic latitude and a dipole term in the Galactic plane that is nearly aligned with the centre-anticentre direction. This is consistent with the known dust foreground SED variation on large angular scales (Fig.~\ref{fig:ratiomaps}), and a residual dipole from the CMB anisotropies. These variations make sense from the Galactic physics point of view: gradients of starlight energy density and opacity with Galactic radius and height lead to an increase of the starlight energy density above and below the narrow Galactic disc. These lead to corrections to the Solar dipole amplitude that are an order of magnitude smaller than the full amplitude of the correction maps: 3.5\microK\ of the solar dipole amplitude for a full amplitude of 34\microK\ of the empirical correction map at 217\,GHz for 60\,\% of the sky.

\subsubsection{Results}

\begin{figure}[htbp!]
\includegraphics[width=\columnwidth]{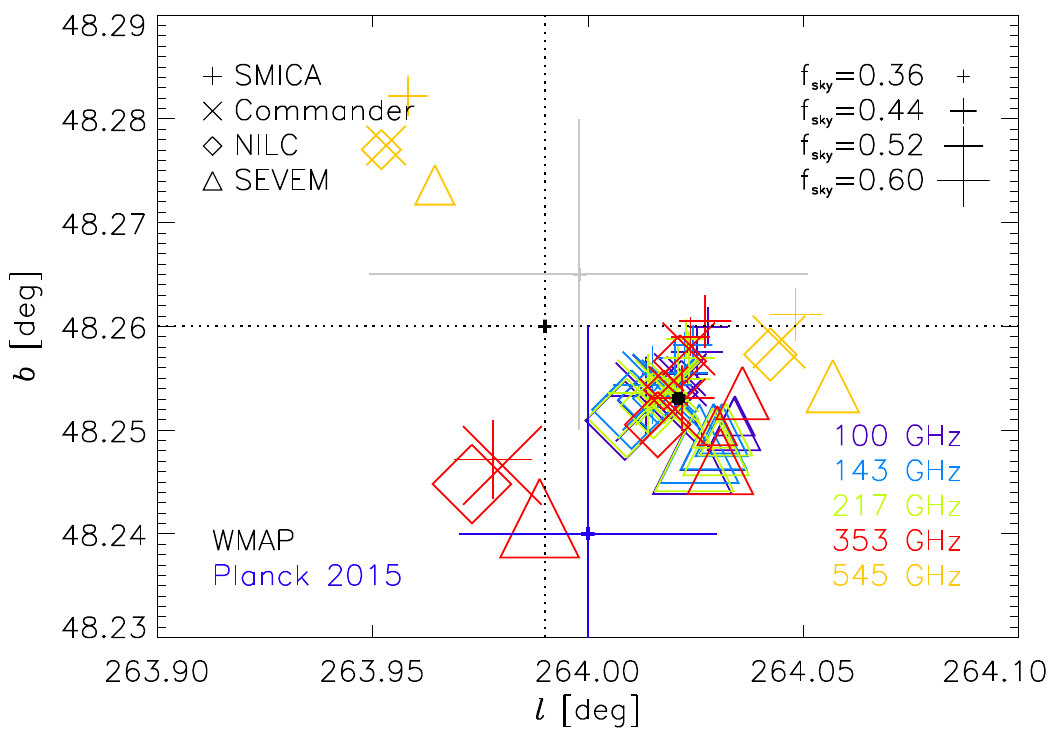}
\includegraphics[width=\columnwidth]{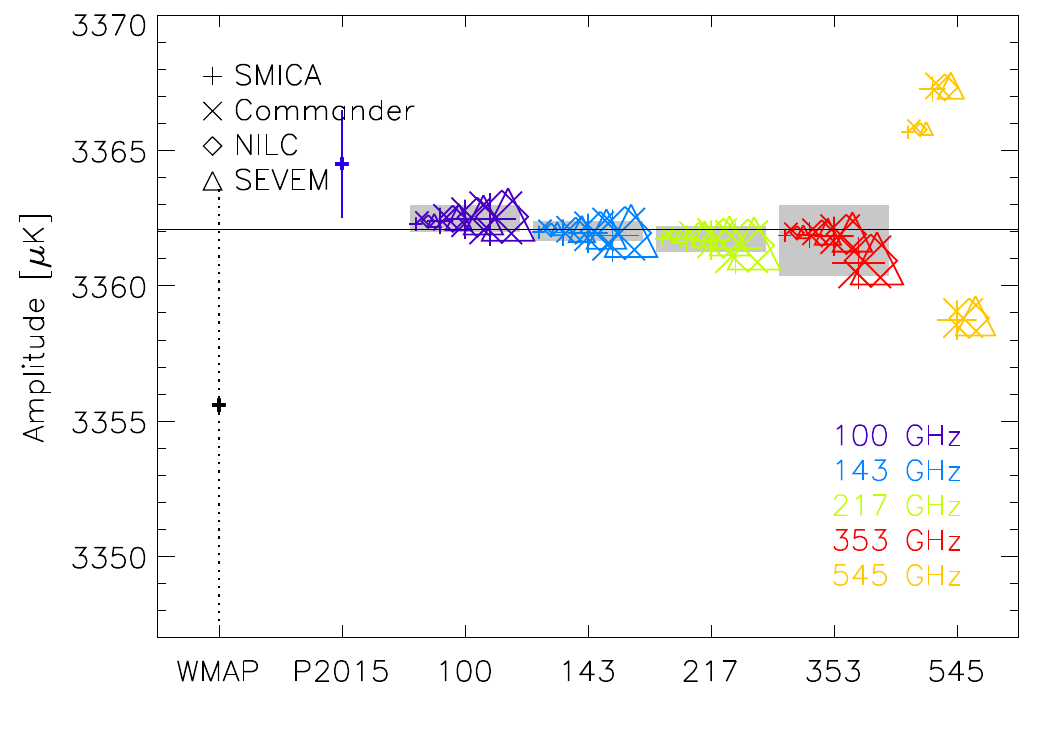}
\caption{Solar dipole directions and amplitudes for the four component-separation methods using different symbols of size, increasing with sky fraction. The colours refer to frequencies for the 2018 release, while the WMAP measurement (see text) is the black dotted plus sign and the \Planck\ 2015 measurement is the blue one. Grey boxes give the absolute bias uncertainties. At 545\,GHz, several points for the largest sky fraction fall outside of the plotted range. The HFI 2018 Solar dipole determination is shown in direction by the black dot and in amplitude by the black horizontal line.} 
\label{fig:Dipole}
\end{figure}

We have extracted the Solar dipole from five frequencies (100--545\GHz), using CMB maps from the four component-separation methods, and for the four Galactic masks chosen to span a large range of sky fraction. The resulting dipole directions and amplitudes are shown in Fig.~\ref{fig:Dipole}, for frequencies from 100 to 545\,GHz.
There is now much better agreement in the dipole amplitude across HFI frequencies (with respect to \citelowell) in the range of sky fraction masks used for the minimization; however, within a given frequency, there is a weaker dependence on the component-separation method used to remove the CMB anisotropies (which would trace foreground residuals and the dipole residual from the CMB map subtraction). We also display for comparison the previous determinations of the Solar dipole direction from WMAP and from the \Planck\ 2015 release. As expected, the fitting process minimizes the angular distance to the 100-GHz dipole directions. All frequencies are in excellent agreement, within 1\arcm\ for all cases for the CMB frequencies and also for 353\,GHz except for the largest sky fraction. Even at 545\,GHz, a similar barycentre is found, but with a larger dispersion (within 6\arcm\ of the best direction found at lower frequencies).

Table~\ref{tab:dipole} summarizes the amplitudes and directions averaged over the 16 cases. The statistical error estimates here are based on the maximum dispersion of the different sky fractions among the four component extractions used. We express the measured amplitudes of the Solar dipole from each of the frequency maps in $\mu$K as calibrated on the orbital dipole adopting the CMB temperature $T_{\rm CMB}=2.72548\,{\rm K}\pm0.57\,{\rm mK}$ \citep{fixsen2009}.
\begin{table}[htbp!]
\newdimen\tblskip \tblskip=5pt
\caption{Amplitudes and directions averaged over the four component-separation methods, with uncertainties given by the rms of the variations as the sky fraction is changed from 36 to 60\,\%.}
\label{tab:dipole}
\vskip -6mm
\footnotesize
\setbox\tablebox=\vbox{
\newdimen\digitwidth
\setbox0=\hbox{\rm 0}
\digitwidth=\wd0
\catcode`*=\active
\def*{\kern\digitwidth}
\newdimen\signwidth
\setbox0=\hbox{+}
\signwidth=\wd0
\catcode`!=\active
\def!{\kern\signwidth}
\newdimen\pointwidth
\setbox0=\hbox{.}
\pointwidth=\wd0
\catcode`?=\active
\def?{\kern\pointwidth}
\halign{\hbox to 1.8 cm{#\leaderfil}\tabskip1.3em&
\hfil#\hfil\tabskip1.0em&
\hfil#\hfil&
\hfil#\hfil\tabskip 0em\cr
\noalign{\doubleline}
\omit\hfil Frequency\hfil&Amplitude&$l$&$b$\cr
\noalign{\vskip 2pt}
\omit\hfil [GHz]\hfil& [\microK]& [deg]& [deg]\cr
\noalign{\vskip 3pt\hrule\vskip 5pt}
\noalign{\vskip 2pt}
100& $3362.48 \pm *0.10$& $264.022 \pm 0.006$& $48.253 \pm 0.003$\cr
143& $3362.02 \pm *0.12$& $264.021 \pm 0.004$& $48.253 \pm 0.002$\cr
217& $3361.73 \pm *0.22$& $264.020 \pm 0.004$& $48.253 \pm 0.002$\cr
353& $3361.68 \pm *0.56$& $264.013 \pm 0.023$& $48.252 \pm 0.006$\cr
545& $3356.59 \pm 15.28$& $263.899 \pm 0.189$& $48.225 \pm 0.052$\cr
\noalign{\vskip 3pt\hrule\vskip 5pt}}}
\endPlancktable
\end{table}

The amplitudes, which are the crucial test, show (with this additional dust correction) a very good agreement between the four CMB-calibrated frequencies, validating the procedure. Furthermore, for each frequency, there is only little apparent trend visible in both direction and in amplitude. The 545-GHz channel, which is calibrated on the giant planets, shows remarkable agreement (${<}\,1\,\%$) with the CMB calibration of the lower frequencies, smaller than the uncertainties coming from the transfer function from point-source calibration to dipole calibration.

\begin{figure}[htbp!]
\includegraphics[width=\columnwidth]{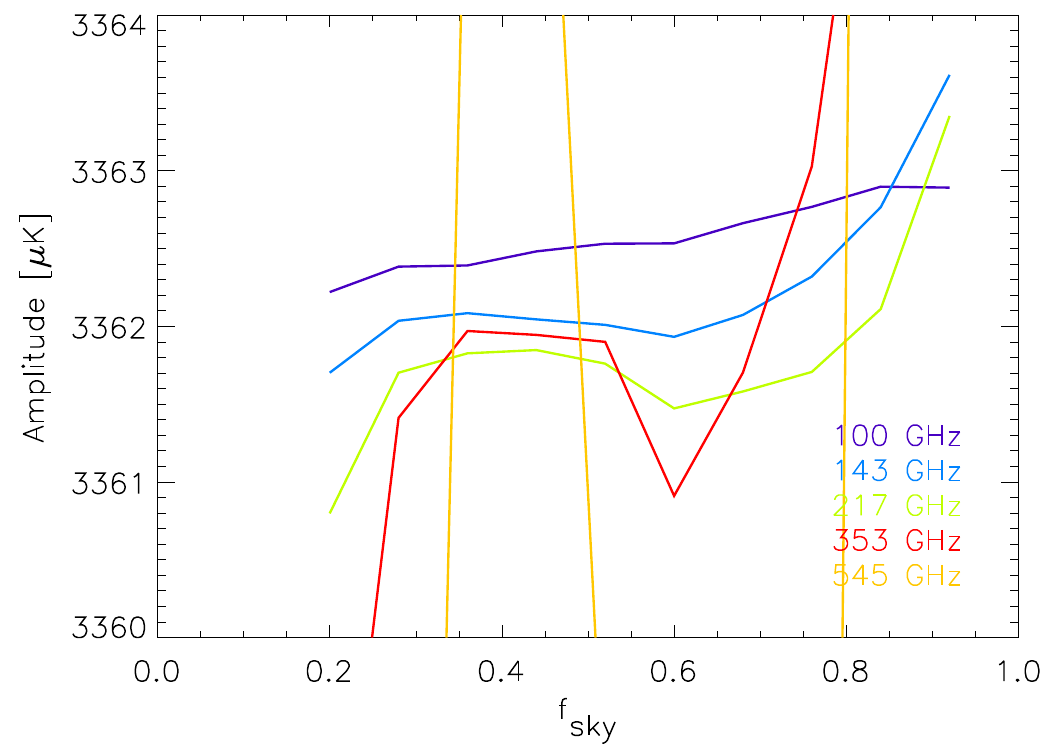}
\caption{HFI 2018 Solar dipole amplitude, for all HFI channels, including 545\,GHz, as a function of sky fraction.}
\label{fig:dipolesvsfsky}
\end{figure}

Figure~\ref{fig:dipolesvsfsky} demonstrates that the Solar dipole amplitude has little trend with the fraction of sky used, over the range $28\,\%<f_{\rm sky}<85\,\%$, for the three lowest frequency channels.
This is a very sensitive test, the result showing that we have fully understood the effect of the dust removal in the determination of a common Solar dipole. The three CMB channels all produce dipole amplitudes within $\pm 0.5\microK$ for sky fractions from 36 to 60\,\%. The 353-GHz results only start to show significant amplitude (and direction) drifts above sky fractions of 68\,\%. This leads us to do a straight average over the three frequencies (100, 143, and 217\,GHz), and between 36 and 60\,\% of the sky, to compute the best HFI 2018 Solar dipole. The noise in the fit of the Solar dipole is negligible compared to the level of systematics, which justifies the straight average.

The coherence of these directions and amplitudes of the Solar dipole with component-separation methods, sky fractions, and frequencies up to 545\GHz, shows that the empirical variable SED dust correction seen in Fig.~\ref{fig:ratiomaps} was contributing significantly in the systematic effects left in the \citelowell\ analysis. 

We construct an estimate of uncertainty on the amplitude, starting from the statistical uncertainties, for a given sky fraction and CMB extraction, using the \sroll\ algorithm; this estimate is presented in Table~\ref{tab:dipole} (0.09\microK\ rms) and referred to as ``stat.''.

However, the dispersion observed with sky fraction and the four component-separation methods is about an order of magnitude larger (0.91\microK\ peak-to-peak). This includes both the effect of the dust removal residuals (traced by sky fraction) and CMB dipole removal residual (traced by the four component-separation methods). This is referred to as ``syst.'', taking the half peak-to-peak amplitude.

Furthermore, the absolute \sroll\ bias, measured on the Solar dipole (columns E of Table~\ref{tab:ratios}), has been applied as a correction to the dipole amplitude. The rms of this correction (column F) is 0.45\microK, which is referred to as ``cal.''.
 
We thus obtain the best HFI 2018 Solar dipole velocity vector and amplitude (which is directly obtained from the orbital dipole). We also give the amplitude in temperature, based on the CMB temperature used in the 2015 release. The description of the velocity vector is:
\begin{eqnarray}
v&=& (369.8160 \pm0.0010)\,{\rm km s}^{-1}; \nonumber\\
A&=& \left[3362.08\pm0.09\,\text{(stat.)}\pm0.45\,\text{(syst.)}\pm0.45\,\text{(cal.)}\right]\,\mu{\rm K}; \nonumber\\
l&=&\text{264\pdeg021} \pm \text{0\pdeg003}\,\text{(stat.)} \pm \text{0\pdeg008}\,\text{(syst.)}; \nonumber\\
b&=&\text{\phantom{0}48\pdeg253} \pm \text{0\pdeg001}\,\text{(stat.)}\pm \text{0\pdeg004}\,\text{(syst.)}.
\end{eqnarray}
Its amplitude and direction are compatible, within their respective uncertainties, with the WMAP ones,\footnote{$(d,l,b) = (3.355 \pm 0.008\,{\rm mK}, \text{263\pdeg99} \pm \text{0\pdeg14}, \text{48\pdeg26} \pm \text{0\pdeg03)}$ \citep{hinshaw2009}, given a CMB monopole temperature of 2.725\,K \citep{Mather1999}} with the \Planck\ 2015 values, and with the 2018 LFI results \citep{planck2016-l02}. The uncertainty on the amplitude does not include the 0.02\,\% uncertainty of the temperature of the CMB monopole.

\subsection{A posteriori inter-calibration within a frequency using the Solar dipole}
\label{sec:intercal}

Using the single-bolometer maps described in Sect.~\ref{sec:monobolomaps}, we can now examine a posteriori using the Solar dipole, the calibration accuracy and gain dispersion of each detector within a frequency band. Because HFI measures polarization by differencing the signals from detectors with different polarization orientations, the relative calibration between these detectors is of the utmost importance. We assume that the Solar dipole is due to motion relative to a pure blackbody spectrum. Expected spectral distortions are negligible at HFI CMB frequencies and sensitivities. Furthermore, fitting the Solar dipole residual amplitudes in single-bolometer maps gives an absolute calibration with respect to the Solar dipole, used as input in the simulations. We explicitly used the one obtained in Sect.~\ref{sec:solardipole}, but this has no first-order effect on the test and just minimizes all non-linear effects.

\begin{figure}[htbp!]
\includegraphics[width=\columnwidth]{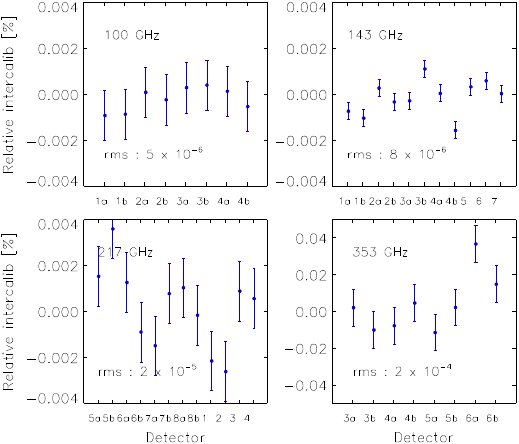}
\caption{Relative calibration measured by the Solar dipole amplitude for all detectors within a frequency channel, with respect to the mean dipole across the frequency band.}
\label{fig:interbol} 
\end{figure}

Figure~\ref{fig:interbol} shows the relative calibration of the different detectors for each frequency (from 100 to 353\,GHz)\footnote{Detectors are identified by a number for SWBs, and adding a letter (``a'' or ``b'') for PSBs.} with respect to the average, and the rms values within each frequency.
The error bars are taken from 100 of the E2E simulations discussed in Sect.~\ref{sec:simcalib}, which are composed of three elements: noise; systematic residuals; and \sroll\ algorithm bias. Intrinsic calibration dispersions from detector to detector, larger than predicted by the error bars, are clearly detected at 143, 217, and 353\,GHz. These induce a small spurious polarization in the temperature anisotropies. This has been propagated in E2E simulations and has been shown in Fig.~B18 of \citelowell\ to be lower than $10^{-6}\microKcarre$ in polarization for the CMB channels, which is negligible.

The calibration depends on residuals left by systematic effects, namely, ADC dipole distortion, low-frequency transfer functions, cross-talk, and polarization-specific parameters. These vary from detector to detector. The improvement in the inter-detector calibration confirms the overall improvement in the correction of these effects in the current data processing. The relative calibration of detectors within a frequency band could be used to give upper limits on the residuals for these systematic effects.

\subsection{Simulations of dipole calibration accuracy}
\label{sec:simcalib}

Bolometer response is known to be very stable and predictable from the bolometer physical parameters. Non-linearity in the electronic amplifiers does not significantly distort the small CMB anisotropy signal, which is affected only by the ADCNL. In the simulations, the parameters of the ADCNL are drawn from the uncertainties estimated for the ADCNL model. These ADCNL uncertainties generate time variations in the linear gain at the level observed. These gain variations do not necessarily average to zero over the mission and leave a small bolometer gain bias with respect to the predicted stable response. It also induces a dispersion between bolometers and, in turn, a bias on the frequency-averaged calibration. The average gain of a bolometer, or a frequency band, can be directly estimated by comparing the input Solar dipole and the \sroll\ solved one in these simulations. The excellent agreement of the amplitudes of the Solar dipole measured at the three CMB frequencies is an indication that the bias is small but still needs to be measured.

\begin{figure}[htbp!]
\includegraphics[width=\columnwidth]{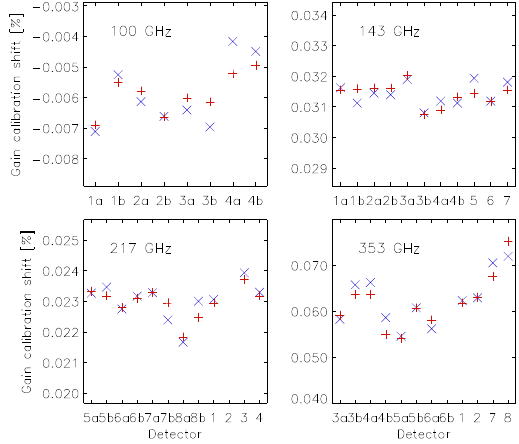}
\caption{Single simulation of detector calibration derived from Solar dipole in single-bolometer maps. Blue is for the absolute calibration bias, while red is for the recovered relative calibration bias.}
\label{fig:calib_accuracy}
\end{figure}

We show in Fig.~\ref{fig:calib_accuracy}, from a single E2E realization for each detector, the absolute calibration bias measured on the strongest signal at CMB frequencies, i.e., the Solar dipole, by comparing the injected Solar dipole and the recovered one. The average value is not null, indicating that there is indeed a small bias induced by the simulation of the ADCNL inaccuracies.
The red crosses in Fig.~\ref{fig:calib_accuracy} are the re-extraction of the inter-calibration, as done in \sroll, but which provides only relative values around zero. Here, we adjust the unknown average to the average value of the blue points. This measures the accuracy of the relative calibration achieved by \sroll\ for each detector in each frequency band.

Table~\ref{tab:ratios} gathers, for CMB-calibrated frequencies, a posteriori Solar dipole calibrations for the data (column~A), for a single simulation (columns~B, C, and D), and for averages over 100 simulations (columns~E and F).
\begin{table*}[htbp!]
\newdimen\tblskip \tblskip=5pt
\caption{A posteriori photometric calibration test using the best Solar dipole estimation (see values in Sect.~\ref{sec:solardipole}) for single detectors within a frequency, and frequency averages. Column~A refers to data, columns~B, C, and D refer to the single fiducial simulation, and columns~E and F refer to 100 E2E simulations where the best Solar dipole was used as input.}
\label{tab:ratios}
\vskip -3mm
\footnotesize
\setbox\tablebox=\vbox{
 \newdimen\digitwidth
 \setbox0=\hbox{\rm 0}
 \digitwidth=\wd0
 \catcode`*=\active
 \def*{\kern\digitwidth}
\newdimen\signwidth
\setbox0=\hbox{+}
\signwidth=\wd0
\catcode`!=\active
\def!{\kern\signwidth}
\newdimen\pointwidth
\setbox0=\hbox{.}
\pointwidth=\wd0
\catcode`?=\active
\def?{\kern\pointwidth}
\halign{\hbox to 3.0cm{#\leaderfil}\tabskip3em&
\hfil#\hfil\tabskip3.5em&
\hfil#\hfil\tabskip1.0em&
\hfil#\hfil\tabskip0.5em&
\hfil#\hfil\tabskip3.5em&
\hfil#\hfil\tabskip 1.25em&
\hfil#\hfil\tabskip 0em\cr
\noalign{\doubleline}
\omit&\sc Data& \multispan3\hfil\sc Single fiducial simulation\hfil& \multispan2\hfil\sc 100 simulations\hfil\cr
\noalign{\vskip -5pt}
\omit&&\multispan3\hrulefill&\multispan2\hrulefill\cr
\noalign{\vskip 3pt}
\omit&&&&\multispan3\hfil Absolute frequency bias\hfil\cr
\noalign{\vskip -4pt}
\omit&\multispan2\hfil Detector rms\hfil&\sroll\ gain&\multispan3\hrulefill\cr
\omit\hfil\sc Frequency\hfil&\multispan2\hrulefill&uncertainty&&Average&rms\cr
\noalign{\vskip 2pt}
\omit\hfil [GHz]\hfil&(A)&(B)&(C)&(D)&(E)&(F)\cr
\noalign{\vskip 3pt\hrule\vskip 5pt}
100& $5\times10^{-6}$& $1.1\times10^{-5}$& $5.8\times10^{-6}$& $-5.9\times10^{-5}$& $8.0\times10^{-5}$& $1.5\times10^{-4}$\cr
143& $8\times10^{-6}$& $3.7\times10^{-6}$& $2.7\times10^{-6}$& $*3.1\times10^{-4}$& $2.1\times10^{-4}$& $1.1\times10^{-4}$\cr
217& $2\times10^{-5}$& $1.3\times10^{-5}$& $3.1\times10^{-6}$& $*2.3\times10^{-4}$& $2.8\times10^{-4}$& $1.4\times10^{-4}$\cr
353& $2\times10^{-4}$& $1.0\times10^{-4}$& $2.8\times10^{-5}$& $*6.0\times10^{-4}$& $2.4\times10^{-4}$& $3.9\times10^{-4}$\cr
\noalign{\vskip 3pt\hrule\vskip 5pt}}}
\endPlancktable
\end{table*}
The dispersion within each frequency is given in column~B and it compares well with column~A for the data, showing that the gain bias modelled by the ADCNL accounts for the dispersion of gain within a frequency. The difference between the red and the blue crosses in Fig.~\ref{fig:calib_accuracy} (column~C in Table~\ref{tab:ratios}) gives the uncertainty introduced by \sroll's determination of the absolute calibration, which is significantly smaller than the dispersion (column~B).

\textit{We thus conclude that the stability in time of the detector-chain gain is affected only by the ADCNL, which does not average to zero over the whole mission, and accounts for the observed dispersion in gain between bolometers. The gains of the bolometers themselves are extremely stable, as expected from their long heritage.}

The bias introduced by \sroll\ on the frequency calibration is obtained by the straight (not noise-weighted, see Sect.~\ref{sec:caveats}) average of the bias of all detectors in that frequency band. For the fiducial simulation, column~D of Table~\ref{tab:ratios} gives, for each frequency, the absolute gain biases, which are small as expected. We neglect the statistical uncertainty on the measurement of the dipole given by \sroll\ (column~C), since it is very small.

The uncertainty of the overall absolute calibration process, based on the orbital dipole, but measured on the recovery of the input Solar dipole, is assessed statistically through 100 E2E simulations. The average bias in the 100 simulations is given in column~E of Table~\ref{tab:ratios} and the rms in column~F. The uncertainty on the average is $1/\sqrt{100}= 0.1$ of the value listed in column~F. This is significantly smaller than column~E, thus this small bias correction was applied to estimate the HFI 2018 Solar dipole amplitude. The accuracy of the absolute calibration based on the orbital dipole and the \sroll\ analysis has thus been tested with 100 simulations by comparing the input values of the Solar dipole and the recovered ones. \textit{The frequency-averaged calibration bias and its uncertainty from these 100 simulations are the ones reported in columns~E and F of Table~\ref{tab:ratios}; the absolute calibration accuracy is better than $3\times10^{-4}$ for the CMB channels and} better than $1.5\times10^{-4}$ \textit{after correction of the bias.}

These biases are removed in the Solar dipole value given in Sect.~\ref{sec:solardipole}, and all have the same sign, amounting to 0.3 to 1\,\microK\ for the three lowest frequencies. Nevertheless, such a correction was not implemented in the released 2018 HFI map calibration, to maintain coherence with the removal of the 2015 Solar dipole common with LFI. This is why we recommend (in Sect.~\ref{sec:freqmaps}) to users of the HFI data, to apply these gain corrections after removing the 2015 dipole, but before subtracting the 2018 dipole.

The small absolute gain correction has a rather large relative uncertainty (column F of Table~\ref{tab:ratios}), which has been added to the error given in Table~\ref{tab:dipole}. This leads to a significant increase in the uncertainty on the amplitude of the Solar dipole (0.3\microK), comparable to the dispersion between frequencies, which leads to the increased error with respect to the estimated errors in Table~\ref{tab:dipole}.

In summary, the new CMB calibration is more accurate than the HFI 2015 one by about an order of magnitude and the best determination to date. Furthermore, through the use of the E2E simulations, we have demonstrated that the calibration dispersion inside a frequency band is due to the ADCNL and we have evaluated the induced calibration bias. This correction has been applied to the Solar dipole amplitude, leading to a consistent picture in Tables~\ref{tab:dipole} and ~\ref{tab:ratios}.

The submillimetre channels are calibrated using giant planet models, for which the uncertainty on the absolute calibration is estimated conservatively at 5\,\%. It is interesting to note that the Solar dipole detected in the 545-GHz channel is within 20\microK\ of the CMB channels amplitude if we take the full range of uncertainty. Figure~\ref{fig:Dipole} shows that the direction of the Solar dipole depends more on the specific component-separation methods and sky fractions used than on the CMB-calibrated frequencies. This gives an a posteriori CMB calibration of the planet model within 1\,\%, which is much smaller than the 5\,\% uncertainty given in \citet{Moreno2010EHSC}. A direct comparison between the dipole CMB calibration and the giant planet calibration (these being nearly point sources for \Planck) requires a knowledge of the transfer function discussed in Sect.~\ref{sec:summaryTF}. We can simply conclude here that the 545-GHz planet calibration is fully in line with the Solar dipole and a transfer function from the dipole to high multipoles at the 6\,\% level, for the range $15< \ell <1000$ (see Fig.~\ref{fig:SMICA}).

\subsection{Intensity inter-frequency band calibration on CMB anisotropies}
\label{sec:interfreq}
\begin{figure}[htbp!]
\includegraphics[width=\columnwidth]{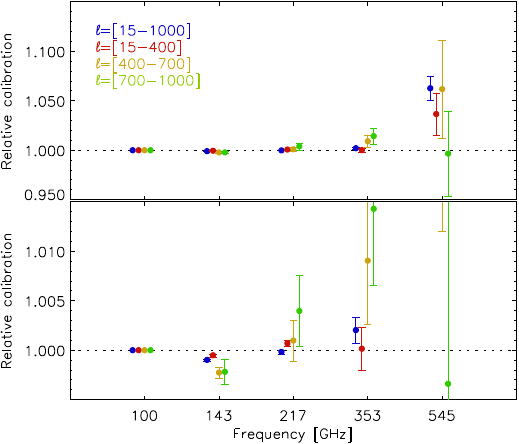}
\caption{Inter-band calibration relative to 100\,GHz, expressed as factors for the maps, measured on power spectra in a broad range ($15<\ell<1000$), and on the three bands around the first ($15<\ell<400$), second ($400<\ell<700$), and third ($700<\ell<1000$) acoustic peaks. The bottom panel is an enlargement of part of the top one.} 
\label{fig:SMICA}
\end{figure}

Following the a posteriori Solar dipole inter-calibration between frequencies reported in Sect.~\ref{sec:solardipole}, we derive relative inter-calibrations based on CMB anisotropies from the {\tt SMICA} component-separation method, on 60\,\% of the sky, for several multipole ranges of the CMB power spectrum. This allows us to test for any multipole-dependent transfer-function residuals. The ranges used ($\ell=15$--400, 400--700, 700--1000, and 15--1000) cover each of the first three acoustic peaks, and the sum of all three. For each of these ranges, the calibration ratio is obtained by performing a noise-weighted ratio of the power spectra between one of the frequencies (143, 217, 353, or 545\,GHz) and a reference frequency taken to be 100\,GHz (the frequency least affected by foregrounds). These ratios are plotted in Fig.~\ref{fig:SMICA} and reported in Table~\ref{tab:dipolechaipa}, together with the corresponding ratio derived from the Solar dipole (also referred to 100\,GHz).
\begin{table*}[htbp!]
\newdimen\tblskip \tblskip=5pt
\caption{Relative differences of CMB response, with respect to 100\,GHz, measured either on the Solar dipole or using the first, second, and third peaks in the CMB power spectrum.}
\label{tab:dipolechaipa}
\vskip -2mm
\footnotesize
\setbox\tablebox=\vbox{
 \newdimen\digitwidth
 \setbox0=\hbox{\rm 0}
 \digitwidth=\wd0
 \catcode`*=\active
 \def*{\kern\digitwidth}
\newdimen\signwidth
\setbox0=\hbox{+}
\signwidth=\wd0
\catcode`!=\active
\def!{\kern\signwidth}
\newdimen\pointwidth
\setbox0=\hbox{.}
\pointwidth=\wd0
\catcode`?=\active
\def?{\kern\pointwidth}
\halign{\hbox to 2.5 cm{#\leaderfil}\tabskip 1em&
\hfil#\hfil&
\hfil#\hfil&
\hfil#\hfil&
\hfil#\hfil\tabskip 0em\cr
\noalign{\doubleline}
\omit\hfil&& \multispan3\hfil First peaks\hfil\cr
\noalign{\vskip -5pt}
\omit\hfil Frequency\hfil&&\multispan3 \hrulefill \cr
\omit\hfil [GHz]\hfil& Solar dipole&$\ell=15$--400& $\ell=400$--700& $\ell=700$--1000\hfil\cr
\noalign{\vskip 3pt\hrule\vskip 5pt}
100& Reference& Reference& Reference& Reference\cr
143& $-8.2\times10^{-6}\pm1.1\times10^{-4}$& $-5.1\times10^{-4}\pm1.5\times10^{-4}$& $-2.3\times10^{-3}\pm5.2\times10^{-4}$& $-2.2\times10^{-3}\pm1.2\times10^{-3}$\cr
217& $-2.5\times10^{-5}\pm1.4\times10^{-4}$& $!6.9\times10^{-4}\pm3.4\times10^{-4}$& $!9.8\times10^{-4}\pm2.1\times10^{-3}$& $!4.0\times10^{-3}\pm3.6\times10^{-3}$\cr
353& $-7.8\times10^{-5}\pm3.9\times10^{-4}$& $!1.5\times10^{-4}\pm2.2\times10^{-3}$& $!9.1\times10^{-3}\pm6.4\times10^{-3}$& $!1.4\times10^{-2}\pm7.8\times10^{-3}$\cr
545& $<1\times10^{-2}\,\textrm{(see text)}$& $!3.7\times10^{-2}\pm2.1\times10^{-2}$& $!6.2\times10^{-2}\pm5.0\times10^{-2}$& $-3.4\times10^{-3}\pm4.3\times10^{-2}$\cr
\noalign{\vskip 3pt\hrule\vskip 5pt}}}
\endPlancktablewide
\end{table*}

The Solar dipole gives upper limits at a few times $10^{-4}$ for inter-calibration at $\ell\,{=}\,1$. Of course the residuals of the beam transfer function will affect the inter-calibration at higher multipoles. From the figure, we can see that there are no highly significant transfer-function residuals revealed by the inter-calibration between the first three acoustic peaks for the CMB channels. A residual transfer function between the three acoustic peaks (red, yellow and green) at 217, 353, and 545\,GHz is marginally detected. The last two frequencies (353 and 545\,GHz) show only marginal calibration differences between the average of the three acoustic peaks (blue dots in Fig.~\ref{fig:SMICA}) and the reference frequency (0.2\,\% and 6\,\%, respectively). These constraints on transfer functions are discussed in Sect.~\ref{sec:summaryTF}.
The transfer function discrepancies between $\ell\,{=}\,1$ and 100 \citep[discussed in][]{planck2013-p01a} have now all been reduced to below $10^{-3}$ for the CMB-calibrated channels.

\subsection{Polarization inter-frequency band calibration on CMB anisotropies}
\label{sec:pol-interfreq}

The {\tt SMICA} method used for intensity in the previous section can also be applied to the CMB polarization data on the first acoustic peaks ($\ell=30$--1000). In the case of polarization, this approach is not expected to provide information on the transfer function associated with the beam window function at the sub-percent level. Indeed, the comparison of the $EE$ CMB acoustic peaks shows larger relative calibration differences between frequencies than was found for intensity; these most likely result from polarization efficiency residuals. These residuals are reported in Table~\ref{tab:calibpolar}, using 143\,GHz as the reference channel, and show significant differences from zero at the percent level with the ground measurements at 100, 143, and 217\,GHz \citep{rosset2010}.
\begin{table*}[htbp!]
\newdimen\tblskip \tblskip=5pt
\caption{Polarization efficiency determination, defined as $\rho$ in Sect.~\ref{sec:instpolpar}. This table gives relative values with respect to 143\,GHz, measured on the {\tt SMICA} $EE$ power spectrum, along with cosmological parameter likelihood values relative to 1, also expressed in terms of map correction. The last column gives the combined residuals.}
\label{tab:calibpolar}
\vskip -2mm
\footnotesize
\setbox\tablebox=\vbox{
 \newdimen\digitwidth
 \setbox0=\hbox{\rm 0}
 \digitwidth=\wd0
 \catcode`*=\active
 \def*{\kern\digitwidth}
\newdimen\signwidth
\setbox0=\hbox{+}
\signwidth=\wd0
\catcode`!=\active
\def!{\kern\signwidth}
\newdimen\pointwidth
\setbox0=\hbox{.}
\pointwidth=\wd0
\catcode`?=\active
\def?{\kern\pointwidth}
\halign{\hbox to 2.5 cm{#\leaderfil}\tabskip 2.0em&
\hfil#\hfil&
\hfil#\hfil\tabskip1.0em&
\hfil#\hfil\tabskip2.0em&
\hfil#\hfil\tabskip 0em\cr
\noalign{\doubleline}
\omit\hfil& & \multispan2\hfil Cosmology driven\hfil\cr
\noalign{\vskip -5pt}
\omit\hfil&\hfil $EE$ first peaks\hfil& \multispan2\hrulefill&\cr
\omit\hfil Frequency\hfil& {\tt SMICA}& {\tt Camspec}& {\tt Plik}&Combined residuals\cr
\omit\hfil [GHz]\hfil& \%& \%& \%&\%\cr
\noalign{\vskip 3pt\hrule\vskip 5pt}
100& $+2.4\pm0.5$& $+1.3\pm0.5$& $+1.0\pm0.5$&$+0.7\pm1.0$\cr
143& !Ref.&$-1.6\pm0.5$& $-1.7\pm0.5$& $-1.7\pm1.0$\cr
217& $+3.6\pm0.5$& $+2.5\pm0.5$& $+2.0\pm0.5$& $+1.9\pm1.0$\cr
\noalign{\vskip 3pt\hrule\vskip 5pt}}}
\endPlancktablewide
\end{table*}

These values can also be compared with the calibration driven by the best $TT$ cosmological model \citep{planck2016-l05}. This provides values of polarization efficiency for each frequency band with respect to intensity calibration, for which uncertainties are better than $3\times10^{-4}$. These values are also reported in Table~\ref{tab:calibpolar} and lead to compatible results with respect to the acoustic peak measurements. This gives confidence that these determinations are realistic (even if not very accurate), and suggests that we should adopt the following scheme. Taking the value for the 143-GHz channel from the {\tt Plik} cosmological parameter likelihood ({\tt Camspec} differs only by 0.2\,\%), we then adjust the other frequencies to the 143-GHz value using the $EE$ acoustic peak comparison. This leads to the combined residuals given in the final column of Table~\ref{tab:calibpolar}. The uncertainties have been linearly combined and are somewhat larger than the estimations of Sect.~\ref{sec:polareff}.

These combined residuals have not been used to correct the delivered HFI frequency maps that are on
the PLA. Because they emerged from a posteriori characterization, they {\it could\/} however be applied to correct the map levels (for example the 100-GHz map is to be multiplied by 1.007), since the two a posteriori residual determinations give a similar pattern, which is compatible with ground and \sroll\ determinations.

\subsection{Point-source calibration}
As noted in previous papers, the scatter in the uncertainty of flux densities of compact sources, for example between different bolometers in the same band, is greater than expected from the uncertainty of the CMB dipole, which is a beam-filling source. This systematic error is discussed at length in section~2.4 of \citet{planck2014-a35}.

\Planck\ is calibrated in brightness on the dipole. The flux density calibration for point sources also requires very good knowledge of the effective beam, which is difficult to obtain in a single survey. Because of the large beams, which include a bright variable background, and because of glitch removal, \Planck\ data are not optimal for good photometry of point sources, and are even less useful for observing moving or variable sources. No improvement over the Second Planck Catalogue of Compact Sources \citep{planck2014-a35} has been identified in this 2018 release, since the accuracy of the point-source photometry is dominated by specific systematic effects, and their correction cannot be improved.

\subsection{Conclusions on calibration}

The improvement obtained through the new mapmaking procedure adopted for this release leads to a much improved dipole calibration stability for polarized bolometers within a frequency band from a few times $10^{-3}$ in 2013 and 15 to better than $2\times 10^{-5}$ for the CMB channels and $2\times 10^{-4}$ for 353\,GHz (column~B of Table~\ref{tab:ratios}). The temporal variability of the bolometer gains has been demonstrated to be due to the ADCNL systematic effect. This effect integrated over the mission induces gain differences between the observed ones and those predicted from the bolometer parameters. This accounts for the associated dispersion of gain within a frequency band, along with the apparent gain variability, and finally the inter-frequency miscalibration.

The 100 E2E simulations give us the uncertainties on the full calibration scheme, which are of the order of 1.0--$1.5\times10^{-4}$ for the CMB channels and $3.9\times10^{-4}$ at 353\,GHz (column~F of Table~{\ref{tab:ratios}). This uncertainty is significantly larger than the \sroll\ statistical uncertainty estimated in Table~\ref{tab:dipole}. Comparing the a posteriori Solar dipole calibration at each frequency from these simulations to the input gain also allows us to estimate the absolute bias per frequency. This bias is corrected for in the amplitude of the Solar dipole, but not in the maps, for a consistent subtraction of the 2015 \Planck\ common Solar dipole with the LFI frequencies.

Polarization-sensitive detectors are calibrated for their response to power input on the unpolarized CMB dipoles, with the same accuracy as the SWBs. Nevertheless the polarization signals also depend linearly on the polarization efficiency, which is known with a much lower accuracy of typically 1\,\%, estimated from the ground measurements, but up to 2\,\% from the data analysis. Furthermore the calibration of signals with angular scales much smaller than the dipole depends on the effective window function. These are discussed in Sects.~\ref{sec:interfreq} and \ref{sec:summaryTF}.

\section{Noise and systematic residuals}
\label{sec:map_characterization}
The accuracy with which systematic effects are removed in the \sroll\ mapmaking has been tested with the E2E simulations, as described in \citelowell. In this section we summarize the results of those earlier tests and also discuss in more detail cases where extraction of instrument parameters has been added to or improved. Furthermore, we investigate how the use of cross-spectra between frequencies helps in removing some systematics. We construct sensitive tests of small residual signals by performing difference tests, i.e., splitting the data into two subsets out of which we can construct maps similar to those released. Such difference maps have been used in many of the tests described in this section. They employ three types of simulations: (i) those that do not include the modelling of the specific systematic effect in the input data; (ii) those with the effect modelled, followed by the full analysis pipeline including correction for that effect; and (iii) the same input, but without correction for that systematic in the processing pipeline. Differences between these maps give either the level of the systematic effects or the level of post-correction residuals that are expected to be present in the data maps. This procedure gives an estimate of the level of the residuals of each systematic effect, which can be compared with the other residuals and with the scientific goals (represented often by the fiducial cosmology power spectra).

Sub-sections~\ref{sec:zodi} to \ref{sec:ADC} discuss each systematic effect in turn,
and shows their residuals. Most of these effects are negligible for the final data products. The last sub-section, Sect.~\ref{sec:syssum}, presents a summary of systematic effects, identifies the main ones, and compares their residuals in a multi-dimensional space, including frequencies and angular scales, based on all of these null tests.

\subsection{Consistency of the zodiacal emission removal}
\label{sec:zodi}
Emission from interplanetary dust is removed from the HFI data, as was already done in the previous 2015 release, using the model from \citet{planck2013-pip88}. The removal of the zodiacal emission was shown to be highly effective through a Survey~1 minus Survey~2 test. That test showed no zodiacal residuals at 545\,GHz (or lower frequencies) and marginal residuals at 857\,GHz, at a level of $10^{-2}$ MJy sr$^{-1}$. The present correction for zodiacal emission applies the same procedure as in the 2015 release, fitting for the emissivities of each component of the zodiacal model. The improvement in the present release comes only from other improvements in the data, which reduce other systematic effects in the maps. We compare the model parameters obtained in table~3 and figure~9 of \citehfimap\ with the updated parameters in Table~\ref{tab:zodi} and in Fig.~\ref{fig:Zodi} of this paper. The improvement is revealed by the much smaller uncertainties, the smoother behaviour with wavelength of the emissivities and the absence of negative emissivities.
\begin{table*}[htbp!]
\newdimen\tblskip \tblskip=5pt
\caption{Zodiacal emissivities for the different components of the Kellsal model \citep{0004-637X-508-1-44}.}
\label{tab:zodi}
\vskip -2mm
\footnotesize
\setbox\tablebox=\vbox{
\newdimen\digitwidth
\setbox0=\hbox{\rm 0}
\digitwidth=\wd0
\catcode`*=\active
\def*{\kern\digitwidth}
\newdimen\signwidth
\setbox0=\hbox{+}
\signwidth=\wd0
\catcode`!=\active
\def!{\kern\signwidth}
\newdimen\pointwidth
\setbox0=\hbox{.}
\pointwidth=\wd0
\catcode`?=\active
\def?{\kern\pointwidth}
\halign{\hbox to 2.5 cm{#\leaderfil}\tabskip2.5em&
\hfil#\hfil\tabskip1.5em&
\hfil#\hfil&
\hfil#\hfil&
\hfil#\hfil\tabskip 0em\cr
\noalign{\doubleline}
\omit\hfil Frequency\hfil\cr
\noalign{\vskip 3pt}
\omit\hfil [GHz]\hfil& Diffuse cloud& Band 1& Band 2& Band 3\cr
\noalign{\vskip 3pt\hrule\vskip 5pt}
\noalign{\vskip 2pt}
857& $0.304\pm0.004$& $1.58\pm0.06$& $0.70\pm0.03$& $2.11\pm0.10$\cr
545& $0.179\pm0.003$& $1.47\pm0.03$& $0.49\pm0.02$& $1.84\pm0.06$\cr
353& $0.082\pm0.002$& $1.52\pm0.02$& $0.35\pm0.02$& $1.77\pm0.05$\cr
217& $0.042\pm0.002$& $1.11\pm0.03$& $0.21\pm0.02$& $1.12\pm0.05$\cr
143& $0.020\pm0.004$& $1.00\pm0.04$& $0.17\pm0.03$& $0.84\pm0.10$\cr
100& $0.018\pm0.006$& $0.54\pm0.10$& $0.07\pm0.04$& $0.19\pm0.12$\cr
\noalign{\vskip 3pt\hrule\vskip 5pt}}}
\endPlancktable
\end{table*}
\begin{figure}[htbp!]
\includegraphics[width=\columnwidth]{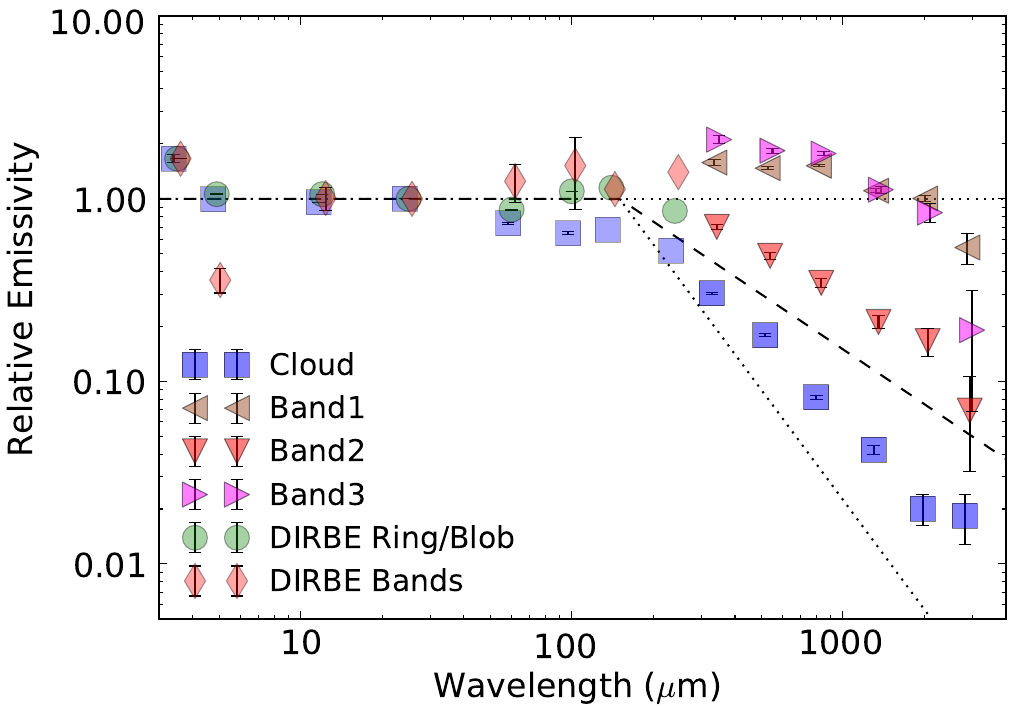}
\caption{\label{fig:Zodi} Zodiacal emissivities as a function of wavelength, combining IRAS, DIRBE, and \Planck-HFI data. For reference, the dotted and dashed lines show emissivities that are unity at wavelengths less than 150\,$\mu$m, and proportional to $\lambda^{-2}$, $\lambda^{-1}$, and $\lambda^{0}$ at longer wavelengths. The emissivities for the cloud and the bands are very similar in level to those reported in \citehfimap, but have smaller errors and show a smoother behaviour.} 
\end{figure}

\subsection{Far sidelobes}

The signal from the far sidelobes (FSL, defined here to be the response of the instrument at angles greater than 5\deg\ from the main beam axis) can introduce spurious polarized signals at large angular scales. In the 2015 release, the FSL contributions were not removed. FSL beam maps over $4\pi$ steradians were computed using {\tt GRASP}\footnote{\url{http://www.ticra.com/products/software/grasp}} software \citep[see figure~14 of][]{planck2013-p03c}. As in \citelowell, their effects on the maps is computed by building HPRs of the FSL beam convolved with an estimate of the sky signal (CMB including dipoles and dust foreground), then running them through the same scan history and destriping procedure as for the real data, to produce FSL map templates for each detector. These templates were subsequently regressed from the final maps as part of the mapmaking procedure.

Table~1 of \citelowell\ and the associated discussion present the direct impact of the FSLs on dipoles and thus on calibration, and show that the very good relative calibrations at 100 and 143\,GHz imply that the FSL corrections are accurate to better than 5 and 2\,\%, respectively. Furthermore, differences between the FSLs of polarized detectors will induce spurious polarization if these differences are not removed. Polarization induced by FSL differences between detectors within a frequency channel (calibration mismatch leakage) can be estimated by taking the rms of the calibration shifts within a frequency. These are all below $2 \times 10^{-5}$ at 143\,GHz and above, and about $10^{-4}$ at 100\,GHz. We use the E2E simulations to propagate these differences to maps and power spectra, as shown in figure~4 of \citelowell. The levels for $EE$ and $BB$ are always below $10^{-6}\microKcarre$ and below $10^{-5}\microKcarre$ for 100 and 143\,GHz, which is much lower than the noise and the dominant residuals of other systematics, as shown below.

\subsection{Warm electronics drifts /second-order non-linearity of bolometers}

Figure~A.2 of \citelowell\ shows that for 143-detset~1 (see definition of detector sets in table~A.1 of \citehfimap), drifts in the warm electronics contribute residuals in the temperature and polarization power spectra that are several orders of magnitude below the noise. Specifically, for $\ell > 10$, $C_{\ell}<10^{-6}\microKcarre$ for $TT$, $EE$, and $BB$, and $C_{\ell}$ reaches $2\times10^{-5}\microKcarre$ at very low multipoles.\footnote{In this paper, we use the term ``very low multipoles'' for multipoles less than 10, which are those relevant for the polarized reionization peak.}

\subsection{Half-ring noise correlation}

Cosmic-ray deglitching is described in \citet{planck2013-p03e}. The deglitching process is based on full ring data. When splitting the data between the first and the second halves of rings for the purposes of carrying out null tests, the deglitching introduces for each glitch in one half, a similar gap in the second half. This effect has been shown to cause correlated noise between the two halves.

Figure~\ref{fig:correlation} shows the effect of deglitching as the glitch detection threshold level is changed.
\begin{figure}[htbp!]
\includegraphics[width=\columnwidth]{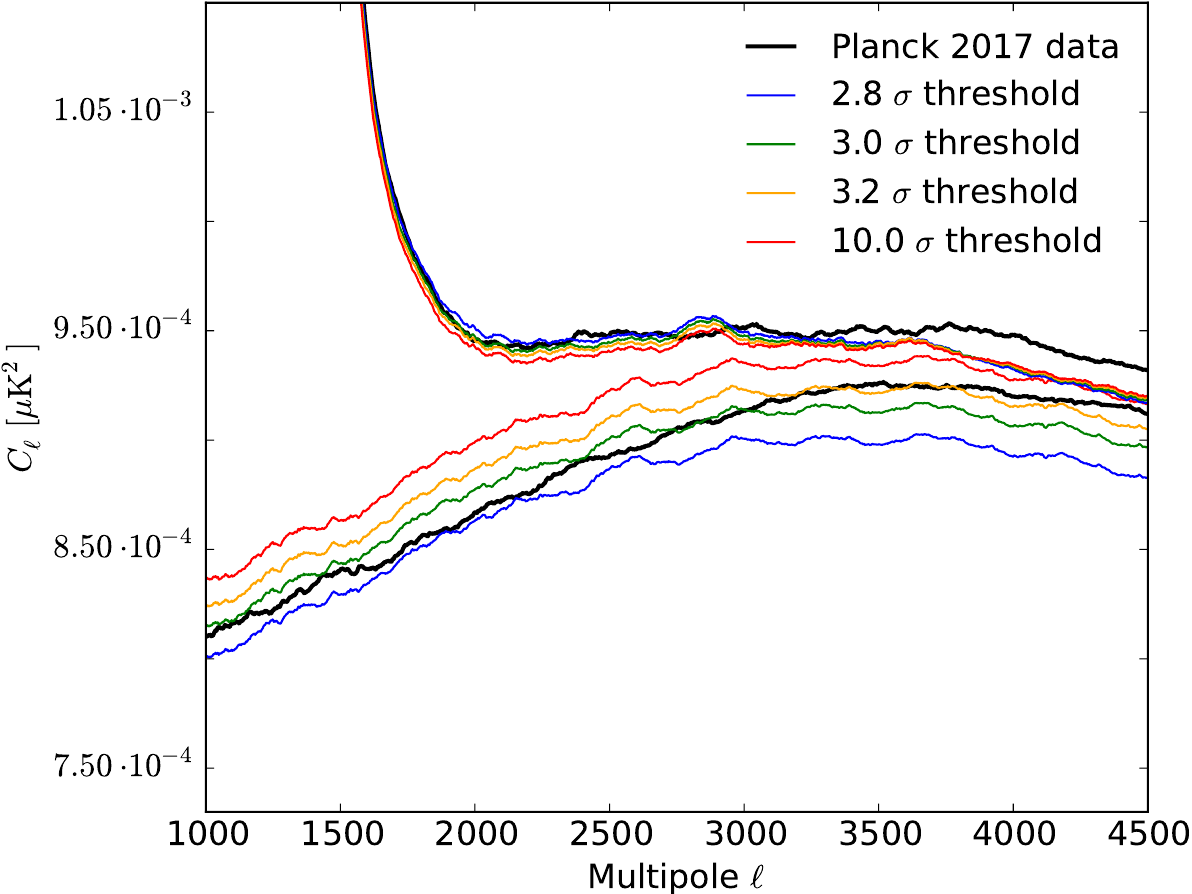}
\caption{Effect of the threshold level on the correlated noise for the 143-1a bolometer. Upper curves are the power spectra of the sums of the half-ring maps, while lower ones are the power spectra of the differences of the half-ring maps.}
\label{fig:correlation} 
\end{figure}
The correlated noise is shown by comparing the behaviour of the sum and the difference of the half-ring maps, colour-coded according to the level of the glitch removal threshold. The black lines represent the data. The CMB signal is clearly seen in the sum at $\ell<3000$; noise dominates above. The best multipole range to study the noise is between 3500 and 4500. For simulations of the sum of the half rings (upper curves), the noise does not depend significantly on the deglitching threshold; all curves are on top of each other in this $\ell$ range. For the half-ring differences (lower curves), the red line ($10\,\sigma$ deglitching threshold) is nearly at the same level as the sum, indicating that there is negligible correlated noise; however, the difference between the sum and the difference increases with decreasing threshold (more glitches masked), which is the sign of an increasing correlated noise fraction. The comparison of the gap between the noise computed for the differences and for the sums, in the data and in the simulations, indicates that the threshold in simulations that corresponds to the data is around $3\,\sigma$, which is close to the expected value for the deglitching parameters.

Because the threshold is set dynamically, and thus is not constant, we cannot accurately evaluate its impact. Nevertheless this simulation has shown that a 3--4\,\% correlation is introduced by the glitch flagging, which is comparable to the correlation detected in the data, even although it cannot be predicted precisely. We thus confirm that half-ring null tests should not be used if sub-percent accuracy is required at high multipoles.

\subsection{4-K lines}

As noted in Sect.~\ref{sec:ouputstosroll}, the HFI's 4-K mechanical cooler induces some noise in the bolometer signal via electromagnetic interference and coupling. These are removed in the TOI processing, as was done for the 2015 release data. The residuals of the 4-K lines are propagated with the E2E simulations to maps and power spectra for all detectors. Figure~\ref{fig:spectrum4Klines} shows, for the case of the 143-3a detector, that the line at $\ell=1800$ reaches $C_{\ell}= 10^{-5}\microKcarre$, which is close to the residual feature identified and discussed in \citet{planck2013-p08}.
\begin{figure}[htbp!]
\includegraphics[width=\columnwidth]{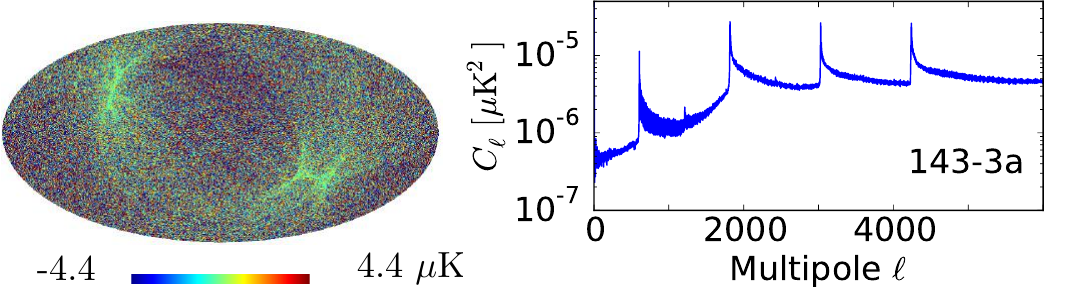}
\caption{Difference map and power spectrum from E2E simulations with and without 4-K lines. The two lower frequency lines at $\ell=600$ and 1200 are smaller than the residual spectra in figure~17 of \citet{planck2014-a13}. Similar figures for all HFI detectors are available in the \citeES.}
\label{fig:spectrum4Klines} 
\end{figure}
The level of the 4-K lines changes by large factors from bolometer to bolometer. The 143-3a results are representative of the average. Nevertheless, given the scatter, it is not possible to model this effect accurately. Hence any weak artefact in the primordial power spectrum that is detected at one of these 4-K line frequencies cannot be interpreted as being meaningful.

\subsection{Compression-decompression}
\label{sec:Compression}

The on-board compression and decompression in the data processing is not lossless. Thus we need to quantify its effect on the CMB and noise through simulations.
\begin{figure}[htbp!]
\includegraphics[width=\columnwidth]{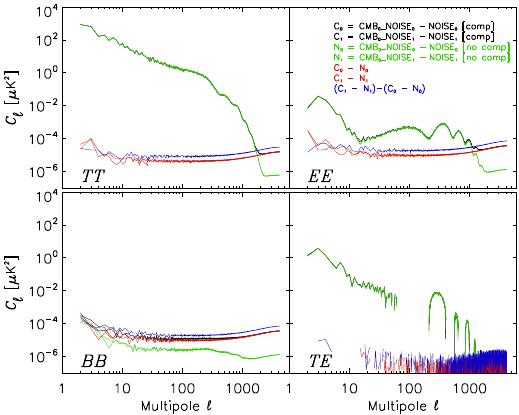}
\caption{Effect of compression/decompression. At 100\,GHz this induces a small noise increase, indicated by the black (with compression-decompression) and green (without) curves, showing the signal (difference between CMB plus noise, with pure noise) for two different noise realizations (NOISE$_0$ and NOISE$_1$). The red curves are the differences between the two noise realizations, which are fully decorrelated, as shown by their difference (blue curve) being higher by a factor of 2. The beam and sub-pixel effects are apparent as a flattening at $\ell > 2000$.}
\label{fig:Compression} 
\end{figure}
We take one CMB and two noise realizations as inputs and propagate the four combinations of these input maps through the E2E simulations with and without compression/decompression. Figure~\ref{fig:Compression} displays all four relevant power spectra, $TT$, $EE$, $BB$, and $TE$. We show differences between different combinations of input maps, specifically:
\begin{itemize}
\item in black, (CMB + NOISE$_0$) $-$ (NOISE$_0$), with compression-decompression, denoted $C_0 $;
\item in black, (CMB + NOISE$_1$) $-$ (NOISE$_1$), with compression-decompression, denoted $C_1 $;
\item in green, (CMB + NOISE$_0$) $-$ (NOISE$_0$), with no compression-decompression, denoted $N_0$;
\item in green, (CMB + NOISE$_1$) $-$ (NOISE$_1$), with no compression-decompression, denoted $N_1$.
\end{itemize}

The black and green curves show these spectra for the difference of an E2E simulation map of the CMB plus noise and the same simulations with only the (identical) noise realization. Both show the CMB spectra and it is thus not surprising that they are nearly on top of each other and not easily distinguishable.

We thus also show, the difference between the two signal spectra $(C_0 - N_0)$ and $(C_0 - N_1)$, both in red As expected, the signal is removed but the noise is only partly removed which reveals the residual effect of the compression/decompression at a level of 10\,\% of the noise. We also examine the effect of two noise realizations by taking their difference $((C_0 - N_0)-(C_0 - N_1))$ in blue, which appears higher than the two noise excess (red) curves by a factor of about 2 for $\ell > 20$, showing almost full decorrelation. At lower multipoles this excess noise is significantly correlated.

In conclusion, compression/decompression affects the noise and not the signal, and induces an order of 10\,\% noise increase. It shows correlation for two independent noise realizations at very low multipoles. Nevertheless, this correlation is too low in amplitude to significantly affect the final results.

\subsection{Beam mismatch leakage and sub-pixel effects}

Mismatch in the size and shape of the main beam between two bolometers can leak CMB and foreground temperature fluctuations into polarization. This is negligible at large angular scales~\citep{tristram2005}, but contributes at small angular scales~\citep{planck2014-a08,planck2014-a13}. This high multipole leakage from $TT$ to $BB$ is also simulated and shown in Fig.~\ref{fig:Compression}. Since this leakage scales with the gradients in the maps, the resulting power spectra of the leakage behave as $\ell^2$ times the temperature signal.

The mapmaking procedure averages all signal samples for which the line of sight falls within the boundaries of a given {\tt HEALPix} pixel. This approximation introduces a small but detectable effect at multipoles corresponding to the pixel size. These subpixel effects are also seen in all cross-spectra signals, as evident in Fig.~\ref{fig:Compression} by the flattening of the green curve (signal) for $\ell > 2000$ at a level of $10^{-6}\microKcarre$.

These effects are very small on the power spectra of diffuse signals. When dealing with very low-level polarized signals, any masks used should include a proper apodization around point sources to mask the strong gradients associated with them. In this 2018 release, we only simulate this effect by computing beam matrices that parameterize subpixel effects, but do not correct for them, given their very low levels.

\subsection{Undetected glitches}
\label{sec:undetected}

Some cosmic-ray hits go undetected in the HFI data. The detected glitch rate mostly depends on the heat capacity and the noise of each detector, and the detected glitch rate is highly variable (by a factor of 4) from detector to detector \citep{planck2013-p03e}. The total (detected and undetected) cosmic-ray hit rate, however, should be nearly constant. A lower rate of detected glitches thus implies more undetected glitches. A model of glitches has been built and is used in the E2E simulations. This model can be used to characterize the undetected glitches. In addition, the tails of detected glitches are not fully corrected for long time constants, and those of order 20 to 30 seconds discussed in Sect.~\ref{sec:xferfunction} are not included at all. Figure~\ref{fig:UndetectedGlitches} shows a noiseless E2E simulation of the residual power spectra of all 143-GHz detectors after removal of detected glitches. Variations from detector to detector are evident. Another simulation also carried out for the 143-GHz detectors on the HPRs, this time including the noise, shows the white noise component above 0.2\,Hz, as displayed in Fig.~\ref{fig:glitches}.
\begin{figure}[htbp!]
\includegraphics[width=\columnwidth]{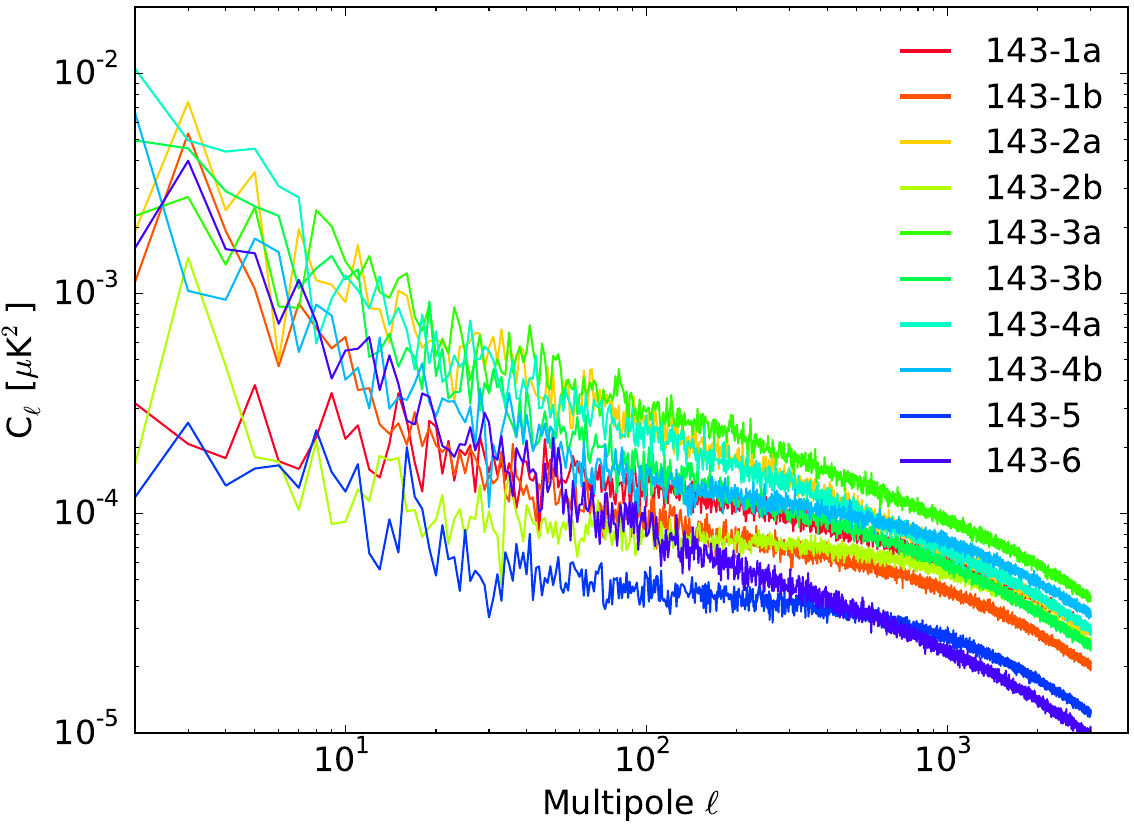}
\caption{For all 143-GHz detectors, simulation of all types of glitches showing power spectra to illustrate the variability both in rate and in tails.}
\label{fig:UndetectedGlitches} 
\end{figure}
\begin{figure}[htbp!]
\includegraphics[width=\columnwidth]{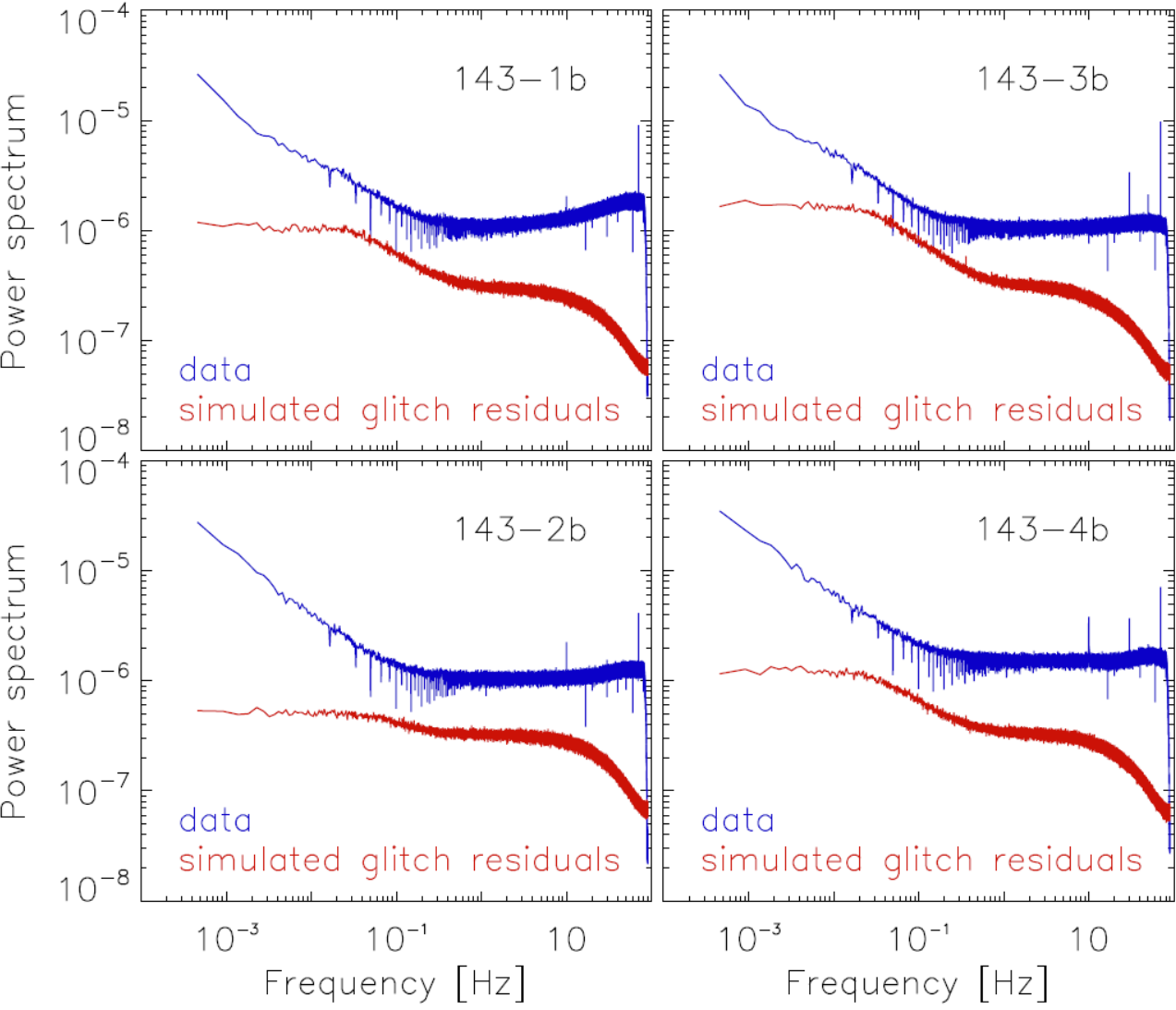}
\caption{TOI noise spectra, in arbitrary units, of four 143-GHz bolometers (blue lines) and simulations of glitch residuals (undetected glitches and remaining tails) in red. The $1/f$ knee frequency of the simulations is identical.} 
\label{fig:glitches}
\end{figure}

The contribution to the noise from the undetected glitches and their remaining tails produces additional white noise from the sharp glitches and a $1/f$ noise component from the tails of glitches (red lines). The knee frequency appears to be stable between 0.2 and 0.3\,Hz. The stability of the knee frequency can be understood as follows: a higher level of noise leads to more undetected glitches and tails, which in turn increases the $1/f$ noise. The simulated knee frequency coincides with the observed knee frequency for the same bolometer (blue line), confirming that the stable knee frequency for very different noise and glitch rates is due to this compensation. This unique feature of a stable knee frequency makes the undetected-glitches model a very likely source candidate for the $1/f$ noise.

The flattening at low frequency in the simulated $1/f$ component (red) is due to the absence of the very long time constant known to be present in the data (see Sect.~\ref{sec:zebra}). Although the level is difficult to predict, this test shows that it is very likely that the undetected glitches and the long time constant tails we ignore are the cause of the $1/f$ noise. This would also account for the Gaussian distribution of this $1/f$ noise component (see figure~3 of \citelowell). This leads to the possibility of using the destriper to remove it \citep[see][]{ashdown2007b}. This was not done yet for this release because the destriping of $1/f$ components above the spin frequency can remove signal at low multipoles and needs to be carefully tested when low multipoles are used for $\tau$ and $r$ determinations.

\subsection{High-energy cosmic-ray showers}

High-energy cosmic-ray showers contribute to the bolometer-plate temperature fluctuations along with the fluctuations of the individual cosmic-ray hits rate on this plate. These temperature fluctuations constitute the correlated noise between all detectors that varies as $1/f^2$ (see figure~2 of \citelowell). This is mostly removed in the TOI processing using the signal of the two dark bolometers and has negligible contribution to the noise. It can also contribute to the low-level correlated white noise observed.

\subsection{Cross-talk and specific instrumental polarization systematics}
\label{sec:instrumentalpolar}
\subsubsection{Cross-talk}
Bolometers show significant bias current cross-talk, which would affect all of the bolometer responses if one (or a few) of these bias currents had been changed. Although the bias currents were adjustable, they was no need to change any of them during the mission. Thus we only need to deal with voltage cross-talk (i.e., signal cross-talk) at constant bias current. The thermal cross-talk signal between bolometer pairs in a PSB can be readily detected thanks to the small time delay (10 to 30 msec) between a strong bolometer glitch and the smaller cross-talk signal in the other bolometers. Wafer glitches are coincident in time and of similar amplitudes, and the separation is very good. The cross-talk between bolometers in different housings is negligible. Conversely, the cross-talk between the two bolometers within the same PSB housing was detected and measured by stacking the nearly coincident strong short glitches (phase-shifted by a few milliseconds) and taking the ratio of the delayed event to the main one. The detected levels are between 0.1 and 0.2\,\%.
\begin{figure}[htbp!]
\includegraphics[width=\columnwidth]{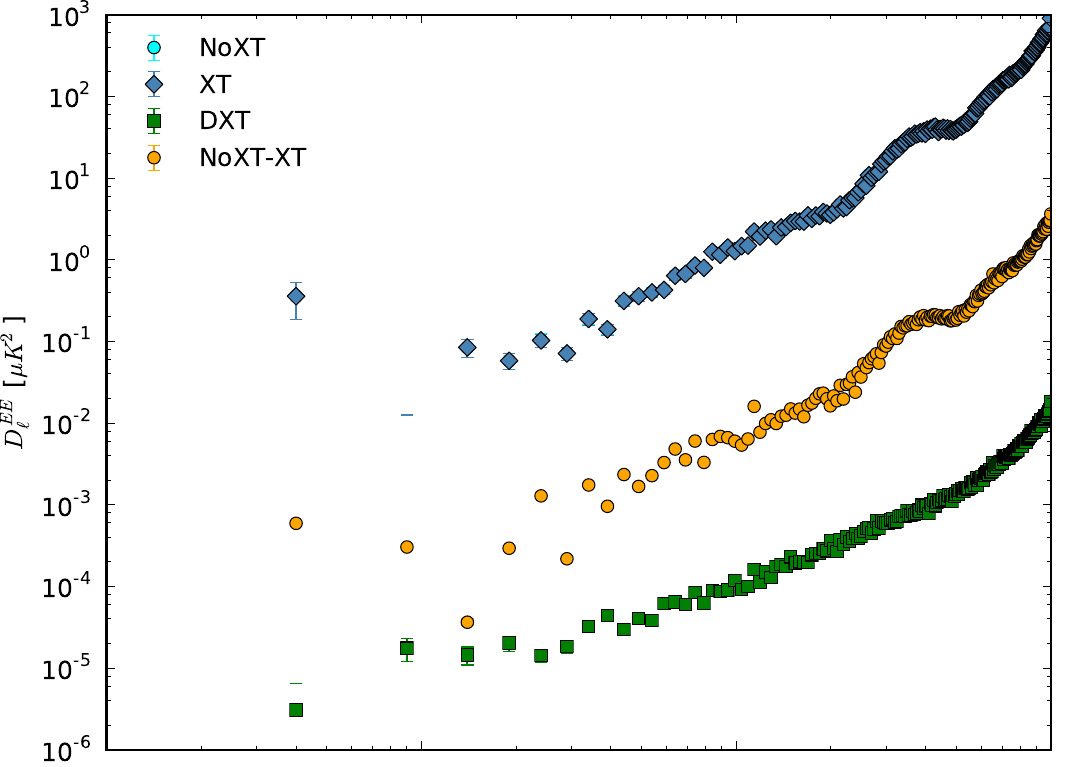}\\
\includegraphics[width=\columnwidth]{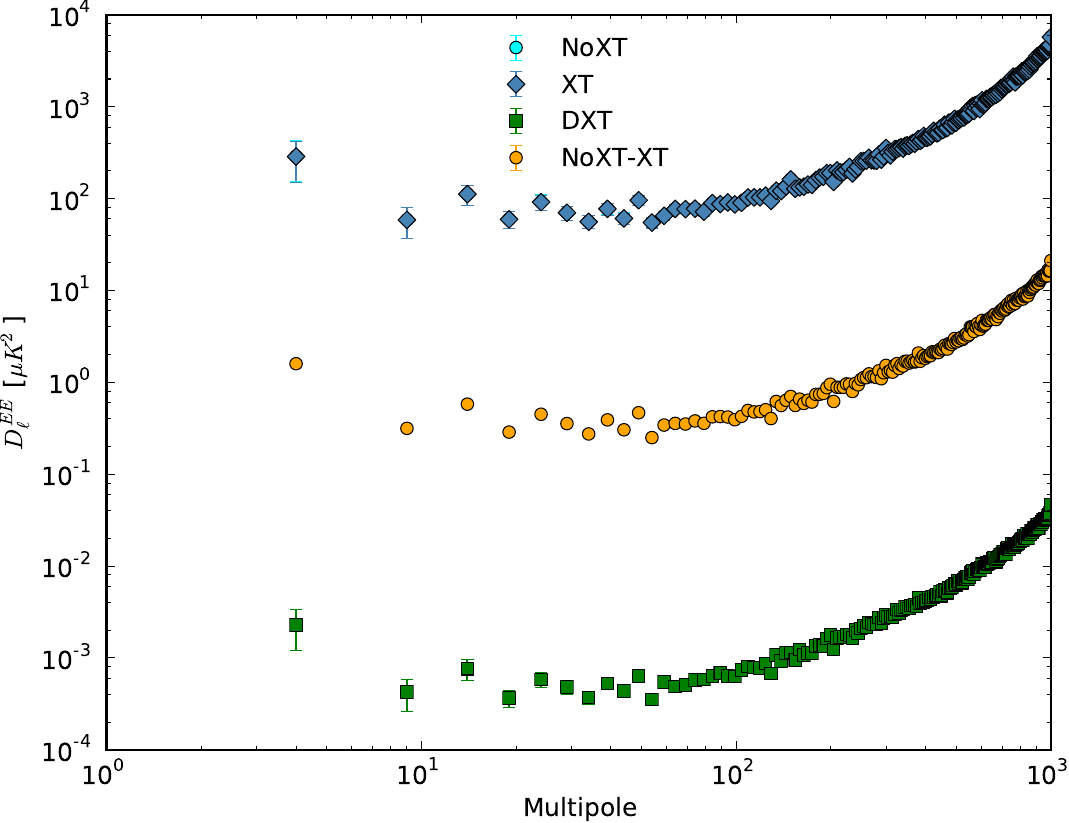}
\caption{$EE$ auto-spectra at 100 (top panel) and 353\,GHz (bottom panel), computed on the simulated maps with (blue) and without cross-talk (light blue). The points are nearly superposed. In orange we plot the difference between auto-spectra obtained with and without cross-talk, while in green we plot the auto-spectra of the difference maps.}
\label{fig:crosstalk} 
\end{figure}
For illustration, Fig.~\ref{fig:crosstalk} shows E2E simulations of the $EE$ auto-spectra at two frequencies.
The blue points show the signal: light blue without cross-talk; and dark blue with cross-talk (the two are on top of each other and not distinguishable in the plot). The orange points show the difference between the two power spectra (with and without cross-talk), which should be approximately equal to the signal plus the noise multiplied by the cross-talk amplitude (around a few times $10^{-3}$ as observed). The green points show the power spectrum of the difference maps made with and without cross-talk. The signal cancels and only the correlated noise appears. It is expected that the cross-talk is approximately equivalent, to first order, to a gain change of the same order, which has been demonstrated (through E2E simulations) to be fully recovered by \sroll\ with an accuracy of $3 \times 10^{-5}$ at 217\,GHz (see table~A1 in \citelowell). The net effect is to induce correlated noise, as shown by the green points at a much lower level than the noise.

The cross-talk is due to thermal conduction between the two bolometers of the same PSB. Its signal is shifted in time by 10 to 30 milliseconds. We also investigate the effect of this time shift on the polarization. We built maps with half-mission data sets, and we take the difference, with and without the time shift. We do the same with the rings data sets. Figure~\ref{fig:Crossshift} shows the cross-spectra between these difference maps for $TT$, $EE$, $BB$, and $TE$. For the half-mission test (blue lines), the effect is symmetric between the odd and the even survey of the same half mission, and the amplitude is negligible. For the ring null test, the effect is not symmetrical, and generates a residual of the polarized signal in $EE$, $EB$ and $BB$. There is no effect in $TE$, since the polarized $E$ residual is not correlated with the $T$ one. This creates an excess noise at the level of a few times $10^{-5}\microKcarre$, which is small but not negligible.
\begin{figure}[htbp!]
\includegraphics[width=\columnwidth]{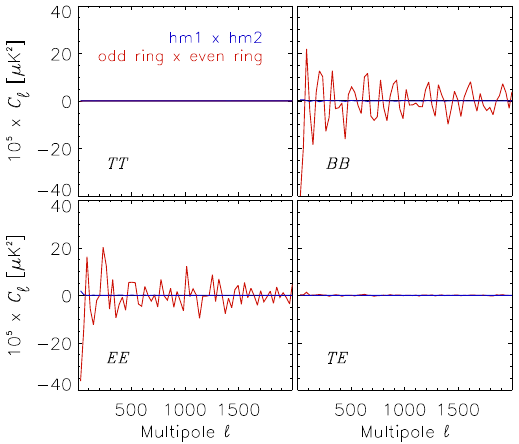}
\caption{Simulation of the effect on polarization of the time shift of the cross-talk between detectors in the same PSB. Cross-spectra between half-mission and odd-even ring differences with and without maps are shown for $TT$, $EE$, $BB$, and $TE$.}
\label{fig:Crossshift} 
\end{figure}

\subsubsection{Instrumental polarization parameters}
\label{sec:instpolpar}

The power seen by a detector is:
\begin{equation}
P_{\mathrm{detector}}=\int \mathrm{gain} \times \left[ I+\rho \left(Q \cos(2\theta) + U \sin(2\theta)\right) \right ] d\Omega , \nonumber
\end{equation}
the polarization efficiency $\rho$ being given by $\rho=(1-\eta)/(1+\eta)$, where $\eta$ is the cross-polarization leakage. Appendix~A.6 of \citelowell\ describes the errors in the polarization parameters found in the ground-based calibration of HFI and their effects on the polarization angular power spectra. Because the primary calibrations are derived from the total intensity of the CMB dipole, uncertainties in these polarization calibration parameters can contribute leakage in $EE$ comparable to the noise level at very low multipoles. The polarization efficiency of the detectors has been measured on the ground with a 0.2\,\% statistical accuracy but there was no evaluation of the systematic effects. The ground measurements of the polarization angles (measurement errors of order 1\deg) were checked with in-flight observations of the Crab Nebula (see \citealt{planck2014-a35}, section~7.4 of \citehfimap, and reference therein), with a null result of a $0\pdeg27\pm0\pdeg22$ shift.

We also need to check, on the sky data, for a global rotation of the focal plane (and corresponding rotation of all the polarization angles), which could be induced by thermo-mechanical effects in flight. To look for such a global rotation, we use the specific leakage induced between the temperature and polarization that leads to non-zero $TB$ and $EB$ power spectra. Section~A.6 of \citelowell\ gives values for polarization angle errors derived from $TB$ and $EB$ power spectra for 100, 143, and 217\,GHz. Using these six measures gives a global rotation angle of $0\pdeg28\pm0\pdeg09$ at the $3\,\sigma$ level. This small rotation is not included in the present data processing. At least for the PSBs, this is within the stated systematic uncertainty of \citet{rosset2010}, and thus consistent with the pre-launch calibrations.

\subsubsection{Polarization angle and polarization efficiency}
\label{sec:polareff}

Polarization efficiency error induces a leakage from $EE$ into $BB$ that is proportional to $\rho ^2$. Simulations of the leakage induced by the errors on the polarization angles are discussed in \citet{rosset2010}, but these did not include foregrounds. The relevant figures, available in the \citeES, show that the angle error affects the $EE$ power spectra at a level of $3\times 10^{-5}$\microKcarre\ on the reionization peak at $\ell=4$.

In the previous section, we have shown that the ground-measured angles used in the data analysis are coherent with the IRAM measurement of the Crab Nebula within $0.3\deg$. The internal HFI $TB$ and $EB$ data gives the same upper limit. This leads to negligible leakage from $E$ to $B$. Levels of leakage from intensity to polarization, due to gain mismatch between detectors, are also negligible, as shown by the quality of the intra-frequency calibration. Finally, the polarization efficiency of each detector has been measured on the ground to be between 0.85 and 0.95, with a statistical error of 0.3\,\% and not much better than $1\,\%$ when systematic effects are considered. This polarization efficiencies are integrated into the mapmaking. The polar efficiency residual induces a gain error in $E$ and a leakage to $B$ that is negligible.

While the effects listed so far are negligible, we still need to check the relative accuracy of the polarization efficiency between bolometers. We build single-bolometer maps (see Sect.~\ref{sec:monobolomaps}), from which we can remove the appropriate bandpass leakage before building the coadded frequency-band maps (see Sect.~\ref{sec:srollscheme}). It is then possible to find the residual polarization efficiency error with respect to ground measurements for each detector within that frequency band. Figure~\ref{fig:CrossPol} shows the residual polarization efficiency values from the data with respect to those measured on the ground. For the 353-GHz PSBs, these residuals could be measured on the strong dust polarized signal, reaching up to 2.5\,\%, and with a 1.2\% rms. This is significantly larger than the statistical uncertainties for the ground measurements but close to the estimates which include the systematic effects \citep{rosset2010}.
\begin{figure}[htbp!]
\includegraphics[width=\columnwidth]{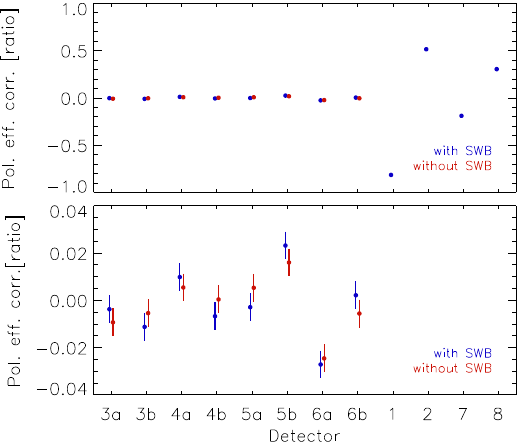}
\caption{Relative polarization efficiency with respect to ground-based measurements, extracted from {\tt SRoll} single-bolometer maps for the 353-GHz bolometers. The bottom panel is an enlargement of part of the top one. It shows the small polarization efficiency difference with respect to ground-based measurements when used with (in blue) the SWBs and without (in red) the SWBs.} 
\label{fig:CrossPol}
\end{figure}

Residual values are plotted in Fig.~\ref{fig:CrossPol} as red points for the case without SWBs and as blue points when including the SWBs. The figure shows in a spectacular way the large relative uncertainties in the low polarization efficiency of the SWBs. This shows that there is a residual systematic effect on the polarization efficiency for PSBs. This residual is comparable to what has been measured on the ground and used in the mapmaking for the SWBs. We thus decided to make public two products for the 353-GHz intensity maps, namely those with and without SWBs.

We want to estimate the effect of the uncertainties in the polarization efficiency demonstrated above (even though these residuals were not included in the processing). To do so, we use two sets of E2E simulations: one without errors in the polarization efficiency, and the other one with a spread in polarization efficiency representative of the error between detectors within one frequency band. We build cross-spectra between two halves of each set and difference those cross-spectra. This is done for three values of the rms of the spread in efficiencies, 0.5, 1, and 2\,\%, to model in a conservative way the errors in the simulations (nominally 0.5\,\% for the PSBs, but showing a larger dispersion in Fig.~\ref{fig:CrossPol}). We test our two main data splits, i.e., rings and half-mission maps. Figure~\ref{fig:Polareff} displays the relative variance within logarithmic bins in the difference of the cross-spectra of the two simulations.
\begin{figure}[htbp!]
\includegraphics[width=\columnwidth]{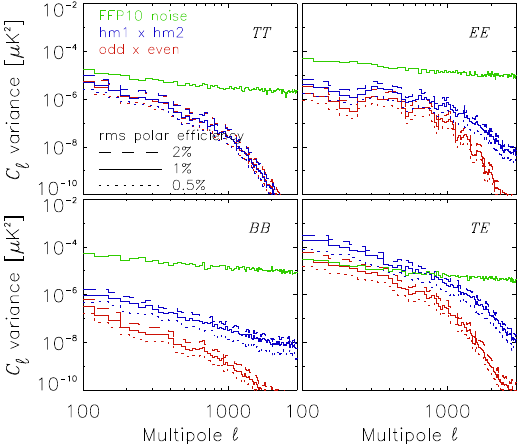}
\caption{Simulation, at 143\,GHz, of the polarization efficiency error propagated to power spectra. Specifically plotted is the relative variance within logarithmic bins in the cross-spectra half-mission~1 $\times$ half-mission~2 (blue curves) and odd $\times$ even rings (red curves) difference without and with polarization efficiency uncertainties (of 0.5, 1.0, and 2\,\%). The green curves show the noise level from the half-mission null test.}
\label{fig:Polareff} 
\end{figure}
The impact is smaller than the noise for $TT$, $EE$, and $BB$, but not for $TE$. The polarization efficiency mismatch causes leakage between temperature and polarization and increases the correlated noise in the $TE$ cross-spectrum. The variance of the half-mission cross-spectra associated with polarization efficiency uncertainties is larger than the odd-even rings cross-spectra. This can be understood, since the scanning strategy is the same from one year to the next, and the half-mission sets have almost the same pixelization.

In conclusion, we have measured small changes in the polarization efficiency compared to the ground-based values, as shown in Fig.~\ref{fig:CrossPol}. These small changes have {\it not\/} been included in the frequency maps in this release. In addition, for 353-GHz polarization studies, one must use the maps based on PSBs only. At other frequencies, when including SWBs, we show in Sect.~\ref{sec:allsyste} that polarized maps are not significantly affected for this 2018 release.

\subsection{Transfer function}
\label{sec:xferfunction}
\subsubsection{Need for an empirical transfer function}
The empirical transfer functions (TFs) are introduced at low harmonics of the spin frequency to account for inaccuracies in the TOI processing step that removes the time constants of the detectors. These corrections are based on the scanning beams measured on planets and the corresponding effective window function derived from a first iteration of the mapmaking with the same TOI-HPR data. Time-constant-induced tails in the effective beams shown in figure~12 of \citet{planck2013-p03c} illustrate these inaccuracies in the transfer functions. This procedure was improved in the 2015 release; nevertheless the accuracy cannot be much better than a few tenths of a percent and the correction cannot detect time constants comparable with the spin frequency (although they are known to be present).

Other residuals are partially degenerate with the time transfer functions. They are associated with the beam ellipticity acting on strong gradients and strong extended signals (CMB dipoles and the Galactic plane) integrated over the FSL. All these effects are different for the same sky pixel when scanned in two opposite directions. The destriper will identify such effects in the differences of signals from the same sky pixel observed by the same bolometer between odd and even surveys. We thus introduced in \sroll\ an empirical complex TF correction in the mapmaking to minimize all these time-like residuals.

\subsubsection{Implementation of the empirical TF}
The empirical TF correction for each bolometer is parameterized with four complex amplitudes for four bins of spin harmonics. These parameters are solved for in the \sroll\ destriper. However, the redundancy and accuracy of the data does not allow us to extract all of these parameters. At all frequencies, we correct for the imaginary part by removing the empirical TF in the ${h=1}$ to ${h=4}$ bins, which show significantly smaller $0.1\,\%$ residuals (figure~11 of \citelowell). 
The real part of the transfer function is not detected accurately at the CMB frequencies (i.e., 100, 143, and 217\,GHz) and is not corrected for at these three frequencies.

At 353\,GHz, and in the submillimetre channels, both the real and imaginary parts are accurately extracted using the strong dust emission signal from the Galactic plane (figure~10 of \citelowell), and are corrected for at this frequency. The phase shifts are at the level of less than $10^{-3}$. The correction at 353\,GHz decreases with spin harmonics, from about $3 \times 10^{-3}$ in the ${h=2}$ bin, to less than $10^{-3}$ in the ${h=4}$ bin. The real part is detected at 353\,GHz on the dust emission in the Galactic ridge and molecular clouds.

As noted, for the CMB channels, the \sroll\ algorithm does detect the phase shift on the dipoles, but does not solve for the real part of the TF (second order on the dipoles). The absence of detection of the real part in the CMB channels is expected, due to the low dust signal.

\subsubsection{Effects of low-multipole TF residuals at 353\,GHz}
\label{sec:zebra}
We use the 353-GHz channel, for which we can extract from \sroll\ both the real and imaginary parts of the empirical transfer function (at least in some range of frequency), to simulate the effects on null-test maps. Assuming that the very long time constants dominate the residuals, we take a simple model of a single time constant of 25 seconds (red line in Fig.~\ref{fig:modeltf}) to represent the low-multipole transfer function extracted by \sroll\ (blue boxes) and used in the processing. We propagate the residuals using the E2E simulations.
\begin{figure}[htbp!]
\includegraphics[width=\columnwidth]{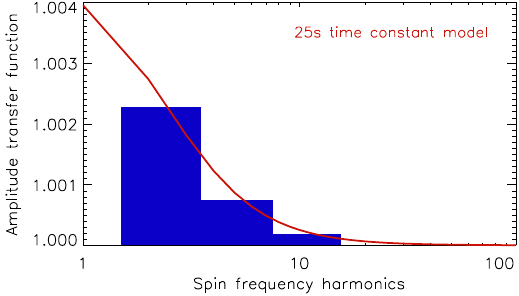}
\caption{Model empirical transfer function at 353\,GHz as a function of spin frequency harmonics. Values are those shown in figure~10 of \citelowell. The four \sroll\ measured bins are shown as blue boxes. The red line is the function used in the simulations.}
\label{fig:modeltf} 
\end{figure}
The survey null-test maps, shown in Fig.~\ref{fig:mapstf}, contain zebra stripes (left figure), with a very specific pattern.
\begin{figure}[htbp!]
\includegraphics[width=0.48\columnwidth]{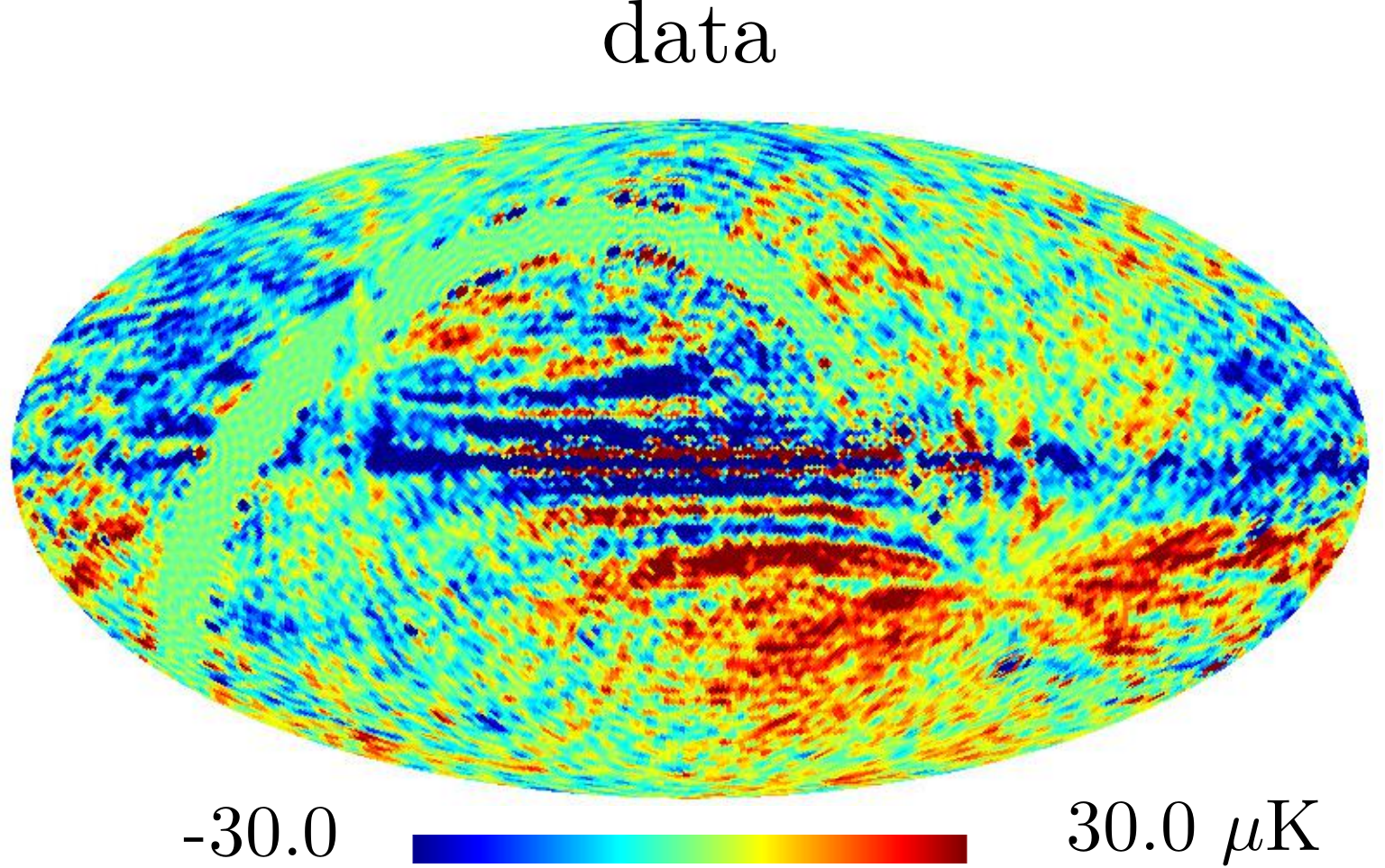}
\includegraphics[width=0.48\columnwidth]{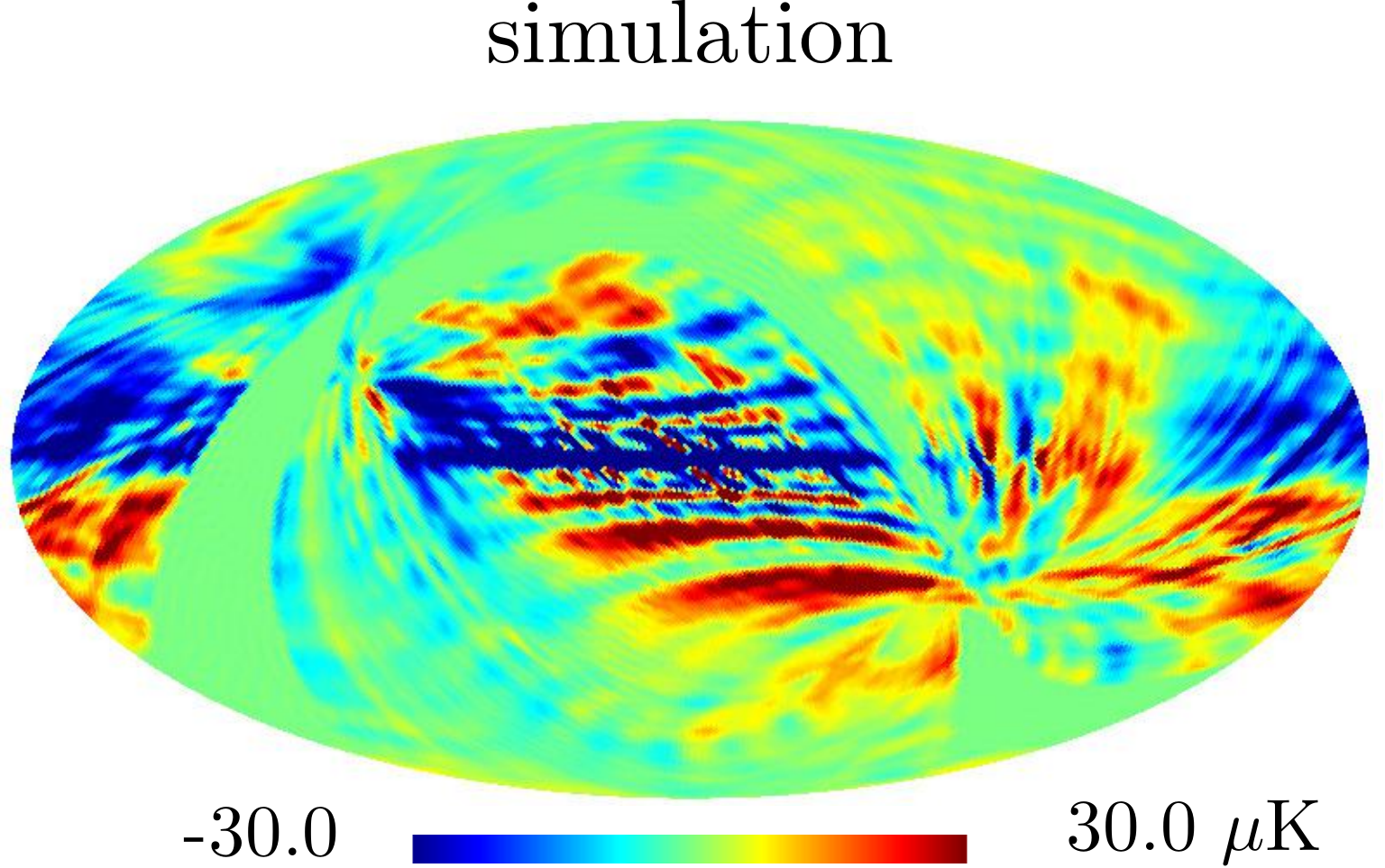}
\caption{353-GHz survey difference maps, with data on the left and simulations on the right. The zebra patterns, due to the difference between the TF implemented (blue boxes in Fig.~\ref{fig:modeltf}) and the very long time constant TF model (red line), are of comparable intensity and shape in the data and in the simulations.} 
\label{fig:mapstf}
\end{figure}
The map on the right shows the simulated residuals associated with the difference between the TF implemented (blue boxes in Fig.~\ref{fig:modeltf}) and the single very long time constant TF model (red line). The pattern due to this difference is strikingly similar to the one observed in the data (left map), demonstrating that the origin of the zebra stripes is indeed in the very long time constants.

The simple model of the empirical transfer function (red line in Fig.~\ref{fig:modeltf}) can also be extrapolated to its ${h=1}$ bin component from the \sroll\ measurements (not corrected in \sroll, as discussed in Sect.~\ref{sec:approx}). We simulate the effect of this time-dependent-dipole in the HPRs, which should be included for consistency, and is expected to show up between odd and even surveys. Figure~\ref{fig:chapochinoi} shows for the data the inter-calibration of all detectors measured survey per survey by the Solar dipole residuals.
\begin{figure}[htbp!]
\includegraphics[width=\columnwidth]{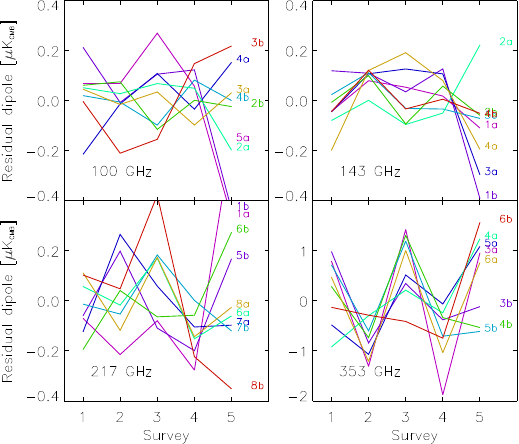}
\caption{Residual Solar dipole amplitude for each bolometer, shown by survey. The average dipole at each frequency has been subtracted.} 
\label{fig:chapochinoi} 
\end{figure}
The 100- and 143-GHz plots do not show any odd-even survey systematic effects, but mostly a noise-like behaviour down to a level of $\pm 0.2\microK$, with an rms calibration dispersion per survey (at the CMB frequencies) between 0.03 and 0.04 and an odd-even survey difference between 0.07 and 0.10, or about $2\sigma$. At 353\,GHz, as expected, a clear oscillatory pattern appears between odd and even surveys, with an amplitude of 1.2 compared to the rms of 0.24 ($5\sigma$). This was seen already in \citelowell\ and the transfer function was mentioned as the probable origin. Figure~\ref{fig:simchapo} shows the E2E simulation of the spin frequency harmonic 1 effect on calibration, which reproduces the odd-even oscillatory pattern with the correct amplitudes at 353\,GHz.
\begin{figure}[htbp!]
\includegraphics[width=\columnwidth]{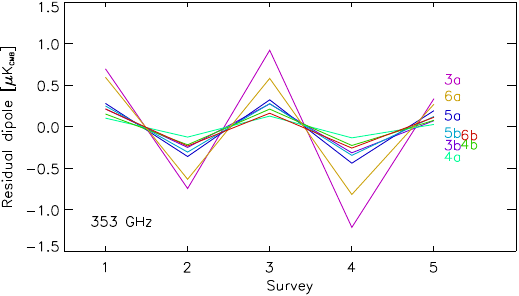}
\caption{Odd-even survey pattern of the Solar dipole calibration for the 353-GHz PSBs. This pattern is observed in the data (see Fig.~\ref{fig:chapochinoi}), especially at 353\,GHz.}
\label{fig:simchapo}
\end{figure}
We have thus demonstrated that the difference between the three bin's empirical transfer function at 353\,GHz with a pure time constant model accounts for both the odd-even survey calibration pattern and the zebra striping. Similar patterns are seen at 545\,GHz in Fig.~\ref{fig:s1234} and also at a lower level at 217\,GHZ. An improvement of the transfer function would require us to fit a single time constant per bolometer to minimize the odd-even calibration pattern and the zebra patterns. This has not been attempted for this paper.

\subsubsection{Summary of constraints on TF residuals}
\label{sec:summaryTF}

Figure~\ref{fig:tf} shows the combination of the constraints on the multipole-dependent transfer function over a broad range of multipoles. The part of the effective beam window functions associated with the intermediate sidelobes induces a loss of power between $\ell=1$ and $\ell=1000$ of order 0.5--0.8\,\% \citep[see figure~C.3 of][]{planck2013-p01a}. This is fully included in the 2015 effective beam. The uncertainties have been estimated to be 0.3\,\% \citep[see figure~21 of][]{planck2014-a13} and are shown in the figure by the grey dash-dotted horizontal line.
\begin{figure}[htbp!]
\includegraphics[width=\columnwidth]{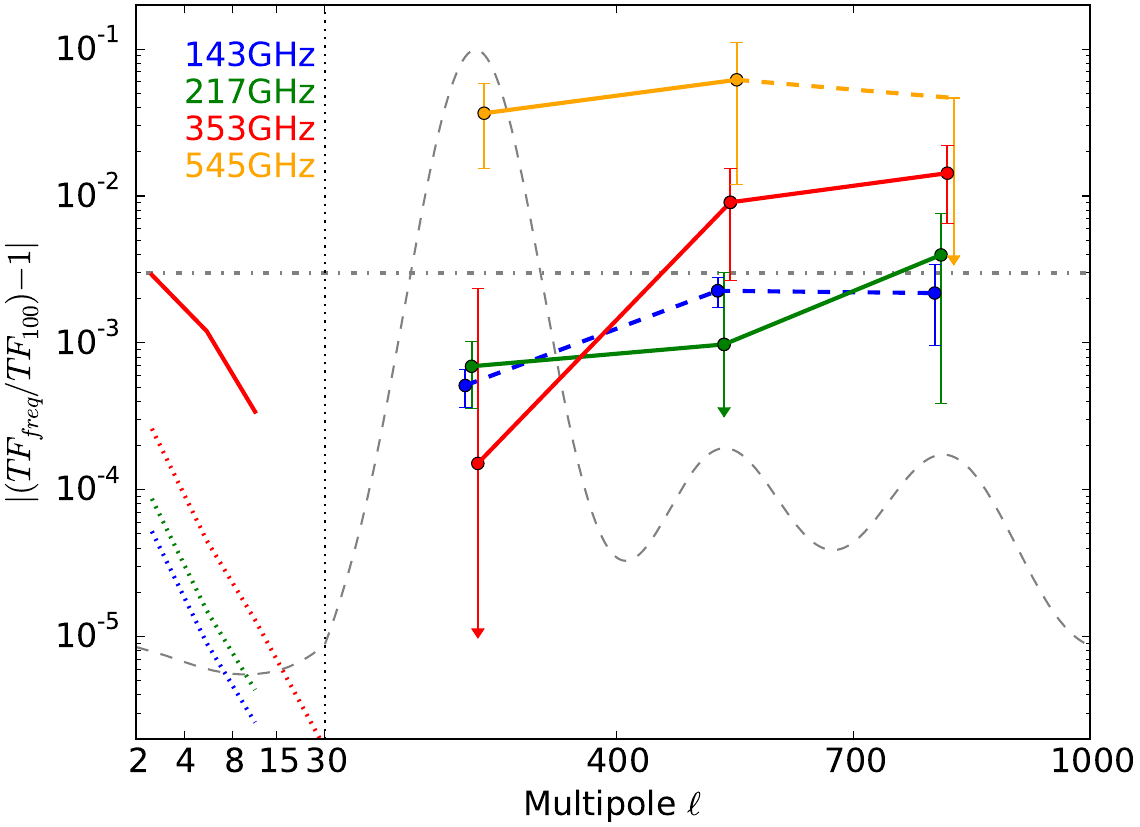}
\caption{Transfer function ratio, referred to 100\,GHz, for various multipoles. For $\ell=2$--30, the solid red line is the measured 353-GHz empirical \sroll\ time TF. The red dotted line shows an upper limit on the 353-GHz residuals after correction, while blue and green dotted lines show an estimate of the residuals at 143 and 217\,GHz. The TF for $\ell=30$--1000 is estimated from CMB anisotropies on the first three acoustic peaks (with negative values displayed by dashed lines). The grey dashed line is a CMB power spectrum, showing the position of the first three acoustic peaks.}
\label{fig:tf} 
\end{figure}

Figure~\ref{fig:tf} shows the relative transfer function with respect to the 100\,GHz one, after including the effective transfer function discussed above. It combines residuals from two corrections:
\begin{itemize}
\item the \sroll\ mapmaking extracts the empirical transfer function at $\ell=2$--15, but we have shown at 353\,GHz that a better model is a single time transfer function around 30 seconds, which leads to an $\ell ^{-2}$ behaviour;
\item at higher multipoles, we can measure the relative calibration of other bands (referred to the 100-GHz calibration), for three ranges of multipoles, centred on the first three acoustic peaks ($\ell=100$--1000), where we can see that at the first acoustic peak ($\ell\simeq200$), the transfer functions within the CMB channels agree to within $10^{-3}$.
\end{itemize}

The 353-GHz TF (full red line) from $\ell=2$ to 15 decreases with multipoles from $3 \times 10^{-3}$ to a very low minimum of order $5 \times10^{-4}$ (figure~10 of \citelowell). This has been corrected in the 2018 release, and the residual is shown by the dotted red line in Fig.~\ref{fig:tf}. At higher multipoles, the difference with 100\,GHz rises to the percent level for the second and third acoustic peaks. This is the multipole range dominated by the effective beam uncertainties.

For the CMB frequencies (143\,GHz in blue, and 217\,GHz in green), the transfer functions are expected to follow a similar behaviour, but were not detected by \sroll. We can estimate upper limits on the changes to the transfer function (shown as dotted lines in Figure~\ref{fig:tf}) as follows. These can be estimated from the better agreement, by a factor of 5, between the odd-even surveys calibration (see Fig.~\ref{fig:chapochinoi}) for the CMB frequencies than at 353\,GHz. The inter-calibration agreement of 100, 143, and 217\,GHz at the first acoustic peak is better than $10^{-3}$, showing that the rise of uncertainties on the transfer functions due to effective beam errors is below this value at $\ell =200$. The transfer function differences rise on the second and third acoustic peaks to a few parts per thousand, and show opposite trends. Nevertheless, as expected for the CMB channels, the level stabilizes around the value of the uncertainty estimated for the effective TF.

Although not fully accurate, these results clearly show two ranges: one is associated with the \sroll\ empirical time transfer function deviations from unity, decreasing with multipoles to very low values at $\ell=10$--30; the other range, increasing with multipoles, is associated with the errors on the TOI transfer function correction (based on effective beams). This regime is measured by the relative calibration on the first three acoustic peaks.

The 545-GHz channel (orange in Fig.~\ref{fig:tf}) shows a transfer function level starting at 0.3\,\% for the dipole, detected at the $4\pm2\,\%$ level on the first acoustic peak, and yielding only upper limits at the two other peaks. The transfer function between dipole and point-source calibrations, at 545\,GHz, is not expected to be very different from the ones at lower frequencies, and thus is at the percent level. This has to be compared with the giant planet model uncertainty, estimated at $5\,\%$, on which this channel is calibrated at high multipoles. This shows that the planet model is compatible with the CMB photometric standard well within the uncertainty of the planet model used for the calibration.

\subsection{Intensity-to-polarization leakage from calibration and bandpass mismatch}
\label{sec:intensity_leakage}

As demonstrated in Sect.~\ref{sec:calibration}, the CMB-based calibration of HFI is exquisite. Because the bandpass of each detector is different, however, response to emission with a non-thermal spectrum will vary from detector to detector. This is turn leads to temperature-to-polarization leakage when the signals from two different detectors in a pair are differenced to extract polarization.

\subsubsection{Polarization leakage from calibration mismatch}

In appendix~ B of \citelowell, we demonstrate that residual polarization leakage arising from calibration mismatch is negligible for the three CMB channels ($<10^{-6}\microKcarre$ at $\ell>3$). At 353\,GHz, the residual power spectrum is higher (at the~$10^{-5}\microKcarre$ level), but the effect on the CMB channels is negligible when the 353\,GHz channel is used to clean them of dust emission. Given the improved calibration for the current release, we expect the residuals from calibration mismatch leakage to be even smaller in the 2018 data.

\subsubsection{Consistency of bandpass leakage coefficients}

The correction for the bandpass-mismatch leakage was discussed in detail in Sect.~2.6 of \citelowell.

\sroll\ allows us to extract the response to dust emission for each individual detector in a given frequency band, relative to the average for that band. The bandpass-mismatch coefficients can then be compared to the estimates from ground-based measurements of the bandpasses of individual detectors. Figure~\ref{fig:bandpassleakage} shows that the sky and ground determinations are in full agreement in most cases.
\begin{figure}[htbp!]
\includegraphics[width=\columnwidth]{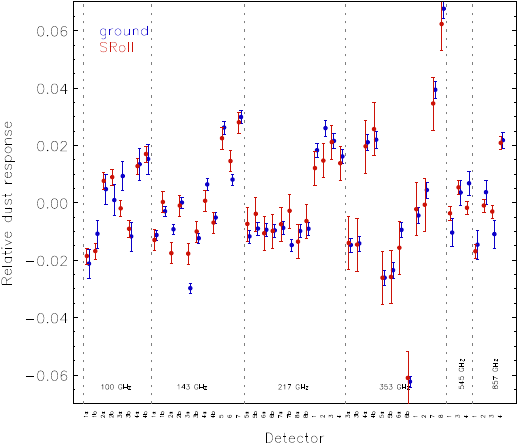}
\caption{Recovery of dust leakage coefficients from the ground measurements (in blue) and those from {\tt SRoll} (in red).}
\label{fig:bandpassleakage} 
\end{figure}
The error bars on the ground measurements shown here are the statistical ones, significantly smaller than the ones reported in figure~14 of \citelowell, which included a very conservative estimate of systematic uncertainties. There are only two exceptions to the good agreement, bolometers 143-2a and 143-3a.

We can also examine the fidelity of the recovery of the bandpass-mismatch coefficients for dust, CO and free-free emission extracted by \sroll\ using simulations. Figure~\ref{fig:bandpassmissmatch} demonstrates that the agreement is excellent, even for 353\,GHz.
\begin{figure}[htbp!]
\includegraphics[width=\columnwidth]{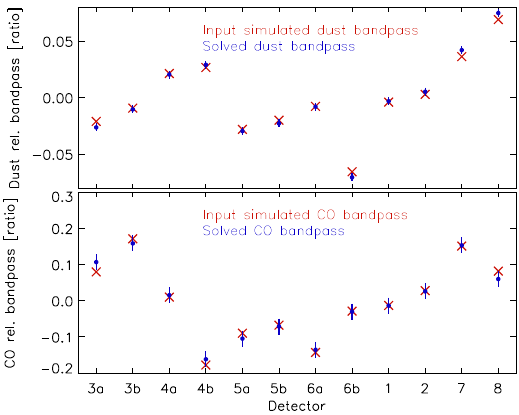}
\caption{Dust (top panel) and CO (bottom panel) bandpass-mismatch leakage coefficients from simulations for the 353-GHz bolometers. Shifts with respect to unity for the ratio of each coefficient to the average of detectors are plotted for the input ones in red and the output ones in blue. The errors bars are estimated from the statistical distribution of a set of simulated realizations.}
\label{fig:bandpassmissmatch} 
\end{figure}
The improvement over earlier results (figure~B7 of \citelowell) arises from a much better model of the dust template used in the current simulations \citep[see][]{2017A&A...603A..62V}. The rms differences between input and extracted coefficients are $3.4 \times 10^{-4}$ and $1.1 \times 10^{-3}$ for dust and CO, respectively.

\subsubsection{Effect of bandpass-mismatch leakage on power spectra}
\label{sec:bpeffect}

The effects of residuals from bandpass-mismatch leakage can be estimated through the difference map between the inputs and outputs of the simulations (as was first done in appendix~B of \citelowell). The corresponding improved power spectra of the \sroll\ solutions are shown for this release in Fig.~\ref{fig:DIFF}.
\begin{figure}[htbp!]
\includegraphics[width=\columnwidth]{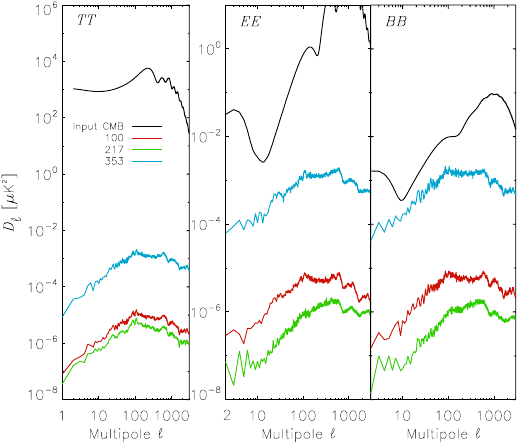}\\
\includegraphics[width=\columnwidth]{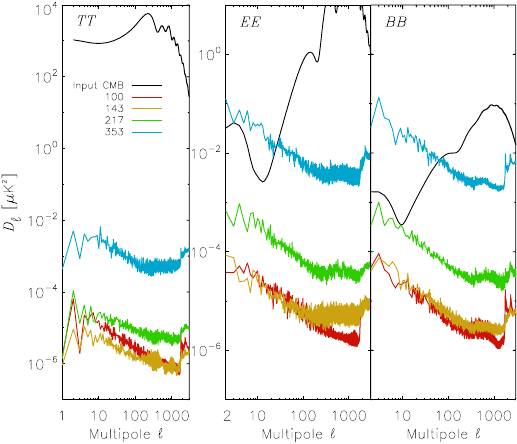}\\
\includegraphics[width=\columnwidth]{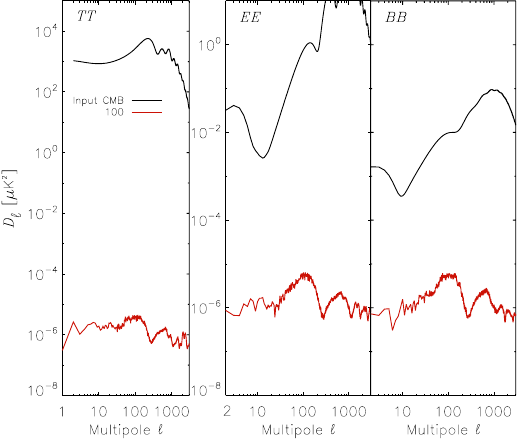}
\caption{Auto-power spectra of the CO (top), dust (middle), and free-free (bottom) bandpass leakage residuals, estimated by E2E simulations.}
\label{fig:DIFF} 
\end{figure}
We note the similarity in the power spectra of the polarization leakage to the power spectrum of intensity, as expected for temperature-to-polarization leakage in the 70\,\% sky fraction used. As stated above, results for 353\,GHz are much improved, but they are still larger than the residual power spectra for the CMB bands. When using the 353-GHz channel to remove polarized dust emission, the scaling factors to 100, 143, and 217\, GHz are $4 \times 10^{-4}$, $2 \times 10^{-3}$, and $2 \times 10^{-2}$, respectively. The induced errors from the 353-GHz leakage scaled down to the CMB channels are thus within the noise.

We also test (through the use of simulation) how the power spectrum depends on the input foreground templates required by \sroll. We introduce a dust template map from the PSM, for which the power spectrum is very different at low multipoles from the standard input sky map based on the 2015 release. The power spectrum of the difference between the input template and the 353-GHz output is shown in Fig.~\ref{fig:dustiter} by the solid blue line.

\begin{figure}[htbp!]
\includegraphics[width=\columnwidth]{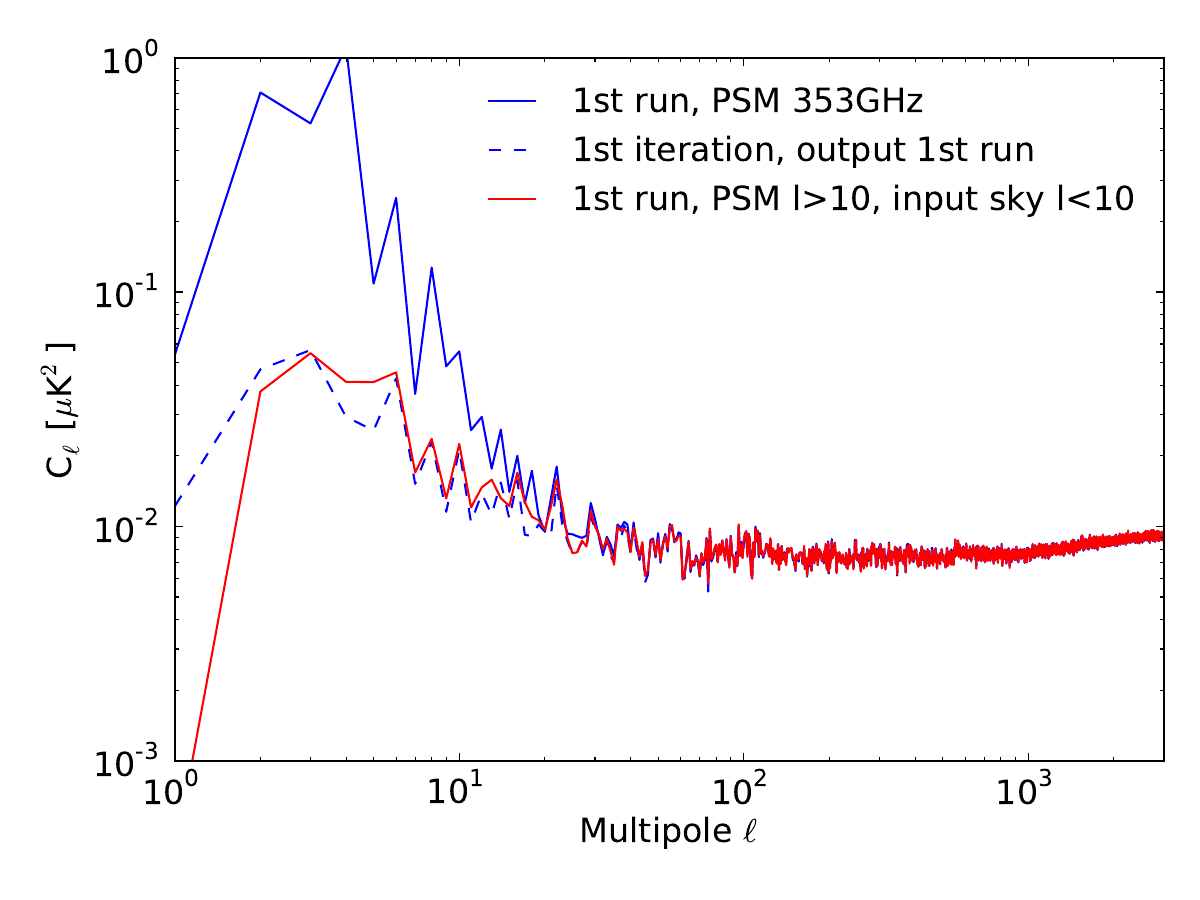}
\caption{Power spectra of the difference between the template dust input and output 353-GHz $Q$ maps. The solid blue line shows a random PSM input dust template. The dashed blue line shows the case where the input dust template to {\tt SRoll} has been taken as the output of the first run. The red curve is obtained by replacing the first ten multipoles in the initial dust template by those of the input sky map.}
\label{fig:dustiter}
\end{figure}
The dashed blue line shows the case where the output of the first iteration has been taken as the input dust template and is close to the noise level. The red curve shows the result of replacing the first ten multipoles in the initial dust input template by those of the input sky map. One iteration, after starting with a non-representative template on large scales, is equivalent to using the input large-scale distribution of the sky map. Of course we already know the large-scale distribution quite well from the previous 2015 release, which is used as the input template in the 2018 release; the test just described demonstrates that we do not need any iterations.

To confirm this, Fig.~\ref{fig:dustiter2} shows the convergence of iterations of the residuals of the half-mission cross-spectra $TT$, $EE$, and $BB$ at 353\,GHz.
\begin{figure}[htbp!]
\includegraphics[width=\columnwidth]{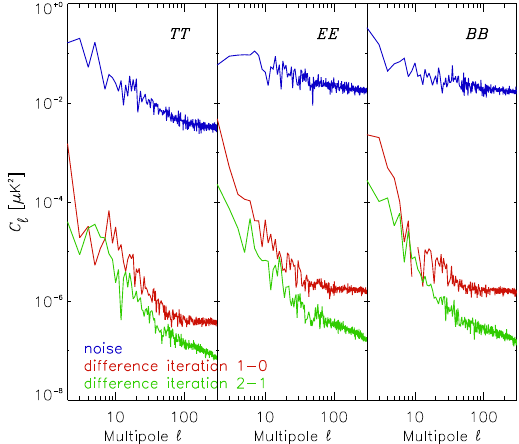}
\caption{Convergence of iterations at 353\,GHz on the dust template input. The residuals (input template minus output 353\,GHz) are computed on power spectra $TT$, $EE$, and $BB$ for a single iteration starting from the 2015 data (in red), and two iterations (in green). The blue curves give the noise from the half-mission differences.}
\label{fig:dustiter2} 
\end{figure}
The convergence is tested through the differences between the dust input template (taken as the 2015 release 353-GHz map) and the 353-GHz output from one run of \sroll\ (red line), and between the output 353-GHz maps first run for iteration 1 and the output (green line). The residuals show clearly that the intensity-to-polarization leakage converges at the level of $10^{-4}\microKcarre$, even at very low multipoles. The drop off at multipoles larger than a few hundred is due to the use of $\Nside\,{=}\,128$ and is not relevant to this discussion.

As discussed in Sect.~\ref{sec:caveats}, the CO template maps, defined at $\Nside\,{=}\,128$, induce a subpixel effect, especially at 100\,GHz. The residual in the power spectrum is shown in Fig.~\ref{fig:fig17}.

\subsection{ADC non-linearities}
\label{sec:ADC}

The non-linearities in the response of the ADCs of each readout electronic chain were not measured accurately enough before flight. However, the ADCNL was measured again during the warm phase of the \Planck\ mission, providing a model of this systematic effect that is removed in the TOI processing. The first-order residual expected from the uncertainties in the TOI correction is a time-dependent linear gain on small signals.
The temporal variation in the linear gain is shown in Fig.~\ref{fig:stimINL} for detector 100-1a (as the blue line).
\begin{figure}[htbp!]
\includegraphics[width=\columnwidth]{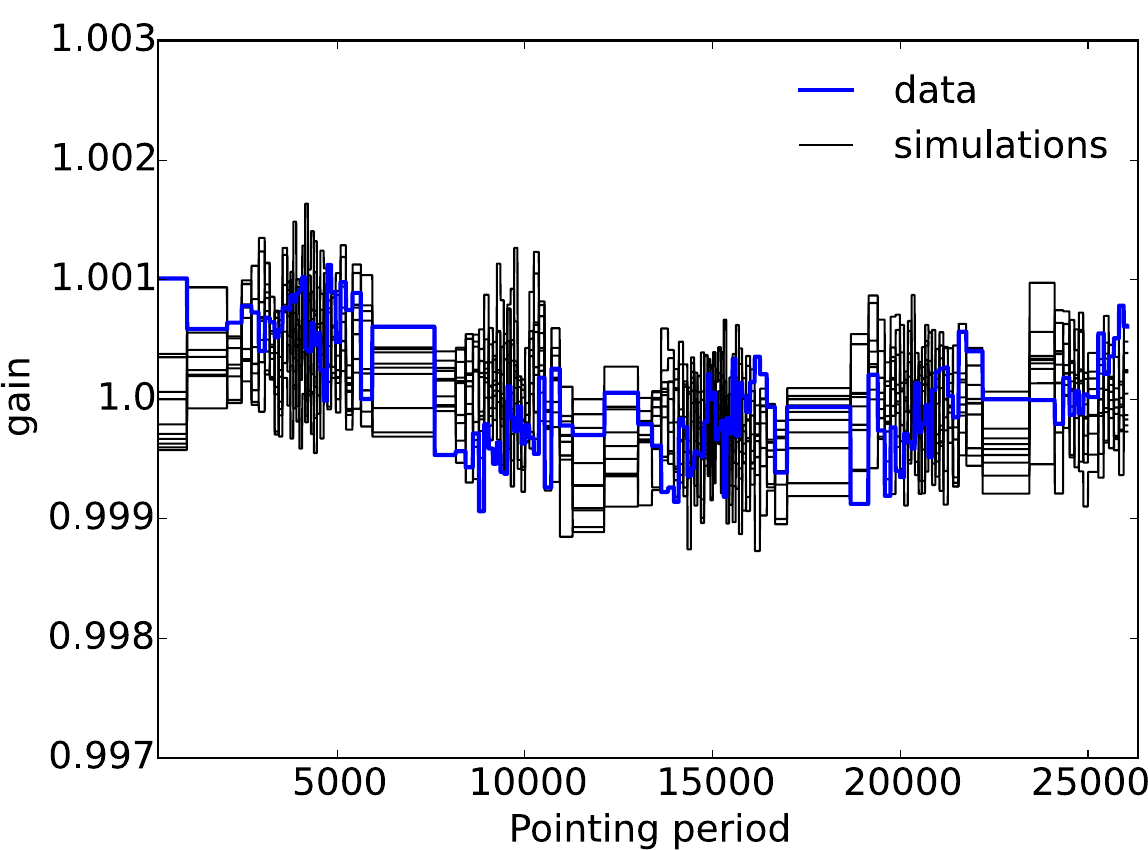}
\caption{ADCNL-induced time variation in gain as a function of pointing periods for bolometer 100-1a. The blue line shows the solved gain variation for the data and the grey ones show ten realizations drawn from the uncertainties of the ADC model. Similar figures are available for all HFI detectors in the \citeES.}
\label{fig:stimINL} 
\end{figure}
This systematic effect was by far the main one in the 2015 release. Using the gain as a function of position of the signal in the relevant ADC range, we can build an empirical parametric model of the ADCNL, using a small number of parameters, leading to a correction of the signal due to the ADCNL as a function of the observed signal; for detail discussion, see sections~2.5, 2.6.1, and B3.3 of \citelowell. Such an empirical model comes out of the \sroll\ extraction, along with error bars on the parameters of the model. We use these uncertainties in the E2E simulations pipeline to simulate statistically significant sets of realizations of the model of the ADCNL. In Fig.~\ref{fig:stimINL} we show (as a set of grey lines) ten realizations drawn from this model. The scatter among these lines is a measure of the uncertainties remaining in the quite-well reconstructed ADCNL correction (the scale of the peak-to-peak variations is only $10^{-3}$).

We built five realizations of the time variation of the gain induced by the ADCNL, then propagated these to maps and power spectra using E2E simulations. In order to make the effects of this systematic visible, we present results both with and without the correction for ADCNL-induced gain variations, and with and without noise. Figure~\ref{fig:ADCNLresidu} displays the Stokes $Q$ maps for frequencies 100 to 353\,GHz. 
\setlength\tabcolsep{1.5pt}
\renewcommand{\arraystretch}{0.5}
\begin{figure}[htbp!]
\begin{tabular}{cccc}
	&Data&Sims with noise&Sims without noise\\
	\\
	\\
	\raisebox{15pt}{\parbox{.06\textwidth}{no\\ corr.}}&
	\includegraphics [width=0.28\columnwidth]{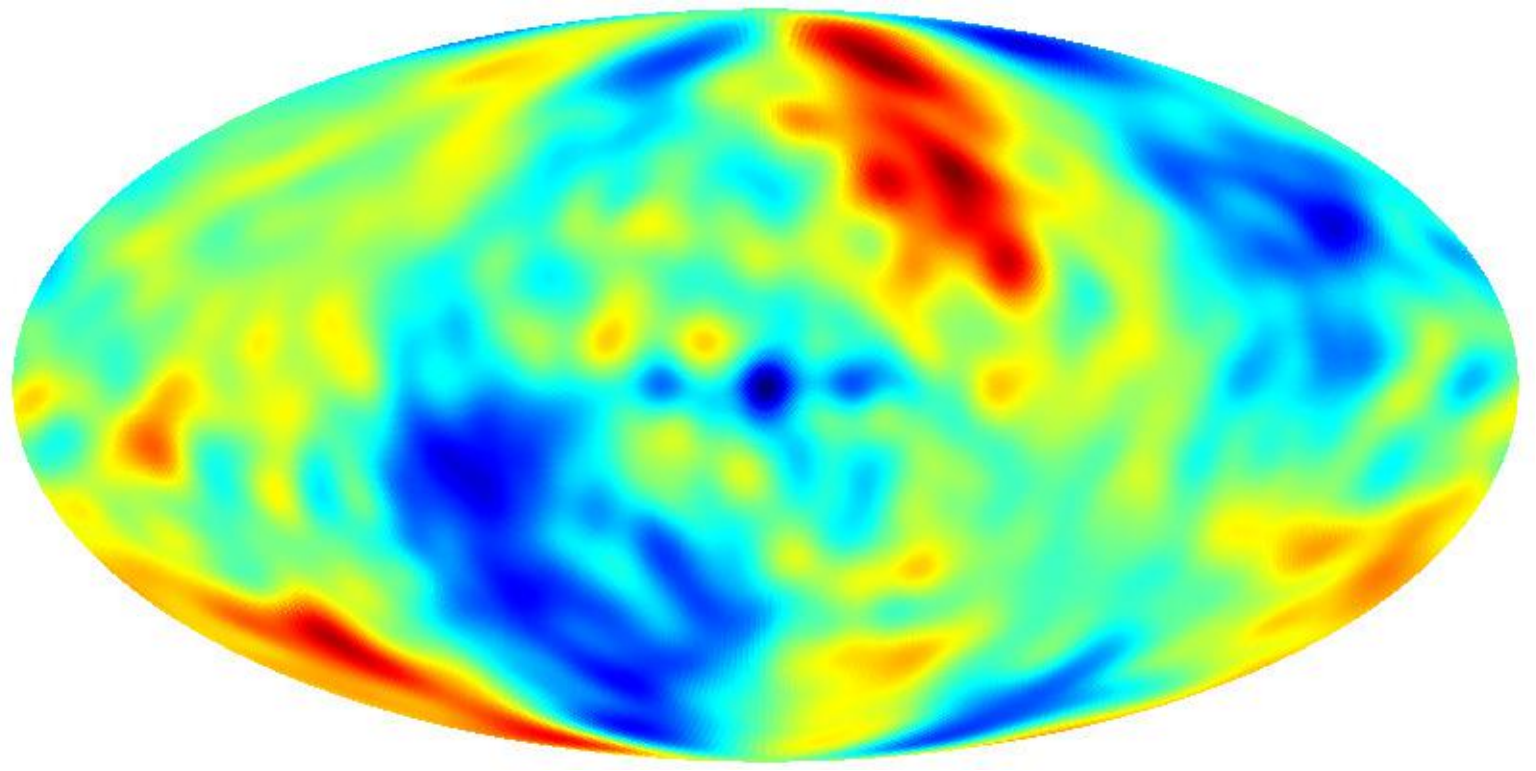}&
	\includegraphics [width=0.28\columnwidth]{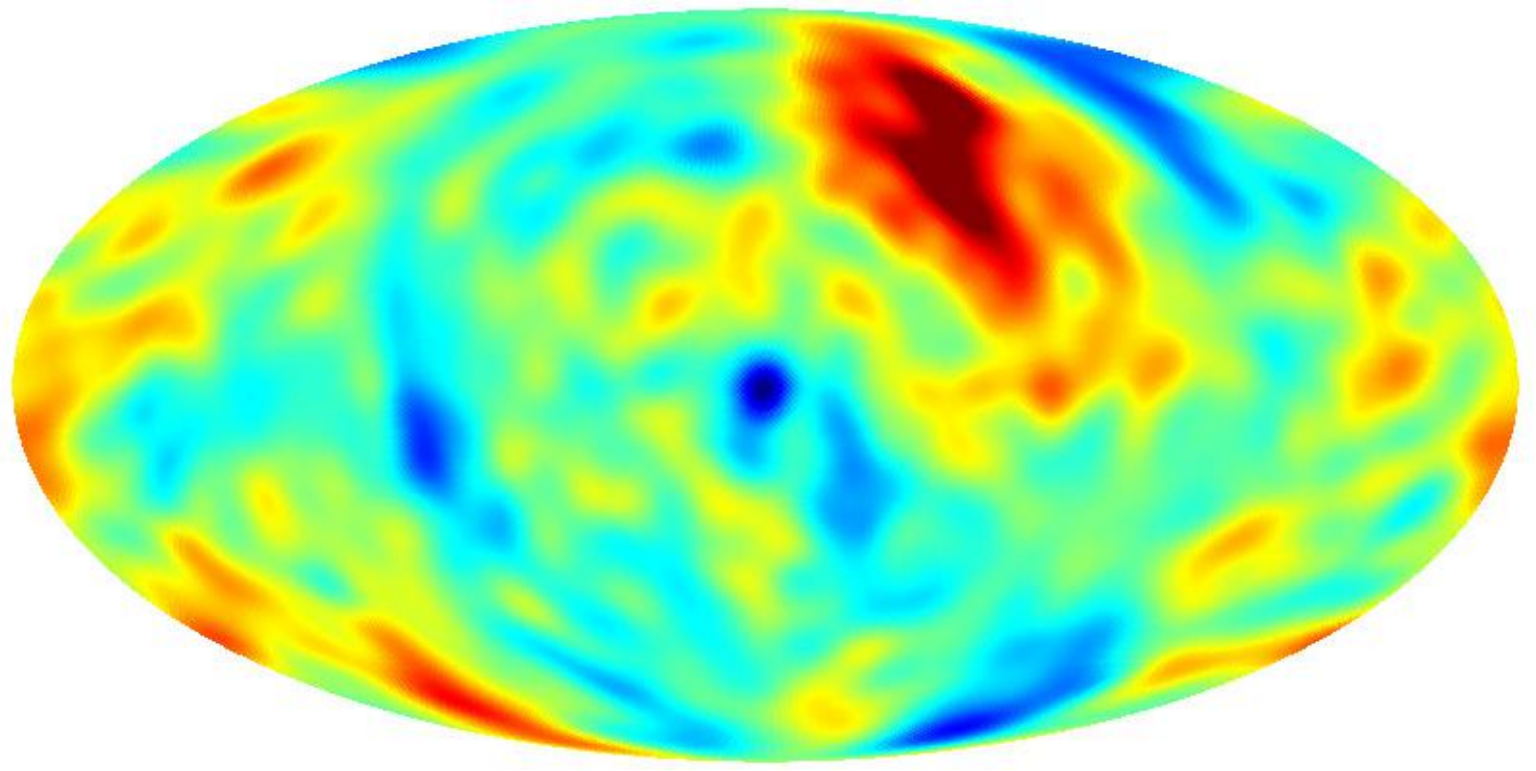}&
	\includegraphics [width=0.28\columnwidth]{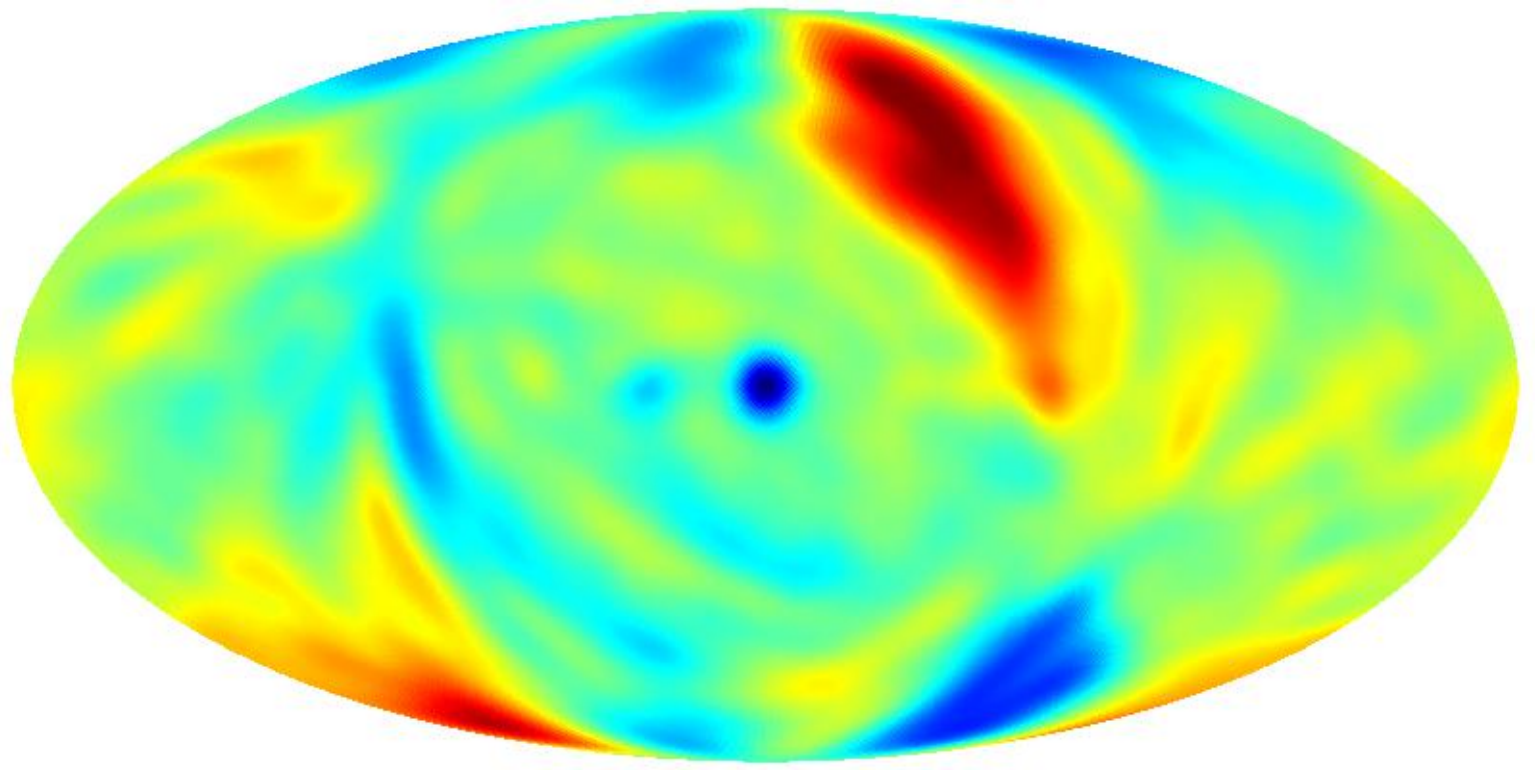}\\
	\raisebox{15pt}{\parbox{.06\textwidth}{linear gain corr.}}&
	\includegraphics [width=0.28\columnwidth]{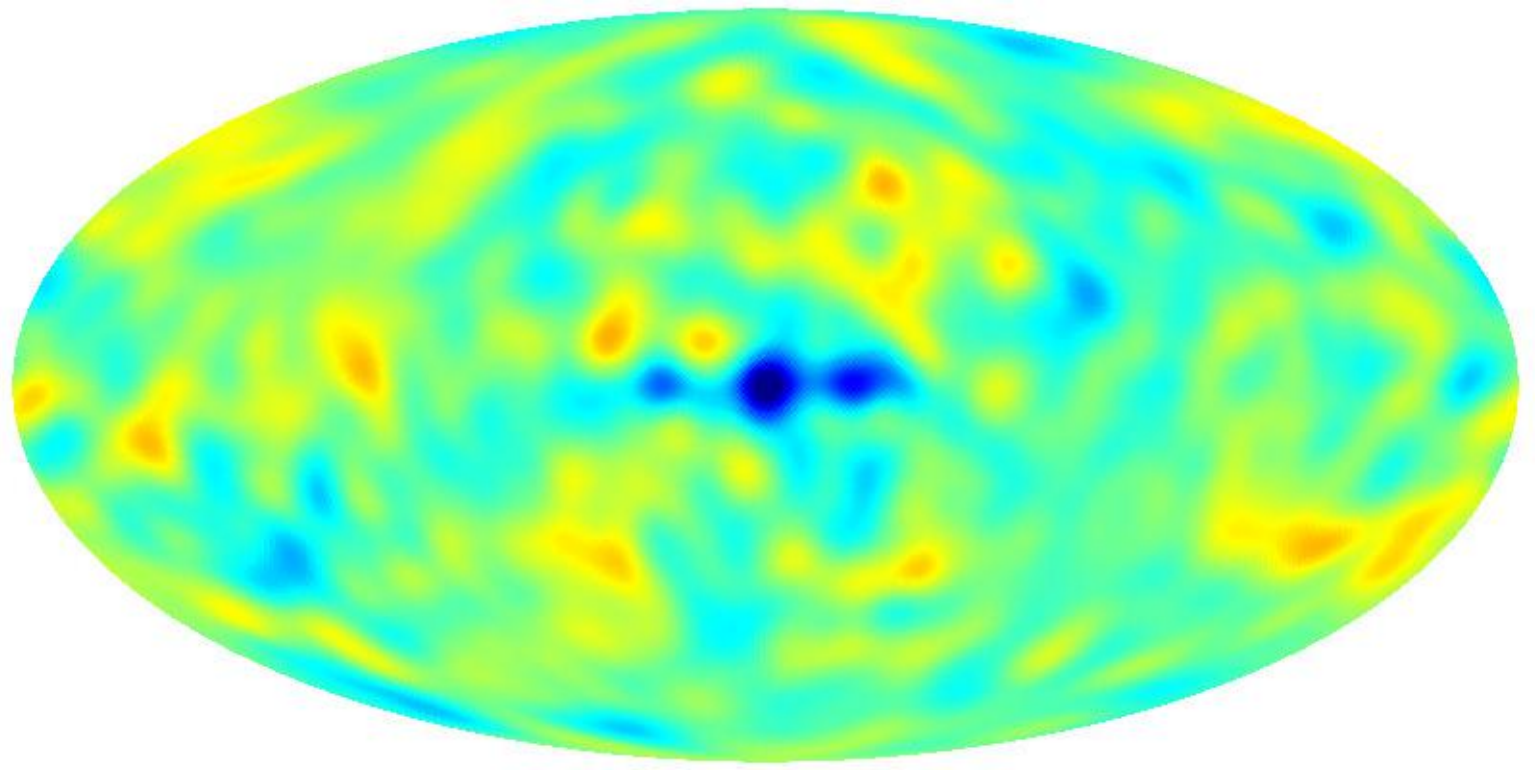}&
	\includegraphics [width=0.28\columnwidth]{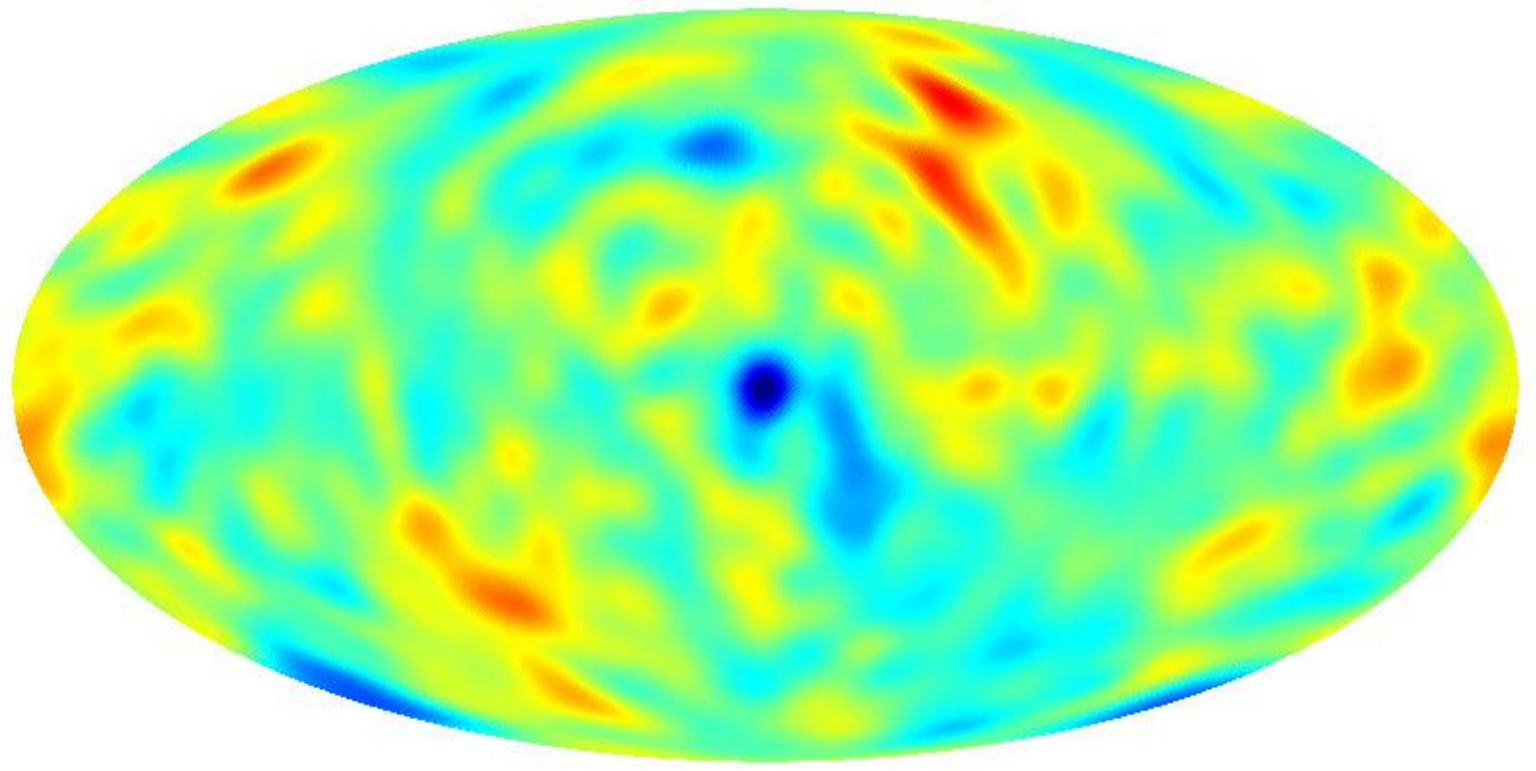}&
	\includegraphics [width=0.28\columnwidth]{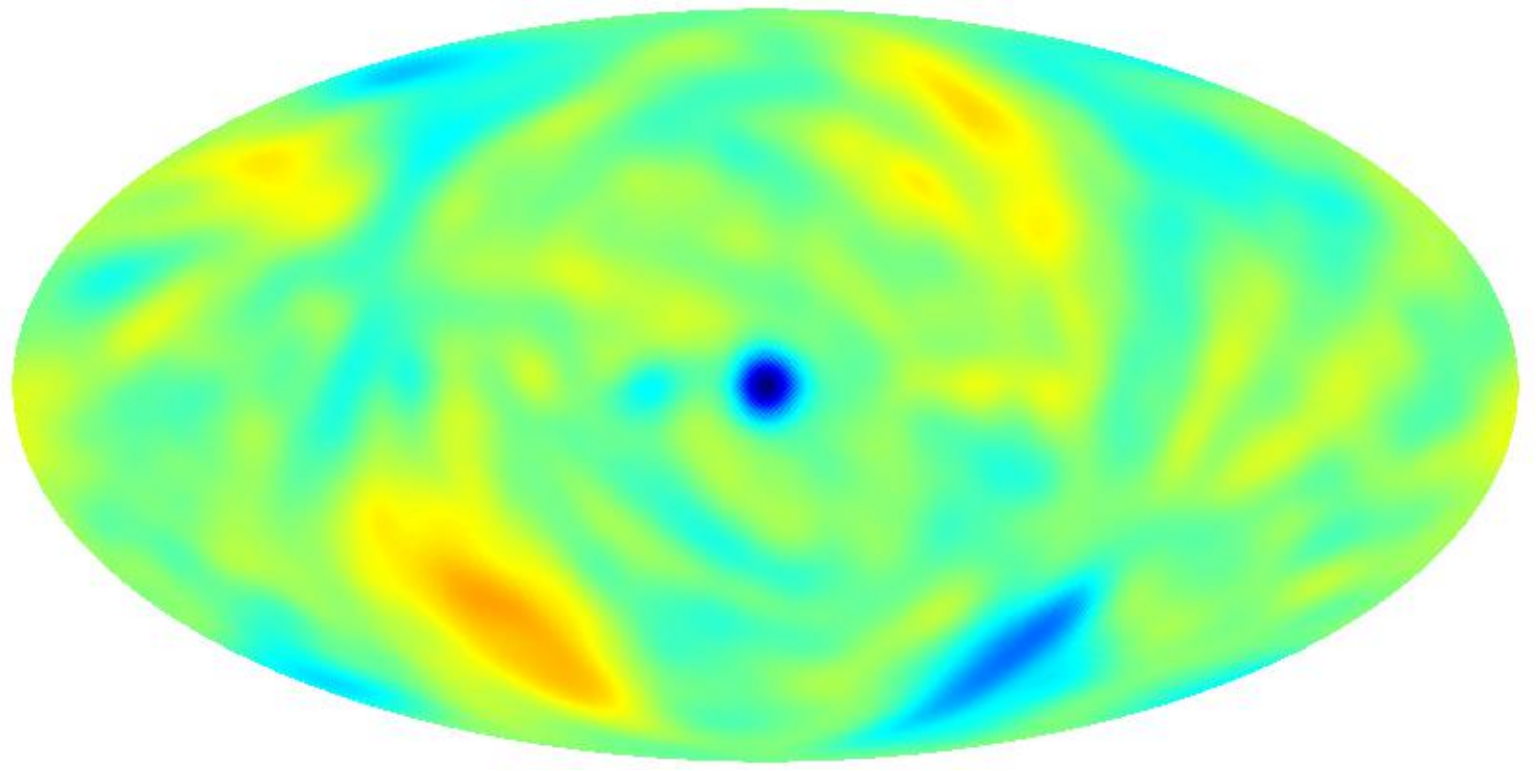}\\
	\raisebox{15pt}{\parbox{.06\textwidth}{dipole distort. corr.}}&
	\includegraphics [width=0.28\columnwidth]{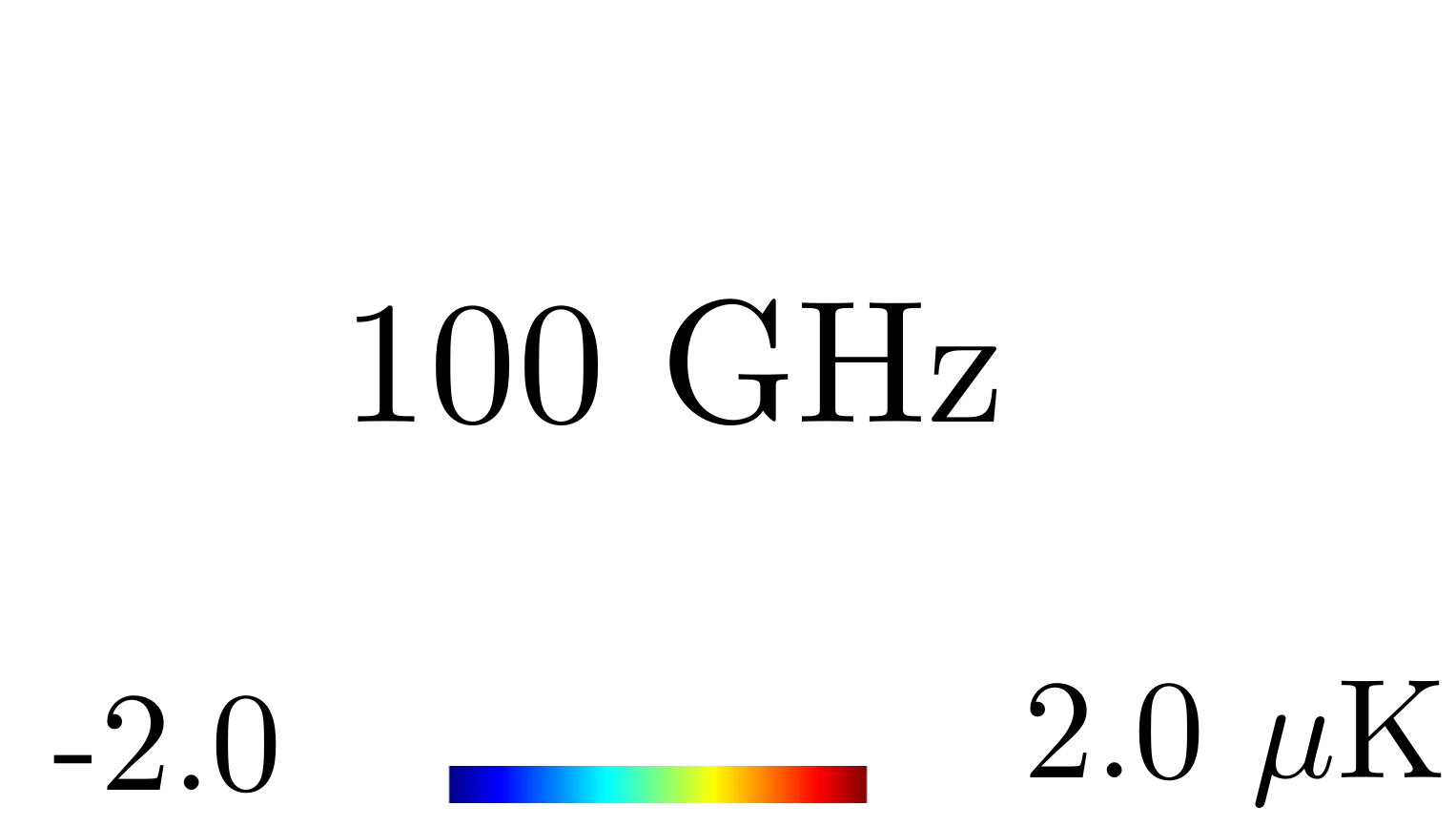}&
	\includegraphics [width=0.28\columnwidth]{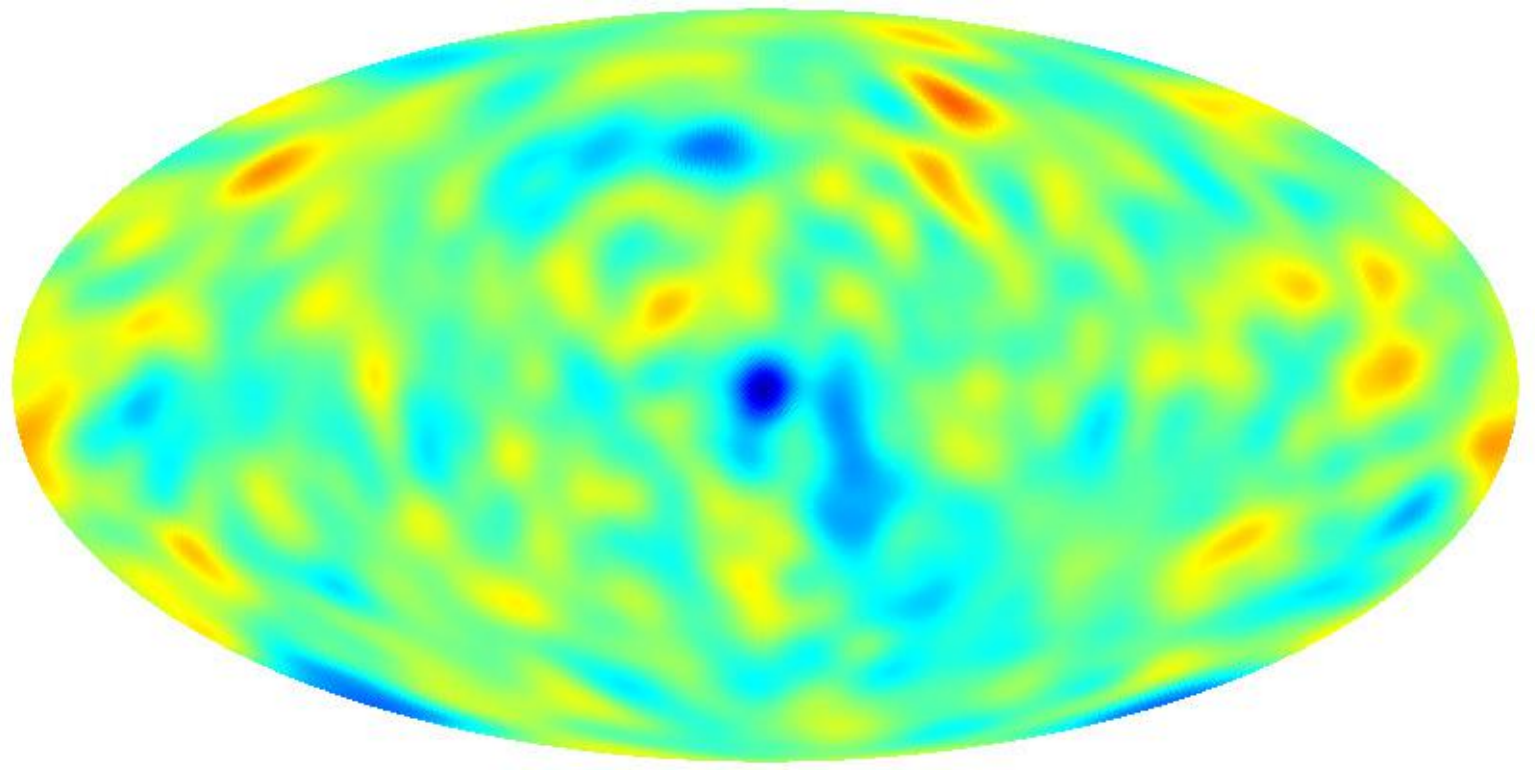}&
	\includegraphics [width=0.28\columnwidth]{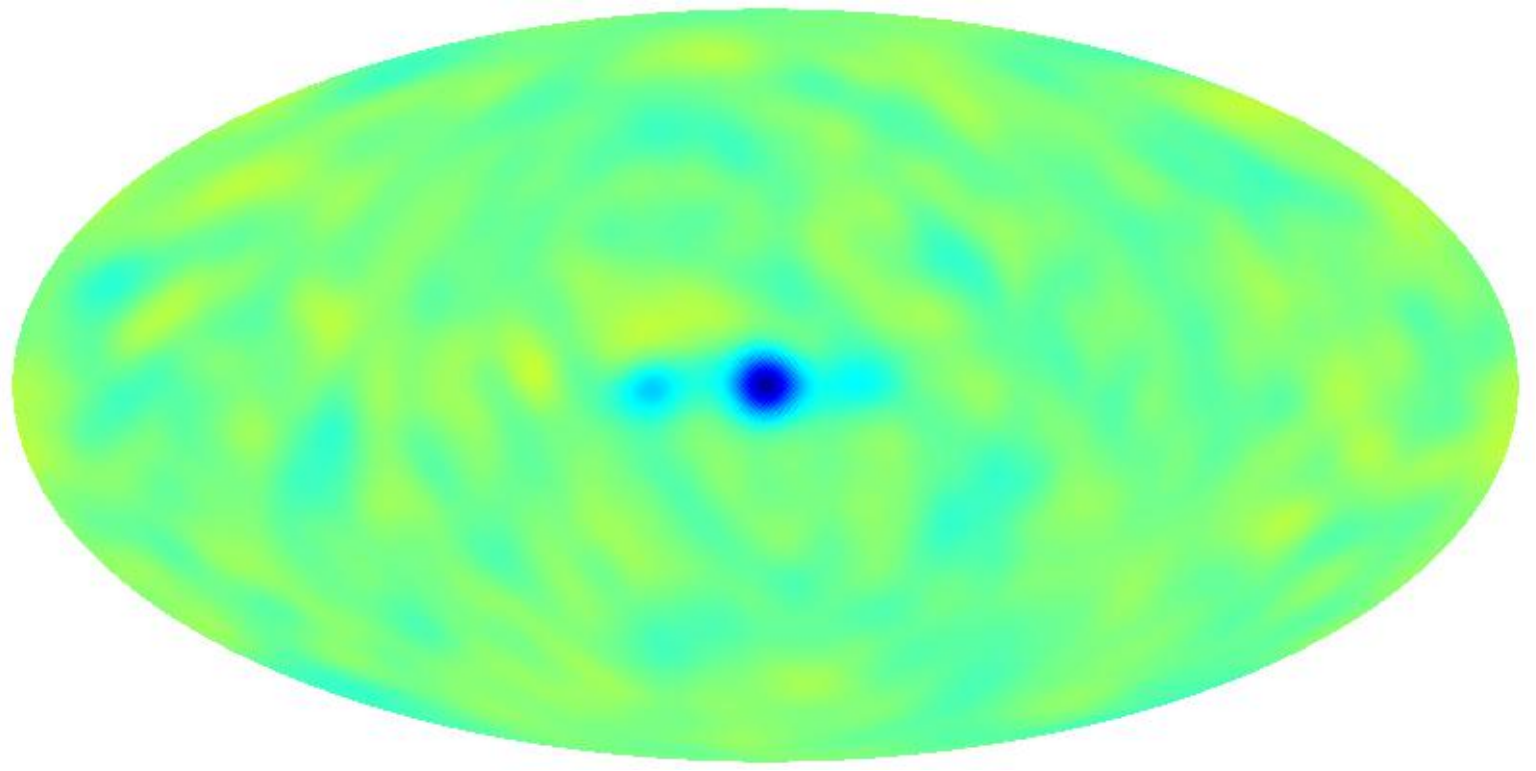}\\
	\\
	\\
	\raisebox{15pt}{\parbox{.06\textwidth}{no\\ corr.}}&
	\includegraphics [width=0.28\columnwidth]{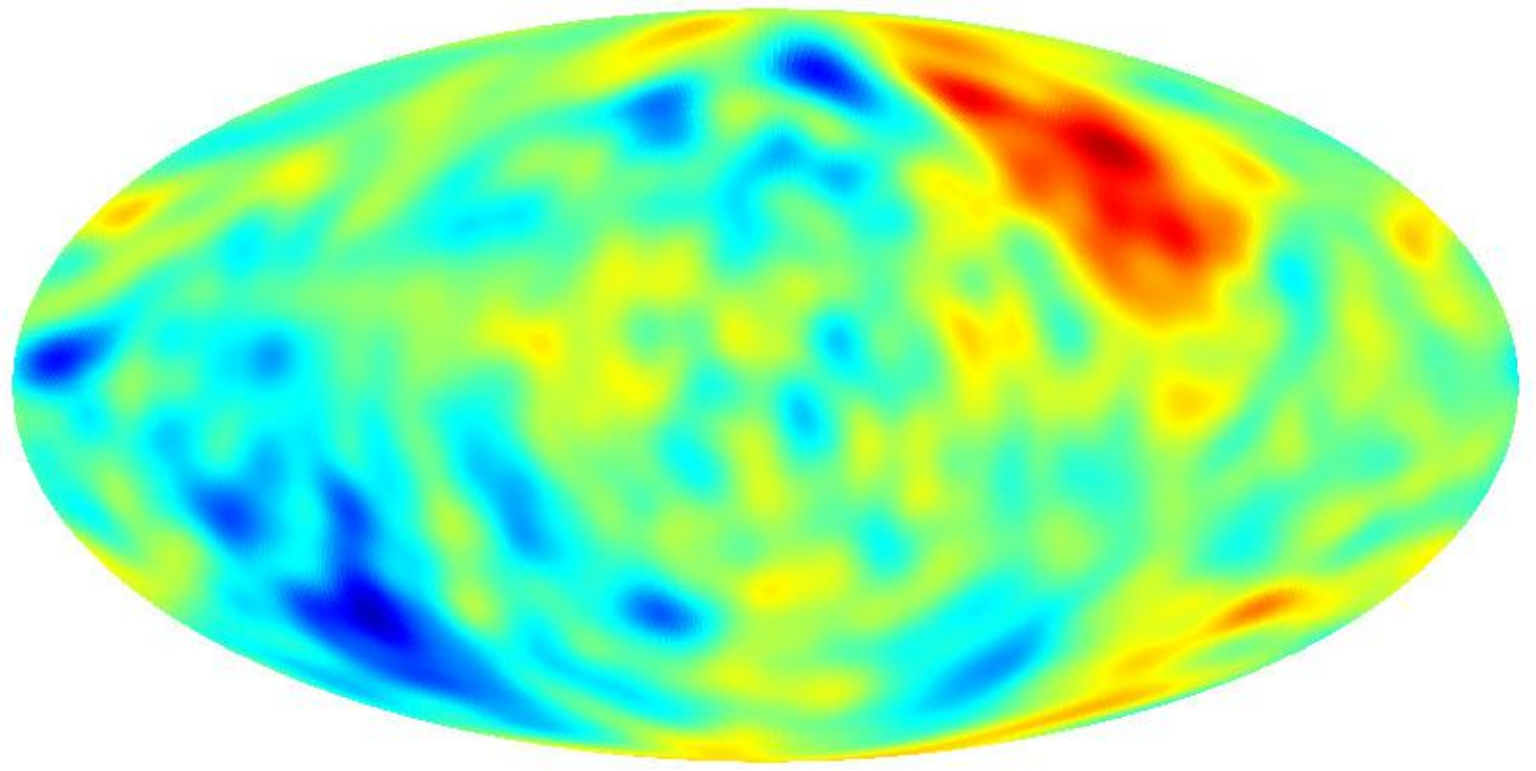}&
	\includegraphics [width=0.28\columnwidth]{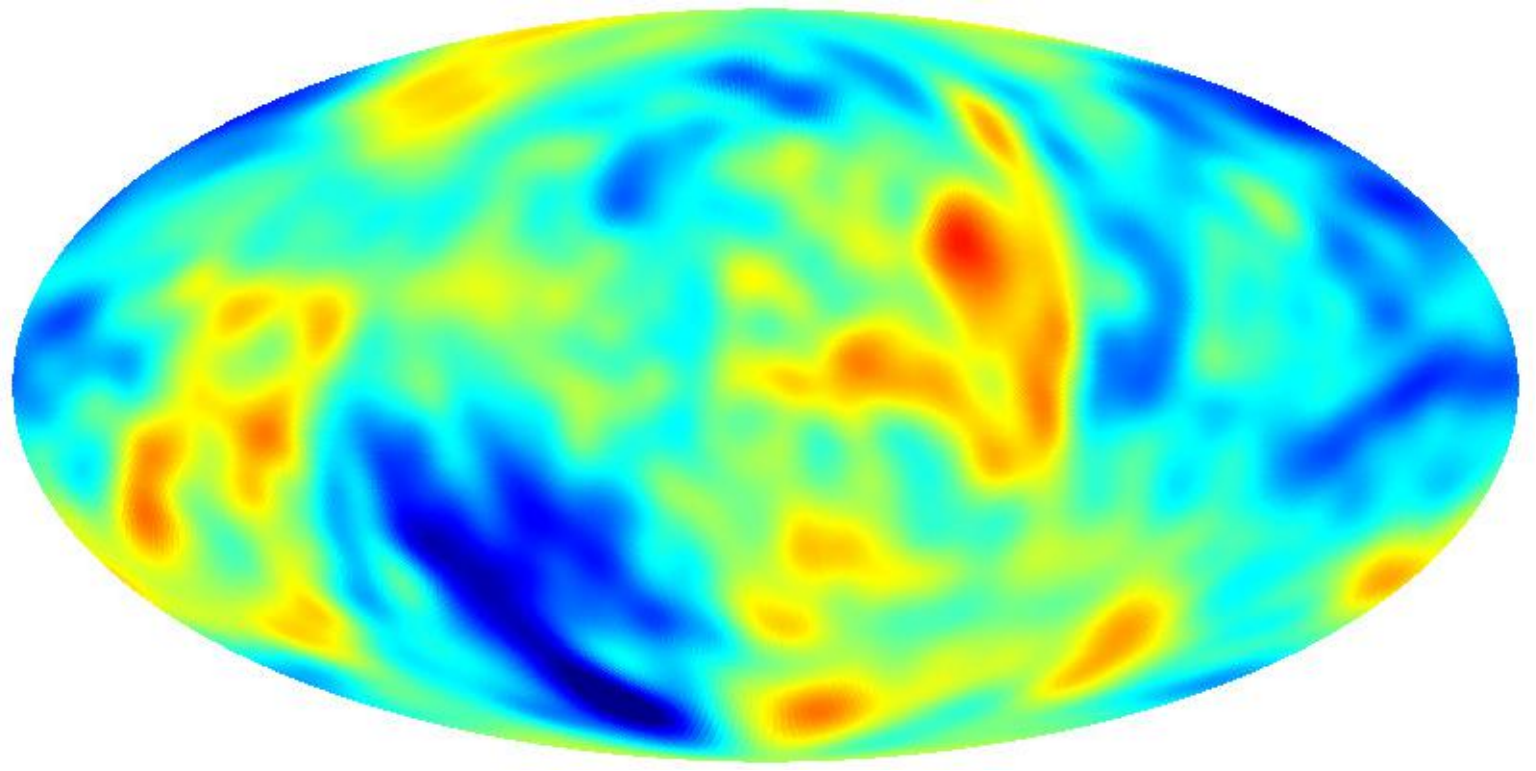}&
	\includegraphics [width=0.28\columnwidth]{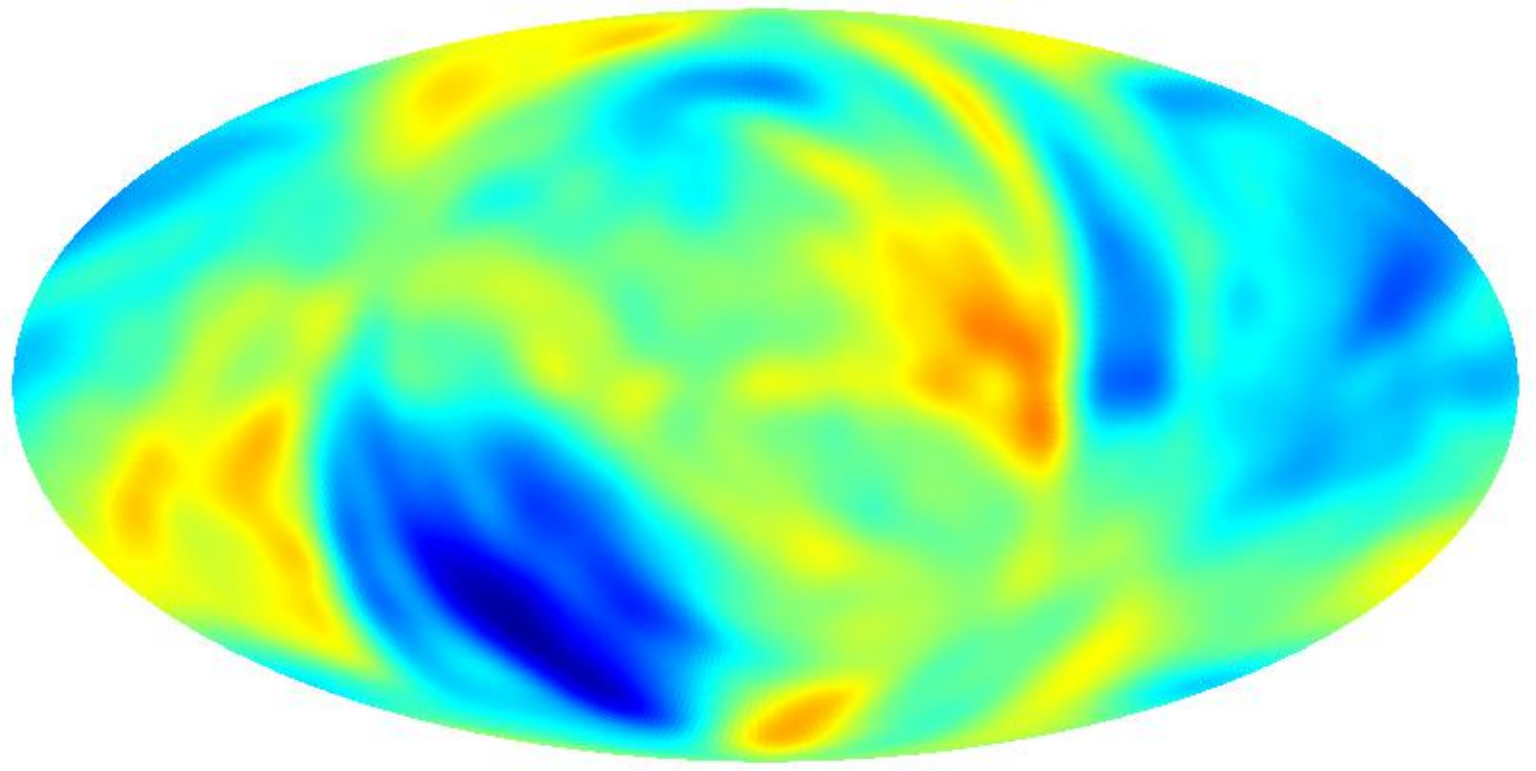}\\
	\raisebox{15pt}{\parbox{.06\textwidth}{linear gain corr.}}&
	\includegraphics [width=0.28\columnwidth]{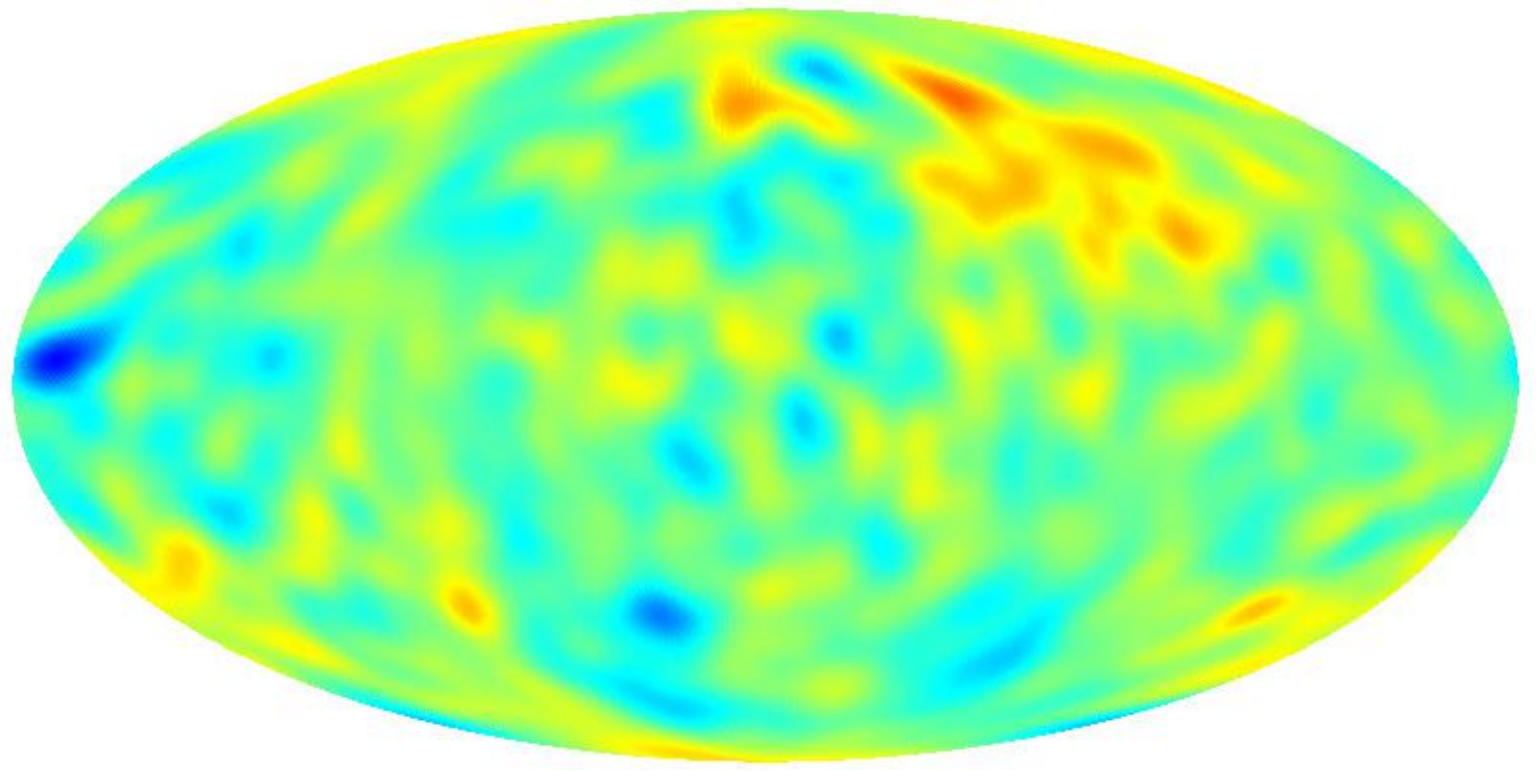}&
	\includegraphics [width=0.28\columnwidth]{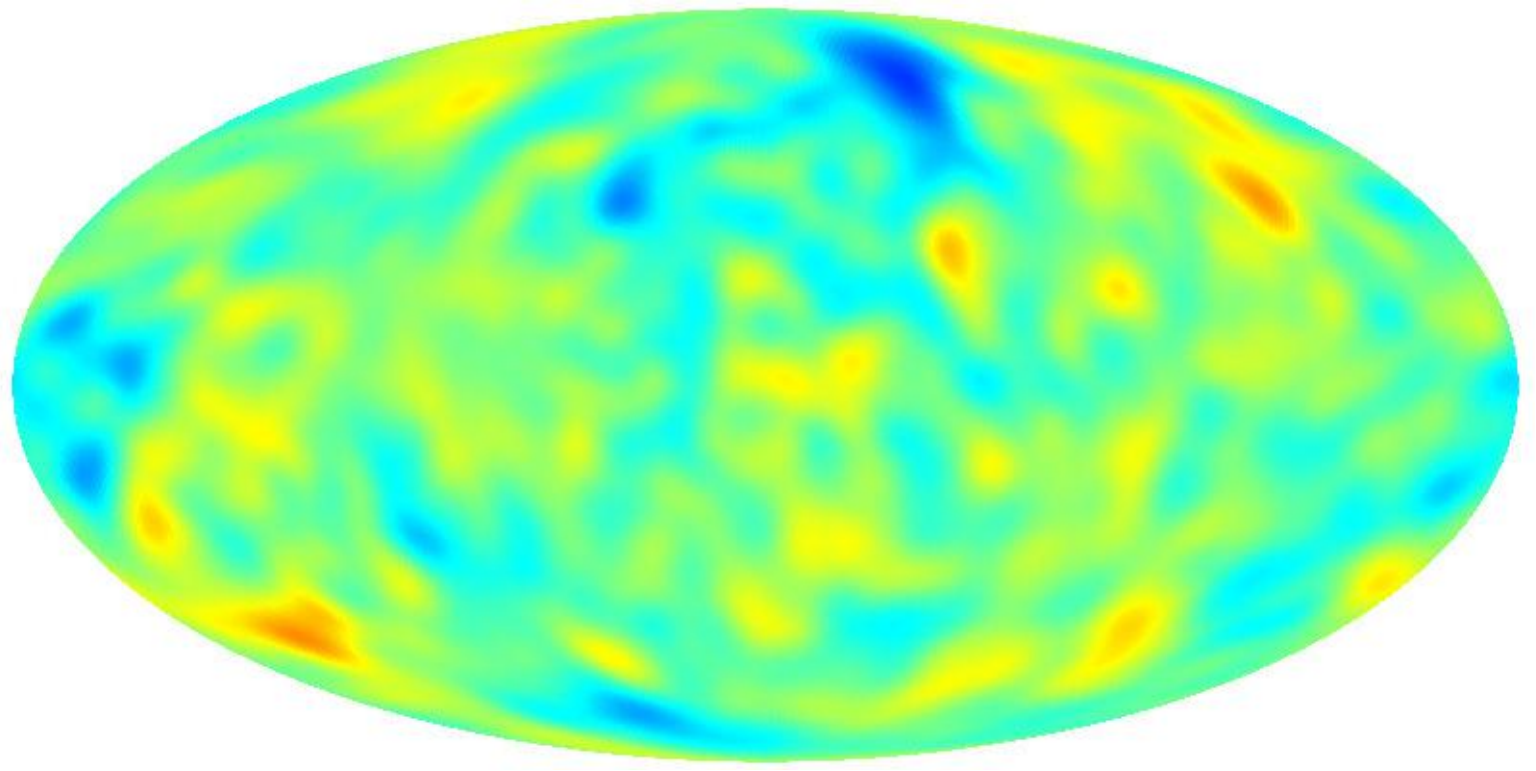}&
	\includegraphics [width=0.28\columnwidth]{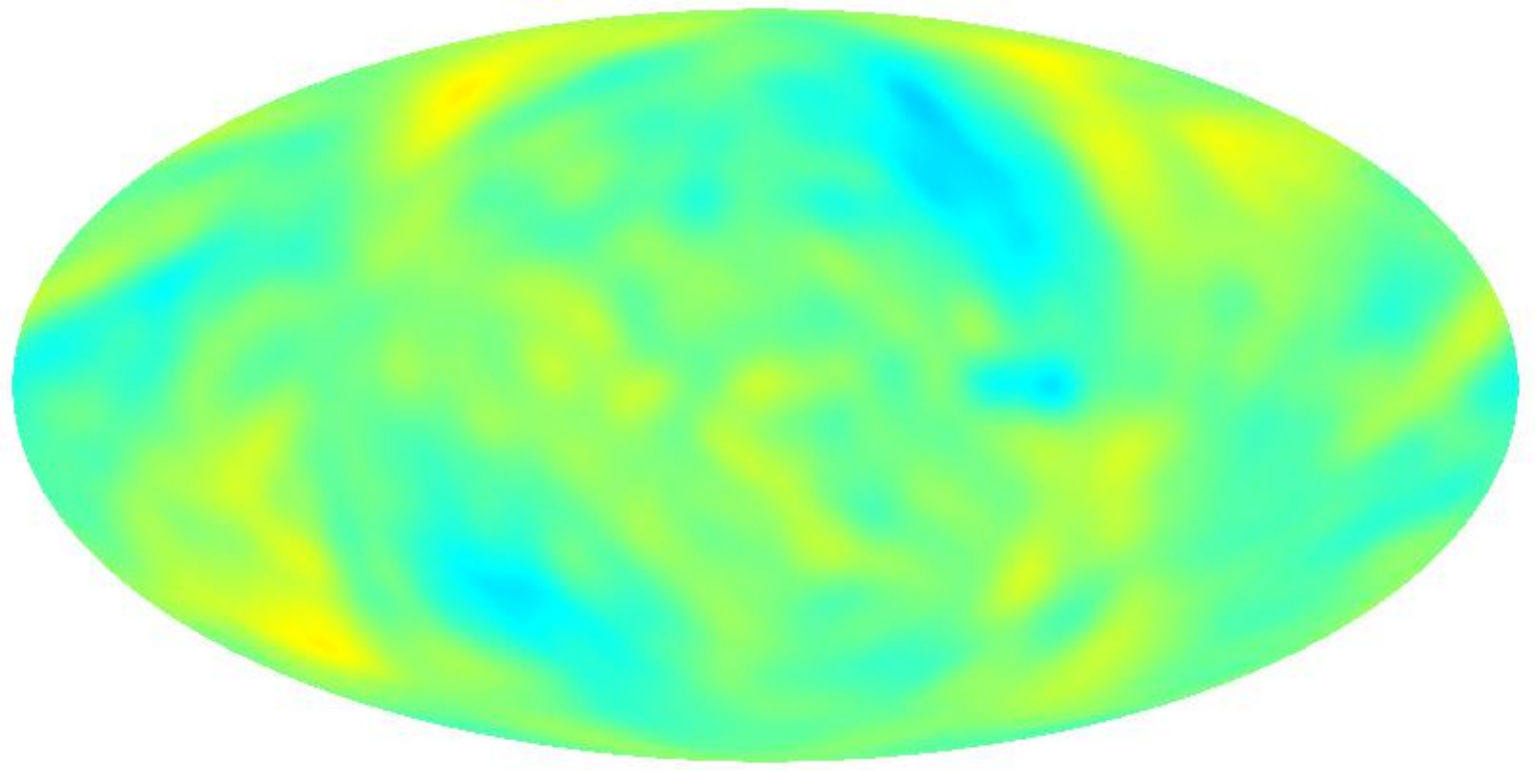}\\
	\raisebox{15pt}{\parbox{.06\textwidth}{dipole distort. corr.}}&
	\includegraphics [width=0.28\columnwidth]{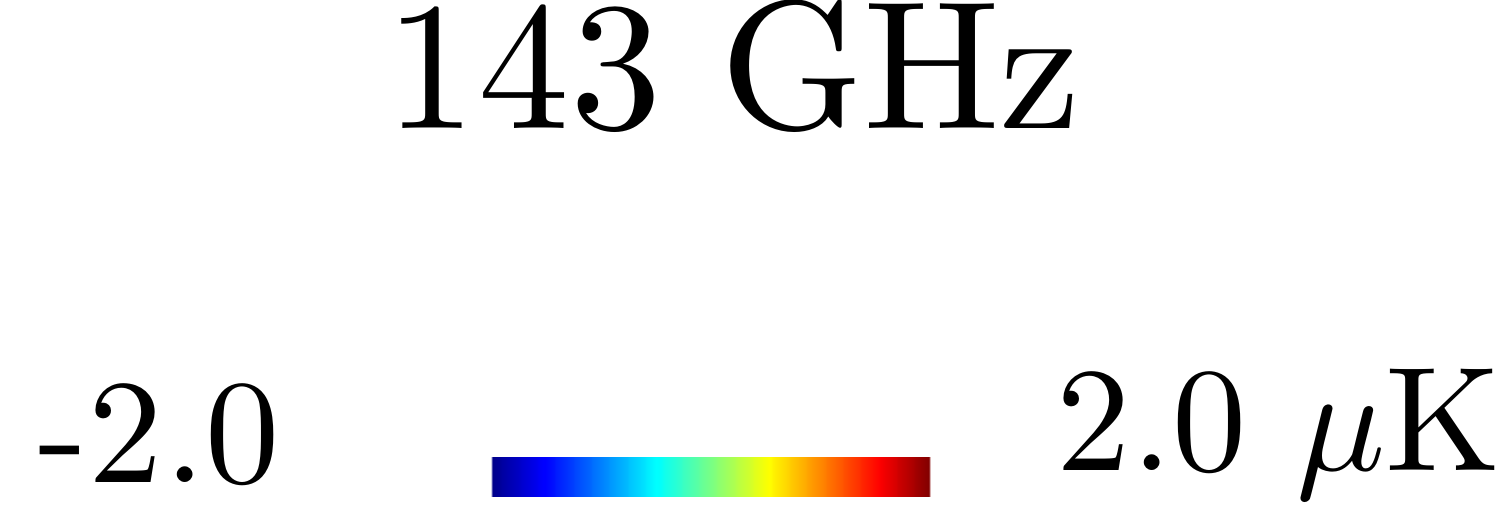}&
	\includegraphics [width=0.28\columnwidth]{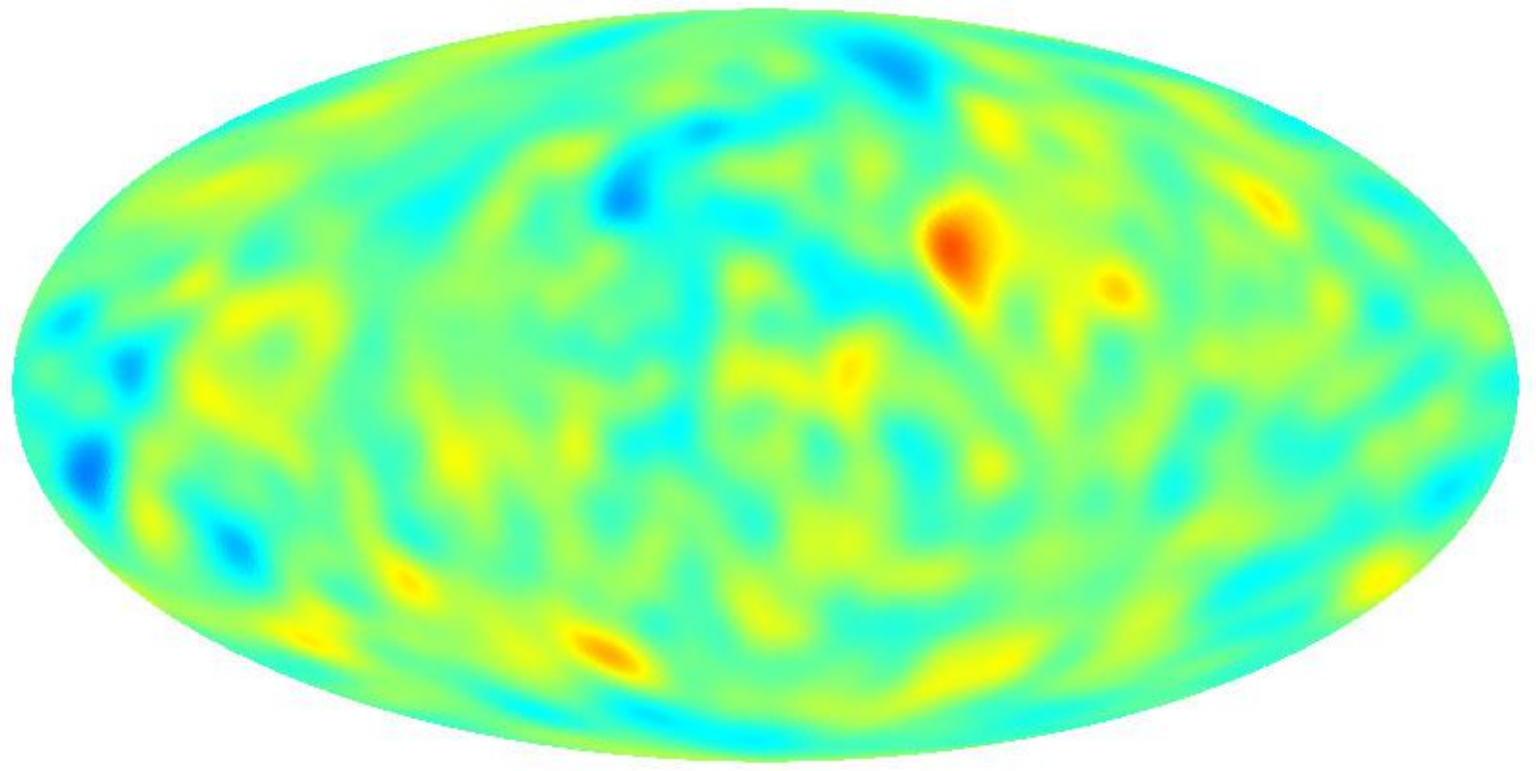}&
	\includegraphics [width=0.28\columnwidth]{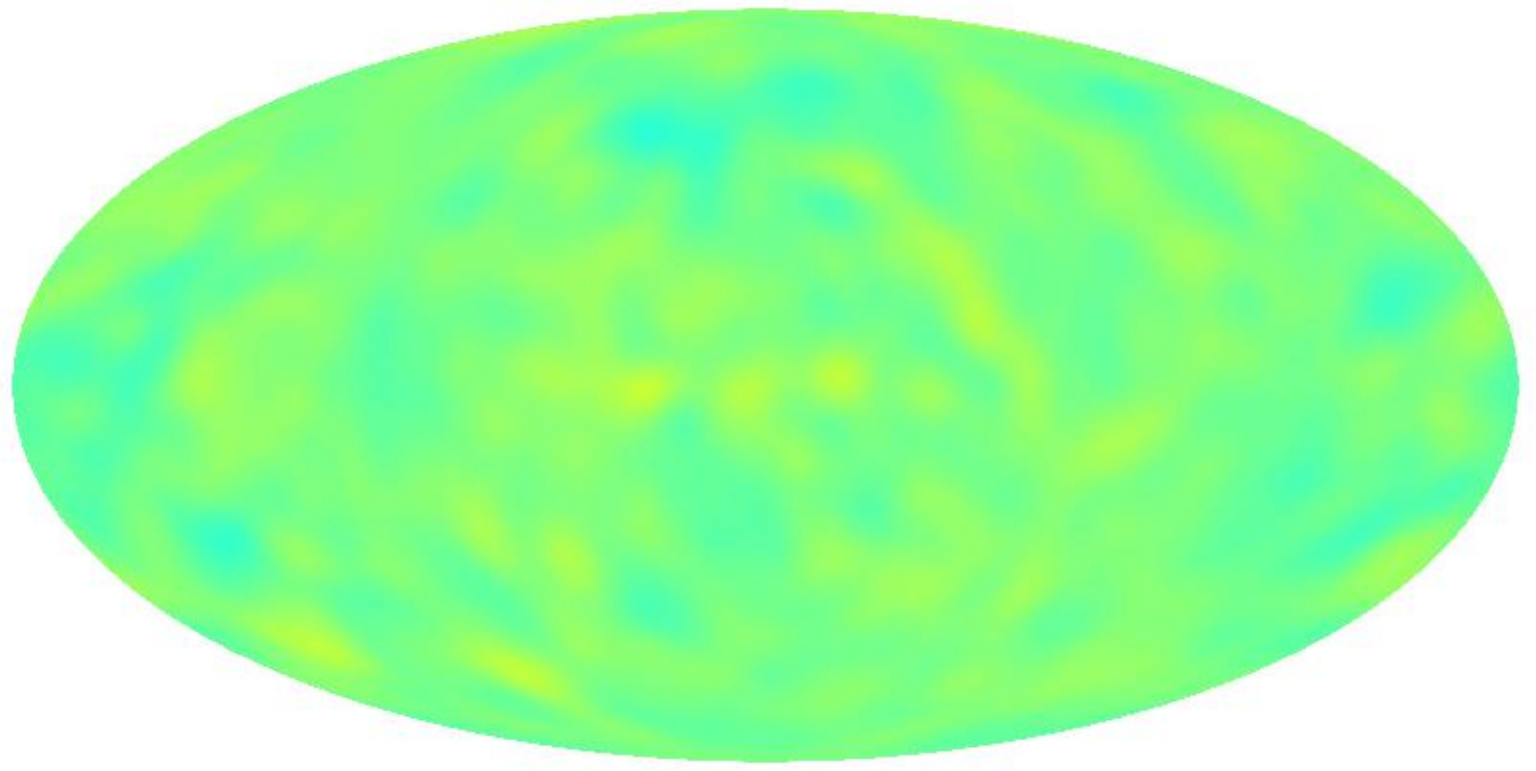}\\
	\\
	\\
	\raisebox{15pt}{\parbox{.06\textwidth}{no\\ corr.}}&
	\includegraphics [width=0.28\columnwidth]{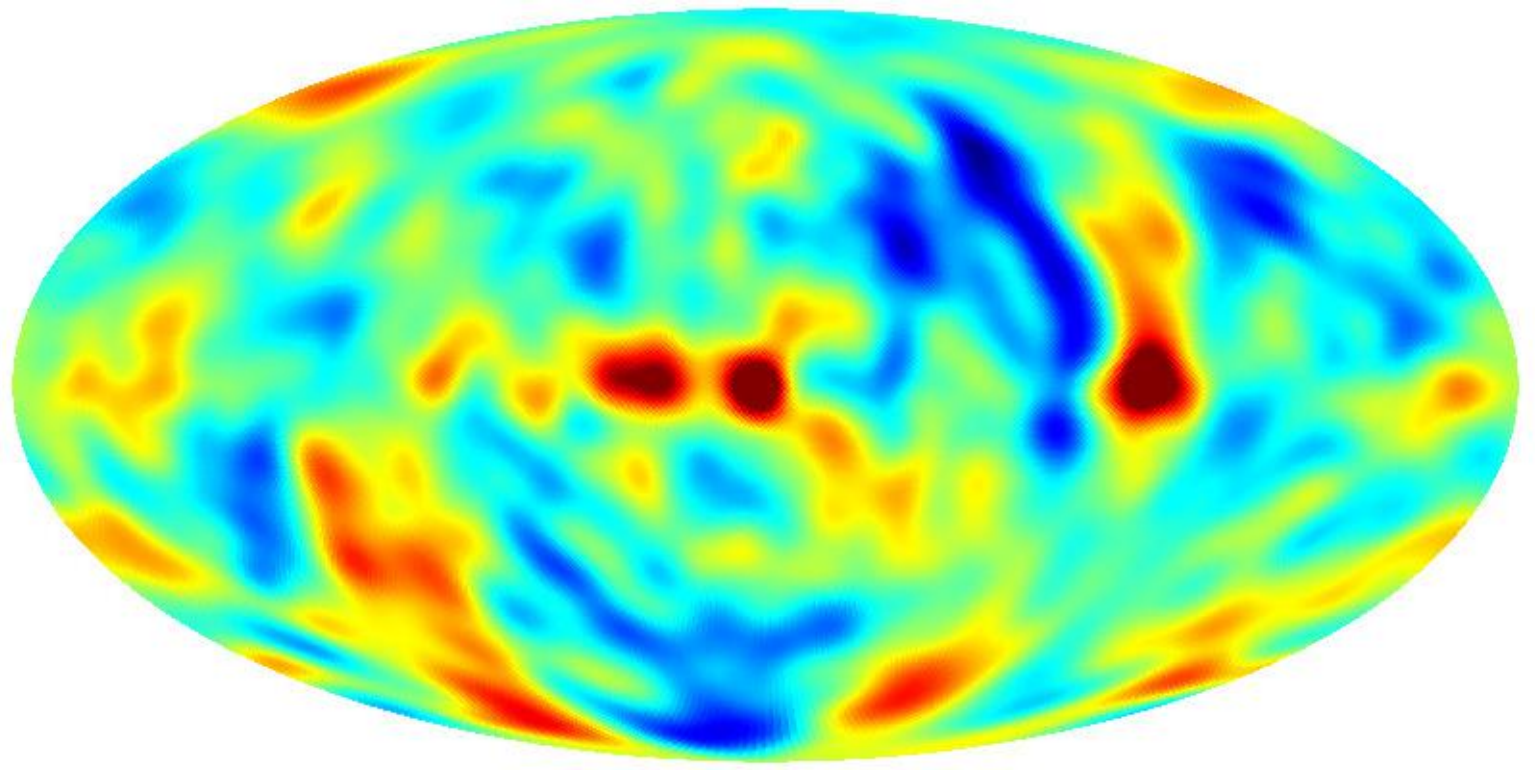}&
	\includegraphics [width=0.28\columnwidth]{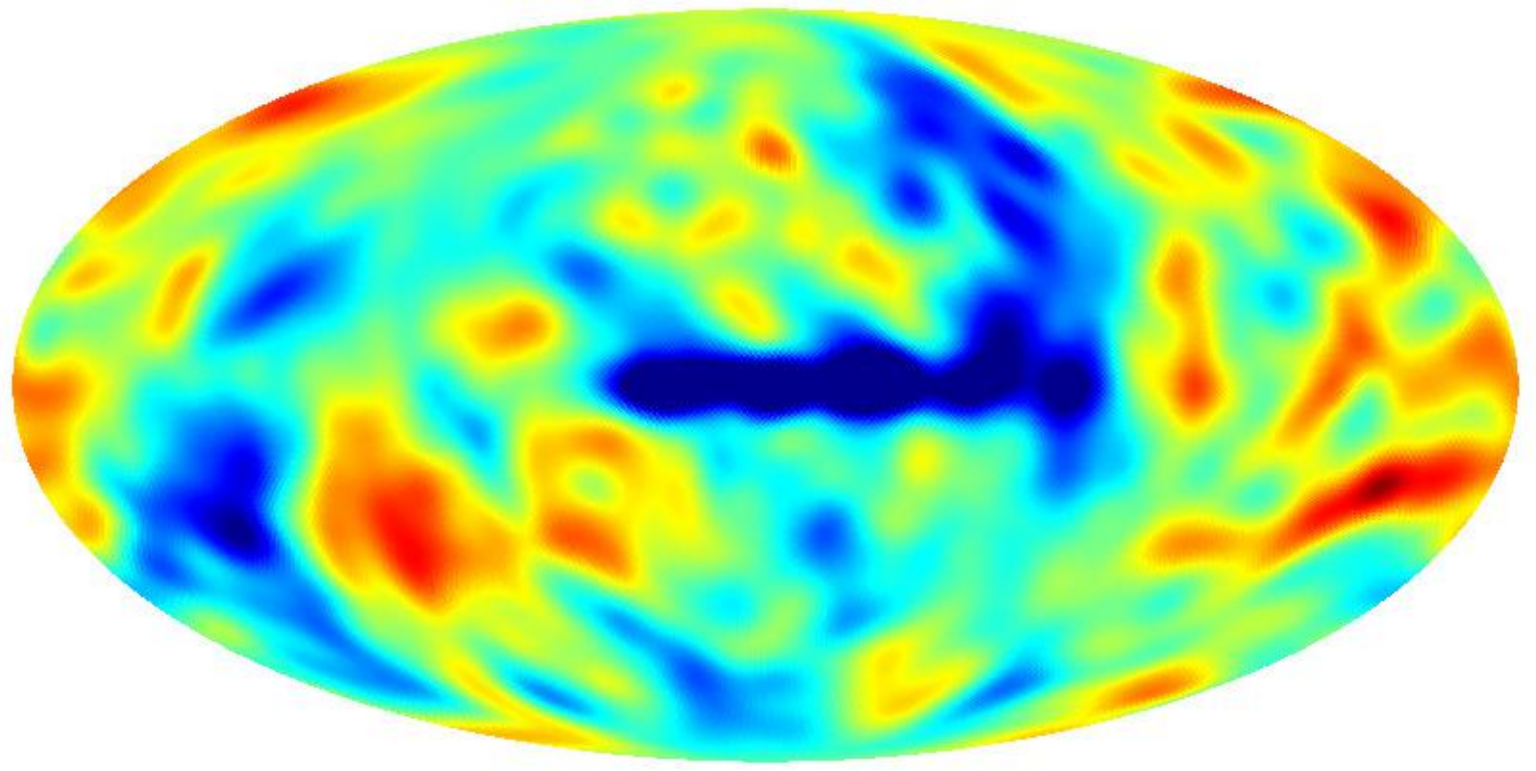}&
	\includegraphics [width=0.28\columnwidth]{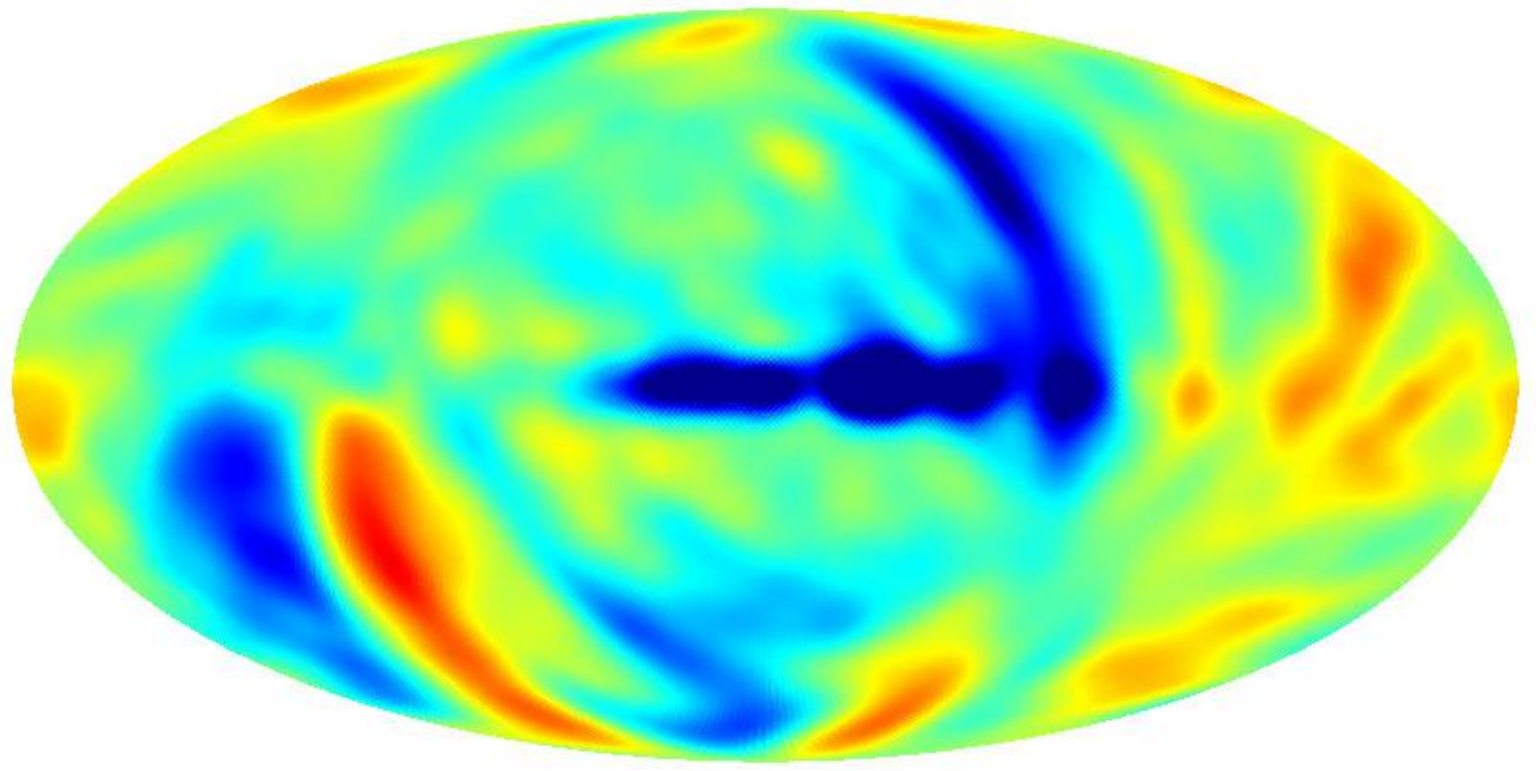}\\
	\raisebox{15pt}{\parbox{.06\textwidth}{linear gain corr.}}&
	\includegraphics [width=0.28\columnwidth]{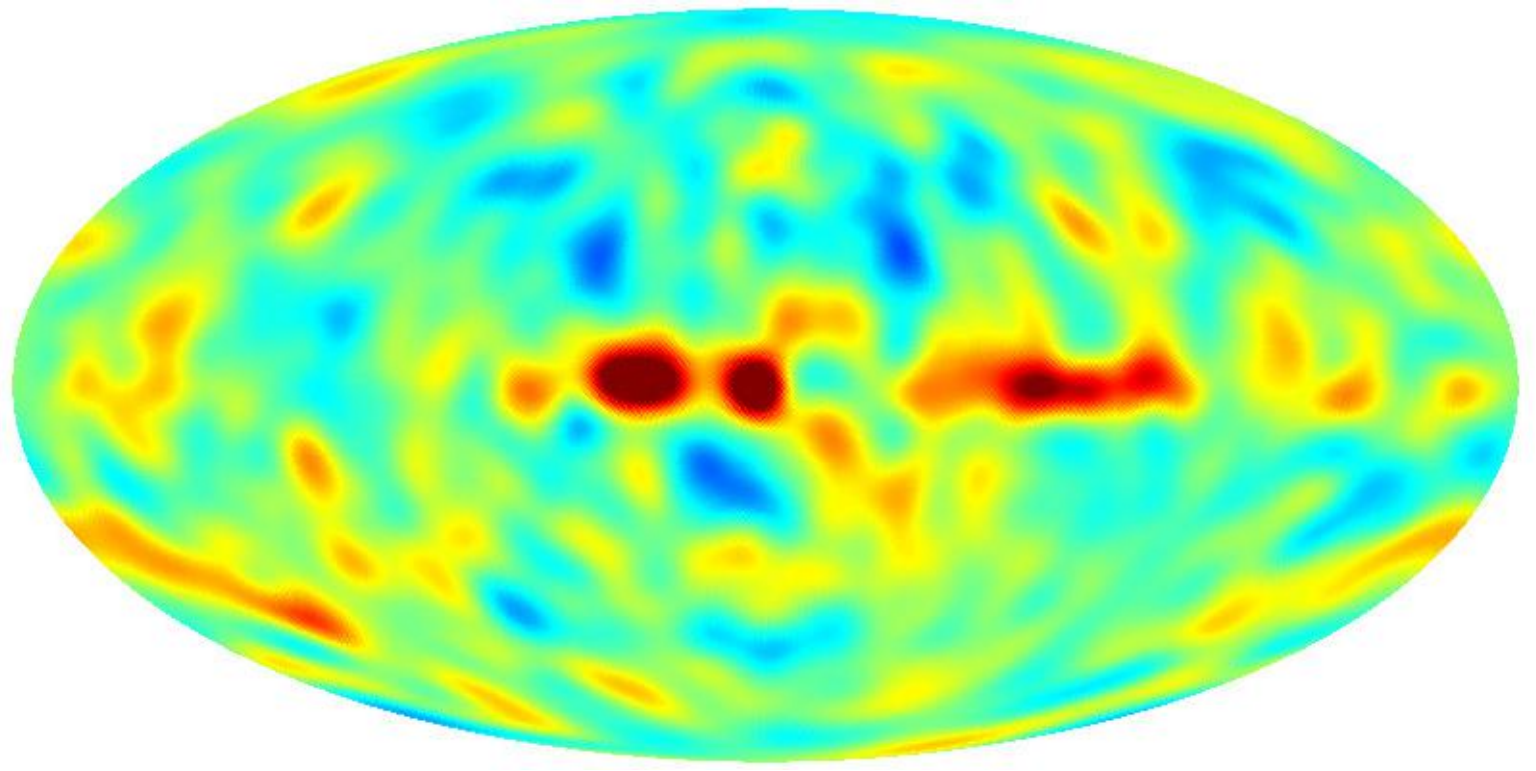}&
	\includegraphics [width=0.28\columnwidth]{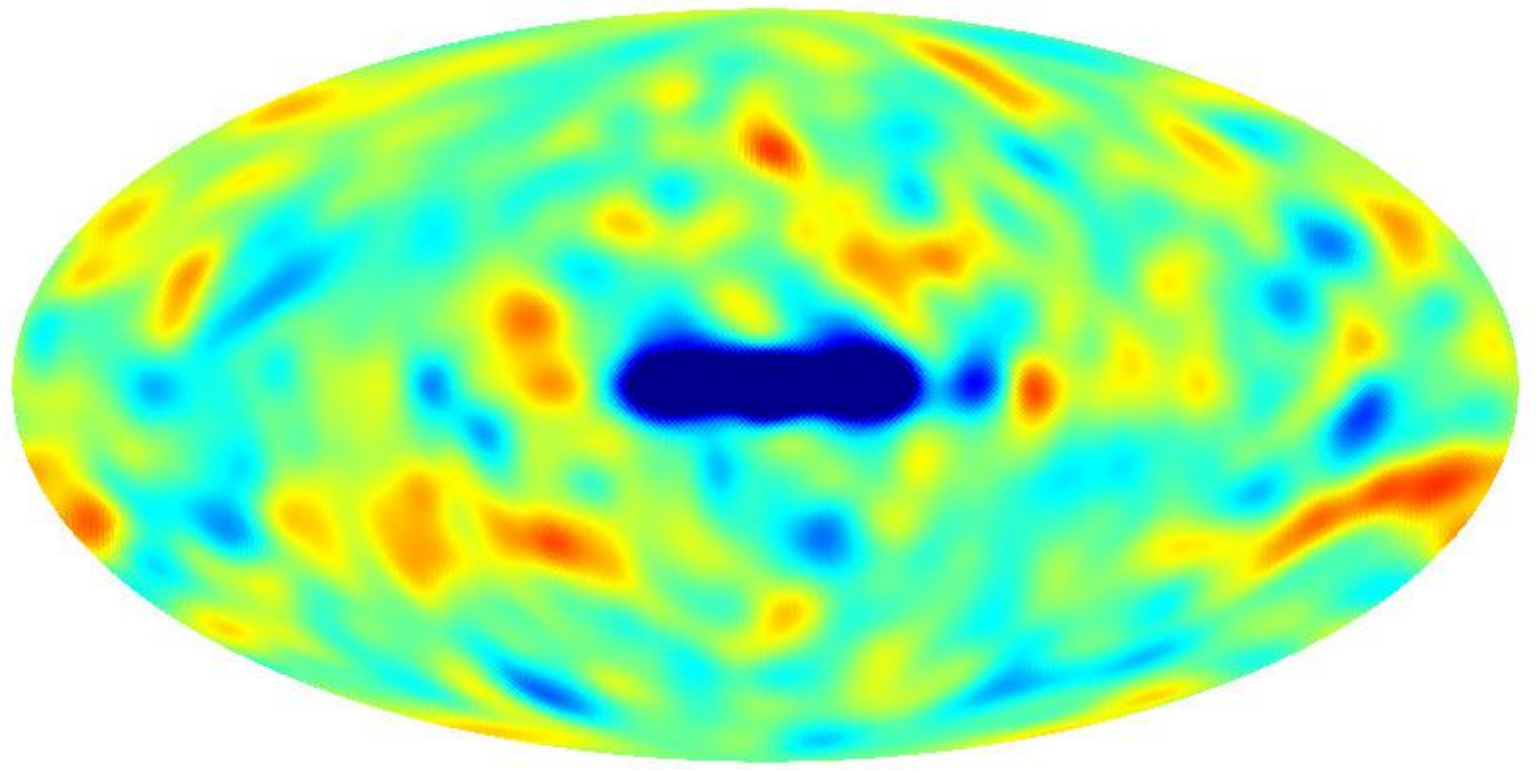}&
	\includegraphics [width=0.28\columnwidth]{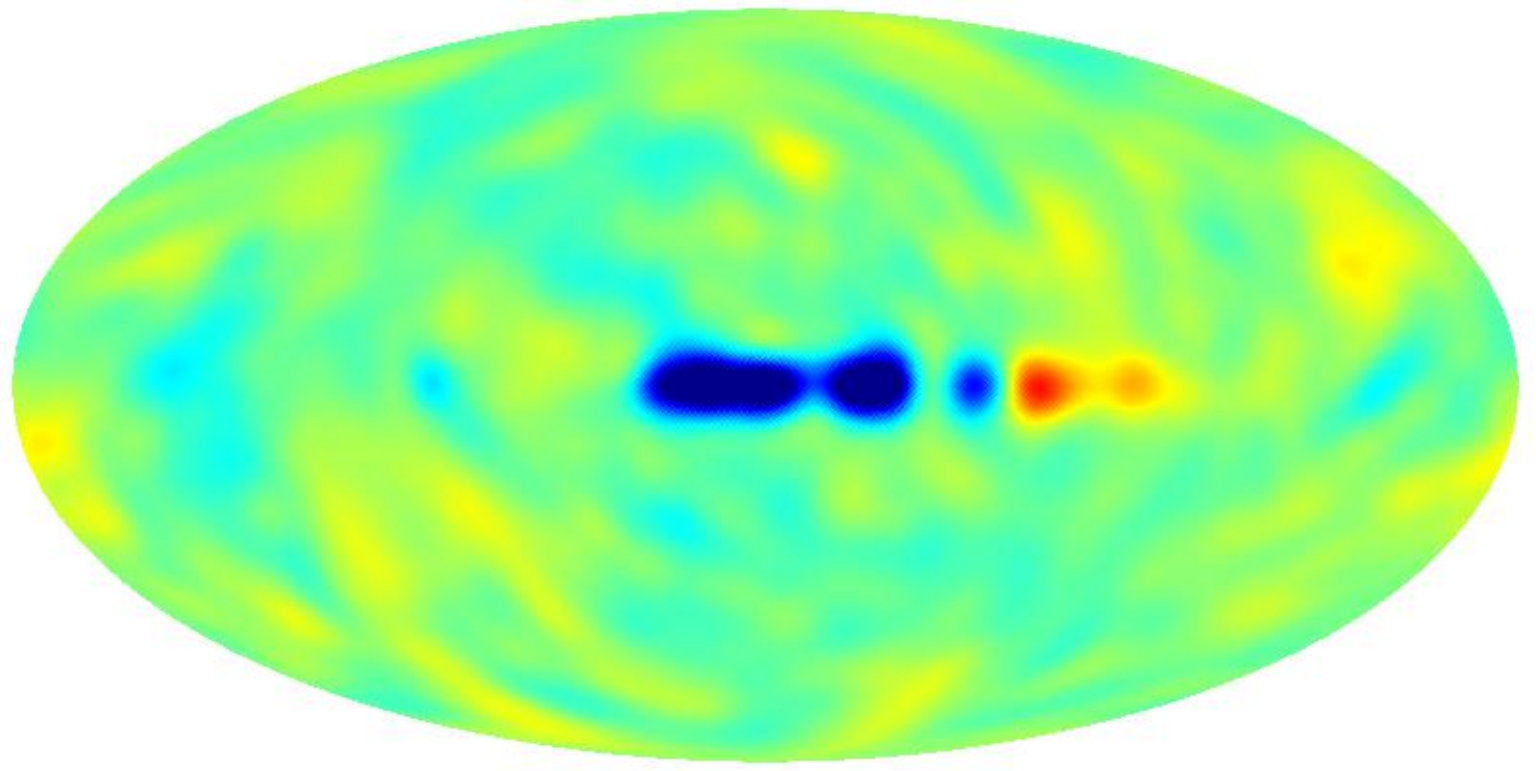}\\
	\raisebox{15pt}{\parbox{.06\textwidth}{dipole distort. corr.}}&
	\includegraphics [width=0.28\columnwidth]{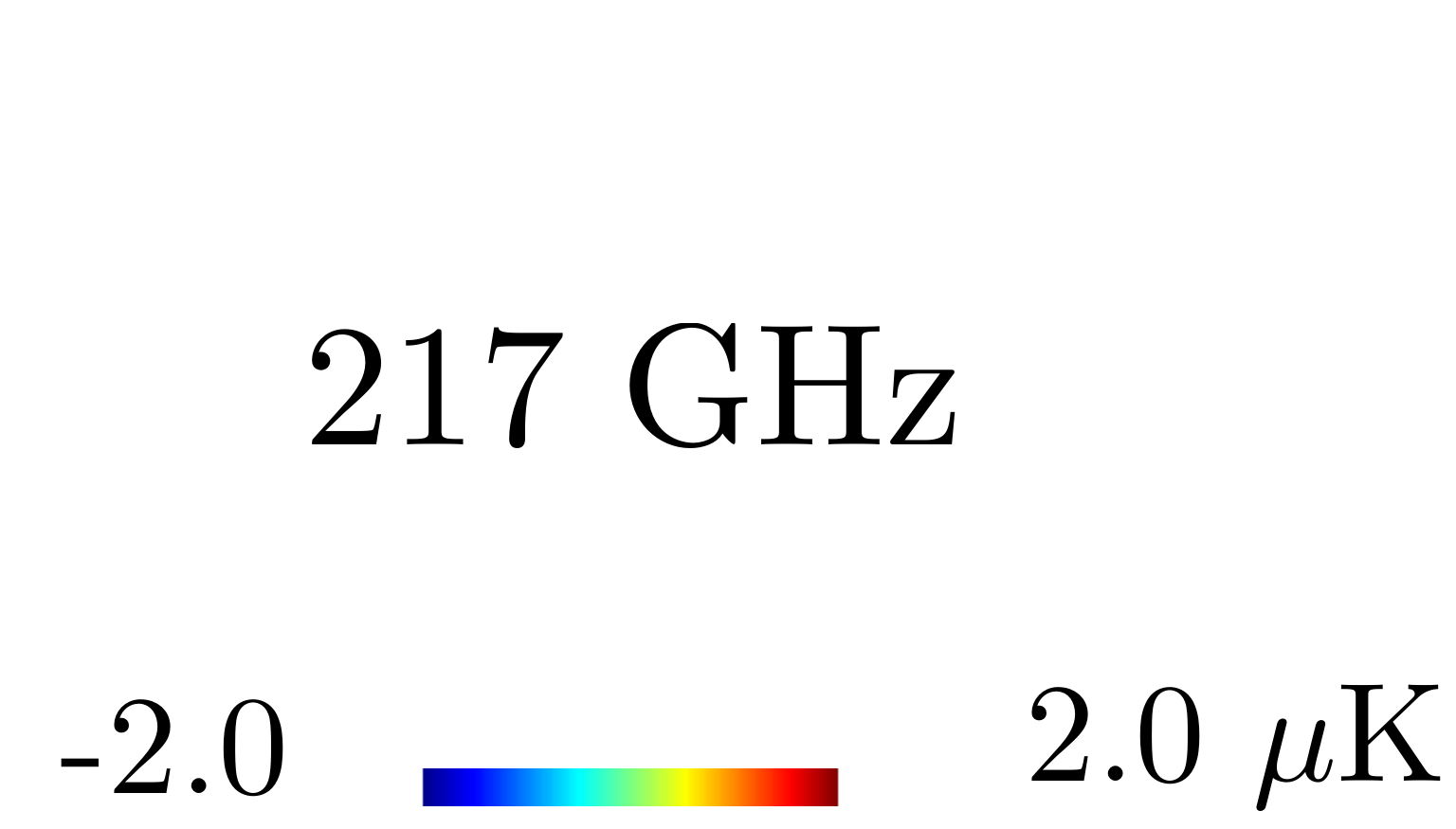}&
	\includegraphics [width=0.28\columnwidth]{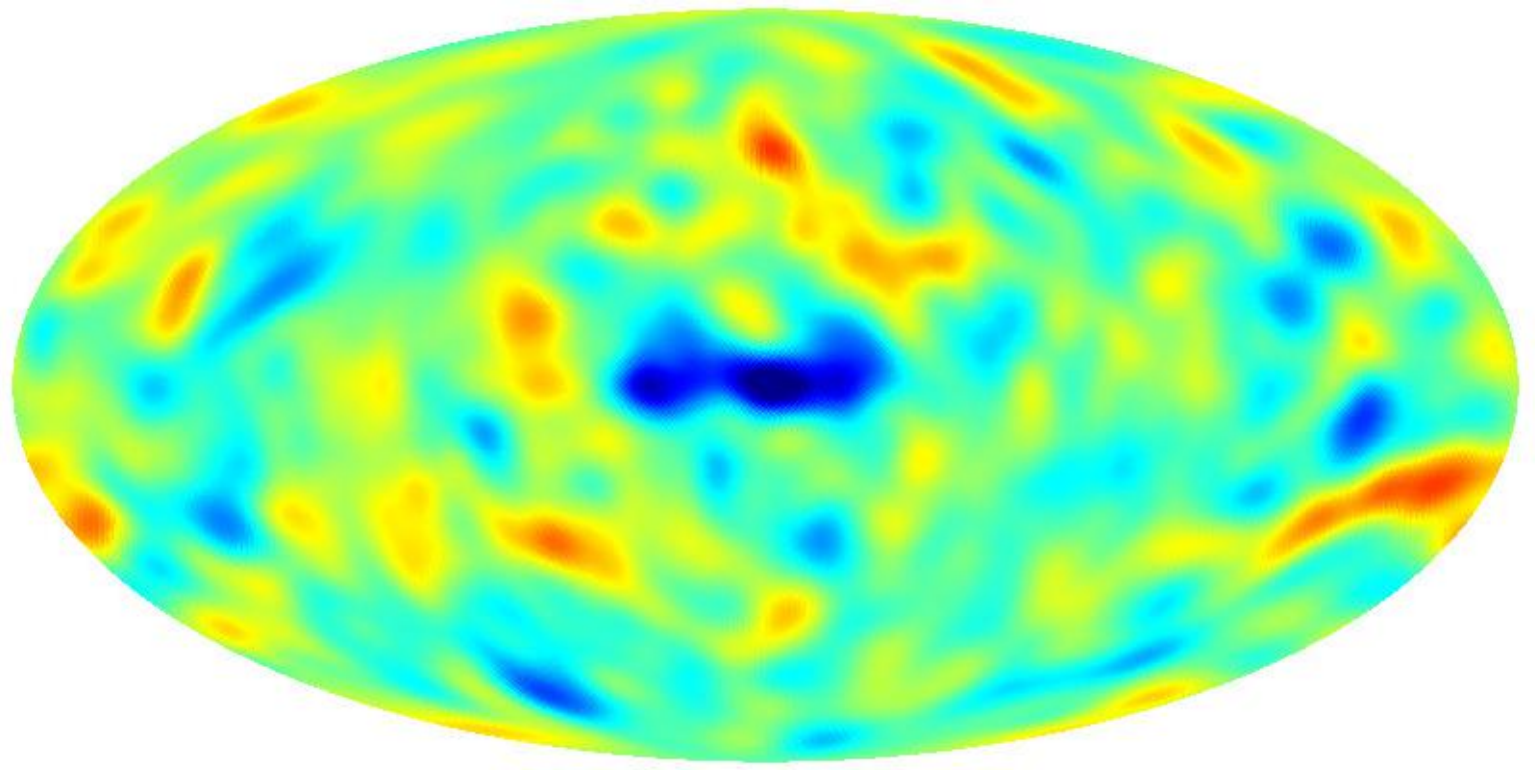}&
	\includegraphics [width=0.28\columnwidth]{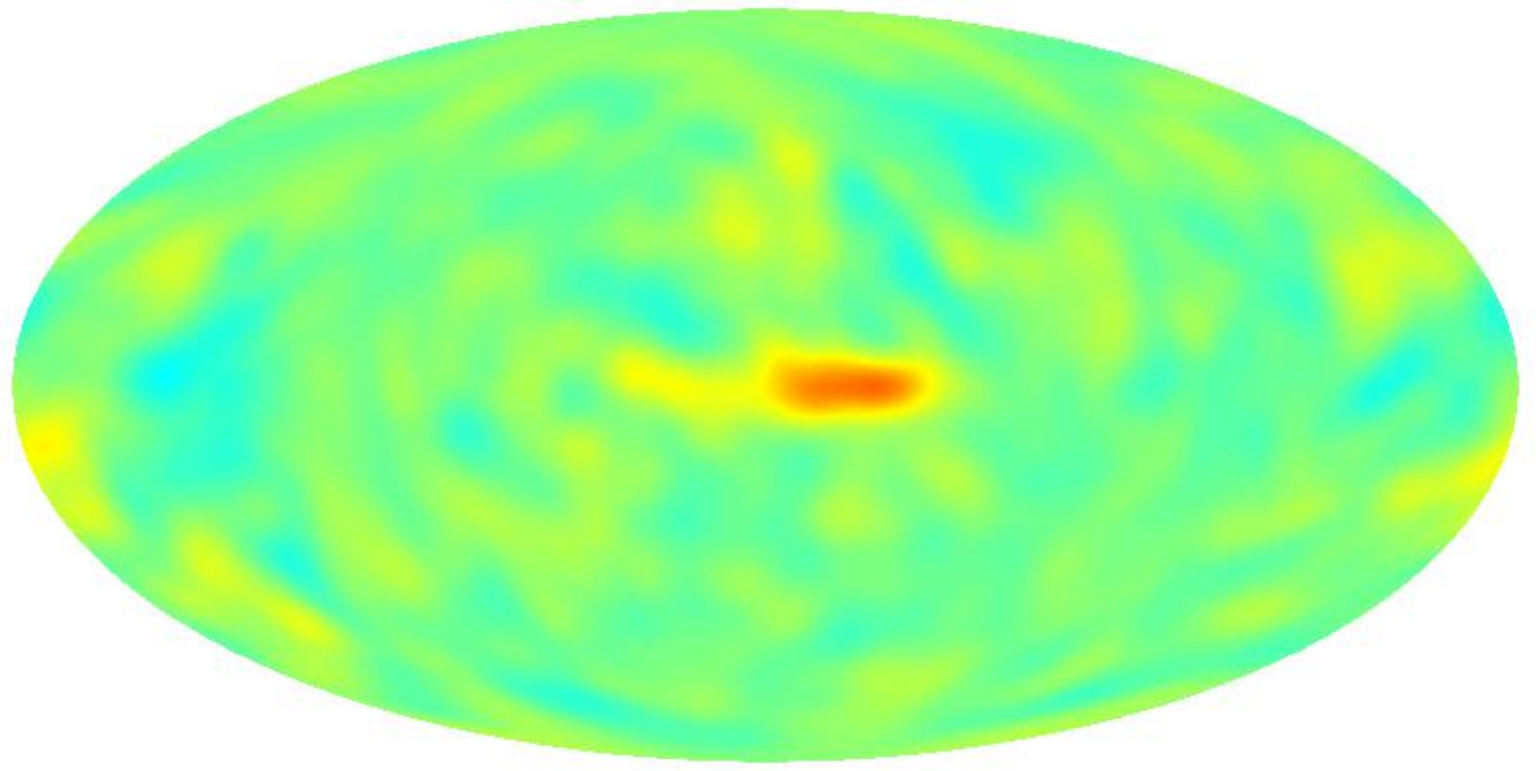}\\
	\\
	\\
	\raisebox{15pt}{\parbox{.06\textwidth}{no\\ corr.}}&
	\includegraphics [width=0.28\columnwidth]{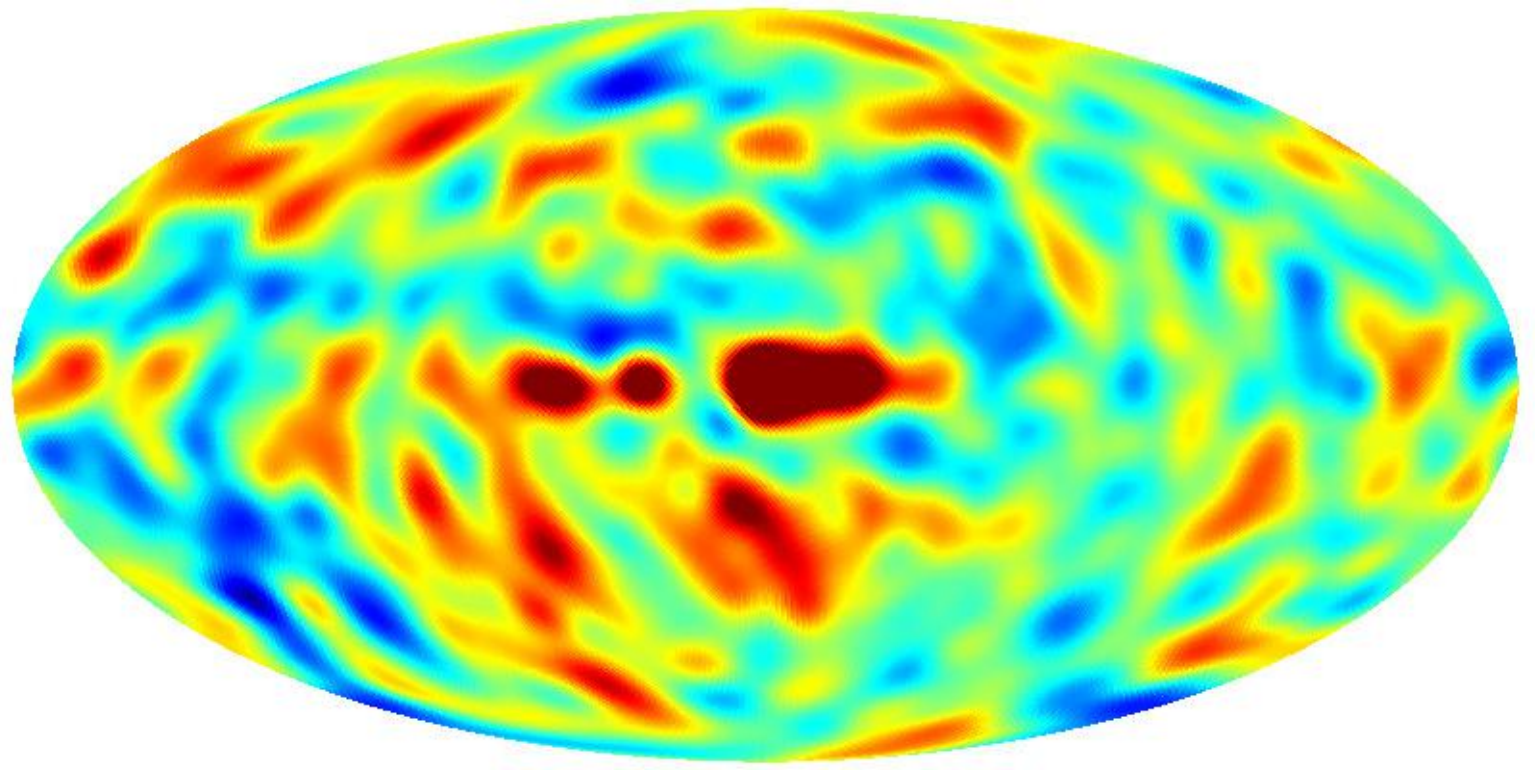}&
	\includegraphics [width=0.28\columnwidth]{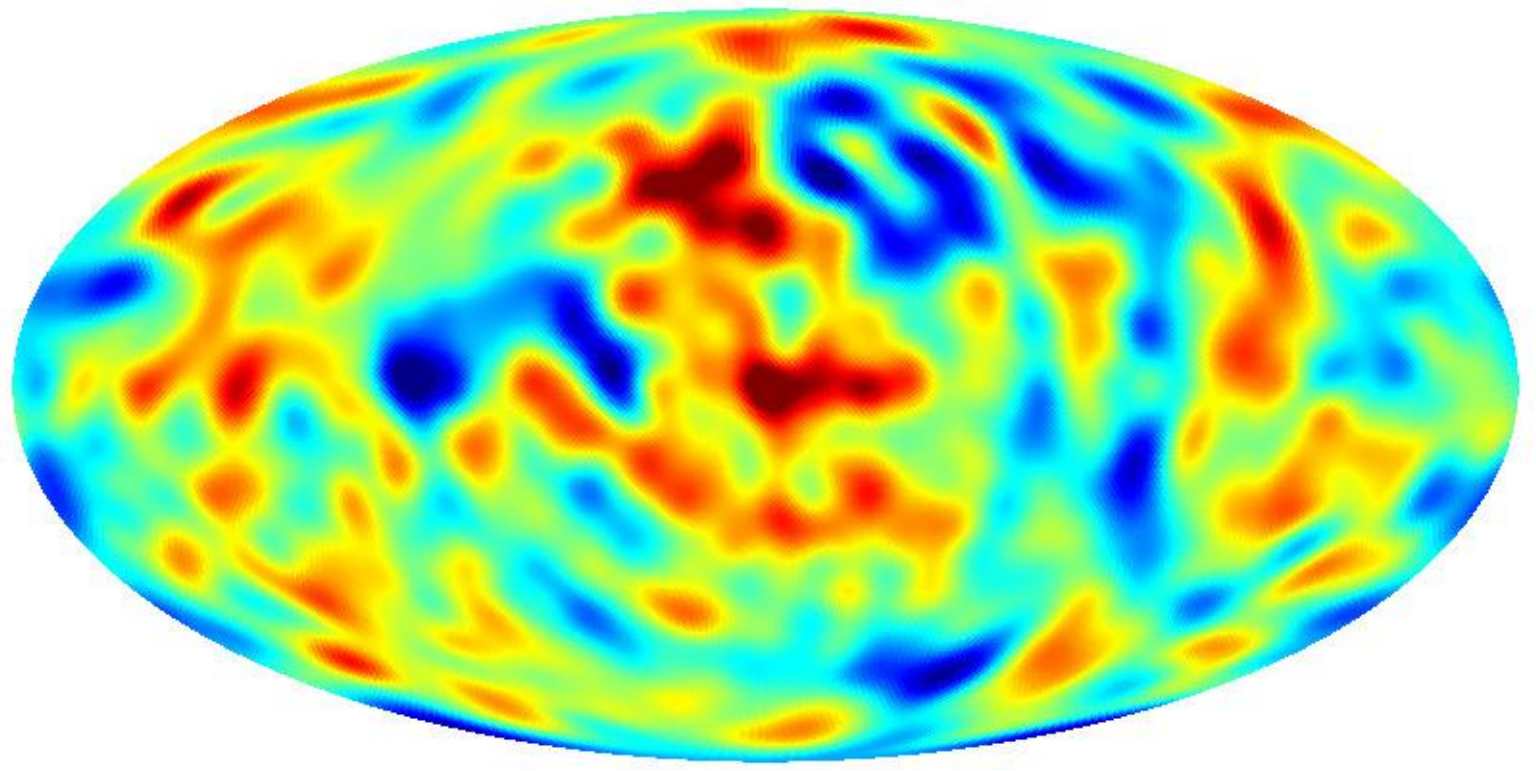}&
	\includegraphics [width=0.28\columnwidth]{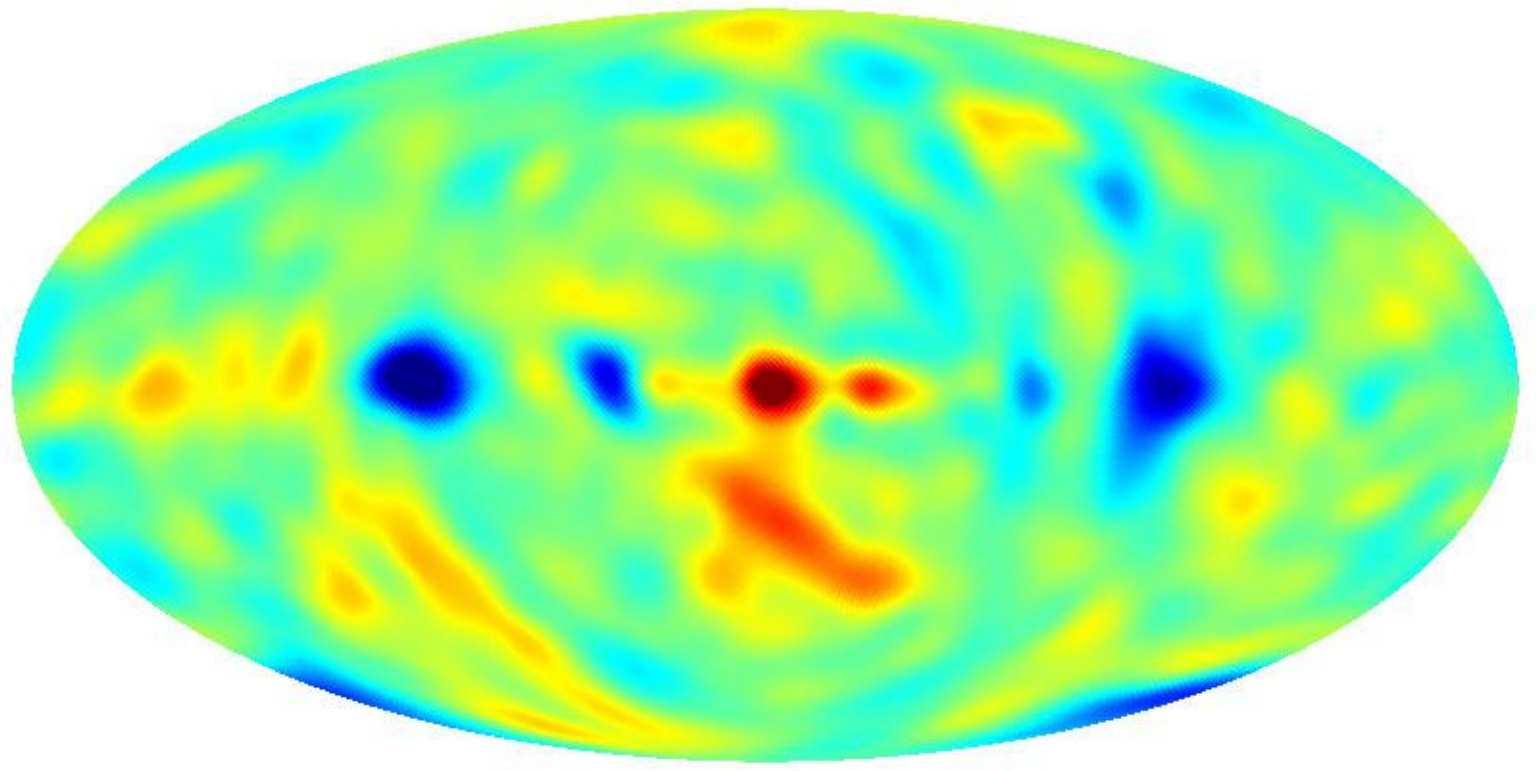}\\
	\raisebox{15pt}{\parbox{.06\textwidth}{linear gain corr.}}&
	\includegraphics [width=0.28\columnwidth]{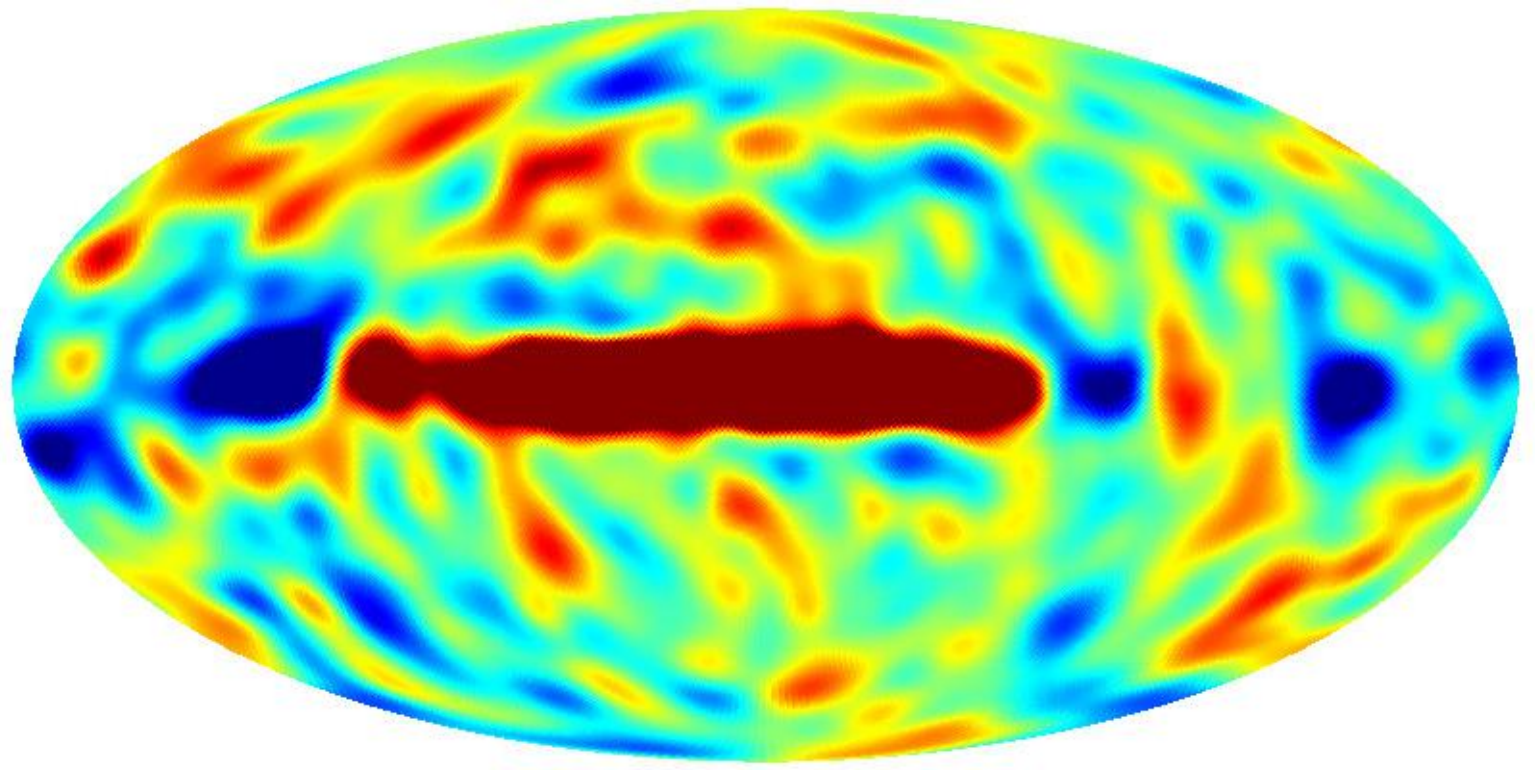}&
	\includegraphics [width=0.28\columnwidth]{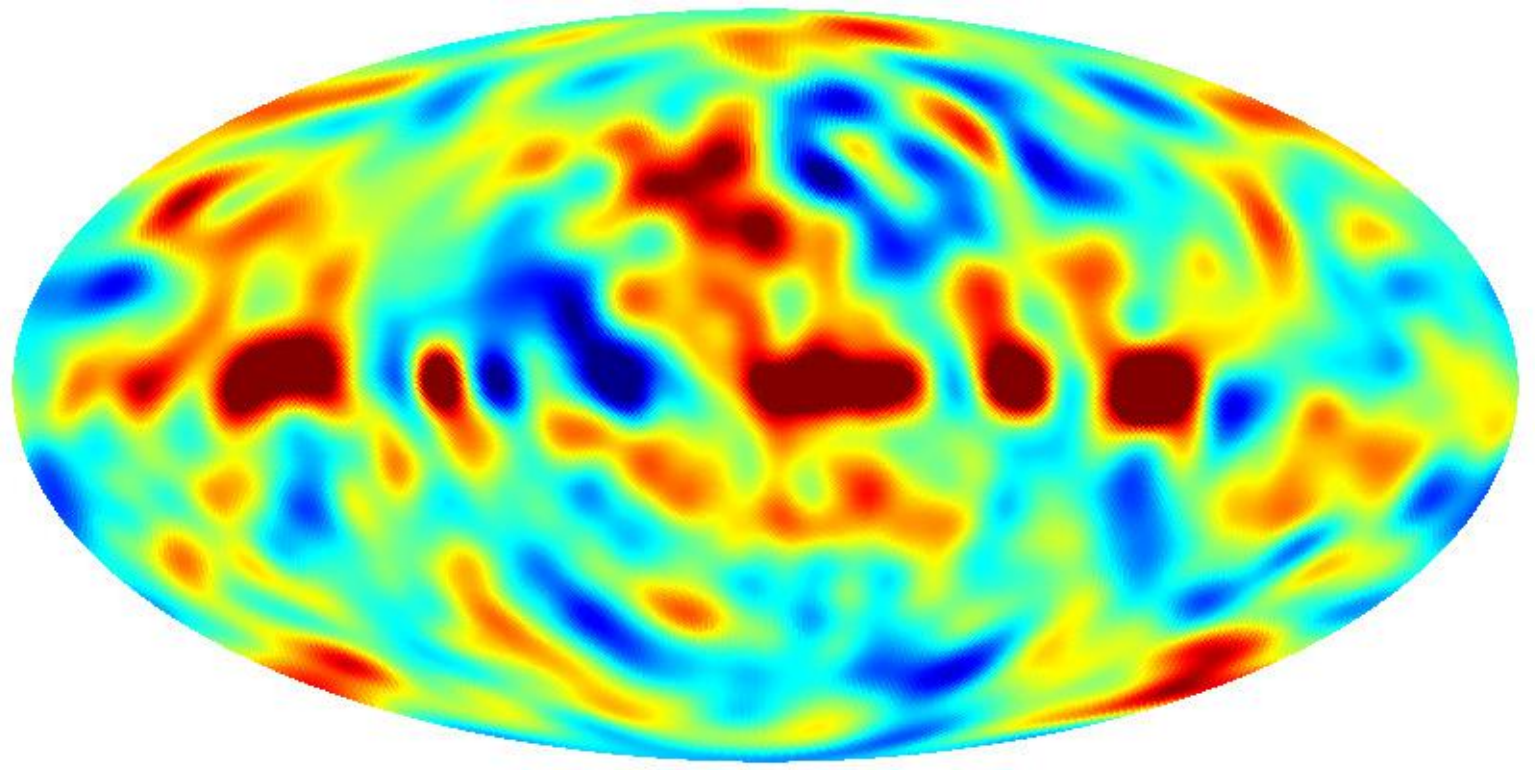}&
	\includegraphics [width=0.28\columnwidth]{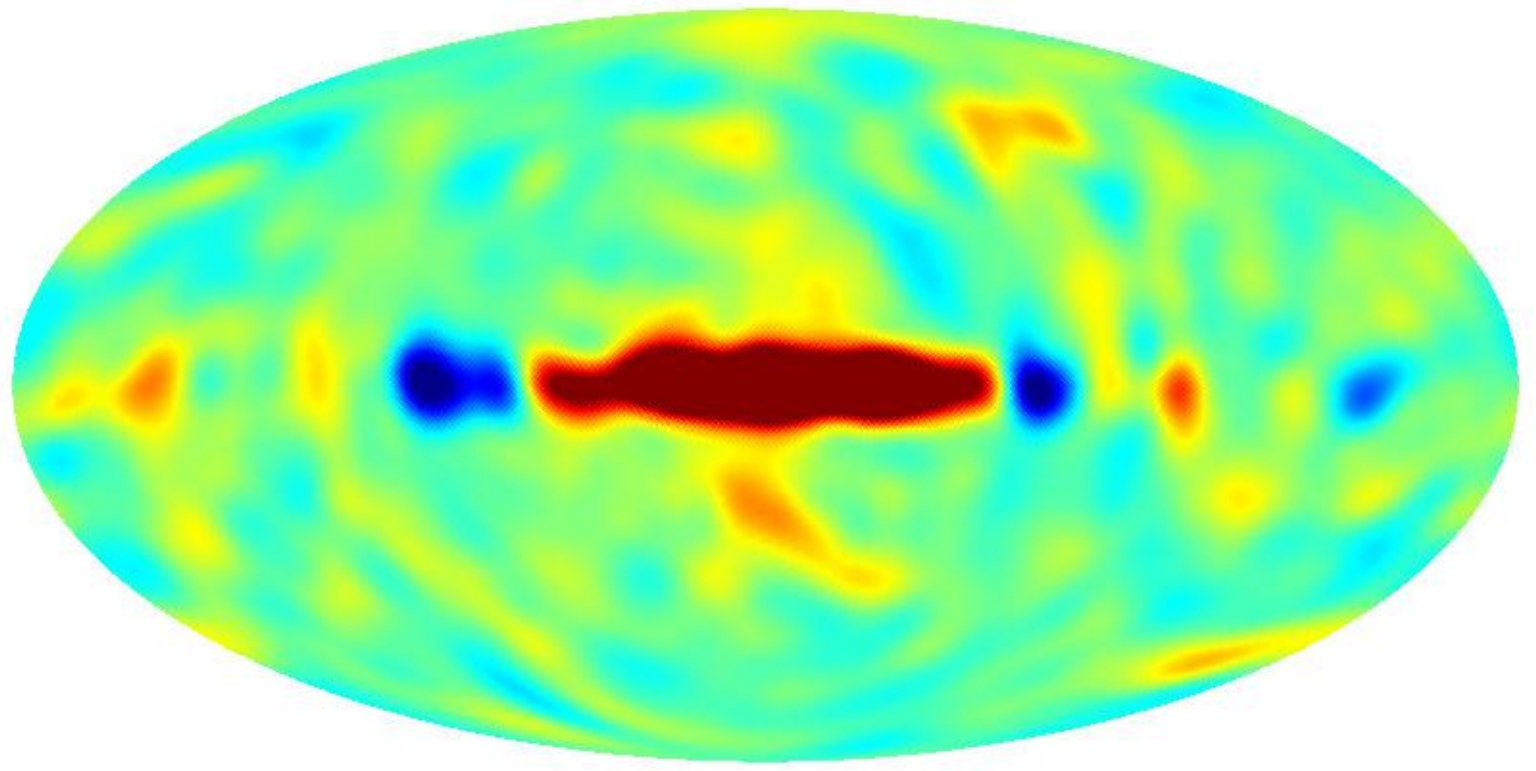}\\
	\raisebox{15pt}{\parbox{.06\textwidth}{dipole distort. corr.}}&
	\includegraphics [width=0.28\columnwidth]{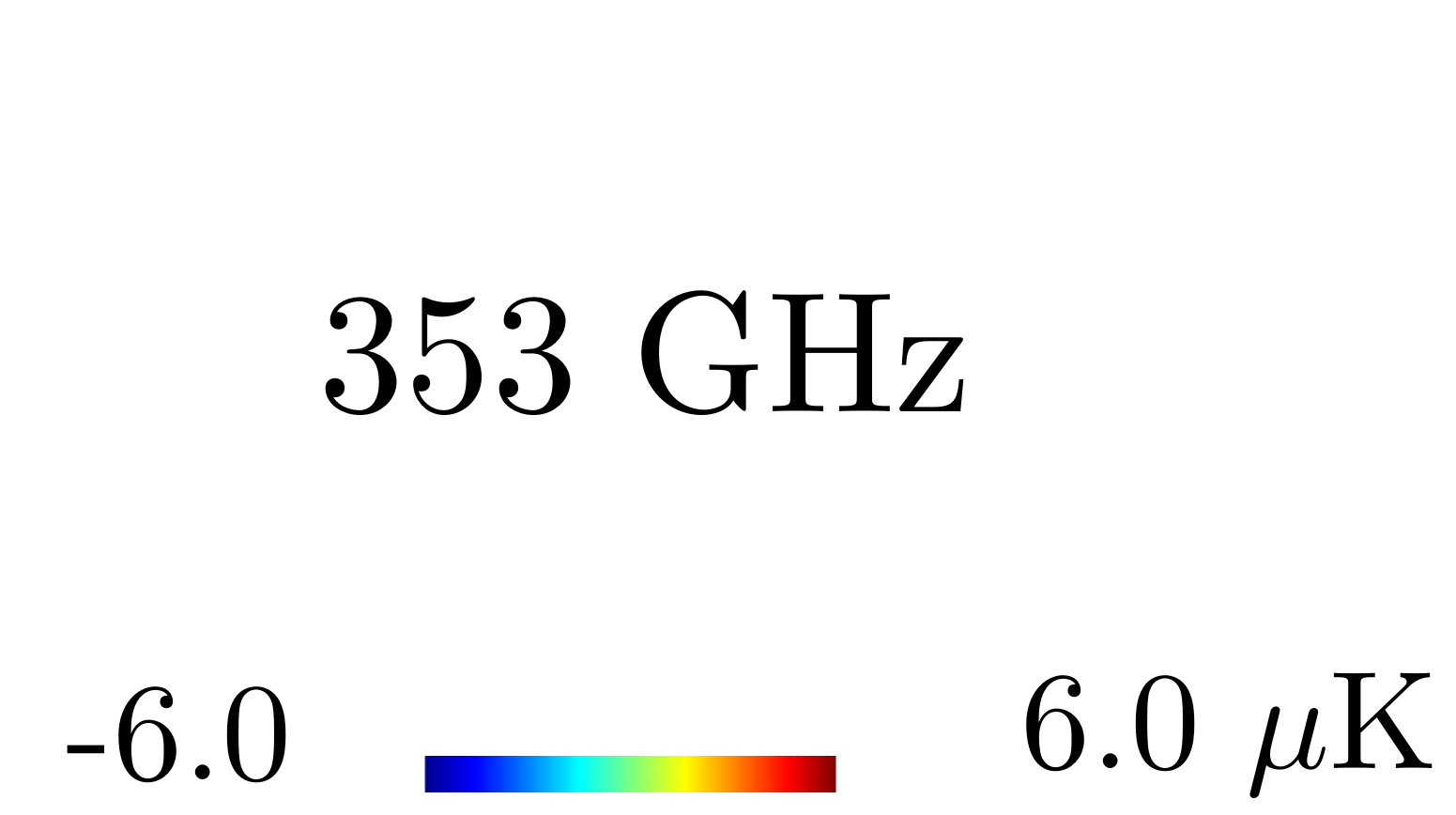}&
	\includegraphics [width=0.28\columnwidth]{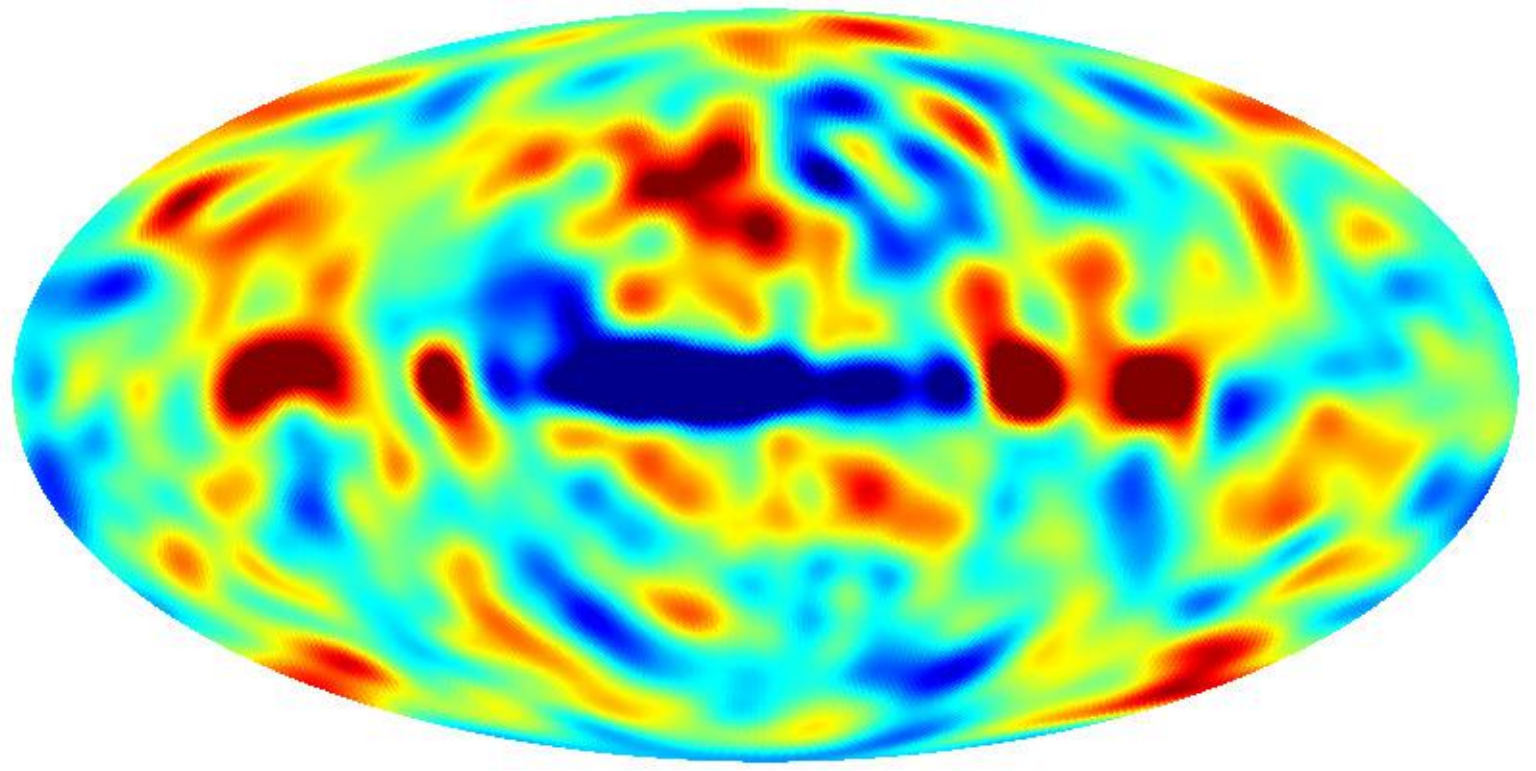}&
	\includegraphics [width=0.28\columnwidth]{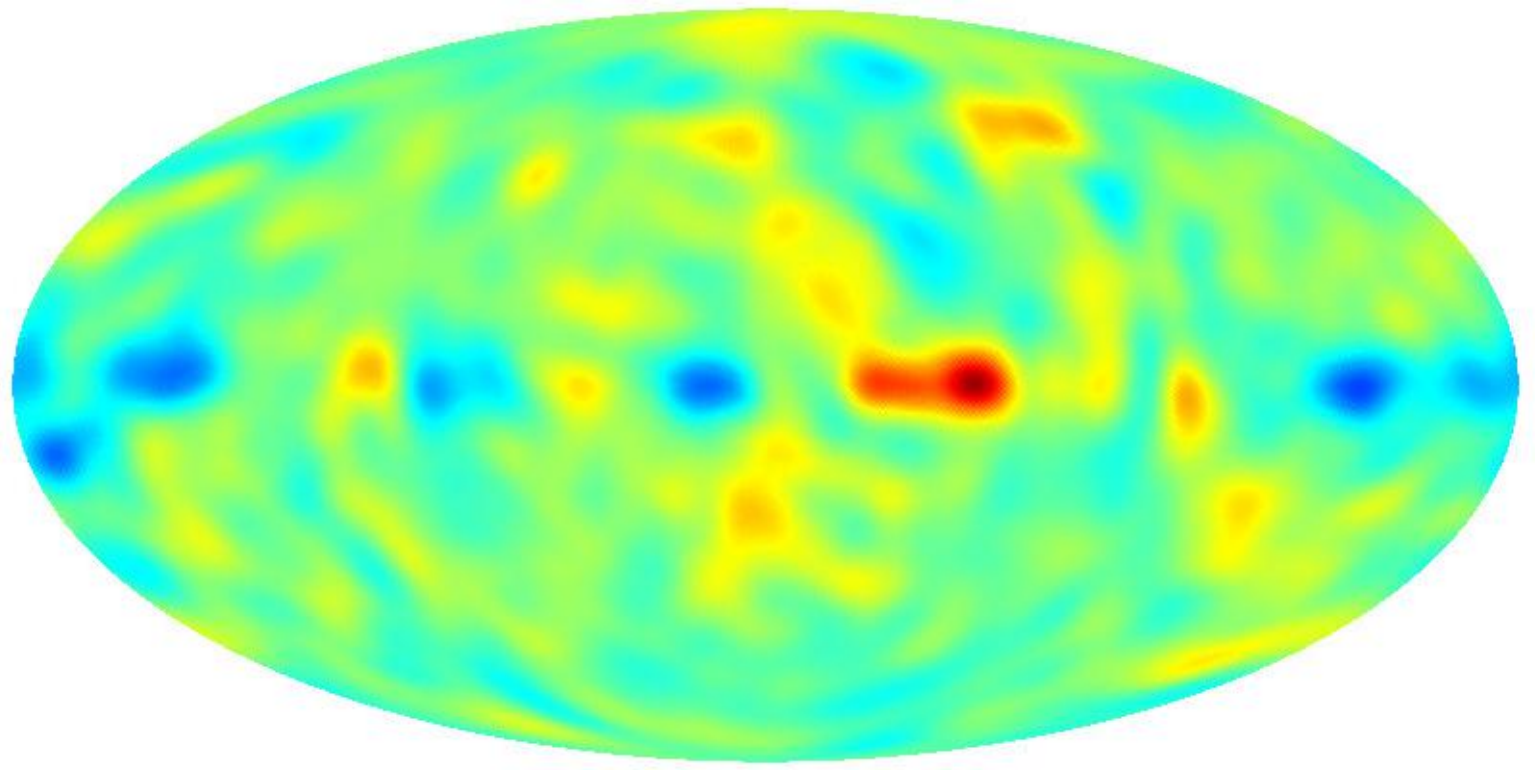}\\
\end{tabular}
\caption{For each of the four frequencies, we show eight $Q$ maps. The first row contains data (left column) and one realization of simulation of the ADCNL with noise (centre column) and without noise included (right column). The second row contains the same after correction for the ADCNL linear gain variation. The third row for each frequency contains simulations carried out without including the ADCNL. With the strong ADCNL signal (especially the dipole part) absent, these show the noise plus other systematics (centre) and just the systematics (right).}
\label{fig:ADCNLresidu} 
\end{figure}
\setlength\tabcolsep{6pt}
\renewcommand{\arraystretch}{1.}

For each frequency, eight maps are displayed, with the actual data shown in the left column, and simulations of the ADCNL effect with and without noise in the centre and right columns, respectively. For each frequency: the top row shows the effects of the ADCNL without any correction; the second row demonstrates the improvement when the correction for the ADCNL-induced gain variation is included; and the third row shows additional correction of the non-linear distortion of strong signals (mostly the dipole). Since this last correction has not been made to the data for this release, the third row shows only simulations.

At 100 and 143\,GHz, maps in the first row show that the ADCNL distortions in the simulations (right maps) can be recognized in the data (left maps), even in the presence of noise. For 217 and 353\,GHz, comparing the centre and right maps of the middle row shows that noise dominates the residual effect after the linear correction for the ADCNL. 

Figure~\ref{fig:adcnl} shows power spectra of both the data and all five simulations (only one of which was mapped in Fig.~\ref{fig:ADCNLresidu}).
\begin{figure*}[htbp!]
\includegraphics[width=0.49\textwidth]{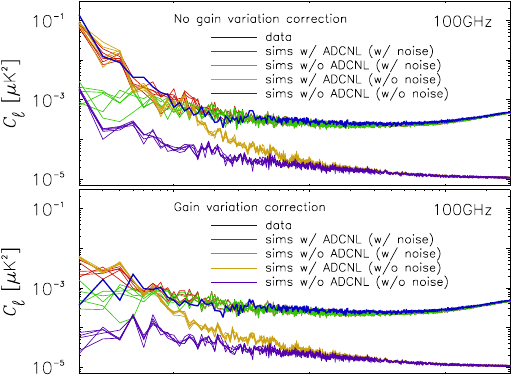}
\includegraphics[width=0.49\textwidth]{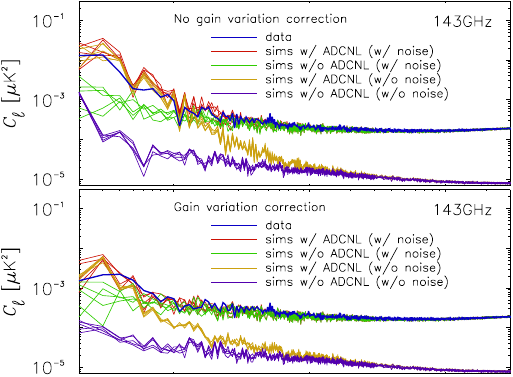}\\
\includegraphics[width=0.49\textwidth]{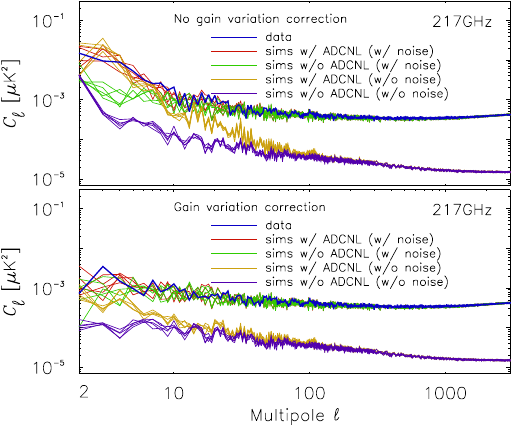}
\includegraphics[width=0.49\textwidth]{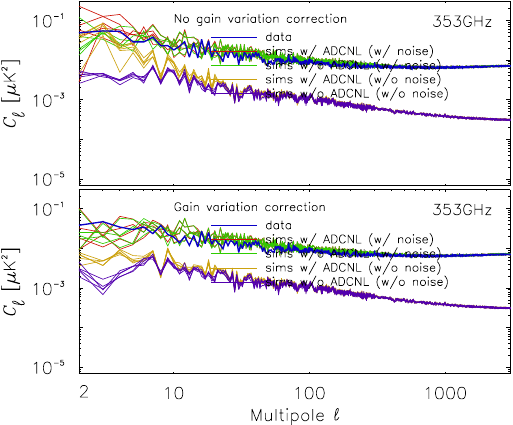}
\caption{For each frequency, $EE$ power spectra are presented for the half-mission null-test difference maps, with and without noise. The plots show the data and five realizations of the ADCNL simulation: red with ADCNL and noise; orange with ADCNL without noise; green without ADCNL, but with noise; and purple without ADCNL, and without noise (residuals from other systematics). In the upper panels, no ADCNL correction has been carried out in the mapmaking, while in the lower panels, the time-varying gain correction has been applied. The simulations are also carried out without ADCNL.}
\label{fig:adcnl} 
\end{figure*}
All plots show the $EE$ power spectra of the half-mission difference maps employed in this test. The differences between the residual power spectra allow us to assess the impact of the ADCNL at each of four HFI frequencies. In all these plots, the half-mission difference results for the data are shown as blue lines.

For each frequency, in the upper panels of Fig.~\ref{fig:adcnl}, along with the data, we show the auto-spectra derived from five simulated realizations of the ADCNL, with noise included in the simulations (as red lines), and with no noise (as orange lines). The red lines are a good approximation to the data. We also show five realizations without the ADCNL included; these show only noise (green lines). The purple lines are simulations with neither noise nor ADCNL included, and hence trace only other systematics and residuals from \sroll.

In the upper panels of Fig.~\ref{fig:adcnl}, no correction for ADCNL-induced gain time variation has been applied in the mapmaking.
By contrast, the lower panels show the same set of power spectra after the time-varying linear gain correction has been applied in the mapmaking. This correction is applied to both the data and the simulations.
The lower panels demonstrate the reduction in the residual power spectra at low multipoles resulting from the correction for time variations in the gain induced by the ADCNL. Again, the simulations including both the ADCNL and noise (red lines) match the data well (blue lines). As before, green lines represent simulation of noise only, and purple lines the remaining systematics (neither noise nor ADCNL).

The plots in the upper panels demonstrate that the ADCNL effect (orange lines) dominates the error at low multipoles for the three CMB channels. In addition, as already noted, the simulations including both noise and ADCNL (red lines) match the data well over the full range of multipoles for all four frequencies. At each frequency, the lower panels show the effectiveness of the correction for the time-varying gain induced by ADCNL included in the 2018 release. It is still the case that residual ADCNL effects dominate all other systematics (purple lines) at $\ell < 100$ for 100 and143\,GHz. It is also the case that the residual ADCNL signal exceeds the noise level at the lowest multipoles even after correction. Additionally, the simulations overestimate the ADCNL effect at 100 and 143\,GHz (the blue lines for the data run below the red and orange lines for the simulations). At 217\,GHZ, the contributions of the ADCNL and the noise (largely $1/f$ at low $\ell$) are comparable. At 353\,GHz, noise is dominant for all multipoles.

At $\ell > 30$, for all four frequencies, noise clearly dominates all systematic effects. This justified the decision to ignore the residual systematic effects in the 2015 release. 

As the plots in the lower panels indicate, at the lowest multipoles ($\ell < 5$) uncorrected non-linear ADCNL effects still dominate, at least for 100 and 143\,GHz. The residual low-multipole power spectrum is at the few times $10^{-3}\microKcarre$ level even in these cases, and is below the noise (although still non-negligible) at 353\,GHz.

Figure~\ref{fig:diffgl} shows the power spectra at 100\,GHz from maps of the five simulations used in Figs.~\ref{fig:ADCNLresidu} and \ref{fig:adcnl}.
\begin{figure}[htbp!]
\includegraphics[width=\columnwidth]{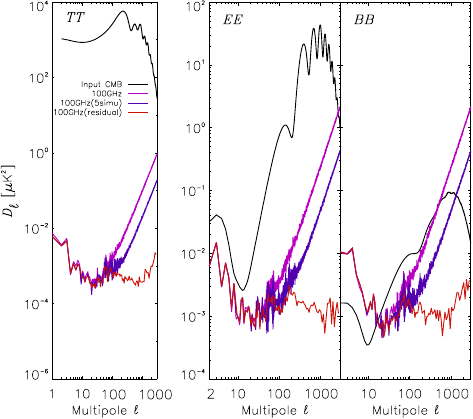}
\includegraphics[width=\columnwidth]{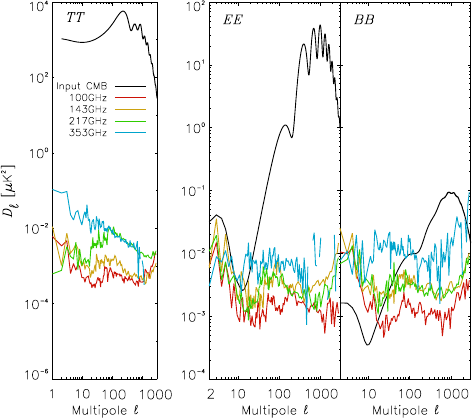}
\caption{{\it Top panel}: residual auto-spectra of the ADCNL at 100\,GHz, for a single simulation (purple line), after correction by the gain variation model. An average over five realizations is shown as the blue line. Using the difference, and scaling it as Gaussian noise for one simulation, we then subtract it from the single simulation (purple). The red curve shows the residuals when removing this estimate of the noise. {\it Bottom panel}: the residuals for the four frequencies.}
\label{fig:diffgl} 
\end{figure}
The figure's upper panel shows for 100 GHz the simulation of the residual effect for $TT$, $EE$, $BB$ in the power spectra of half-mission difference maps for one simulation of the ADCNL after linear gain variation correction (purple line), but not the second-order ADCNL effect that is not corrected in the 2018 data. The average of this, with five different noise realizations, is shown as the dark blue line. The difference between these allows us to separate the systematic effect from the noise. The red curve shows the residual of the ADCNL systematic effect after removing this estimate of the noise. It takes out the rise at $\ell <100$ where the noise is dominant. At low multipoles, the residual ADCNL systematic dominates and the noise is negligible. The lower panel in the figure shows this ADCNL auto-spectra for the four frequencies. The red curve for 100\,GHz, obtained in the upper panels, is also shown in this panel. At $\ell>1000$ the rise is due to a small additional digitalization noise.

In summary, the apparent variation in the linear gain is a good correction for the first-order approximation of the ADCNL systematic effect. However, large signals, especially the dipoles, are distorted by the second-order ADCNL effect. This is apparent at very low multipoles, and is not corrected for in the present 2018 data release. This is the main systematic residual at 100 and 143\,GHz at very large scales; while we can model it statistically (with a slight overestimate at 143\,GHz), these estimates are not accurate enough to use to correct the maps.

\subsection{Systematic effects summary}
\label{sec:syssum}
The inaccuracies in the residuals after correction for systematics effects were already discussed in \citelowell, the first paper using the \sroll\ mapmaking products. These residuals have been updated in this section and an overview is presented here in Table~\ref{tab:syst}.

\subsubsection{Systematic effects that do not project directly onto the sky maps}
The TOI processing remains basically the same as in 2015. We rely on this step to correct any part of the TOI signal that does not project onto the sky maps of Galactic and extragalactic emission. This includes Solar system emissions, time transfer functions, ADCNL, and beam asymmetry effects.

The zodiacal emission is removed in the HPRs. The emissivity of each component of the COBE model of zodiacal emission has been re-estimated. The emissivity model is significantly better than in the 2015 release (smaller uncertainties and smoother behaviour of the emissivities with frequency). There is no sign of the typical signatures of zodiacal residuals in the maps (i.e., diffuse ecliptic emission or zodiacal bands) for the CMB channels, and only very weak ones in the submillimetre channels. This is consistent with the expected level of the residuals after correction seen from the E2E simulations. The zodiacal model is confirmed by the good correction in the submillimetre channels and thus the prediction of much smaller and negligible residuals at CMB frequencies is also verified.

The far sidelobes are very asymmetrical and thus contribute different signals depending on the orientation of the beam around its fiducial axis. The FSL pattern has now been convolved with the sky maps of the previous release and removed from the TOIs. The 2015 maps (with the FSL not removed) show a clear FSL signal in the null tests, but none are apparent in 2018 (as can be seen in Fig.~\ref{fig:s1234}).

The time transfer functions used to correct the TOI processing are based on the scanning beams. For HFI beams these are measured on strong point sources, namely Jupiter, Saturn, and Mars and allows us to correct time constants up to 3\,s. The scanning beams are unchanged from the 2015 release.

Similarly, the ADCNL effects do not project onto the sky maps and are corrected to first order in the TOI processing, based on measurements made on the ground and during the warm phase of the mission (see detailed discussions in \citelowell). The \sroll\ mapmaking detects an apparent gain variation due to residuals of the ADCNL TOI correction. This linear gain variation is corrected in \sroll, but the associated second-order term non-linear distortion of strong signals (including dipoles) is not corrected in the 2018 release. This procedure leaves residuals after correction that are detectable in half-mission or odd-even survey null tests due to the long-term time dependence of the ADCNL (see \citelowell\ and Sect.~\ref{sec:ADC}).

Survey null-test maps are very sensitive, but in isolation are not a specific test of any of these three systematic effects by themselves, although they do provide a good global test of the improvement between releases. The separation of these three effects has been achieved in combination with E2E simulations in Sect.~\ref{sec:xferfunction}.

\subsubsection{Cross-spectra}
\label{sec:crossspectra}

Figure~\ref{fig:crossspectra} shows that the $100 \times 143$, $100 \times 217$ and $143 \times 217$ cross-spectra of the residual ADCNL effect are at a lower level than the $143 \times 143$ auto-spectrum, which we show as a reference (yellow line). 
\begin{figure}[htbp!]
\includegraphics[width=\columnwidth]{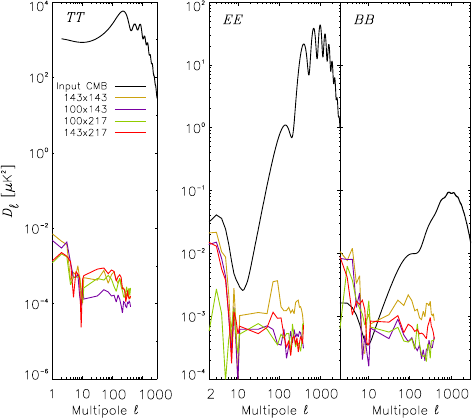}
\caption{Cross-spectra between the three CMB channels, showing the non-linear distortion of strong signals (not corrected in \sroll). The auto-spectrum of the lowest noise 143-GHz channel is shown to illustrate the reduction of the systematic level brought by the use of cross-spectra. At $\ell>300$, quantization noise generates a rise not related to this test; hence that range of multipoles is not shown.}
\label{fig:crossspectra} 
\end{figure}
The power spectra are taken from the simulations of the residuals of the ADCNL in the 2018 release maps shown in Fig.~\ref{fig:ADCNLresidu}. At $\ell<5$ the reduction from the auto-spectra (Fig.~\ref{fig:diffgl}) is a factor 2 to 3; at $\ell =200$ it reaches a factor of 10, showing that the ADCNL is not well correlated between HFI frequencies, a property that can be taken advantage of in the science analysis. Furthermore it should also be recalled that Sect.~\ref{sec:ADC} and Fig.~\ref{fig:crossspectra} show that the simulations tend to overestimate this systematic effect at very low multipoles (2, 3, and 4) at 100 and 143\,GHz. The ADCNL residuals in the cross-spectra are also well below the minimum in $EE$ predicted by the cosmological models to fall between the reionization peak and the first recombination peak at $10<\ell<30$.

\subsubsection{All systematic effects summary figure and table}
\label{sec:allsyste}

Figure~\ref{fig:fig17} displays the $EE$ power spectra of the residuals from E2E simulations of each of the main systematic effects discussed in Sects.~\ref{sec:intensity_leakage} and \ref{sec:ADC}.
\begin{figure*}[htbp!]
\includegraphics[width=\textwidth]{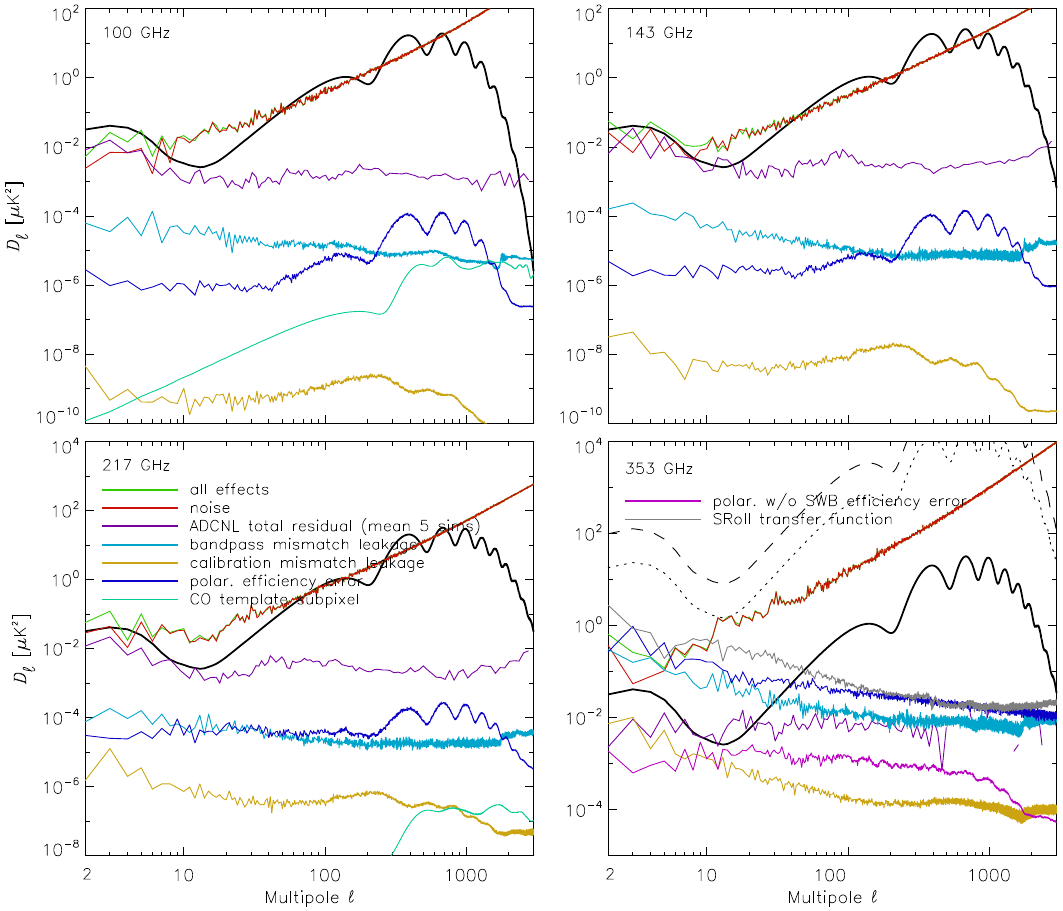}
\caption{Polarization power spectra in $D_{\ell}^{EE}$, showing the noise (red) and the main systematic residuals: ADCNL remaining dipole distortion after variable gain correction (purple); bandpass-mismatch leakage (light blue); leakage from calibration mismatch (orange); and the sum of all these (green). Polarization efficiency (dark blue), \sroll\ residual empirical transfer function (grey), and CO template subpixel effect (turquoise) have not been included in the sum. The fiducial CMB power spectrum is shown in full, dotted and dashed black lines (see text).}
\label{fig:fig17} 
\end{figure*}
The $EE$ fiducial power spectrum (black line) is shown to be compared with the noise and the residuals of each systematic effect. As could be expected, systematic effects and noise exceed at 353\,GHz (and is comparable to at 217\,GHz) this fiducial spectrum. Another comparison relevant for dust foreground removal from the 100- and 143-GHz CMB channels, is given by the $EE$ fiducial CMB power spectrum shifted upwards by the dust SED scaling factor between 353\,GHz and 100\,GHz (dashed black line) or 143\,GHz (dotted black line). The 353-GHz noise and systematics, if used to remove dust, can be compared directly to these shifted fiducial CMB. This shows that noise and systematic effects introduced in the best CMB channels (100 and 143\,GHz) through the dust-removal process using the dust foreground at 353 or 217\,GHz is lower by one order of magnitude than the noise and systematics at these frequencies.
 
The $1/f$ Gaussian component of the noise dominates over the white noise at $\ell <10$ for all frequency channels. Nevertheless, at 100, 143, and 217\,GHz, the ADCNL residual is the most dominant systematic effect at all multipoles and it is comparable to the noise for $2<\ell<10$. The passband leakage and polar efficiency are the next systematic effects covering all multipoles. Subpixel effects are very small, except for the one induced by the CO template at 100\,GHz.

At 353\,GHz, the bandpass leakage residual from dust at low multipoles reaches the noise level of $10^{-1}\microKcarre$ at $\ell=4$, and exceeds it at lower multipoles. The transfer function residuals that generate the zebra striping exceed the noise level around $\ell=10$ and is the dominant systematic effect at $\ell<10$, an order of magnitude above the $1/f$ noise at $\ell=3$--4. The 1\,\% error on polarization efficiency would reach a similar level at low multipoles if we included the SWBs. As discussed in Sect.~\ref{sec:caveats}, we recommend the use of the 353\,GHz map constructed without the SWBs; this brings the residuals to a negligible level. All other systematic effects are well below the noise.

The improvement in the CMB calibration reduces the leakage due to calibration mismatch to a subdominant systematic effect at all frequencies and multipole ranges. This was already true in \citelowell.
It is useful to compare the results in that paper (in its section B.4.1) in which the bandpass-mismatch and calibration-mismatch leakage terms were dominant and not completely negligible in polarization at low multipoles. Details of the analysis have been given in Fig.~\ref{fig:DIFF}. In CMB channels the corrections are comparable (for free-free at 100\,GHz) or reduced (for CO and dust in all CMB channels). At 353\,GHz, the improvement is much larger than one order of magnitude. This was expected, since the transfer-function effects were identified but not well corrected in the previous 2015 release. These improvements come from better foreground templates, partially based on the \sroll\ analysis.

Table~\ref{tab:syst} summarizes the levels of the systematic effects for three multipole ranges: very low multipoles relevant for the reionization peak; intermediate multipoles; and high multipoles. 
\begin{table*}[htbp!]
\newdimen\tblskip \tblskip=5pt
\caption{Residual levels in \microKcarre\ for the main systematic effects at each CMB frequency for three notable multipole values ($\ell=4$--5, 100, and 2000). We highlight in bold type the most significant effect for each multipole range.}
\label{tab:syst}
\vskip -2mm
\footnotesize
\setbox\tablebox=\vbox{
\newdimen\digitwidth
\setbox0=\hbox{\rm 0}
\digitwidth=\wd0
\catcode`*=\active
\def*{\kern\digitwidth}
\newdimen\signwidth
\setbox0=\hbox{+}
\signwidth=\wd0
\catcode`!=\active
\def!{\kern\signwidth}
\newdimen\lesswidth
\setbox0=\hbox{$\,{<}\,$}
\lesswidth=\wd0
\catcode`?=\active
\def?{\kern\lesswidth}
\halign{\hbox to 5.3 cm{#\leaderfil}\tabskip 2em&
\vtop{\hsize 4.5cm\hangindent=1em\hangafter=1\strut#\hfil\strut\par}\tabskip1em&
\hfil#\hfil&
\hfil#\hfil&
\hfil#\hfil&
\hfil#\hfil\tabskip 0em\cr
\noalign{\doubleline}
\omit\hfil Effect\hfil& \omit\hfil Estimation method\hfil& ?100\,GHz& ?143\,GHz& 217\,GHz& ?353\,GHz\cr
\noalign{\vskip 3pt\hrule\vskip 10pt}
\omit\bf Multipoles $\mathbf{\ell=4}$--5\hfil& & & & & \cr
\noalign{\vskip 3pt}
\hglue 1.3em Compression/decompression&Additional noise& ?$3\times10^{-5}$& ?\dots& \dots& ?\dots\cr
\hglue 1.3em Calibration mismatch&& $<1\times10^{-9}$& $<1\times10^{-8}$& $1\times10^{-6}$& ?$1\times10^{-3}$\cr
\hglue 1.3em Bandpass mismatch (total)& Simulations& ?$3\times10^{-5}$& ?$1\times10^{-4}$& $1\times10^{-4}$& ?$\mathbf{5\times10^{-2}}$\cr
\hglue 1.3em Bandpass mismatch free-free&Simulations& ?$1\times10^{-6}$& ?\dots& \dots& ?\dots \cr
\hglue 1.3em Bandpass mismatch CO&Simulations& ?$3\times10^{-7}$& ?\dots& $3\times10^{-8}$& ?$1\times10^{-4}$\cr
\hglue 1.3em Bandpass mismatch dust&Simulations& ?$3\times10^{-5}$& ?$3\times10^{-5}$& $3\times10^{-4}$& ?$3\times10^{-2}$\cr
\hglue 1.3em Polarization efficiency errors&Residuals from half mission& ?\dots& ?\dots& \dots& ?\dots\cr
\omit&Residuals from odd-even rings& ?\dots& ?\dots& \dots& ?\dots\cr
\hglue 1.3em Transfer function&Odd-even surveys oscillation& ?\dots& ?\dots& \dots& ?\dots\cr
\hglue 1.3em ADCNL&In cross-spectra& ?$\mathbf{3\times10^{-3}}$& ?$\mathbf{5\times10^{-3}}$& $\mathbf{5\times10^{-3}}$& ?$5\times10^{-3}$\cr
\hglue 1.3em Beam leakage&$TT$ or $EE$ to $B$& ?$5\times10^{-5}$& ?\dots& \dots& ?\dots\cr
\noalign{\vskip 14pt}
\omit\bf Multipoles $\mathbf{\ell\simeq100}$\hfil& & & & & \cr
\noalign{\vskip 3pt}
\hglue 1.3em Compression/decompression&Additional noise& ?$1\times10^{-5}$& ?\dots& \dots&?\dots\cr
\hglue 1.3em Calibration mismatch& & $<1\times10^{-9}$& $1\times10^{-8}$& $3\times10^{-7}$& ?$2\times10^{-4}$\cr
\hglue 1.3em Bandpass mismatch (total)& total& ?$1\times10^{-5}$& $2\times10^{-5}$& $3\times10^{-4}$& ?$\mathbf{1\times10^{-2}}$\cr
\hglue 1.3em Bandpass mismatch free-free& Simulations& ?$3\times10^{-6}$& ?\dots& \dots& ?\dots\cr
\hglue 1.3em Bandpass mismatch CO& Simulations& ?$1\times10^{-5}$& \dots& $1\times10^{-5}$& ?$1\times10^{-3}$\cr
\hglue 1.3em Bandpass mismatch dust& Simulations& ?$5\times10^{-6}$& $5\times10^{-6}$& $5\times10^{-5}$& ?$5\times10^{-3}$\cr
\hglue 1.3em Polarization efficiency errors& Residuals from half mission from $EE$ ($TE$)&& $4\times10^{-6}$ ($2\times10^{-4}$)& \dots&?\dots\cr
\omit&Residuals from odd-even rings from $EE$ ($TE$)& ?\dots& $2\times10^{-6}$ ($3\times10^{-5})$& \dots& ?\dots\cr
\hglue 1.3em Transfer function&Odd-even surveys oscillation& ?\dots& ?\dots& \dots& ?\dots\cr
\hglue 1.3em ADCNL&In cross-spectra& ?$\mathbf{1\times10^{-3}}$& $\mathbf{3\times10^{-3}}$& $\mathbf{3\times10^{-3}}$& ?$5\times10^{-3}$\cr
\hglue 1.3em Beam leakage&$TT$ or $EE$ to $B$& $<5\times10^{-5}$& \dots& \dots& ?\dots\cr
\noalign{\vskip 14pt}
\omit\bf Multipoles $\mathbf{\ell\simeq2000}$\hfil& & ?\dots& ?\dots& \dots& ?\dots\cr
\noalign{\vskip 3pt}
\hglue 1.3em Compression/decompression& Additional noise& ?$2\times10^{-5}$&?\dots&?\dots& ?\dots\cr
\hglue 1.3em Calibration mismatch& & $<1\times10^{-9}$& $<3\times10^{-10}$& $1\times10^{-7}$& ?$1\times10^{-4}$\cr
\hglue 1.3em Bandpass mismatch (total)& Simulations& ?$5\times10^{-6}$& ?$1\times10^{-5\phantom{0}}$& $3\times10^{-4}$& ?$\mathbf{1\times10^{-2}}$\cr
\hglue 1.3em Bandpass mismatch free-free& Simulations& ?$1\times10^{-6}$& ?\dots& \dots& ?\dots\cr
\hglue 1.3em Bandpass mismatch CO& Simulations& ?$1\times10^{-5}$& ?\dots& $1\times10^{-5}$& ?$5\times10^{-4}$\cr
\hglue 1.3em Bandpass mismatch dust& Simulations& ?$2\times10^{-6}$& ?$3\times10^{-6\phantom{0}}$& $5\times10^{-4}$& ?$3\times10^{-3}$\cr
\hglue 1.3em Polarization efficiency errors& Residuals from half mission from $EE$ ($TE$)& ?\dots& $5\times10^{-8}$ ($3\times10^{-7}$)& \dots& ?\dots\cr
\omit&Residuals from odd-even rings from $EE$ ($TE$)& & $1\times10^{-9}$ ($1\times10^{-9}$)& \dots& ?\dots\cr
\hglue 1.3em Transfer function& Odd-even surveys oscillation& ?\dots& ?\dots& \dots& ?\dots\cr
\hglue 1.3em ADCNL&In cross-spectra& ?$\mathbf{1\times10^{-3}}$& ?$\mathbf{5\times10^{-3}}$& $\mathbf{3\times10^{-3}}$& $<3\times10^{-3}$\cr
\hglue 1.3em Beam leakage&$TT$ or $EE$ to $B$& $<1\times10^{-6}$& ?\dots& \dots& ?\dots\cr
\noalign{\vskip 3pt\hrule\vskip 5pt}}}
\endPlancktable
\end{table*}

\section{Conclusions}
\label{sec:conclusion}

The HFI maps in the 2018 release are improved with respect to the 2015 ones in several ways. The \sroll\ mapmaking destriper fits for more systematic effects than was done and discussed in the intermediate \Planck\ paper \citelowell, improves on some of them, and now includes the two submillimetre channels. The intensity-to-polarization leakage is improved, both for the calibration and bandpass mismatch effects.

Some systematic effects residuals are still significant at low multipoles, specifically, the very long time constants (of order half a minute), which could not be detected in the scanning beams extracted from the planet observations, as well as the ADC non-linearity, are not fully corrected and are still contributing above the $1/f$ component of the TOI noise (although they are smaller than the TOI noise for multipoles higher than 20).

We have succeeded in making first estimates of the polarization efficiencies from the sky data at 353\,GHz; these estimates lead to rms dispersion that agree with the ground measurements, but do not improve on them, thus they were not used in this release. In all the cases described above, this opens the way for future corrections of such systematic effects in the polarization data. No estimate could be extracted for the CMB frequencies (100 and 217\,GHz). {\tt SMICA} estimates and cosmology-based likelihoods lead to combined results that are up to 1.9\,\% at 217\,GHz, somewhat bigger than the 1\,\% ground-based estimate.

The Solar dipole measurements in \citelowell\ showed obvious residuals of the Galactic (mostly dust) foreground removal, in the form of drifts in direction and amplitude with increasing frequency and sky fraction used. This has been understood as requiring a dust removal model that includes SED corrections of the 857-GHz dust template on large scales, with a similar pattern to correct the drifts observed at 100 to 545\,GHz. The main effect of dust SED large-scale variations in latitude have been on the diffuse foreground map ratio.

The Solar dipole is now stable in direction within 1\arcm, and amplitude within 0.5\microK, for frequencies 100 to 353\,GHz and sky fractions from 30 to 75\,\%. This provides a legacy of \Planck-HFI for future CMB observations that have only limited sky coverage. The channels calibrated on the
orbital CMB dipole have an absolute map calibration accuracy, measured a posteriori on the Solar dipole, better than $4\times 10^{-4}$ at $\ell=1$. For smaller scales the transfer functions have been shown to be smaller than $3\times10^{-3}$ from the dipole up to the first three acoustic peaks ($\ell\,{<}\,1000$), which brings the inter-frequency calibration to the same level. The calibration of the submillimetre channels, based on giant planet models, has been shown to be in agreement with the CMB photometry through the 545-GHz channel at a level better than 5\,\% (the uncertainty of the planet model).

Table~\ref{tab:mapcharacteristics} gives the main characteristics of the full-mission maps. The scanning beams of the maps are the same as those from 2015. The effective beams are not exactly the same because of the 1000 rings removed at the end of the mission, but the difference is negligible. The table also provides the 2018 release high-multipole sensitivity for the detector-noise-dominated scales ($\ell\,{>}\,100$), and compares it with the expectation based on the TOI noise. The TOI noise level is the same in the 2015 and 2018 releases and all differences with 2015 are due to the mapmaking improvements. This performance is close to the preflight expectations \citep[see for example][]{planck2005-bluebook}. The sensitivities have been converted to $C_{\ell}$ for the full mission, full sky, and we report the $TT$ and $EE$ values in the table. The improvement is due to the destriping at \Nside=2048 versus 512 (Fig.~\ref{fig:diffdx11rd12}), combined with a degradation due to the 1000 rings removed at the end of the mission. The $TT$ and $EE$ spectra quoted in the table are not directly comparable to the 2015 values \citep[reported in table~5 of][]{planck2014-a09} due to different sky masks being used for different frequencies and a scaling for the full sky done without hit-counts weighting.

We also report in Table~\ref{tab:mapcharacteristics} uncertainties on absolute calibration on the Solar dipole, based on simulations. These are an order of magnitude smaller than the 2015 ones, now being down to a few times $10^{-4}$. This improvment is useful for limiting the intensity-to-polarization leakage. Nevertheless, the calibration uncertainties at higher multipoles are dominated by the uncertainties on the transfer function (at the level of a few times $10^{-3}$), which are measured by comparing the first three acoustic peaks of the CMB power spectra. Polarization data calibration is further affected by the dominant uncertainty on the calibration efficiency of the PSBs at the percent level.

The monopole of each frequency map is set to be representative of the absolute one (not measured by \Planck). It is adjusted to a CIB model. The zero level of the Galactic dust component is obtained by scaling to the lowest \ion{H}{i} column densities. Zodiacal emission is not included in the frequency maps because of its dependence on the observation time during the year, but a zodiacal monopole is adjusted on the minimum emission. This ensures the best colour ratio between frequency maps.\footnote{For work separating CMB and diffuse Galactic components from HFI frequency maps, the CIB and Zodiacal emission monopole should be removed.}

Simulations have also demonstrated that the \sroll\ mapmaking does not affect the CMB sky input at the level of $3\times10^{-5}\microKcarre$ in the $D_{\ell}$ power spectra (figure~B5 of \citelowell). Figure~\ref{fig:Compression} also shows another example of the difference of CMB input and ouput, showing no effect up to $\ell=2000$, where subpixel effects start to dominate at the level of 1\,\% of the noise.

In the future, the correction of the relative bandpass response measured from the sky data will have profound implications on the way that foregrounds are removed from broadband experiments. In this paper, the bandpass-mismatch coefficients are fitted using foreground template maps, and the coefficients found are consistent with those measured on the ground, but with higher accuracy. The example of the CO line maps demonstrates that component separation could be better done using the information on the single-detector response to different foregrounds, for which a good template exists. This capability has not been fully exploited yet, and could lead to an integration of the mapmaking and component-separation procedures.

Furthermore, we have shown that the simulations characterize the data very well (although not completely), using different null tests, comparisons of the input versus output skies, and the effect of pixelization of the foreground templates.

Finally the full E2E simulations have been shown to be a powerful common tool for verifying consistency, investigating discrepancies, and providing the only way to build meaningful likelihoods when instrumental systematic effects and Galactic foregrounds become significant with respect to detector sensitivity.

\begin{table*}[htbp!]
\newdimen\tblskip \tblskip=5pt
\caption{Main characteristics of HFI full-mission maps.}
\label{tab:mapcharacteristics}
\vskip -5mm
\footnotesize
\setbox\tablebox=\vbox{
\newdimen\digitwidth
\setbox0=\hbox{\rm 0} 
\digitwidth=\wd0 
\catcode`*=\active 
\def*{\kern\digitwidth}
\newdimen\signwidth
\setbox0=\hbox{+}
\signwidth=\wd0
\catcode`!=\active
\def!{\kern\signwidth}
\newdimen\pointwidth
\setbox0=\hbox{.}
\pointwidth=\wd0
\catcode`?=\active
\def?{\kern\pointwidth}
\halign{\hbox to 10.5cm{#\leaderfil}\tabskip 1em&
\hfil#\hfil\tabskip 0.94em&
\hfil#\hfil&
\hfil#\hfil&
\hfil#\hfil&
\hfil#\hfil&
\hfil#\hfil&
\hfil#\hfil\/\tabskip=0pt\cr 
\noalign{\doubleline}
\omit&\multispan6\hfil Reference Frequency [GHz]\tablefootmark{\rm a1}\hfil\cr
\noalign{\vskip -3pt}
\omit&\multispan6\hrulefill\cr
\omit\hfil Quantity \hfil& 100& 143& 217& 353& 545& 857& Notes\cr
\noalign{\vskip 3pt\hrule\vskip 5pt}
Number of bolometers& 8& 11& 12& 12& 3& 4& \tablefootmark{{\rm a2}}\cr
\noalign{\vskip 12pt} 
Effective beam solid angle $\Omega$ [arcmin$^2$]& 106.22& 60.44& 28.57& 27.69& 26.44& 24.37& \tablefootmark{\rm b1}\cr
Error in solid angle $\sigma_\Omega$ [arcmin$^2$]& **0.14& *0.04& *0.04& *0.02& *0.02& *0.02& \tablefootmark{\rm b2}\cr
Spatial variation (rms) $\Delta\Omega$ [arcmin$^2$]& **0.20& *0.20& *0.19& *0.20& *0.21& *0.12& \tablefootmark{\rm b3}\cr
Effective beam FWHM$_1$ [arcmin]& **9.68& *7.30& *5.02& *4.94& *4.83& *4.64& \tablefootmark{\rm b4}\cr
Effective beam FWHM$_2$ [arcmin]& **9.66& *7.22& *4.90& *4.92& *4.67& *4.22& \tablefootmark{\rm b5}\cr
Effective beam ellipticity $\epsilon$& *1.186& 1.040& 1.169& 1.166& 1.137& 1.336& \tablefootmark{\rm b6}\cr
Variation (rms) of the ellipticity $\Delta\epsilon$&*0.024& 0.009& 0.029& 0.039& 0.061& 0.125& \tablefootmark{\rm b7}\cr
\noalign{\vskip 12pt}
$C_{\ell}^{TT}$ expected for full-mission map sensitivity [$10^{-4}\,\mu\mathrm{K_{CMB}}^2$]& 1.93& 0.48& 1.11& 13.0& & & \tablefootmark{\rm c1}\cr
\phantom{$C_{\ell}^{TT}$ expected for full-mission map sensitivity} [$10^{-4}$kJy$^2$.sr]& & & & &1.38&1.21&\tablefootmark{\rm c1}\cr
$C_{\ell}^{TT}$ map sensitivity from the 2018 release [$10^{-4}\,\mu\mathrm{K_{CMB}}^2$]& 1.52& 0.36& 0.78& 11.6& & & \tablefootmark{\rm c2}\cr
\phantom{$C_{\ell}^{TT}$ map sensitivity from the 2018 release} [$10^{-4}$kJy$^2$.sr]& & & & &2.9--8& 2.7--8& \tablefootmark{\rm c3}\cr
$C_{\ell}^{EE}$ map expected sensitivity from TOI white noise [$10^{-4}\,\mu\mathrm{K_{CMB}}^2$]& 5.01& 2.70& 2.77& 51.1& & & \tablefootmark{\rm c1}\cr
$C_{\ell}^{EE}$ map sensitivity from the 2018 release [$10^{-4}\,\mu\mathrm{K_{CMB}}^2$]& 2.94& 1.61& 3.25& *7.0& & & \tablefootmark{\rm c2}\cr
\noalign{\vskip 12pt}
Dipole absolute calibration accuracy [\%]& $0.008$& $0.021$& $0.028$& $0.024$& $\simeq1$& & \tablefootmark{\rm d1}\cr
Planet submm intercalibration accuracy [\%]& & & & & & $\simeq3$& \tablefootmark{\rm d2}\cr
Intensity transfer function uncertainty ($700<\ell<1000$) [\%]&Ref.&0.12& 0.36& 0.78& 4.3& & \tablefootmark{\rm d3}\cr
Polarization efficiency residual errors [\%]& 0.7& $-$1.7& 1.9& & & & \tablefootmark{\rm d4}\cr
\noalign{\vskip 12pt} 
Galactic emission zero level uncertainty [\MJysr]& 0.0008& 0.0010& 0.0024& 0.0067& 0.0165& 0.0147& \tablefootmark{\rm e1}\cr
CIB monopole assumption [\MJysr]& 0.0030& 0.0079& 0.033& 0.13& 0.35& 0.64& \tablefootmark{\rm e2}\cr
CIB monopole uncertainty [\%]& 100& 100& 40& 20& 20& 20& \tablefootmark{\rm e3}\cr
Zodiacal emission monopole level [$\mu$K$_{\mathrm{CMB}}$]& 0.43& 0.94& 3.8& 34& & & \tablefootmark{\rm e4}\cr
\phantom{Zodiacal emission monopole level} [\MJysr]& & & & & 0.04& 0.12& \tablefootmark{\rm e4}\cr
\noalign{\vskip 5pt\hrule\vskip 3pt}}}
\endPlancktablewide
\raggedright
\tablenote {{\rm a1}} {~Channel-map reference frequency, and channel identifier.}\par
\tablenote {{\rm a2}} {~Number of bolometers whose data were used in producing the channel map. At 353\,GHz, only eight PSBs are used for polarization maps.}\par
\tablenote {{\rm b1}} {~Mean value over bolometers at the same frequency.}\par
\tablenote {{\rm b2}} {~As given by simulations. }\par
\tablenote {{\rm b3}} {~Variation (rms) of the solid angle across the sky. }\par
\tablenote {{\rm b4}} {~FWHM of the Gaussian whose solid angle is equivalent to that of the effective beams.}\par
\tablenote {{\rm b5}} {~Mean FWHM of the elliptical Gaussian fit.}\par
\tablenote {{\rm b6}} {~Ratio of the major to minor axis of the best-fit Gaussian averaged over the full sky. }\par
\tablenote {{\rm b7}} {~Variability (rms) on the sky. }\par
\tablenote {{\rm c1}} {~Estimate of the $C_{\ell}$ map noise for the full mission, derived from the TOI white noise (see Sect.~\ref{sec:preprocessing}), \citep[table~5 of ][]{planck2014-a08}.}\par
\tablenote {{\rm c2}} {~Estimate of the $C_{\ell}$ map noise for the full mission, derived from the odd-even ring null-test noise (Fig.~\ref{fig:newpte}, multipole range $\ell=200$--1000).}\par
\tablenote {{\rm c3}} {~Estimate of the $C_{\ell}$ map noise for the full mission, derived from the odd-even ring null-test noise (Fig.~\ref{fig:diffdx11dx12}, multipole range $\ell=200$--2000).}\par
\tablenote {{\rm d1}} {~Absolute calibration accuracy from simulations with the Solar dipole (Table~\ref{tab:ratios}). The 545-GHz channel retains the 2015 planet calibration, and the accuracy is calculated a posteriori on the Solar dipole.}\par
\tablenote {{\rm d2}} {~The 857\,GHz channel retains the 2015 planet calibration, and the accuracy is calculated a posteriori using the planet model \citep{planck2016-LII} and the 545-GHz data.}\par
\tablenote {{\rm d3}} {~Derived upper limits of the transfer function on the first three acoustic peaks (Table~\ref{tab:dipolechaipa}).}\par
\tablenote {{\rm d4}} {~The polarization efficiencies residual errors exceed somewhat the estimated uncertainty (1\,\%). These are a posteriori tests, and have not been applied to the maps, but taken into account in the likelihoods (Table~\ref{tab:calibpolar}).}\par
\tablenote {{\rm e1}} {~The monopoles of the maps are built using a Galactic dust model extrapolated to a zero level for \ion{H}{i}. Uncertainties are discussed in \citet{planck2013-p03f} and presented in table~5 of that paper.}\par
\tablenote {{\rm e2}} {~The monopole of the~\cite{bethermin2012} CIB model (table 6 of \citet{planck2014-a09}).}\par
\tablenote {{\rm e3}} {~The CIB uncertainties are estimated by combining the absolute measurements of FIRAS \citep{1996A&A...308L...5P, 1998ApJ...508..123F} and the anisotropies from \Planck\ HFI \citep{planck2013-pip56}, assuming the same SED for the absolute value and the anisotropies.}\par
\tablenote {{\rm e4}} {~Monopole contribution of the zodiacal emission, adjusted for high Galatic and ecliptic latitudes \citep[table~6 of ][]{planck2014-a09}.}\par
\end{table*}

\begin{acknowledgements}
The Planck Collaboration acknowledges the support of: ESA; CNES and CNRS/INSU-IN2P3-INP (France); ASI, CNR, and INAF (Italy); NASA and DoE (USA); STFC and UKSA (UK); CSIC, MINECO, JA, and RES (Spain); Tekes, AoF, and CSC (Finland); DLR and MPG (Germany); CSA (Canada); DTU Space (Denmark); SER/SSO (Switzerland); RCN (Norway); SFI (Ireland); FCT/MCTES (Portugal); ERC and PRACE (EU). A description of the Planck Collaboration and a list of its members, indicating which technical or scientific activities they have been involved in, can be found at \href{http://www.cosmos.esa.int/web/planck/planck-collaboration}{\texttt{http://www.cosmos.esa.int/web/planck/planck-collaboration}}. 
\end{acknowledgements}

\bibliographystyle{aat}
\bibliography{L03_HFI_Data_Processing,Planck_bib}

\appendix

\section{HFI FFP10 simulations for product characterization}
\label{sec:end2end}

The cosmology analysis requires simulations that allow to estimate chance coincidences between CMB anisotropies, foregrounds, and noise. For the previous 2015 \Planck\ release, this was performed using the FFP8 simulations \citep{planck2014-a14} by using a fixed foreground, and swapping the CMB and noise realizations. In this 2018 release, we also want to statistically investigate the effects of systematic residuals. It has been shown (see appendix~B.3.1 of \citelowell) that whether the CMB map is included in the inputs or added after {\tt SRoll} processing, leads to differences for the power spectra in the CMB channels that are below $10^{-4}\microKcarre$. This justifies the use of this ``CMB swapping'' procedure, even when non-Gaussian systematic effects dominate over the TOI detector noise. For statistical analysis, we thus have built 1000 CMB map realizations, together with 300 E2E simulations that have one fiducial CMB realization and one set of foregrounds but variable noise and parameters describing the systematic effects.

The basic scheme for the creation of simulations is shown in Fig.~\ref{fig:SimulationFlow}. The following sections detail these processes.  We also use this E2E pipeline with a single sky (CMB plus foregrounds or CMB only), and with or without noise, to characterize the level of the systematic effects and verify that they are representative of the data.

\begin{figure}[htbp!]
\includegraphics[width=\columnwidth]{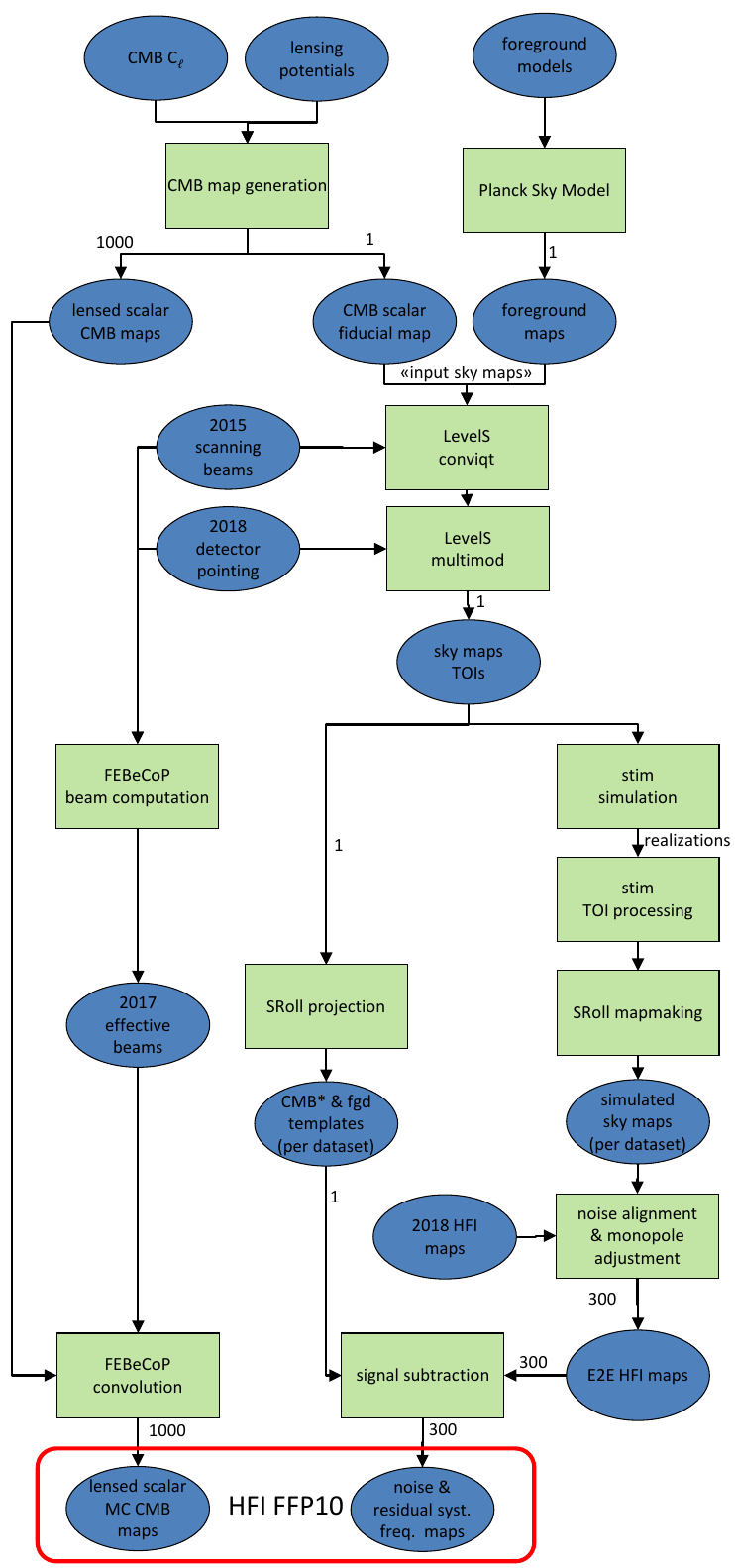}
\caption{Schematic of the HFI simulation pipeline. The numbers are the number of realizations. The red frames show the products available in the Planck Legacy Archive.}
\label{fig:SimulationFlow} 
\end{figure}

\subsection{Building the CMB Monte Carlo maps}

\subsubsection{Generating the CMB maps}

The FFP10 lensed CMB maps are generated in the same way as for the previous FFP8 release and described in detail (including the cosmological model parameters) in \citet{planck2014-a14} and in the \citeES. The FFP10 simulations only contain the scalar parts, lensed with independent lensing potential realizations. 

The FFP10 simulation set contains 1000 CMB maps, intended for swapping with the noise plus systematic residual maps. An additional CMB map realization, called the ``fiducial,'' is used for building the input sky for the E2E simulations.

\subsubsection{Convolving the CMB maps with effective beams}

The 1000 CMB maps are then convolved with the {\tt FEBeCoP} effective beams computed as described in Sect.~\ref{sec:freqmaps}. These maps are available in the PLA.

\subsection{Building the foreground maps}

Tests and validation of the data analysis pipelines are performed using full focal-plane simulations of the \Planck\ data streams. One single foreground sky model is used, referred to as ``the FFP10 sky model'' in this paper. Foregrounds in the FFP10 sky model include the following components:
\begin{itemize}
\item Galactic thermal dust, spinning dust, synchrotron, free-free, and CO line emission;
\item the cosmic infrared background;
\item Galactic and extragalactic faint point sources (radio and infrared);
\item thermal and kinetic Sunyaev-Zeldovich effects from Galaxy clusters.
\end{itemize}
The foreground maps do not include resolved compact sources.

For all of the sky components except dust polarization, we use the latest version of the Planck Sky Model \citep[PSM,][]{delabrouille2012} described in \citet{planck2014-a14}.

\subsubsection{Diffuse Galactic components}

The dust model maps are built as follows. 
The Stokes $I$ map at 353\,GHz is the dust total intensity \Planck\ map obtained by applying the Generalized Needlet Internal Linear Combination ({\tt GNILC}) method of \citet{Remazeilles11} to the 2015 release of \Planck\ HFI maps (PR2), as described in \citet{planck2016-XLVIII}, and subtracting the monopole of the cosmic infrared background \citep{planck2014-a09}. For the Stokes $Q$ and $U$ maps at 353\,GHz, we started with one realization of the statistical model of \citet{2017A&A...603A..62V}. The portions of the simulated Stokes $Q$ and $U$ maps near Galactic plane were replaced by the \Planck\ 353-GHz PR2 data. The transition between data and the simulations was made using a Galactic mask with a $5^\circ$ apodization, which leaves 68\,\% of the sky unmasked at high latitude. Furthermore, on the full sky, the large angular scales in the simulated Stokes $Q$ and $U$ maps were replaced by the \Planck\ data. Specifically, the first ten multipoles came from the \Planck\ data, while over $\ell=10$--20 the simulations were introduced smoothly using the function $\left(1+\sin\left[\pi\left(15-\ell\right)/10\right]\right)/2$.

To scale the dust Stokes maps from the 353-GHz templates to other \Planck\ frequencies, we follow the FFP8 prescription \citep{planck2014-a14}. A different modified blackbody emission law is used for each of the $N_{\rm side}=2048$ {\tt HEALPix} pixels. The dust spectral index used for scaling in frequency is different for frequencies above and below 353\,GHz. For frequencies above 353\,GHz, the parameters come from the modified blackbody fit of the dust spectral energy distribution (SED) for total intensity obtained by applying the {\tt GNILC} method to the PR2 HFI maps \citep{planck2016-XLVIII}. These parameter maps have a variable angular resolution that decreases towards high Galactic latitudes. Below 353\,GHz, we also use the dust temperature map from \citet{planck2016-XLVIII}, but with a distinct map of spectral indices from \citet{planck2013-p06b}, which has an angular resolution of 30\arcm. These maps introduce significant spectral variations over the sky at high Galactic latitudes, and between the dust SEDs for total intensity and polarization. The spatial variations of the dust SED for polarization in the FFP10 sky model are quantified in \citet{planck2016-l11A}.

Synchrotron intensity is modelled by scaling in frequency the 408-MHz template map from \citet{haslam1982}, as reprocessed by \citet{2015MNRAS.451.4311R} using a single power law per pixel. The pixel-dependent spectral index is derived from an analysis of WMAP data by \citet{2008A&A...490.1093M}. The generation of synchrotron polarization follows the prescription of \citet{delabrouille2012}. 

Free-free, spinning dust models, and Galactic CO emissions are essentially the same as used for the FFP8 sky model \citep{planck2014-a14}, but the actual synchrotron and free-free maps used for FFP10 are obtained with a different realization of small-scale fluctuations of the intensity. CO maps do not include small-scale fluctuations, and are generated from the spectroscopic survey of \citet{dame2001}. None of these three components is polarized in the FFP10 simulations.

\subsubsection{Unresolved point sources and the cosmic infrared background}

Catalogues of individual radio and low-redshift infrared sources are generated in the same way as for FFP8 simulations \citep{planck2014-a14}, but use a different seed for random number generation. Number counts for three types of galaxies (early-type proto-spheroids, and more recent spiral and starburst galaxies) are based on the model of \citet{Cai2013}. The entire Hubble volume out to redshift $z=6$ is cut into 64 spherical shells and for each shell we generate a map of density contrast integrated along the line of sight between $z_{\rm min}$ and $z_{\rm max}$, such that the statistics of these density contrast maps (i.e., power spectrum of linear density fluctuations, and cross-spectra between adjacent shells, as well as with the CMB lensing potential\footnote{The CMB and its lensing potential used here are not the CMB fiducial one, and thus, in the E2E simulations, the CIB is not correlated with the CMB and its lensing.}), obey statistics computed using the Cosmic Linear Anisotropy Solving System ({\tt CLASS}) code \citep{Blas2011,DiDio2013}. For each type of galaxy, a catalogue of randomly-generated galaxies is generated for each shell, following the appropriate number counts. These galaxies are then distributed in the shell to generate a single intensity map at a given reference frequency, which is scaled across frequencies using the prototype galaxy SED at the appropriate redshift.

\subsubsection{Galaxy clusters}

A full-sky catalogue of galaxy clusters is generated based on number counts following the method of \citet{Delabrouille2002}. The mass function of \citet{Tinker2008} is used to predict number counts. Clusters are distributed in redshift shells, proportionally to the density contrast in each pixel with a bias $b(z,M)$, in agreement with the linear bias model of \citet{MoWhite1996}. For each cluster, we assign a universal profile based on XMM observations, as described in \citet{Arnaud2010}. Relativistic corrections are included to first order following the expansion of \citet{Nozawa1998}. To assign an SZ flux to each cluster, we use a mass bias of $M_{\rm Xray}/M_{\rm true}=0.63$ to match actual cluster number counts observed by \Planck\ for the best-fit cosmological model coming from CMB observations. We use the specific value $\sigma_8 = 0.8159$.

The kinetic SZ effect is computed by assigning to each cluster a radial velocity that is randomly drawn from a centred Gaussian distribution, with a redshift-dependent standard deviation that is computed from the power spectrum of density fluctuations. This neglects correlations between cluster motions, such as bulk flows or pairwise velocities of nearby clusters.

\subsection{Building the sky map TOIs}

As for the 2015 data release, the frequency simulated maps are built using the {\tt LevelS} software package \citep{reinecke2006} and its modules {\tt conviqt} and {\tt multimod}. The generated TOIs are convolved with the same scanning beam as for the 2015 data release, but with an updated 2018 scanning strategy omitting the 1000 pointing periods from the end of the mission (see Sect.~\ref{sec:1000rings}). Scanning beams are the 2015 intensity-only scanning beams issued from the 2015 maps, to which a fake polarization is added using a simple model based on each bolometer polarization angle and leakage.

\subsection{{\tt stim} simulation}

The main new aspect of the HFI 2018 simulations is the production of E2E simulations. These include all significant systematic effects, and are used to produce maps of noise plus systematic effect residuals. The {\tt stim} pipeline adds the modelled instrumental systematic effects at the timeline level. It includes noise only up to the time response convolution step, after which the signal is added and the systematics simulated. It was shown in appendix~B.3.1 of \citelowell\ that, including the CMB map in the inputs or adding it after {\tt SRoll} processing, leads to differences for the power spectra in CMB channels below the $10^{-4}\microKcarre$ level. This justifies the use of CMB swapping even when non-Gaussian systematic effects dominate over the TOI detector noise.

We now describe each of the main systematic effect ingredients of the E2E simulations.

\textbf{White noise:} the noise is based on a physical model composed of photon noise, phonon noise, and electronic noise. The time-transfer functions are different for these three noise sources. A timeline of noise only is created, with the level adjusted to agree with the observed TOI white noise after removal of the sky signal averaged in a ring.

\textbf{Bolometer signal time-response convolution:} the photon white noise is convolved with the bolometer time response using the same code and same parameters as in the 2015 TOI processing. A second white noise contribution is added to the convolved photon white noise to simulate the electronics noise.

\textbf{Noise auto-correlation due to deglitching:} it has been found that the deglitching step in the TOI processing creates noise auto-correlation by flagging samples that are synchronous with the sky. Nevertheless, since we do not simulate the cosmic-ray glitches, we mimic this behaviour by adjusting the noise of samples above a given threshold to simulate their flagging.

\textbf{Time response deconvolution:} the timeline containing the photon and electronic noise contributions is then deconvolved with the bolometer time response and low-pass filtered to limit the amplification of the high-frequency noise, using the same parameters as in the 2015 TOI processing.

The input sky signal timeline is added to the convolved/deconvolved noise timeline and is then put through the instrument simulation. The sky signal is not convolved/deconvolved with the bolometer time response, since it is already convolved with the scanning beam extracted from the 2015 TOI processing output, and thus already contains the low-pass filter associated with the time-response deconvolution.

\textbf{Simulation of the signal non-linearity:} the first step of electronics simulation is the conversion of the input sky plus noise signal from K$_{\textrm{CMB}}$ units to analogue-to-digital units (ADU) using the detector response measured on the ground and assumed to be very stable in time. The ADU signal is then fed through a simulator of a non-linear analogue-to-digital converter. 

This step is the one introducing complexity into the signal, inducing time variation of the response, and causing gain difference with respect to the ground-based measurements. This corresponds to specific new modules of correction in the mapmaking.

The ADCNL transfer-function simulation is based on the TOI processing, with correction from the ground measurements, combined with in-flight measurements carried out during the warm extension of the mission. A reference simulation is built for each bolometer, which minimizes the difference between the simulation and the data gain variations, measured in a first run of the \sroll\ mapmaking. Realizations of the ADCNL are then drawn to mimic the variable behaviour of the gains seen in the 2018 data.

\textbf{Compression/decompression:} the signal is then compressed by the lossy algorithm required by the telemetry rate allocated to the HFI instrument. While very close to the compression algorithm used on-board, the one used in the simulation pipeline differs slightly, due to the non-simulation of the cosmic-ray glitches, together with the use of the average of the signal in the compression slice.

The number of compression steps, the signal mean of each compression slice and the step value for each sample are then used by the decompression algorithm to reconstruct the modulated signal.

\subsection{{\tt stim} TOI processing}

The TOIs issued from the steps outlined above are then processed in the same way as the flight TOI data. Because of the granularity needed and the required computational performance, the TOI processing pipeline applied to the simulated data is not exactly the same as the one applied the data. The specific steps are the following.

\textbf{ADCNL correction:} the ADCNL correction is carried out with the same parameters as the 2015 data TOI processing, and with the same algorithm. 

\textbf{Demodulation:} signal demodulation is also performed in the same way as the flight TOI processing. First, the signal is converted from ADU to volts. Next, the signal is demodulated by subtracting from each sample the average of the modulated signal over 1\,hour and then taking the opposite value for negative parity samples.

\textbf{Conversion to watts and thermal baseline subtraction:} the demodulated signal is then converted to watts (ignoring the conversion non-linearity of the bolometers and amplifiers, which has been shown to be negligible). Finally, a thermal baseline is subtracted; this is derived from the flight signals of the two dark bolometers, smoothed over 1\,minute.

\textbf{\textit{1/f} noise:} a $1/f$-type noise component is then added to each signal ring, with parameters (slope and knee frequency) adjusted on the flight data.

\textbf{Projection to HPR:} the signal is then projected to HPRs, after removal of flight-flagged data (unstable pointing periods, glitches, Solar system objects, planets, etc.).

\textbf{4-K line residuals:} a HPR of the 4-K lines residuals for each bolometer, built by stacking the 2015 TOI, is added to the simulation output HPR.

\textbf{List of modules and effects not included in the E2E simulations of TOI and HPR processing:}
\begin{itemize}
\item no discrete point sources;
\item no glitching/deglitching, only deglitching-induced noise auto-correlation;
\item no 4-K line simulation and removal, only the simulation of their residuals;
\item no bolometer volts-to-watts conversion non-linearity from the bolometers and amplifiers,\footnote{These are only important for strong glitches and planets that are not simulated. The ad hoc planet TOI processing used to derive the scanning beams is a separate pipeline and the rings containing planets are flagged.}
\item no far sidelobes (FSLs) are added or removed;
\item reduced simulation pipeline at 545- and 857-GHz.
\end{itemize}
To be more specific about this last item, the processing uses a reduced simulation pipeline without electronics simulation. This contains only photon and electronic noise, deglitching noise auto-correlation, and time-response convolution/deconvolution, and $1/f$ noise. Bolometer by bolometer baseline addition and thermal baseline subtraction, compression/decompression, and 4-K line residuals are not included.

\subsection{Mapmaking} 

The next stage is the processing of {\tt stim}-projected HPRs by the \sroll\ mapmaking. The following \sroll\ parameters are all the same for simulation mapmaking as for the data:
\begin{itemize}
\item thermal dust, CO, and free-free map templates;
\item detector NEP and polarization parameters;
\item bad rings list and sample flagging.
 \end{itemize}
The FSL removal performed in the \sroll\ destriper is not activated (since no FSL effect is included in the input). The total dipole removed by \sroll\ is the same as the input in the sky TOIs generated by {\tt LevelS}.

\subsection{Post-processing}

\textbf{Noise alignment:} an additional noise component is added to align the noise levels of the simulations with the noise estimate from the 2018 odd-even rings. Of course, this adjustment of the noise level does not satisfy all the other noise null tests (see Sect.~\ref{sec:highmultipoles}).

To follow the structure in frequency of the TOI noise (white noise, $1/f$ noise, and frequency transfer-function effects), this additional TOI noise is built up of six spin-frequency harmonic wavelets, shown in Fig.~\ref{fig:plnoisewave}. Each noise TOI associated with a wavelet is projected according to the scanning strategy, producing templates for which coefficients are then adjusted to fit the data noise level.
\begin{figure}[htbp!]
\includegraphics[width=\columnwidth]{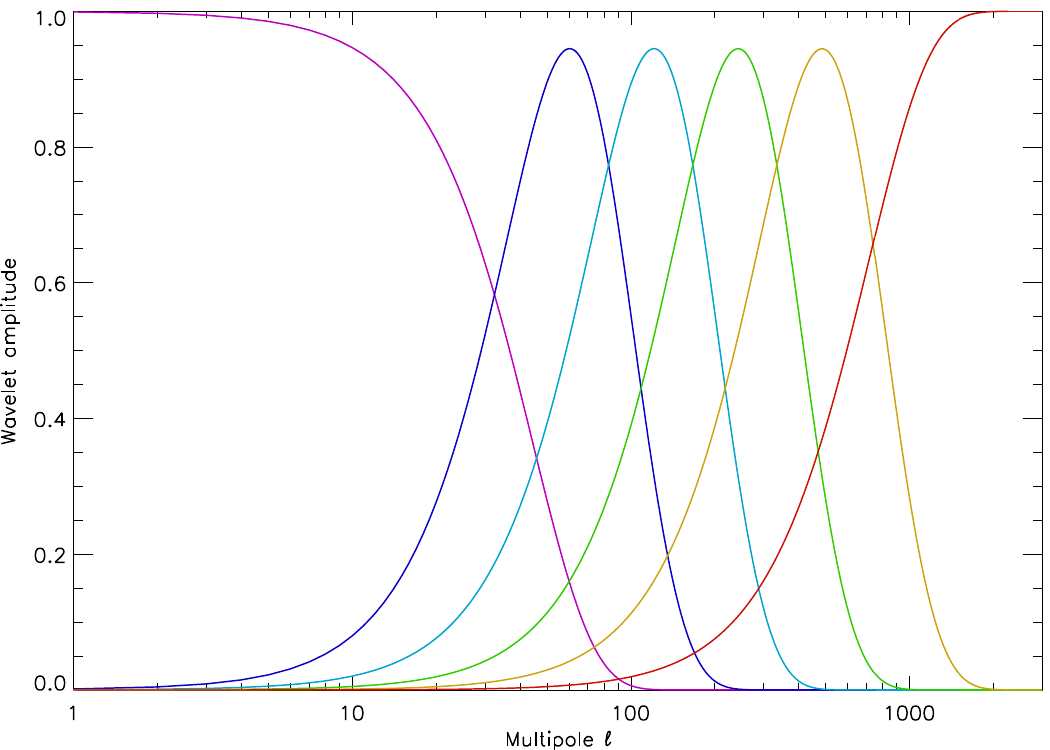}
\caption{Profile of the wavelets in temporal frequency rescaled over the corresponding multipole range on a ring (and thus on a map).}
\label{fig:plnoisewave} 
\end{figure}

This alignment is different for temperature and for polarization maps, in order to simulate the effect of the noise correlation between detectors within a PSB. The noise alignment model follows
\begin{equation}
\begin{bmatrix}
I\\ 
Q\\ 
U
\end{bmatrix}=\sum_{i=0}^{i=5}A_i W_i * \tens{P}.N . \nonumber
\end{equation}
The components of this equation are:
\begin{itemize}
\item $ A_i=\begin{bmatrix} a_i & 0 & 0\\ 0 & b_i & 0\\ 0 & 0 & b_i \end{bmatrix}$, the noise weight fitted on ring null tests for each frequency range associated with a wavelet $W_i$, with $a_i$ and $b_i$ being the coefficients reported in Table~\ref{tab:wavelets};
\item $\tens{P}$, the $II$, $IQ$, \dots $UU$ variance;
\item $N$, the white noise. 
\end{itemize}
\begin{table*}[htbp!]
\newdimen\tblskip \tblskip=5pt
\caption{For each wavelet, $a$ and $b$ bolometer coefficients averaged per frequency, in percentage}
\label{tab:wavelets}
\vskip -3mm
\footnotesize
\setbox\tablebox=\vbox{
\newdimen\digitwidth
\setbox0=\hbox{\rm 0}
\digitwidth=\wd0
\catcode`*=\active
\def*{\kern\digitwidth}
\newdimen\signwidth
\setbox0=\hbox{+}
\signwidth=\wd0
\catcode`!=\active
\def!{\kern\signwidth}
\newdimen\pointwidth
\setbox0=\hbox{.}
\pointwidth=\wd0
\catcode`?=\active
\def?{\kern\pointwidth}
\halign{\hbox to 3cm{#\leaderfil}\tabskip 2.0em&
\hfil#\hfil\tabskip2.0em&
\hfil#\hfil&
\hfil#\hfil&
\hfil#\hfil&
\hfil#\hfil&
\hfil#\hfil\tabskip 0em\cr
\noalign{\doubleline}
\omit\hfil Quantity\hfil& Wavelet 0& Wavelet 1& Wavelet 2& Wavelet 3& Wavelet 4& Wavelet 5\cr
\noalign{\vskip 3pt\hrule\vskip 5pt}
\noalign{\vskip 2pt}
$(a/b)_{100}$& *2.9/*0.0& 1.4/0.6& 2.2/1.1& 1.7/0.8& 1.4/1.1& 1.8/1.2\cr
$(a/b)_{143}$& *8.6/*0.0& 4.3/1.7& 3.5/0.8& 2.2/0.8& 2.9/1.3& 1.1/0.6\cr
$(a/b)_{217}$& 15.8/14.6& 7.9/0.1& 8.0/0.3& 3.3/0.8& 3.6/1.4& 1.3/0.8\cr
$(a/b)_{353}$& *8.9/*0.0& 4.5/0.0& 4.4/0.0& 2.7/0.0& 2.8/0.0& 0.4/0.1\cr
$(a)_{545}$ [\%]& 12.8& 0.1& 19.0& *7.4& 6.7& 6.8\cr
$(a)_{857}$ [\%]& 19.4& 0.1& *6.3& 16.0& 3.4& 5.0\cr
\noalign{\vskip 3pt\hrule\vskip 5pt}}}
\endPlancktable
\end{table*}
The higher amplitudes correspond to the first wavelet, which covers the low frequencies, dominated by the residuals of the strong systematic effects. The amplitudes in the range of the quasi-white noise range are much smaller, less than 8\,\%. At 545 and 857\,GHz, the noise description is more complicated (no quasi-white noise), and thus, the noise alignment correction amplitudes are larger.

This noise alignment procedure provides a good representation of the noise at high multipoles, with residuals being within the uncertainties of the TOI noise model. At low multipoles, which are dominated by systematic effect residuals, the noise cannot be described by a smooth model in multipoles. A noise adjustment of the data using one wavelet only is a good approximation of the statistics of the data (for $\ell<30$), as demonstrated by the PTE (see Fig.~\ref{fig:pteLuca}).

\textbf{Monopole adjustment:} a constant is added to each simulated map to bring its monopole to the same value as the corresponding 2018 maps.

The above steps lead to the production of the 300 HFI E2E simulations that are used extensively in this paper. 

\textbf{Signal subtraction:} from each map, the input sky (CMB and foreground) is subtracted to build the ``noise and residual systematics frequency maps.'' The systematics include additional noise and residuals induced by sky-signal distortion. Those maps are part of the FFP10 data release.

\end{document}